\documentclass[12pt]{iopart}
\usepackage[utf8]{inputenc}
\usepackage{iopams}
\usepackage{overpic,bbm,bm,epsfig}
\usepackage{multirow}
\usepackage{color}
\usepackage{braket}

\renewcommand{\theequation}{\thesection.\arabic{equation}}
\renewcommand{\thefigure}{\thesection.\arabic{figure}}

\begin{document}

\title[Zhi-zhong Xing $\&$ Zhen-hua Zhao]{The minimal seesaw and leptogenesis
models}

\author{Zhi-zhong Xing$^{1}$ {\rm and} Zhen-hua Zhao$^{2}$\footnote{Corresponding author}}

\address{\footnotesize $^1$Institute of High Energy Physics and
School of Physical Sciences, University of Chinese Academy of
Sciences, Beijing 100049, China \\
$^2$Department of Physics, Liaoning Normal University, Dalian
116029, China}
\ead{zhaozhenhua@lnnu.edu.cn}
\vspace{10pt}
\begin{indented}
\item[]August 2020
\end{indented}

\begin{abstract}
Given its briefness and predictability, the minimal seesaw --- a simplified
version of the canonical seesaw mechanism with only two right-handed neutrino fields
--- has been studied in depth and from many perspectives, and now it is being pushed
close to a position of directly facing experimental tests. This article is intended
to provide an up-to-date review of various phenomenological aspects of the minimal
seesaw and its associated leptogenesis mechanism in neutrino physics and
cosmology. Our focus is on possible flavor structures of such benchmark
seesaw and leptogenesis scenarios and confronting their predictions with current
neutrino oscillation data and cosmological observations. In this connection
particular attention will be paid to the topics of lepton number violation, lepton
flavor violation, discrete flavor symmetries, CP violation and antimatter
of the Universe.
\end{abstract}

%

%
%
%
%

\tableofcontents

\def\thefootnote{\arabic{footnote}}
\setcounter{footnote}{0}


\setcounter{equation}{0}
\section{Introduction}
\label{section 1}

\subsection{Massive neutrinos: known and unknown}
\label{section 1.1}

Since 1998 a number of impressive underground experiments have convincingly
verified the long-standing hypothesis that a type of neutrino flavor eigenstate
$\nu^{}_\alpha$ (for $\alpha = e, \mu, \tau$) as a superposition of the neutrino
mass eigenstates $\nu^{}_i$ (for $i = 1, 2, 3$) travelling in space can
spontaneously and periodically convert to another type of neutrino flavor
eigenstate $\nu^{}_\beta$ (for $\beta = e, \mu, \tau$)
via a pure quantum effect --- {\it flavor oscillation} \cite{Tanabashi:2018oca}.
This achievement is marvelous because flavor oscillations definitely
mean that $\nu^{}_1$, $\nu^{}_2$ and $\nu^{}_3$ have different but tiny masses
and there exists a mismatch between the mass and flavor eigenstates of three
neutrinos or three charged leptons
--- the so-called lepton flavor mixing \cite{Pontecorvo:1957cp,Maki:1962mu},
a phenomenon analogous to the well-established quark flavor mixing
\cite{Cabibbo:1963yz,Kobayashi:1973fv}.

To be explicit, the mismatch between $\nu^{}_\alpha$ and $\nu^{}_i$
is described by the $3\times 3$ Pontecorvo-Maki-Nakagawa-Sakata (PMNS) matrix
$U$ \cite{Pontecorvo:1957cp,Maki:1962mu} appearing in the weak charged-current
interactions of charged leptons and massive neutrinos:
\begin{eqnarray}
-{\cal L}^{}_{\rm cc} = \frac{g}{\sqrt{2}} \
\overline{\pmatrix{e & \mu & \tau}^{}_{\rm L}} \ \gamma^\mu \ U
\pmatrix{ \nu^{}_{1} \cr \nu^{}_{2} \cr
\nu^{}_{3} \cr}^{}_{\hspace{-0.1cm} \rm L} W^-_\mu + {\rm h.c.} \; ,
\label{eq:1.1}
\end{eqnarray}
in which ``L" stands for the left chirality of a fermion field, and
$e$, $\mu$ and $\tau$ stand respectively for  the mass eigenstates of electron, muon
and tau. In the basis where the flavor eigenstates of three charged leptons are
identical with their mass eigenstates, the link between $\nu^{}_\alpha$ and $\nu^{}_i$
is straightforward:
\begin{eqnarray}
\pmatrix{ \nu^{}_{e} \cr \nu^{}_{\mu} \cr \nu^{}_{\tau}
\cr}^{}_{\hspace{-0.1cm} \rm L} =
U \pmatrix{ \nu^{}_{1} \cr \nu^{}_{2} \cr \nu^{}_{3}
\cr}^{}_{\hspace{-0.1cm} \rm L} =
\pmatrix{
U^{}_{e 1} & U^{}_{e 2} & U^{}_{e 3} \cr U^{}_{\mu 1} & U^{}_{\mu 2}
& U^{}_{\mu 3} \cr U^{}_{\tau 1} & U^{}_{\tau 2} & U^{}_{\tau 3} \cr}
\pmatrix{ \nu^{}_{1} \cr \nu^{}_{2} \cr \nu^{}_{3}
\cr}^{}_{\hspace{-0.1cm} \rm L} \; .
\label{eq:1.2}
\end{eqnarray}
If $U$ is assumed to be unitary, it can always be parameterized in terms of three
real two-dimensional rotation matrices and two complex phase matrices. The most
commonly used or ``standard" parametrization of $U$ takes the form of
\begin{eqnarray}
U = O^{}_{23} (\theta^{}_{23}) \otimes P^{}_\delta \otimes O^{}_{13} (\theta^{}_{13})
\otimes P^{\dagger}_\delta \otimes O^{}_{12} (\theta^{}_{12}) \otimes P^{}_\nu \; ,
\label{eq:1.3}
\end{eqnarray}
where
\begin{eqnarray}
O^{}_{12} (\theta^{}_{12}) = \pmatrix{ c^{}_{12} & s^{}_{12} & 0 \cr
-s^{}_{12} & c^{}_{12} & 0 \cr 0 & 0 & 1 \cr} \; ,
\nonumber \\
O^{}_{13} (\theta^{}_{13}) = \pmatrix{ c^{}_{13} & 0 & s^{}_{13} \cr 0 & 1 & 0
\cr -s^{}_{13} & 0 & c^{}_{13} \cr} \; ,
\nonumber \\
O^{}_{23} (\theta^{}_{23}) = \pmatrix{ 1 & 0 & 0 \cr 0 & c^{}_{23} & s^{}_{23} \cr
0 & -s^{}_{23} & c^{}_{23} \cr} \; ,
\label{eq:1.4}
\end{eqnarray}
together with the diagonal phase matrices
$P^{}_\delta = {\rm Diag}\{1, 1, e^{{\rm i}\delta}\}$ and
$P^{}_\nu = {\rm Diag}\{e^{{\rm i}\rho}, e^{{\rm i}\sigma}, 1\}$. In
Eq.~(\ref{eq:1.4}) we have defined $s^{}_{ij} \equiv \sin{\theta^{}_{ij}}$
and $c^{}_{ij} \equiv \cos{\theta^{}_{ij}}$ (for $ij = 12, 13, 23$).
Note that $\delta$ is usually referred to as the Dirac CP phase, while
$\rho$ and $\sigma$ are the so-called Majorana CP phases which are physical
only when massive neutrinos are the Majorana particles.
More explicitly, the expression of $U$ reads
\begin{eqnarray}
U = \pmatrix{
c^{}_{12} c^{}_{13} & s^{}_{12} c^{}_{13} & s^{}_{13} e^{-{\rm i} \delta} \cr
-s^{}_{12} c^{}_{23} - c^{}_{12} s^{}_{13} s^{}_{23}  e^{{\rm i} \delta}
& c^{}_{12} c^{}_{23} - s^{}_{12} s^{}_{13} s^{}_{23} e^{{\rm i} \delta}
& c^{}_{13} s^{}_{23} \cr
s^{}_{12} s^{}_{23} - c^{}_{12} s^{}_{13} c^{}_{23}  e^{{\rm i} \delta}
& -c^{}_{12} s^{}_{23} - s^{}_{12} s^{}_{13} c^{}_{23} e^{{\rm i} \delta}
& c^{}_{13} c^{}_{23}} P^{}_\nu \; .
\label{eq:1.5}
\end{eqnarray}
The strength of CP violation in neutrino oscillations is
measured by the well-known Jarlskog invariant $\cal J$ \cite{Jarlskog:1985ht},
which is defined through
\begin{eqnarray}
{\rm Im} \left( U^{}_{\alpha i} U^{}_{\beta j} U^{*}_{\alpha j} U^{*}_{\beta i} \right)
= {\cal J} \sum^{}_\gamma \epsilon^{}_{\alpha \beta \gamma}
\sum^{}_k \epsilon^{}_{i j k} \; ,
\label{eq:1.6}
\end{eqnarray}
where the Greek and Latin subscripts run respectively over $(e, \mu, \tau)$ and
$(1,2,3)$, and $\epsilon^{}_{\alpha \beta \gamma}$ or $\epsilon^{}_{i j k}$
denotes the  three-dimensional Levi-Civita symbol. Given
the parametrization of $U$ in Eq.~(\ref{eq:1.5}),
one has ${\cal J} = c^{}_{12} s^{}_{12} c^2_{13} s^{}_{13} c^{}_{23} s^{}_{23} \sin\delta$,
which depends only on the Dirac CP phase $\delta$.
In comparison, the phase parameters $\rho$ and $\sigma$ have nothing to do with
neutrino oscillations but they are important in those lepton-number-violating
processes such as the neutrinoless double-beta ($0\nu 2\beta$) decays \cite{Furry:1939qr}.
\begin{table}[t]
\caption{The best-fit values, 1$\sigma$ and 3$\sigma$ intervals of six neutrino
oscillation parameters extracted from a global analysis of current
experimental data \cite{Esteban:2018azc}.} \vspace{0.2cm}
\label{Table:1}
\begin{indented}
\item[]\begin{tabular}{cccc} \br Parameter & Best fit &
1$\sigma$ interval & 3$\sigma$ interval \\ \mr \multicolumn{4}{c}{Normal
mass ordering $(m^{}_1 < m^{}_2 < m^{}_3$)} \\ \mr \vspace{0.1cm}
$\Delta m^2_{21}/10^{-5} ~{\rm eV}^2$ & $7.39$  & 7.19 --- 7.60 &
6.79 --- 8.01 \\ \vspace{0.1cm}
$\Delta m^2_{31}/10^{-3} ~ {\rm eV}^2$~ & $2.525$ & 2.493 --- 2.558 &
2.427 --- 2.625 \\ \vspace{0.1cm}
$\sin^2\theta_{12}/10^{-1}$ & $3.10$ & 2.98 --- 3.25 & 2.75 --- 3.50
\\ \vspace{0.1cm}
$\sin^2\theta_{13}/10^{-2}$ & $2.241$ & 2.176 --- 2.306 & 2.045 --- 2.439
\\ \vspace{0.1cm}
$\sin^2\theta_{23}/10^{-1}$ & $5.80$  & 5.59 --- 5.97 & 4.18 ---
6.27 \\ \vspace{0.1cm}
$\delta/\pi$ &  $1.19$ & 1.03 --- 1.42 & 0.69
--- 2.18 \\ \br
\multicolumn{4}{c}{Inverted mass ordering $(m^{}_3 < m^{}_1 <
m^{}_2$)} \\ \mr \vspace{0.1cm}
$\Delta m^2_{21}/10^{-5} ~{\rm eV}^2$ & $7.39$  & 7.19 --- 7.60 &
6.79 --- 8.01 \\ \vspace{0.1cm}
$-\Delta m^2_{31}/10^{-3} ~ {\rm eV}^2$~ & $2.438$ & 2.404--- 2.470 &
2.336 --- 2.534 \\ \vspace{0.1cm}
$\sin^2\theta_{12}/10^{-1}$ & $3.10$ & 2.98 --- 3.25 & 2.75 --- 3.50
\\ \vspace{0.1cm}
$\sin^2\theta_{13}/10^{-2}$ & $2.264$ & 2.198 --- 2.330 & 2.068 --- 2.463
\\ \vspace{0.1cm}
$\sin^2\theta_{23}/10^{-1}$ & $5.84$  & 5.64 --- 6.00 & 4.23 ---
6.29 \\ \vspace{0.1cm}
$\delta/\pi$ &  $1.58$ & 1.42 --- 1.73 & 1.09
--- 2.00 \\ \br
\end{tabular}
\end{indented}
\end{table}

It is well known that the behaviors of three-flavor neutrino oscillations
are governed by six independent parameters: three flavor mixing angles $\theta^{}_{12}, \theta^{}_{13}$ and $\theta^{}_{23}$, the Dirac CP phase $\delta$, and two distinctive
neutrino mass-squared differences $\Delta m^2_{21} \equiv m^2_2 - m^2_1$
and $\Delta m^2_{31} \equiv m^2_3 - m^2_1$ (or $\Delta m^2_{32} \equiv m^2_3 - m^2_2$).
So far $\theta^{}_{12}$, $\theta^{}_{13}$, $\theta^{}_{23}$, $\Delta m^2_{21}$
and $|\Delta m^2_{31}|$ (or $|\Delta m^2_{32}|$)  have been determined,
to a good degree of accuracy, from solar, atmospheric, reactor and accelerator
neutrino oscillation experiments \cite{Tanabashi:2018oca}. Several groups have
performed a global analysis of the existing experimental data to extract or
constrain the values of these six neutrino oscillation parameters \cite{Esteban:2018azc,Capozzi:2018ubv,deSalas:2017kay,Capozzi:2020qhw,deSalas:2020pgw}.
In Table~\ref{Table:1} we list the results obtained by Esteban {\it et al}
\cite{Esteban:2018azc} as the reference values for the subsequent numerical
illustration and discussions. It is then straightforward to obtain
the $3\sigma$ intervals of the magnitudes of nine PMNS matrix elements from
Table~\ref{Table:1}:
\begin{eqnarray}
\left| U \right| \simeq \pmatrix{ 0.797 - 0.842 & 0.518
- 0.585 & 0.143 - 0.156 \cr 0.233 - 0.495 & 0.448
- 0.679 & 0.639 - 0.783 \cr 0.287 - 0.532 & 0.486
- 0.706 & 0.604 - 0.754 \cr} \; .
\label{eq:1.7}
\end{eqnarray}
A few immediate comments are in order.
\begin{itemize}
\item  Given $\Delta m^2_{21} >0$ from the solar neutrino oscillation
experiment, the unfixed sign of $\Delta m^2_{31}$ in atmospheric and accelerator neutrino
oscillation experiments allows for two possible neutrino mass ordering cases as illustrated
in Fig.~\ref{Fig:1}: the normal ordering (NO) $m^{}_1 < m^{}_2 <
m^{}_3$ or the inverted ordering (IO) $m^{}_3 < m^{}_1 < m^{}_2$. The values of
$\theta^{}_{13}$, $\theta^{}_{23}$ and $\delta$ extracted from current neutrino
oscillation data are more or less sensitive to such an ambiguity,
as one can clearly see in Table~\ref{Table:1}.
\begin{figure}[t]
\begin{center}
\includegraphics[width=4.5in]{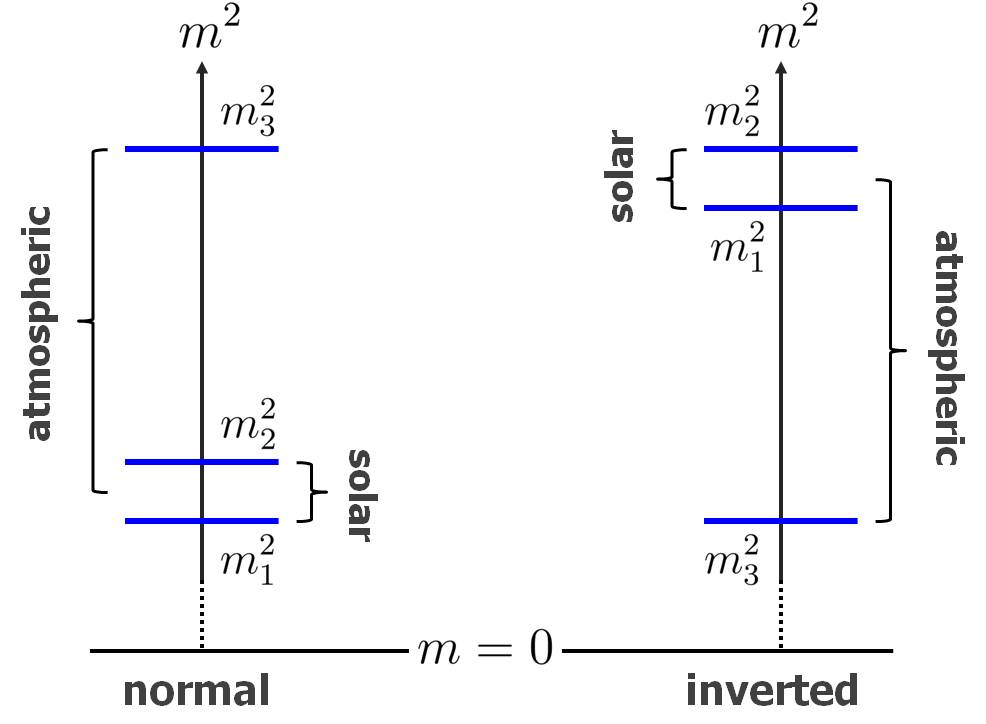}
\vspace{-0.07cm}
\caption{A schematic illustration of normal or inverted neutrino mass
ordering, where the smaller and larger mass-squared differences are responsible
for the dominant oscillations of solar and atmospheric neutrinos, respectively.}
\label{Fig:1}
\end{center}
\end{figure}

\item At present the NO case is found to be favored over the IO case at the
$3\sigma$ level. The sign of $\Delta m^2_{31}$ will hopefully be pinned down by
some ongoing and upcoming neutrino oscillation experiments in the near future \cite{Ayres:2004js,An:2015jdp,Acciarri:2015uup,Abe:2018uyc}.

\item As opposed to the small quark flavor mixing angles, here $\theta^{}_{12}$ and
$\theta^{}_{23}$ are very large. Moreover, they lie around special values (i.e., $\sin^2\theta^{}_{12} \sim 1/3$ and $\sin^2\theta^{}_{23} \sim 1/2$).
This may be suggestive of an underlying flavor symmetry in the lepton sector.
In this connection many flavor symmetries have been examined to interpret the
observed pattern of lepton flavor mixing \cite{Altarelli:2010gt,King:2013eh}.

\item The best-fit value of $\delta$ is around $1.5\pi$, which implies the existence
of large CP violation in the lepton sector. Of course, the significance of this
observation remains weak, but there is no reason why CP symmetry would be conserved in
leptonic weak interactions. Note that $\delta \sim 3\pi/2$ and $\sin^2\theta^{}_{23}
\sim 1/2$ may interestingly point towards an approximate $\mu$-$\tau$ reflection
symmetry \cite{Xing:2015fdg}. The latter is also revealed by the striking
relations $|U^{}_{\mu i}| \sim |U^{}_{\tau i}|$ as shown in Eq.~(\ref{eq:1.7}).
\end{itemize}
In short, the next-generation neutrino oscillation experiments will answer
two important questions in neutrino physics: the sign of $\Delta m^2_{31}$
and the value of $\delta$.

Unfortunately, neutrino oscillations have nothing to do with the absolute neutrino
mass scale and the Majorana CP phases. In order to acquire the knowledge of $m^{}_i$,
$\rho$ and $\sigma$, one has to resort to some non-oscillation experiments.
There are three kinds of promising experiments for this purpose.
\begin{itemize}
\item     In a given beta decay the neutrino masses will slightly affect the energy
spectrum of emitted electrons. By carefully measuring this spectrum's endpoint
where the effect of neutrino masses may develop to be observable, one finds that
the tritium beta decay experiments is capable of probing the effective
electron-neutrino mass
\begin{eqnarray}
\langle m\rangle^{}_e = \sqrt{ m^{2}_1 \left|U^{}_{e1}\right|^2 + m^{2}_2
\left|U^{}_{e2}\right|^2 + m^{2}_3 \left|U^{}_{e3}\right|^2} \; .
\label{eq:1.8}
\end{eqnarray}
While the present upper limit for $\langle m\rangle^{}_e$ is about 1 eV
\cite{Aker:2019uuj}, the future KATRIN experiment may reach a sensitivity
down to 0.2 eV \cite{Osipowicz:2001sq}.

\item     Since the Universe is flooded with a huge number of cosmic background
neutrinos, the tiny neutrino masses should have played a crucial role in the
cosmic evolution and left some imprints on the cosmic microwave background anisotropies
and large-scale structure formation. In this regard the cosmological observations
allow us to probe the sum of three neutrino masses
$\Sigma^{}_\nu = m^{}_1+ m^{}_2 + m^{}_3$ \cite{Bilenky:2002aw}. For the time being
the Planck 2018 results put the most stringent bound $\Sigma^{}_\nu \le 0.12$ eV
at the $95\%$ confidence level \cite{Aghanim:2018eyx, Vagnozzi:2017ovm}.

\item     If massive neutrinos are the Majorana particles,
then they can mediate the $0\nu 2\beta$ decays of some even-even nuclei (e.g.,
$^{76}_{32}{\rm Ge} \to  ^{76}_{34}{\rm Se} + 2 e^-$) \cite{Rodejohann:2011mu}.
The rates of such lepton-number-violating processes are governed by the magnitude
of the effective Majorana electron-neutrino mass
\begin{eqnarray}
\langle m\rangle^{}_{ee} =  m^{}_1 U^2_{e1} + m^{}_2 U^2_{e2} + m^{}_3 U^2_{e3} \;.
\label{eq:1.9}
\end{eqnarray}
The present upper limit for $|\langle m\rangle^{}_{ee}|$ is $0.06$ eV to $0.2$ eV at
the $95\%$ confidence level \cite{Agostini:2013mzu,Albert:2014awa,KamLAND-Zen:2016pfg},
where the large uncertainty originates from the inconclusiveness of relevant nuclear
physics calculations. It is expected that the next-generation $0\nu 2\beta$
experiments may bring the sensitivity down to the ${\cal O}(10)$ meV level.
\end{itemize}
So far these three kinds of experiments have not yet placed any lower constraint on
the absolute neutrino mass scale, nor any constraints on the Majorana CP phases.

There are certainly many other unknowns about massive neutrinos, such as whether
there exist extra (sterile) neutrino species of different mass scales,
whether low-energy CP violation in the lepton sector is related to the observed
matter-antimatter asymmetry of the Universe, why neutrino masses are so tiny,
what kind of flavor symmetry is behind large leptonic flavor mixing angles, and so on.

\subsection{Seesaw mechanisms with Occam's razor}
\label{section 1.2}

The facts that neutrinos are massive and lepton flavors are mixed
provide us with the first solid evidence that the standard model (SM)
of particle physics is incomplete, at least in its neutrino sector.
How to partly but wisely modify the SM turns out to be a burning question
in particle physics, simply because the true origin of finite neutrino masses
is definitely a window of new physics beyond the SM. In this connection
``the considerations have always
been qualitative, and, despite some interesting attempts, there has never
been a convincing quantitative model of the neutrino masses", as argued
by Edward Witten \cite{Witten:2000dt}. The popular seesaw mechanisms are
just a typical example of this kind --- they can offer a
{\it qualitative} explanation of smallness of three neutrino masses with
the help of some new degrees of freedom, but they are in general unable to
make any {\it quantitative} predictions unless a very specific flavor structure
associated with those new particles is assumed \cite{Xing:2019vks}.

Given the $\rm SU(2)^{}_{\rm L} \times U(1)^{}_{\rm Y}$ gauge symmetry and
field contents of the SM, there is no dimension-four operator that
can render the neutrinos massive. If the requirements of renormalizability
and lepton number conservation are loosened, a unique dimension-five operator
--- known as the Weinberg operator \cite{Weinberg:1979sa} will emerge to
give rise to finite but suppressed neutrino masses:
\begin{eqnarray}
{\cal O}^{}_\nu = \frac{1}{2} \cdot \frac{y^{}_{\alpha \beta}}{\Lambda}
\overline{\ell^{}_{\alpha \rm L}} \widetilde{H} \widetilde{H}^{T} \ell^{c}_{\beta \rm L} \;,
\label{eq:1.10}
\end{eqnarray}
where $\widetilde{H} \equiv {\rm i} \sigma^{}_2 H^*$ with $\sigma^{}_2$ being
the second Pauli matrix and $H$ being the Higgs doublet,
$\ell^{}_{\alpha\rm L} = \pmatrix{\nu^{}_{\alpha \rm L}, l^{}_{\alpha \rm L}}^{T}$
represent the left-handed lepton doublets, $y^{}_{\alpha \beta}$ (for $\alpha, \beta
= e, \mu, \tau$) are some dimensionless
coefficients, and $\Lambda$ stands for the typical cut-off scale of new physics
responsible for the origin of neutrino masses. Note that $l^{}_e$, $l^{}_\mu$ and
$l^{}_\tau$ denote the flavor eigenstates of
three charged leptons, whose mass eigenstates have been denoted as $e$, $\mu$ and
$\tau$ in Eq.~(\ref{eq:1.1}). Once the electroweak symmetry
is spontaneously broken by the vacuum expectation value of the neutral
Higgs component (i.e., $v \equiv \langle H^0 \rangle  \simeq 174$ GeV), the above
operator will yield an effective neutrino mass matrix $M^{}_\nu$
with the elements $(M^{}_\nu)^{}_{\alpha\beta} = y^{}_{\alpha \beta} v^2/\Lambda$.
One can see that the Weinberg operator violates lepton number by two units
and thus the massive neutrinos are of the Majorana nature.
In order to achieve sub-eV neutrino masses, it is compulsory to take either
extremely small $y^{}_{\alpha \beta}$ or extremely large $\Lambda$.
Since $y^{}_{\alpha \beta} \sim {\cal O}(1)$ is a natural expectation from the
point of view of model building, $\Lambda$ should be around ${\cal O}(10^{14})$
GeV --- an energy scale which happens to be not far from the presumable scale
of grand unification theories (GUTs)
\footnote{With the help of additional suppression mechanisms, the new physics
scale can be naturally lowered to an experimentally accessible scale.
There are a few typical ways to do so. (1) In a model where neutrino masses are
generated radiatively \cite{Cai:2017jrq}, the loop integrals will supply the
required additional suppression. (2) Given the Weinberg operator with lepton number
violation, the smallness of neutrino masses can be attributed to the smallness of
lepton-number-violating parameters (e.g., in the inverse seesaw mechanism \cite{Wyler:1982dd,Mohapatra:1986bd}).
(3) In a model where the Weinberg operator is forbidden and the neutrino masses
can only stem from certain higher-dimension operators, additional $v/\Lambda$
suppression factors will contribute (e.g., in the multiple and cascade seesaw
mechanisms \cite{Xing:2009hx,Bonnet:2009ej,Liao:2010cc}).}.
In such an intriguing scenario the smallness of neutrino masses is ascribed to the
largeness of $\Lambda$ as compared with the electroweak scale
$\Lambda^{}_{\rm EW} \sim v$, and hence it works like a {\it seesaw}.

The unique Weinberg operator can be derived from the Yukawa interactions
mediated by heavy particles in certain renormalizable extensions of the SM.
At the tree level there are three and only three ways to realize this idea,
which are known as the type-I \cite{Fritzsch:1975sr,Minkowski:1977sc,Yanagida:1979as,
GellMann:1980vs,Glashow:1979nm,Mohapatra:1979ia}, type-II \cite{Konetschny:1977bn,Magg:1980ut,Schechter:1980gr,Cheng:1980qt,Lazarides:1980nt,
Mohapatra:1980yp} and type-III \cite{Foot:1988aq,Ma:1998dn} seesaw mechanisms.
Here let us outline their main features as follows \cite{Xing:2009in}.

(1) Type-I seesaw: three heavy right-handed neutrino fields $N^{}_{\alpha \rm R}$
(for $\alpha = e, \mu, \tau$) are introduced into the SM and lepton number conservation
is violated by their Majorana mass term. In this case the leptonic mass terms
can be written as
\begin{eqnarray}
-{\cal L}^{}_{\ell + \nu} = \overline{\ell^{}_{\rm L}}
Y^{}_l H E^{}_{\rm R} + \overline{\ell^{}_{\rm L}} Y^{}_\nu
\widetilde{H} N^{}_{\rm R} + \frac{1}{2} \overline{N^{c}_{\rm R}}
M^{}_{\rm R} N^{}_{\rm R} + {\rm h.c.} \; ,
\label{eq:1.11}
\end{eqnarray}
where $N^{}_{\rm R} = \pmatrix{N^{}_{e \rm R} , N^{}_{\mu \rm R} ,
N^{}_{\tau \rm R}}^{T}$, and $M^{}_{\rm R}$ is a symmetric Majorana
mass matrix. After integrating out the heavy degrees of freedom in Eq.~(\ref{eq:1.11})
\cite{Xing:2011zza}, we are left with the effective Weinberg operator
\begin{eqnarray}
{\cal O}^{}_\nu = \frac{1}{2} \left(Y^{}_\nu M^{-1}_{\rm R}
Y^T_\nu\right)^{}_{\alpha\beta} \overline{\ell^{}_{\alpha \rm L}}
\widetilde{H} \widetilde{H}^T \ell^{c}_{\beta \rm L} \; .
\label{eq:1.12}
\end{eqnarray}

(2) Type-II seesaw: a heavy Higgs triplet $\Delta$ is introduced
into the SM and lepton number conservation is violated by the interactions
of $\Delta$ with both the lepton doublet and the Higgs doublet. In this case,
\begin{eqnarray}
-{\cal L}^{}_{\ell + \nu} = \overline{\ell^{}_{\rm L}}
Y^{}_l H E^{}_{\rm R} + \frac{1}{2} \overline{\ell^{}_{\rm L}}
Y^{}_\Delta \Delta {\rm i}\sigma^{}_2 \ell^{c}_{\rm L} -
\lambda^{}_\Delta M^{}_\Delta H^T {\rm i}\sigma^{}_2 \Delta H + {\rm h.c.} \; ,
\label{eq:1.13}
\end{eqnarray}
where $Y^{}_\Delta$ stands for the neutrino Yukawa coupling matrix, $\lambda^{}_\Delta$
represents the scalar coupling coefficient, and $M^{}_\Delta$ is the mass scale
of $\Delta$. After the heavy degrees of freedom are integrated out, one obtains
\begin{eqnarray}
{\cal O}^{}_\nu = -\frac{\lambda^{}_\Delta}{M^{}_\Delta}
\left(Y^{}_\Delta\right)^{}_{\alpha\beta}
\overline{\ell^{}_{\alpha \rm L}} \widetilde{H} \widetilde{H}^T
\ell^{c}_{\beta \rm L} \; .
\label{eq:1.14}
\end{eqnarray}

(3) Type-III seesaw: three heavy fermion triplets are introduced into
the SM and lepton number conservation is violated by their Majorana mass term.
In this case,
\begin{eqnarray}
-{\cal L}^{}_{\ell + \nu} \; = \; \overline{\ell^{}_{\rm L}}
Y^{}_l H E^{}_{\rm R} + \overline{\ell^{}_{\rm L}} \sqrt{2}
Y^{}_\Sigma \Sigma^{c} \widetilde{H} + \frac{1}{2} {\rm Tr} \left(
\overline{\Sigma} M^{}_\Sigma \Sigma^{c} \right) + {\rm h.c.} \; ,
\label{eq:1.15}
\end{eqnarray}
where $Y^{}_\Sigma$ and $M^{}_\Sigma$ stand for the Yukawa coupling matrix
and the mass scale of $\Sigma$, respectively. After integrating out
the heavy degrees of freedom, we have
\begin{eqnarray}
{\cal O}^{}_\nu = \frac{1}{2} \left(Y^{}_\Sigma M^{-1}_{\rm \Sigma}
Y^T_\Sigma\right)^{}_{\alpha\beta} \overline{\ell^{}_{\alpha \rm
L}} \widetilde{H} \widetilde{H}^T \ell^{c}_{\beta \rm L} \; .
\label{eq:1.16}
\end{eqnarray}

It is obvious that Eqs.~(\ref{eq:1.12}), (\ref{eq:1.14}) and (\ref{eq:1.16})
lead us to the same effective Majorana mass term for three light
neutrinos at the electroweak scale, after spontaneous gauge symmetry breaking:
\begin{eqnarray}
-{\cal L}^{\prime}_{\ell+\nu} = \overline{l^{}_{\rm L}} M^{}_l E^{}_{\rm R}
+ \frac{1}{2} \overline{\nu^{}_{\rm
L}} M^{}_\nu \nu^{c}_{\rm L} + {\rm h.c.} \; ,
\label{eq:1.17}
\end{eqnarray}
in which
$M^{}_l = Y^{}_l v$ is the charged-lepton mass matrix, and
the symmetric Majorana neutrino mass matrix $M^{}_\nu$ is given by one
of the seesaw formulas
\begin{eqnarray}
M^{}_\nu \simeq \left\{ \begin{array}{lcl}
\displaystyle - M^{}_{\rm D} M^{-1}_{\rm R}
M^T_{\rm D} && ({\rm type ~ I}) \; , \vspace{0.15cm} \\
\displaystyle 2 v^2 \lambda^{}_\Delta Y^{}_\Delta M^{-1}_\Delta
&& ({\rm type ~ II}) \; , \vspace{0.15cm} \\
\displaystyle - M^{}_{\rm D} M^{-1}_\Sigma M^{T}_{\rm D}
&& ({\rm type ~ III}) \; ,
\end{array}
\right .
\label{eq:1.18}
\end{eqnarray}
where $M^{}_{\rm D} = Y^{}_\nu v$ (type-I seesaw) or
$M^{}_{\rm D} = Y^{}_\Sigma v$ (type-III seesaw). It becomes transparent that the
smallness of $M^{}_\nu$ can be naturally attributed to the largeness of $M^{}_{\rm R}$,
$M^{}_\Delta$ or $M^{}_\Sigma$ as compared with $v$ in such seesaw mechanisms.
However, this qualitative observation does not mean that one can figure out
the values of neutrino masses from Eq.~(\ref{eq:1.18}) because each seesaw formula
involves quite a lot of free parameters.

There are two possibilities of enhancing predictive power of the
most popular type-I seesaw mechanism. One of them is to determine the flavor structures of
$M^{}_{\rm D}$ and $M^{}_{\rm R}$ with the help of a kind of flavor symmetry,
which is certainly model-dependent. The other possibility, which is independent
of any model details, is to reasonably simplify this seesaw mechanism with the
so-called principle of Occam's razor --- ``entities must not be multiplied beyond necessity"
\footnote{It was William of Ockham, an English philosopher and theologian in the
14th century, who invented this law of briefness and expressed it in Latin as
``Entia non sunt multiplicanda praeter necessitatem".}.
Namely, one may cut off one of the three heavy right-handed neutrino fields
with Occam's razor such that $M^{}_{\rm R}$ is of rank two and thus $M^{}_\nu$
is also of rank two \cite{Kleppe:1995zz,Ma:1998zg}, predicting one of the three light
Majorana neutrinos to be massless at the tree level. The resultant scenario is commonly
referred to as the {\it minimal} (type-I) seesaw mechanism. It is not only
compatible with current neutrino oscillation data but also able to interpret
the observed baryon-antibaryon asymmetry of the Universe \cite{Frampton:2002qc}
via thermal leptogenesis \cite{Fukugita:1986hr}. Similarly, a minimal version
of the type-II (or type-III) seesaw mechanism can be achieved by taking
$Y^{}_\Delta$ (or $M^{}_\Sigma$) to be of rank two, leading us to a massless
Majorana neutrino at the tree level.

We find that the minimal seesaw mechanism deserves particular attention because
it can serve as a predictive benchmark seesaw scenario which will be
confronted with the upcoming precision measurements in both neutrino physics and
cosmology. In fact, several hundreds of papers have been published in the past
twenty years to explore this economical but viable and testable mechanism of neutrino
mass generation in depth and from many perspectives,
especially since the seminal work done by Frampton, Glashow and
Yanagida appeared in 2002 \cite{Frampton:2002qc}. So it is highly timely and
important today to review the theoretical aspect of the minimal seesaw mechanism
and its various phenomenological consequences, including those consequences in
cosmology.

The purpose of the present article is just to provide an up-to-date review of all the
important progress that has so far been made in the studies of the minimal seesaw
and leptogenesis models. Our focus is on possible flavor structures of such models
and confronting their predictions with current experimental measurements. We are
going to pay special attention to the topics of lepton number violation, discrete
flavor symmetries, CP violation, leptogenesis and antimatter in this connection.

The remaining parts of this review article are organized as follows. In section 2 we
outline salient features of the minimal seesaw mechanism, discuss the
stability of $m^{}_1 =0$ or $m^{}_3 =0$ against quantum corrections, and
introduce the minimal thermal leptogenesis mechanism. We also make some
brief comments on the minimal versions of a few other
seesaw scenarios. Section 3 is devoted to a number of generic
descriptions of the flavor structures of $M^{}_{\rm D}$ and $M^{}_{\rm R}$ in
the minimal seesaw case, including an exact Euler-like parametrization. We
explore some striking scenarios of the minimal leptogenesis mechanism in
section 4, such as the vanilla leptogenesis, flavored leptogenesis, resonant leptogenesis, and possible contributions of $N^{}_2$ to leptogenesis. Quantum corrections to the minimal seesaw relation and their effects on leptogenesis will also be discussed. In section 5 we
study some particular textures of $M^{}_{\rm D}$ and $M^{}_{\rm R}$ which contain
one or more zero entries, and confront their phenomenological consequences with
current neutrino oscillation data.
Section 6 is devoted to some simple but instructive flavor symmetries that can be embedded in the minimal seesaw models. The typical examples of this kind include the $\mu$-$\tau$ reflection symmetry and the $\rm S^{}_4$ symmetry. The so-called littlest seesaw model, which is actually a special form of the minimal seesaw scenario, will also be introduced in some detail.
Some other aspects of the minimal seesaw model, including the lepton-number-violating and lepton-flavor-violating processes mediated by both light and heavy Majorana neutrinos, are discussed in section 7, where some comments on the low-scale seesaw models are also made. Finally, we give some concluding remarks and outlooks in section 8.

\setcounter{equation}{0}
\setcounter{figure}{0}
\section{The minimal seesaw and thermal leptogenesis}
\label{section 2}

\subsection{The minimal seesaw and its salient features}
\label{section:2.1}

As argued above, a straightforward way of enhancing predictability of the canonical seesaw
mechanism is to reduce the number of the hypothetical right-handed neutrino fields. But the
minimal number of right-handed neutrino fields that can be consistent with the two
experimentally-observed neutrino mass-squared differences (i.e.,
$\Delta m^2_{21} \ll |\Delta m^2_{31}|$) is {\it two} instead
of one, because the so-called ``seesaw fair play rule" requires that the number of
heavy Majorana neutrinos $N^{}_i$ should exactly match that of light
Majorana neutrinos $\nu^{}_j$ in a type-I seesaw scenario with arbitrary $N^{}_i$
(for $i,j = 1, 2, \cdots$) \cite{Xing:2007uq}
\footnote{This point can easily be seen from the seesaw formula in Eq.~(\ref{eq:1.18}),
where the rank of $M^{}_\nu$ must be equal to that of $M^{}_{\rm R}$. If there were
only a single right-handed neutrino field, then two of the three light neutrinos would
be massless as a consequence of the seesaw relation --- a tree-level result that is
definitely in conflict with current neutrino oscillation data.}.
Moreover, at least two right-handed neutrino fields are needed for a successful realization
of thermal leptogenesis \cite{Frampton:2002qc}, so as to interpret the observed
baryon-antibaryon asymmetry of the Universe. That is why the {\it minimal} seesaw
mechanism is defined to contain {\it two} right-handed neutrinos
and allow for lepton number violation.

This simple but intriguing scenario nicely conforms with the philosophy of Occam's razor.
There are actually several good reasons to consider and study such a simplified seesaw
mechanism. (1) Although $N^{}_{\alpha \rm R}$ are called the right-handed neutrino fields,
they are in fact the gauge-singlet fermions which carry no gauge quantum number of the SM.
So it is {\it not} unnatural that the number of $N^{}_{\alpha \rm R}$ does not match
that of $\nu^{}_{\beta \rm L}$ (for $\alpha, \beta = e, \mu, \cdots$), at least
from a purely theoretical point of view. (2) Even if there are three right-handed
neutrino fields in a given model, it is still possible to arrive at an effective minimal
seesaw scenario in the case that one of them is essentially decoupled
from the other two (e.g., if such a heavy neutrino is much heavier than the other two
or its associated Yukawa couplings are vanishing
or vanishingly small). (3) The number of free parameters in the minimal seesaw
mechanism can be significantly reduced as compared with that in an ordinary seesaw model,
making its predictive power accordingly enhanced. (4) One of the most salient features of
the minimal seesaw paradigm is that one of the three light neutrinos must be massless
\footnote{Note that this statement is valid at the tree and one-loop levels in the SM
framework \cite{Mei:2003gn}, and the two-loop quantum corrections may give rise to a
vanishingly small mass for the lightest neutrino in this case (see
Refs.~\cite{Davidson:2006tg,Xing:2020ezi} and section~\ref{section 2.2} for some
discussions).},
and thus their mass spectrum can be fixed after current neutrino oscillation
data on $\Delta m^2_{21}$ and $\Delta m^2_{31}$ are taken into account. Moreover,
the Majorana CP phase associated with the massless neutrino is not well defined
and hence does not have any physical impacts. These two features make the minimal
seesaw very economical and easily testable among various viable seesaw
scenarios on the market.

Without loss of generality, let us take the flavor eigenstates of two heavy
(essentially right-handed) neutrinos to be identical with their mass eigenstates
in the minimal seesaw mechanism. In this basis the neutrino mass terms can be expressed as
\begin{eqnarray}
-{\cal L}^{}_{\nu} & = & \overline{\pmatrix{\nu^{}_{e} & \nu^{}_{\mu} &
\nu^{}_{\tau}}^{}_{\rm L}} \ M^{}_{\rm D} \pmatrix{N^{}_1 \cr
N^{}_2}^{}_{\hspace{-0.05cm} \rm R} + \frac{1}{2} \ \overline{\pmatrix{N^{c}_{1}
& N^{c}_{2}}^{}_{\rm L}} \ D^{}_{N} \pmatrix{N^{}_1 \cr
N^{}_2}^{}_{\hspace{-0.05cm} \rm R} + {\rm h.c.}
\nonumber \\
& = & \frac{1}{2} \ \overline{\pmatrix{\nu^{}_{e} & \nu^{}_{\mu} &
\nu^{}_{\tau} & N^{c}_{1} & N^{c}_{2}}^{}_{\rm L}} \ \pmatrix{0 & M^{}_{\rm D} \cr
M^T_{\rm D} & D^{}_N} \pmatrix{\nu^{c}_e \cr \nu^{c}_\mu \cr \nu^{c}_\tau \cr
N^{}_1 \cr N^{}_2}^{}_{\hspace{-0.1cm} \rm R} + {\rm h.c.} \; ,
\label{eq:2.1}
\end{eqnarray}
where the $3\times 2$ Dirac mass matrix $M^{}_{\rm D}$ and the diagonal $2\times 2$
Majorana mass matrix $M^{}_{\rm R}$ are denoted respectively as
\begin{eqnarray}
M^{}_{\rm D} = \pmatrix{ a^{}_{1} & b^{}_{1} \cr
a^{}_{2} & b^{}_{2} \cr a^{}_{3} & b^{}_{3}} \; ,
\quad
D^{}_{N} = \pmatrix{ M^{}_1 & 0 \cr 0 & M^{}_2} \;
\label{eq:2.2}
\end{eqnarray}
with $M^{}_{1,2}$ being the masses of heavy Majorana neutrinos $N^{}_{1,2}$.
Although the elements of $M^{}_{\rm D}$ are all complex in general, it is
definitely possible to remove three of their six phase parameters by redefining the
phases of three left-handed neutrino fields. Note, however, that the three
phase differences ${\rm arg}(b^{}_i) - {\rm arg}(a^{}_i)$ (for $i=1,2,3$)
can always survive rephasing of the left-handed neutrino fields. Therefore, the
minimal seesaw mechanism only contains a total of {\it eleven} real parameters.

By definition, the rank of the overall $5\times 5$ neutrino mass matrix in
Eq.~(\ref{eq:2.1}) is the number of nonzero rows in the reduced row echelon form of
this matrix, which can be calculated with the method of Gauss elimination
\cite{Xing:2007uq}. Since the upper-left $3\times 3$ submatrix is a zero matrix, it
is straightforward to convert the upper-right $3\times 2$ submatrix (namely, $M^{}_{\rm D}$)
into a reduced row echelon form where the first row is full of zero elements.
In comparison, the lower-right $2\times 2$ submatrix (that is, $D^{}_N$) is of
rank two. The total rank of the symmetric $5\times 5$ neutrino mass matrix turns out
to be $3 - 1 + 2 = 4$, corresponding to four massive neutrino eigenstates. As a
result, one of the three light Majorana neutrinos must be exactly massless
at the tree level.

This observation can easily be seen from the approximate type-I seesaw
formula obtained in Eq.~(\ref{eq:1.18}), simply because the rank of $M^{}_\nu$
must be equal to that of the mass matrix with the lowest rank on the right-hand
side of Eq.~(\ref{eq:1.18}) --- the rank of $M^{}_{\rm R}$ in the minimal
seesaw scenario under discussion. To be more explicit, let us calculate
$M^{}_\nu$ with the help of Eqs.~(\ref{eq:1.18}) and (\ref{eq:2.2}).
The expression of $M^{}_\nu$ is found to be
\begin{eqnarray}
M^{}_\nu \simeq -\pmatrix{ \vspace{0.2cm}
\displaystyle \frac{a^2_1}{M^{}_1} + \frac{b^2_1}{M^{}_2}
& ~ \displaystyle \frac{a^{}_1 a^{}_2}{M^{}_1} + \frac{b^{}_1 b^{}_2}{M^{}_2} ~
& \displaystyle \frac{a^{}_1 a^{}_3}{M^{}_1} + \frac{b^{}_1 b^{}_3}{M^{}_2} \cr
\vspace{0.2cm}
\displaystyle \frac{a^{}_1 a^{}_2}{M^{}_1} + \frac{b^{}_1 b^{}_2}{M^{}_2}
& \displaystyle \frac{a^{2}_2 }{M^{}_1} + \frac{b^{2}_2 }{M^{}_2}
& \displaystyle \frac{a^{}_2 a^{}_3}{M^{}_1} + \frac{b^{}_2 b^{}_3}{M^{}_2} \cr
\displaystyle \frac{a^{}_1 a^{}_3}{M^{}_1} + \frac{b^{}_1 b^{}_3}{M^{}_2}
& \displaystyle \frac{a^{}_2 a^{}_3}{M^{}_1} + \frac{b^{}_2 b^{}_3}{M^{}_2}
& \displaystyle \frac{a^{2}_3}{M^{}_1} + \frac{ b^{2}_3}{M^{}_2} \cr } \;.
\label{eq:2.3}
\end{eqnarray}
Then it is easy to show $\det(M^{}_\nu) = 0$, implying the existence of a
massless light Majorana neutrino. It is also easy to see that $M^{}_1$ and $M^{}_2$
can be absorbed by making the rescaling transformations $a^{}_i \to a^{}_i/\sqrt{M^{}_1}$
and $b^{}_i \to b^{}_i/\sqrt{M^{}_2}$, implying that these two heavy degrees of
freedom are actually redundant in producing the masses and flavor mixing parameters of
three light Majorana neutrinos.

Given the neutrino mass spectrum $m^{}_1 < m^{}_2 < m^{}_3$ (NO) or
$m^{}_3 < m^{}_1 < m^{}_2$ (IO) as constrained by current neutrino oscillation data,
we are left with either $m^{}_1 = 0$ (NO) or $m^{}_3 = 0$ (IO) in the minimal
seesaw mechanism. As an immediate consequence, one of the Majorana CP phases of
the PMNS matrix $U$ is not well defined in this case and can therefore
be removed. To see this point clearly,
let us take the basis in which the flavor eigenstates of three charged leptons are
identical with their mass eigenstates (i.e., $M^{}_l = D^{}_l \equiv
{\rm Diag}\{ m^{}_e, m^{}_\mu, m^{}_\tau\}$), and reconstruct the symmetric
Majorana neutrino mass matrix $M^{}_\nu = U D^{}_\nu U^T$ with $D^{}_\nu \equiv
{\rm Diag}\{m^{}_1, m^{}_2, m^{}_3\}$ in terms of $m^{}_i$ and $U^{}_{\alpha i}$
(for $i = 1,2,3$ and $\alpha = e, \mu, \tau$). Then we arrive at
\begin{eqnarray}
\langle m\rangle^{}_{\alpha\beta} \equiv \left(M^{}_\nu\right)^{}_{\alpha\beta}
=  m^{}_1 U^{}_{\alpha 1} U^{}_{\beta 1} +
m^{}_2 U^{}_{\alpha 2} U^{}_{\beta 2} + m^{}_3 U^{}_{\alpha 3} U^{}_{\beta 3} \;,
\label{eq:2.4}
\end{eqnarray}
and thus $m^{}_1 = 0$ or $m^{}_3 = 0$ will eliminate one of the three terms on
the right-hand side of Eq.~(\ref{eq:2.4}). That is why one of
the Majorana CP phases in $P^{}_\nu$ of $U$ in Eq.~(\ref{eq:1.3}) can always be
removed in the $m^{}_1 \to 0$ (or $m^{}_3 \to 0$) limit.
To be more specific, $\rho$ will automatically disappear
in the $m^{}_1 = 0$ case; and it can also be eliminated in the $m^{}_3 = 0$ case
by a global rephasing of the three left-handed neutrino fields
(i.e., $\nu^{}_{i \rm L} \to e^{-{\rm i}\rho} \nu^{}_{i \rm L}$) and
a redefinition of $\sigma - \rho$ as $\sigma$ \cite{Mei:2003gn}.
So we simply take $P^{}_\nu = {\rm Diag}\{1, e^{{\rm i}\sigma}, 1\}$ hereafter.
In other words, the PMNS matrix $U$ only contains two nontrivial CP-violating phases
($\delta$ and $\sigma$) in the minimal seesaw mechanism.

Now let us fix the neutrino mass spectrum in the minimal seesaw mechanism with the
help of current neutrino oscillation data as listed in Table~\ref{Table:1}.
\begin{itemize}
\item     In the $m^{}_1 =0$ case, we are left with
\begin{eqnarray}
m^{}_2 = \sqrt{\Delta m^2_{21}} \simeq 8.60^{+0.12}_{-0.12} \times 10^{-3} \ {\rm eV} \;,
\nonumber \\
m^{}_3 = \sqrt{\Delta m^2_{31}} \simeq 5.02^{+0.03}_{-0.03} \times 10^{-2} \ {\rm eV} \;.
\label{eq:2.5}
\end{eqnarray}
The effective electron-neutrino mass of a beta decay and the sum of
three neutrino masses are then given as
\begin{eqnarray}
\langle m\rangle^{}_e \simeq 8.88^{+0.12}_{-0.12} \times 10^{-3} \ {\rm eV} \;, \quad
\Sigma^{}_\nu \simeq 5.88^{+0.04}_{-0.03} \times 10^{-2} \ {\rm eV} \;.
\label{eq:2.6}
\end{eqnarray}

\item     In the $m^{}_3 =0$ case, we arrive at
\begin{eqnarray}
m^{}_1 = \sqrt{|\Delta m^2_{31}|} \simeq 4.94^{+0.03}_{-0.03} \times 10^{-2} \ {\rm eV} \;, \nonumber \\
m^{}_2 = \sqrt{\Delta m^2_{21} + |\Delta m^2_{31}|} \simeq 5.01^{+0.03}_{-0.03}
\times 10^{-2} \ {\rm eV} \;.
\label{eq:2.7}
\end{eqnarray}
Accordingly, the values of $\langle m\rangle^{}_e$ and $\Sigma^{}_\nu$ are found to be
\begin{eqnarray}
\langle m\rangle^{}_e \simeq 4.90^{+0.03}_{-0.03} \times 10^{-2} \ {\rm eV} \;, \quad
\Sigma^{}_\nu \simeq 9.95^{+0.06}_{-0.07} \times 10^{-2} \ {\rm eV} \;.
\label{eq:2.8}
\end{eqnarray}
\end{itemize}
In either case the observable quantities $\langle m\rangle^{}_e$ and $\Sigma^{}_\nu$,
together with the neutrino mass spectrum, are fully determined.

We proceed to take a look at the simplified result of the effective electron-neutrino
mass $\langle m\rangle^{}_{ee}$ for a lepton-number-violating $0\nu 2\beta$ decay mode,
which has already been defined in Eq.~(\ref{eq:1.9}). In the minimal seesaw framework,
\begin{itemize}
\item     $m^{}_1 =0$ leads us to
\begin{eqnarray}
\left|\langle m\rangle^{}_{ee}\right| = \sqrt{ m^2_2 s^4_{12} c^4_{13} + m^2_3 s^4_{13} + 2 m^{}_2 m^{}_3 s^2_{12} c^2_{13} s^2_{13} \cos{2\left(\sigma + \delta\right)} } \;,
\label{eq:2.9}
\end{eqnarray}
from which the upper and lower bounds of $|\langle m\rangle^{}_{ee}|$ are found to be
\begin{eqnarray}
\left|\langle m\rangle^{}_{ee}\right|^{}_{\rm max} = m^{}_2 s^2_{12} c^2_{13}+ m^{}_3 s^{2}_{13} \simeq 3.73^{+0.14}_{-0.11} \times 10^{-3} \ {\rm eV} \;,
\nonumber \\
\left|\langle m\rangle^{}_{ee}\right|^{}_{\rm min} = m^{}_2 s^2_{12} c^2_{13} - m^{}_3 s^{2}_{13} \simeq 1.48^{+0.14}_{-0.11} \times 10^{-3} \ {\rm eV} \;,
\label{eq:2.10}
\end{eqnarray}
corresponding to $\sigma+\delta =0$ and $\pi$, respectively;

\item    $m^{}_3 =0$ leads us to
\begin{eqnarray}
\left|\langle m\rangle^{}_{ee}\right| = c^2_{13} \sqrt{ m^2_1 c^4_{12}
+ m^2_2 s^4_{12} + 2 m^{}_1 m^{}_2 c^2_{12} s^2_{12}  \cos{2\sigma} } \;,
\label{eq:2.11}
\end{eqnarray}
from which the maximal and minimal values of $|\langle m\rangle^{}_{ee}|$ are found to be
\begin{eqnarray}
\left|\langle m\rangle^{}_{ee}\right|^{}_{\rm max} = \left(m^{}_1 c^2_{12} + m^{}_2 s^{2}_{12}\right)c^2_{13} \simeq 4.85^{+0.03}_{-0.03} \times 10^{-2} \ {\rm eV} \;,
\nonumber \\
\left|\langle m\rangle^{}_{ee}\right|^{}_{\rm min} = \left(m^{}_1 c^2_{12} - m^{}_2 s^{2}_{12}\right)c^2_{13} \simeq  1.81^{+0.15}_{-0.12} \times 10^{-2} \ {\rm eV} \;,
\label{eq:2.12}
\end{eqnarray}
corresponding to $\sigma =0$ and $\pi$, respectively.
\end{itemize}
In either case Eq.~(\ref{eq:2.4}) can be simplified to describe an effective Majorana mass
triangle in the complex plane, as discussed in Refs.~\cite{Xing:2015uqa,Xing:2015wzz}.
\begin{figure*}[t]
\centering
\includegraphics[width=6in]{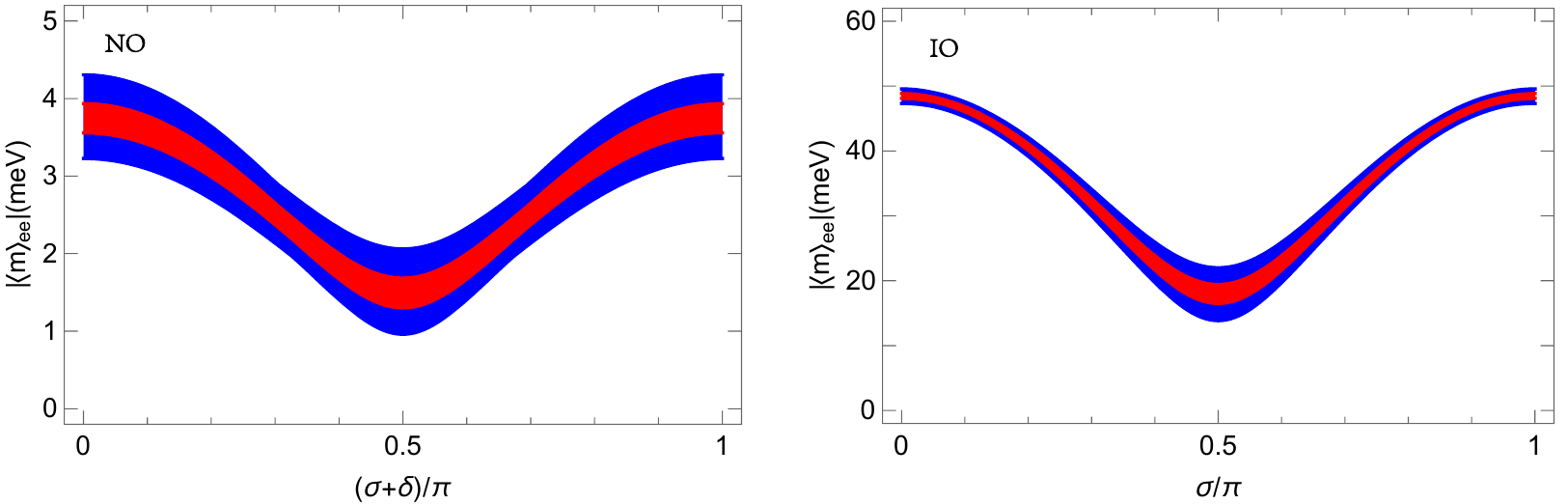}
\caption{The allowed range of $|\langle m\rangle^{}_{ee}|$ as a function of
$\sigma+\delta$ in the $m^{}_1=0$ (NO) case or of $\sigma$ in the $m^{}_3=0$ (IO) case, where the red and blue bands correspond to inputting
the $1\sigma$ and $3\sigma$ intervals of current neutrino oscillation
data listed in Table~\ref{Table:1}, respectively.}
\label{Fig:2}
\end{figure*}

In Fig.~\ref{Fig:2} the allowed range of $|\langle m\rangle^{}_{ee}|$ is illustrated as a
function of $\sigma+\delta$ in the $m^{}_1 =0$ case or of $\sigma$ in the $m^{}_3=0$ case,
where the red and blue bands are obtained by inputting the
$1\sigma$ and $3\sigma$ intervals of current neutrino oscillation parameters listed
in Table~\ref{Table:1}, respectively. It is found that the value of $|\langle m\rangle^{}_{ee}|$
is strongly sensitive to $\sigma$, implying that a measurement of the former will
enable us to determine the latter in the minimal seesaw scenario.

So far we have not made any specific assumptions about flavor structures of the minimal
seesaw mechanism itself, but some striking phenomenological consequences have been
achieved even in this general case. Once the flavor structures of $M^{}_{\rm D}$ and
$M^{}_{\rm R}$ are fixed with the help of certain flavor symmetries or empirical conditions, the
resulting minimal seesaw models will have much more predictive power. We shall elaborate
on such model building issues in sections 5 and 6.

\subsection{On the stability of $m^{}_1 =0$ or $m^{}_3 =0$}
\label{section 2.2}

Although $m^{}_1$ (or $m^{}_3$) is dictated to be vanishing in the minimal seesaw
mechanism under discussion, this result is only valid at the tree level.
Given the fact that the three charged leptons have quite different Yukawa coupling
eigenvalues (i.e., $y^{}_e \ll y^{}_\mu \ll y^{}_\tau$ with
$y^{}_\alpha \equiv m^{}_\alpha/v$ in the SM for $\alpha = e, \mu, \tau$), no fundamental
symmetry demands that the massless neutrino should stay massless. In other words,
quantum corrections at the loop level are expected to make the massless neutrino
massive. A careful study shows that $m^{}_1 =0$ (or $m^{}_3 =0$) still holds well if
the one-loop renormalization-group equation (RGE) is considered for the
evolution of neutrino masses from the seesaw scale down to the electroweak scale
\cite{Mei:2003gn}, but a tiny departure of $m^{}_1$ (or $m^{}_3$) from zero will
come into being when the two-loop RGE effect is taken into account \cite{Davidson:2006tg}.

At the one-loop level, the RGE of the effective Majorana neutrino coupling matrix
$\kappa = Y^{}_\nu M^{-1}_{\rm R} Y^T_\nu$ in the type-I seesaw scenario is given
by \cite{Chankowski:1993tx,Babu:1993qv,Antusch:2001ck}
\begin{eqnarray}
16\pi^2 \frac{{\rm d}\kappa}{{\rm d}t} = \alpha^{}_\kappa \kappa -
\frac{3}{2} \left[ \left( Y^{}_l Y^{\dagger}_l \right) \kappa +
\kappa \left( Y^{}_l Y^\dagger_l \right)^{T} \right] \; ,
\label{eq:2.13}
\end{eqnarray}
in which $t \equiv \ln \left(\mu/\Lambda_{\rm EW}\right)$ with $\mu$
being an arbitrary renormalization scale between the Fermi scale
$\Lambda^{}_{\rm EW} \sim 10^2$ GeV and the seesaw scale
$\Lambda^{}_{\rm SS}$, and $\alpha^{}_\kappa \simeq \lambda -3g^2_2 + 6y^2_t$
with $\lambda$, $g^{}_2$ and $y^{}_t$ standing respectively for the Higgs
self-coupling constant, the ${\rm SU(2)^{}_{L}}$ gauge coupling and the
top-quark Yukawa coupling eigenvalue in the SM. Note that the flavor structure
of the RGE on the right-hand side of Eq.~(\ref{eq:2.13}) does not change the
rank of $\kappa$, and thus the rank of the effective Majorana neutrino mass matrix
$M^{}_\nu = - v^2 \kappa$ is insensitive to the one-loop quantum correction. To
be more explicit, the derivative of $m^{}_i$ against $t$ is proportional to
$m^{}_i$ itself (for $i=1,2,3$) at the one-loop level, implying that
$m^{}_1 =0$ (or $m^{}_3 =0$) at $\Lambda^{}_{\rm SS}$ will simply lead us to
$m^{}_1 =0$ (or $m^{}_3 =0$) at $\Lambda^{}_{\rm EW}$ \cite{Mei:2003gn}.

\begin{figure*}[t]
\centering
\includegraphics[width=2in]{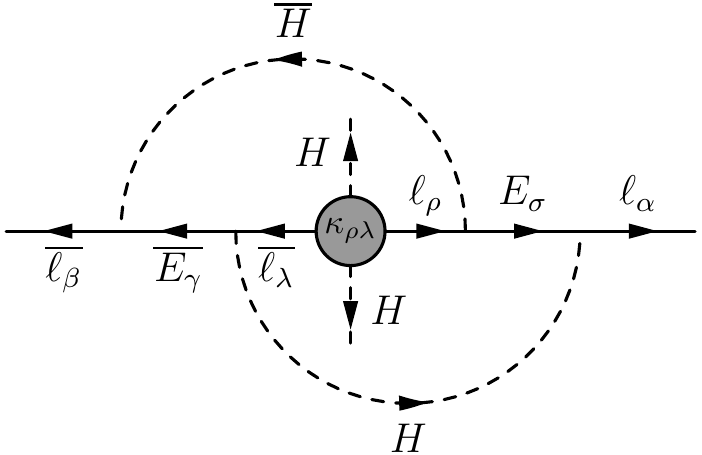}
\caption{The dominant two-loop Feynman diagram that can increase the rank of
$\kappa$ from two to three in the SM extended with the minimal seesaw mechanism.}
\label{Fig:3}
\end{figure*}

This situation will change after the two-loop RGE of $\kappa$ is taken into
consideration:
\begin{eqnarray}
16\pi^2 \frac{{\rm d}\kappa}{{\rm d}t} & = & \alpha^{}_\kappa \kappa -
\frac{3}{2} \left[ \left( Y^{}_l Y^{\dagger}_l \right) \kappa +
\kappa \left( Y^{}_l Y^\dagger_l \right)^{T} \right]
\nonumber \\
& & + \frac{1}{8\pi^2} \left( Y^{}_l Y^{\dagger}_l \right) \kappa
\left( Y^{}_l Y^{\dagger}_l \right)^T + \cdots \; ,
\label{eq:2.14}
\end{eqnarray}
where the last term on the right-hand side is the dominant effect that can
increase the rank of $\kappa$, and its corresponding Feynman diagram is shown
in Fig.~\ref{Fig:3}; and the dots denote other possible contributions at the
two-loop order and higher, but they do not give rise to any qualitatively
new effects \cite{Davidson:2006tg,Xing:2020ezi}. Working in the flavor basis with
$Y^{}_l$ being diagonal (i.e., $Y^{}_l = {\rm Diag}\{y^{}_e, y^{}_\mu, y^{}_\tau\}
\simeq \{0, 0, y^{}_\tau\}$ as a good approximation), we integrate Eq.~(\ref{eq:2.14})
from $\Lambda^{}_{\rm SS}$ to $\Lambda^{}_{\rm EW}$ and obtain
\begin{eqnarray}
\kappa \left( \Lambda^{}_{\rm EW} \right) \simeq I^{}_0
\pmatrix{ \kappa^{}_{ee} \left( \Lambda^{}_{\rm SS} \right) &
\kappa^{}_{e\mu} \left( \Lambda^{}_{\rm SS} \right) &
I^{}_\tau \kappa^{}_{e\tau} \left( \Lambda^{}_{\rm SS} \right) \cr
\kappa^{}_{e\mu} \left( \Lambda^{}_{\rm SS} \right) &
\kappa^{}_{\mu\mu} \left( \Lambda^{}_{\rm SS} \right) &
I^{}_\tau \kappa^{}_{\mu\tau} \left( \Lambda^{}_{\rm SS} \right) \cr
I^{}_\tau \kappa^{}_{e\tau} \left( \Lambda^{}_{\rm SS} \right) &
I^{}_\tau\kappa^{}_{\mu\tau} \left( \Lambda^{}_{\rm SS} \right) &
I^2_\tau I^\prime_\tau \kappa^{}_{\tau\tau} \left( \Lambda^{}_{\rm SS}
\right) \cr} \; ,
\label{eq:2.15}
\end{eqnarray}
where
\begin{eqnarray}
I^{}_0 & = & \exp{\left[ -\frac{1}{16\pi^2} \int^{\ln{(\Lambda^{}_{\rm SS}
/\Lambda^{}_{\rm EW})}}_{0} \alpha^{}_\kappa (t) \hspace{0.05cm} {\rm d}t \right]} \; ,
\nonumber \\
I^{}_\tau & = & \exp{\left[ +\frac{3}{32\pi^2} \int^{\ln{
(\Lambda^{}_{\rm SS}/\Lambda^{}_{\rm EW})}}_{0} y^{2}_\tau (t)
\hspace{0.05cm} {\rm d}t \right]} \; ,
\nonumber \\
I^{\prime}_\tau & = & \exp{\left[ -\frac{1}{128\pi^4} \int^{\ln{
(\Lambda^{}_{\rm SS}/\Lambda^{}_{\rm EW})}}_{0} y^{4}_\tau (t)
\hspace{0.05cm} {\rm d}t \right]} \; .
\label{eq:2.16}
\end{eqnarray}
It is obvious that the departures of $I^{}_\tau$ and $I^\prime_\tau$ from
unity measure the one- and two-loop RGE-induced corrections to the texture
of $\kappa$, respectively. Taking $\Lambda^{}_{\rm SS}
\simeq 10^{14}$ GeV for example, one has $I^{}_0 \simeq 0.76$,
$I^{}_\tau -1 \simeq 2.80 \times 10^{-5}$ and $1 - I^\prime_\tau \simeq
2.52 \times 10^{-11}$ at $\Lambda^{}_{\rm EW}$ in the SM \cite{Xing:2020ezi}.
So the two-loop effect is really tiny, but it is crucial to make the initially
vanishing singular value of $\kappa$ at $\Lambda^{}_{\rm SS}$ become nonzero
at $\Lambda^{}_{\rm EW}$.

After the above two-loop correction to $\kappa$ is taken into account, a detailed
numerical calculation shows that the smallest neutrino mass and its associated Majorana CP
phase at $\Lambda^{}_{\rm EW}$ are given by \cite{Xing:2020ezi}
\begin{eqnarray}
m^{}_1 (\Lambda^{}_{\rm EW}) \simeq \left\{\begin{array}{l}
1.38 \times 10^{-13} ~{\rm eV} \hspace{0.5cm} {\rm for} ~ \sigma (\Lambda^{}_{\rm EW})
= 0 \; , \\
1.25 \times 10^{-13} ~{\rm eV} \hspace{0.5cm} {\rm for} ~ \sigma (\Lambda^{}_{\rm EW})
= \pi/4 \; , \\
1.06 \times 10^{-13} ~{\rm eV} \hspace{0.5cm} {\rm for} ~ \sigma (\Lambda^{}_{\rm EW})
= \pi/2 \; ,
\end{array} \right. \nonumber \\
\rho (\Lambda^{}_{\rm EW}) \simeq \left\{\begin{array}{l}
81.5^\circ \hspace{0.5cm} {\rm for} ~ \sigma (\Lambda^{}_{\rm EW})
= 0 \; , \\
85.7^\circ \hspace{0.5cm} {\rm for} ~ \sigma (\Lambda^{}_{\rm EW})
= \pi/4 \; , \\
82.6^\circ \hspace{0.5cm} {\rm for} ~ \sigma (\Lambda^{}_{\rm EW})
= \pi/2 \; ,
\end{array} \right.
\label{eq:2.17}
\end{eqnarray}
in the NO case; or
\begin{eqnarray}
m^{}_3 (\Lambda^{}_{\rm EW}) \simeq \left\{\begin{array}{l}
2.97 \times 10^{-13} ~{\rm eV} \hspace{0.5cm} {\rm for} ~ \sigma (\Lambda^{}_{\rm EW})
= 0 \; , \\
2.78 \times 10^{-13} ~{\rm eV} \hspace{0.5cm} {\rm for} ~ \sigma (\Lambda^{}_{\rm EW})
= \pi/4 \; , \\
1.48 \times 10^{-13} ~{\rm eV} \hspace{0.5cm} {\rm for} ~ \sigma (\Lambda^{}_{\rm EW})
= \pi/2 \; ,
\end{array} \right. \nonumber \\
\varrho (\Lambda^{}_{\rm EW}) \simeq \left\{\begin{array}{l}
89.8^\circ \hspace{0.5cm} {\rm for} ~ \sigma (\Lambda^{}_{\rm EW})
= 0 \; , \\
301.9^\circ \hspace{0.5cm} {\rm for} ~ \sigma (\Lambda^{}_{\rm EW})
= \pi/4 \; , \\
344.9^\circ \hspace{0.5cm} {\rm for} ~ \sigma (\Lambda^{}_{\rm EW})
= \pi/2 \; ,
\end{array} \right.
\label{eq:2.18}
\end{eqnarray}
in the IO case, where $\rho$ (or $\varrho$) is defined to be directly associated
with $m^{}_1$ (or $m^{}_3$) in the standard parametrization of $U$.

It is worth mentioning that two-loop RGE-induced corrections to the effective
Majorana neutrino coupling matrix $\kappa$ have also been calculated in the minimal
supersymmetric standard model (MSSM) \cite{Antusch:2005gp}, but they do not change the
rank of $\kappa$ unless supersymmetry is broken at an energy scale just above the
electroweak scale \cite{Davidson:2006tg}. Given $m^{}_1 = 0$ or $m^{}_3 =0$ at
$\Lambda^{}_{\rm SS}$, the finite value of $m^{}_1$ or $m^{}_3$ at $\Lambda^{}_{\rm EW}$
originating from quantum corrections is vanishingly small in any case. That is why
it is absolutely safe to explore various phenomenological consequences of the minimal
seesaw mechanism by simply taking $m^{}_1 = 0$ or $m^{}_3 =0$ at any energy scales.

\subsection{The minimal leptogenesis in a nutshell}

As is known, the CPT theorem (a profound implication of the relativistic quantum field theory) dictates that every kind of fundamental fermion has a corresponding
antiparticle with the same mass and lifetime but the opposite charge. Given such a
particle-antiparticle symmetry, it is expected that there should be equal amounts of
baryons and antibaryons in the Universe, at least in the very beginning or soon after the
Big Bang. But something must have happened during the evolution of the early Universe,
because today's Universe is found to be predominantly composed of baryons instead of
antibaryons. In other words, the primordial antibaryons have mysteriously disappeared.
This is just the ``missing antimatter" puzzle.

To be more explicit, a careful analysis of recent Planck measurements of the cosmic
microwave background (CMB) anisotropies leads us to the baryon number density
$\Omega^{}_{\rm b} h^2 = 0.0224 \pm 0.0001$ at the $68\%$ confidence level
\cite{Aghanim:2018eyx}, which can be translated into the baryon-to-photon ratio
\begin{eqnarray}
\eta \equiv \frac{n^{}_{\rm B}}{n^{}_\gamma} \simeq 273 \times 10^{-10} ~
\Omega^{}_{\rm b} h^2 \simeq \left(6.12 \pm 0.03\right)
\times 10^{-10} \; ,
\label{eq:2.19}
\end{eqnarray}
a result consistent very well with the result
$5.8 \times 10^{-10} < \eta < 6.6 \times 10^{-10}$ that has been extracted
from current observational data on the primordial abundances of light element
isotopes based on the Big Bang nucleosynthesis (BBN) theory
\cite{Tanabashi:2018oca}. Note that the BBN and CMB formation took place at
two very different time points of the Universe: $t^{}_{\rm BBN} \gtrsim 1 ~ {\rm s}$
but $t^{}_{\rm CMB} \sim 3.8 \times 10^5 ~ {\rm yr}$. So a good agreement
between the values of $\eta$ determined from the above two events signifies a
great success of the Big Bang cosmology.

A viable {\it baryogenesis} mechanism, which is able to successfully account for the
observed value of $\eta$ as shown above, dictates the Universe to satisfy the three
``Sakharov conditions" \cite{Sakharov:1967dj}: (a) baryon number violation; (b) C
and CP violation; and (c) departure from thermal equilibrium. Fortunately,
even the SM itself can accommodate C and CP violation and allow for baryon
number violation at the non-perturbative regime \cite{tHooft:1976rip,tHooft:1976snw}.
As compared with a few other popular baryogenesis scenarios
\cite{Kuzmin:1985mm,Cohen:1993nk,Trodden:1998ym,Affleck:1984fy,Riotto:1999yt,Dine:2003ax},
the thermal {\it leptogenesis} mechanism \cite{Fukugita:1986hr} is of particular interest
because it is closely related to lepton number violation and neutrino mass generation
(see, e.g., Refs.~\cite{Xing:2011zza,Buchmuller:2004nz,Buchmuller:2005eh,Davidson:2008bu}
for comprehensive reviews). This elegant mechanism especially works well when it is
combined with the canonical seesaw mechanism. Here we concentrate on the
{\it minimal} leptogenesis scenario, a simplified version of thermal leptogenesis based
on the {\it minimal} seesaw scenario that has been described in section~\ref{section:2.1}
\cite{Frampton:2002qc}. The three key points of the minimal thermal leptogenesis mechanism
can be summarized as follows.
\begin{figure*}[t]
\centering
\includegraphics[width=6in]{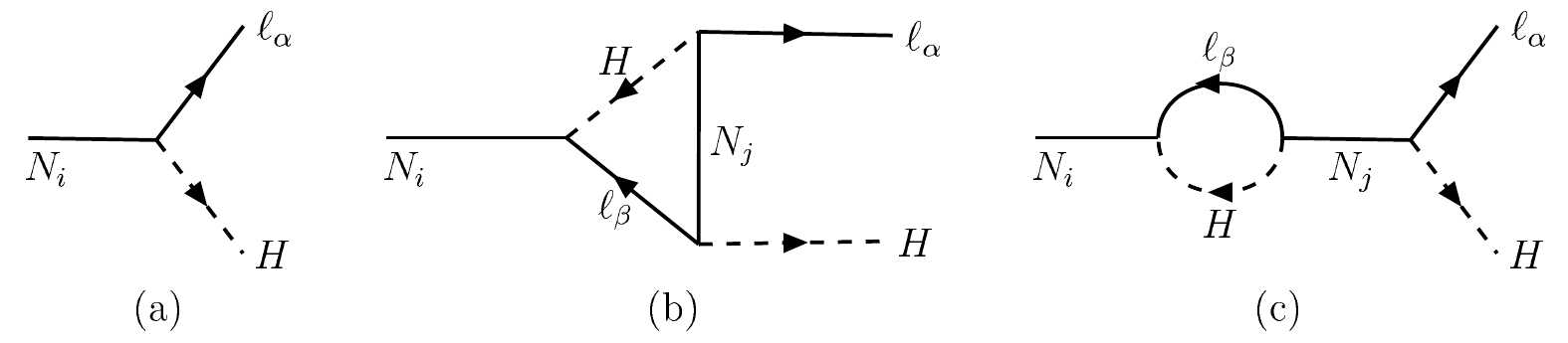}
\caption{Feynman diagrams for the decay modes $N^{}_i \to \ell^{}_\alpha + H$ at the
tree and one-loop levels in the minimal seesaw scenario, where the Latin and Greek
subscripts run over $(1, 2)$ and $(e, \mu, \tau)$, respectively.}
\label{Fig:4}
\end{figure*}

(a) The two heavy Majorana neutrinos $N^{}_1$ and $N^{}_2$ decay into
the lepton doublet $\ell^{}_\alpha$ (for $\alpha = e, \mu, \tau$) and the Higgs
doublet $H$ via the Yukawa interactions at both the tree level and the one-loop level,
as shown in Fig.~\ref{Fig:4}. Such a decay mode is lepton-number-violating,
since $N^{}_1$ and $N^{}_2$ are their own antiparticles. It is also CP-violating,
because the interference between the tree and one-loop amplitudes may give rise to an
observable asymmetry between the decay rates of $N^{}_i \to \ell^{}_\alpha + H$ and
its CP-conjugate process $N^{}_i \to \overline{\ell^{}_\alpha} + \overline{H}$,
defined as
\begin{eqnarray}
\varepsilon^{}_{i \alpha} \equiv \frac{\Gamma(N^{}_i \to
\ell^{}_\alpha + H) - \Gamma(N^{}_i \to \overline{\ell^{}_\alpha} +
\overline{H})}{\displaystyle \sum_\alpha \left[\Gamma(N^{}_i \to
\ell^{}_\alpha + H) + \Gamma(N^{}_i \to \overline{\ell^{}_\alpha} +
\overline{H})\right]} \; .
\label{eq:2.20}
\end{eqnarray}
The explicit expressions of $\varepsilon^{}_{1 \alpha}$ and $\varepsilon^{}_{2 \alpha}$
in the minimal seesaw case can be read off from the more general result obtained in
the type-I seesaw mechanism \cite{Luty:1992un,Covi:1996wh,Plumacher:1996kc}. Namely,
\begin{eqnarray}
\varepsilon^{}_{1 \alpha} & = & \frac{+1}{8\pi (Y^\dagger_\nu
Y^{}_\nu)^{}_{11}} \left\{ {\rm Im}\left[(Y^*_\nu)^{}_{\alpha 1} (Y^{}_\nu)^{}_{\alpha 2}
(Y^\dagger_\nu Y^{}_\nu)^{}_{12}\right] {\cal F}(x) \right.
\nonumber \\
& & + \left. {\rm Im}\left[(Y^*_\nu)^{}_{\alpha 1} (Y^{}_\nu)^{}_{\alpha 2} (Y^\dagger_\nu Y^{}_\nu)^*_{12}\right] {\cal G}(x) \right\} \; ,
\nonumber \\
\varepsilon^{}_{2 \alpha} & = & \frac{-1}{8\pi (Y^\dagger_\nu
Y^{}_\nu)^{}_{22}} \left\{ {\rm Im}\left[(Y^*_\nu)^{}_{\alpha 1} (Y^{}_\nu)^{}_{\alpha 2}
(Y^\dagger_\nu Y^{}_\nu)^{}_{12}\right] {\cal F}(z) \right.
\nonumber \\
& & + \left. {\rm Im}\left[(Y^*_\nu)^{}_{\alpha 1} (Y^{}_\nu)^{}_{\alpha 2} (Y^\dagger_\nu Y^{}_\nu)^*_{12}\right] {\cal G}(z) \right\} \; ,
\label{eq:2.21}
\end{eqnarray}
where $x \equiv M^2_2/M^2_1 \equiv 1/z$, ${\cal F}(x) = \sqrt{x} \left\{(2-x)/(1-x)
+ (1+x) \ln [x/(1+x)] \right\}$ and ${\cal G}(x) = 1/(1-x)$. A sum of the three
{\it flavored} CP-violating asymmetries leads us to the {\it unflavored} (or
flavor-independent) CP-violating asymmetry
\begin{eqnarray}
\varepsilon^{}_{1} = \sum_\alpha \varepsilon^{}_{1 \alpha} & = &
\frac{+1}{8\pi (Y^\dagger_\nu Y^{}_\nu)^{}_{11}}
{\rm Im}\left[(Y^\dagger_\nu Y^{}_\nu)^{2}_{12}\right] {\cal F}(x) \; ,
\nonumber \\
\varepsilon^{}_{2} = \sum_\alpha \varepsilon^{}_{2 \alpha} & = &
\frac{-1}{8\pi (Y^\dagger_\nu Y^{}_\nu)^{}_{22}}
{\rm Im}\left[(Y^\dagger_\nu Y^{}_\nu)^{2}_{12}\right] {\cal F}(z) \; ,
\label{eq:2.22}
\end{eqnarray}
in which the loop functions ${\cal F}(x)$ and ${\cal F}(z)$ have been defined above.
One can easily see that $\varepsilon^{}_{1 \alpha}$ and $\varepsilon^{}_{2 \alpha}$ (or
$\varepsilon^{}_{1}$ and $\varepsilon^{}_{2}$) depend on the same phase combination
of $Y^{}_\nu$, but their magnitudes are different in general.

Note that Eqs.~(\ref{eq:2.21}) and (\ref{eq:2.22}) will become invalid if the
values of $M^{}_1$ and $M^{}_2$ are very close. In the latter case the one-loop
self-energy corrections in Fig.~\ref{Fig:4}(c) may significantly enhance
the relevant CP-violating asymmetries
\cite{Pilaftsis:1997dr,Pilaftsis:1997jf,Pilaftsis:2003gt,Anisimov:2005hr}, leading
us to the following special results in the minimal seesaw mechanism:
\begin{eqnarray}
\varepsilon^{\prime}_1 = -\frac{{\rm Im}\left[(Y^\dagger_\nu
Y^{}_\nu)^2_{12}\right]}{(Y^\dagger_\nu Y^{}_\nu)^{}_{11}
(Y^\dagger_\nu Y^{}_\nu)^{}_{22}} \cdot \frac{\Delta M^2_{21} M^{}_1
\Gamma^{}_2}{(\Delta M^2_{21})^2 + M^2_1 \Gamma^2_2} \; ,
\nonumber \\
\varepsilon^{\prime}_2 = -\frac{{\rm Im}\left[(Y^\dagger_\nu
Y^{}_\nu)^2_{12}\right]}{(Y^\dagger_\nu Y^{}_\nu)^{}_{11}
(Y^\dagger_\nu Y^{}_\nu)^{}_{22}} \cdot \frac{\Delta M^2_{21} M^{}_2
\Gamma^{}_1}{(\Delta M^2_{21})^2 + M^2_2 \Gamma^2_1} \; ,
\label{eq:2.23}
\end{eqnarray}
where $\Delta M^2_{21} \equiv M^2_2 - M^2_1$ is defined, $\Gamma^{}_{1,2}$
denote the decay widths of $N^{}_{1,2}$, and the conditions $M^{}_1 \sim M^{}_2$
and $|M^{}_1 - M^{}_2| \sim \Gamma^{}_1 \sim \Gamma^{}_2$ are satisfied.
As a consequence, the approximate equality
$\varepsilon^{\prime}_1 \sim \varepsilon^{\prime}_2$ is expected to hold.

(b) The primordial CP-violating asymmetries $\varepsilon^{}_{i \alpha}$ between
$N^{}_i \to \ell^{}_\alpha + H$ and
$N^{}_i \to \overline{\ell^{}_\alpha} + \overline{H}$ decay modes (for $i = 1, 2$ and
$\alpha = e, \mu, \tau$) make it possible to
generate a net lepton-antilepton number asymmetry (i.e.,
$n^{}_{\rm L} \neq n^{}_{\overline{\rm L}}$) in the early Universe. To prevent
such an asymmetry from being washed out by the inverse decays of $N^{}_{1,2}$
and various $\Delta L =1$ and $\Delta L =2$ scattering processes,
the decays of $N^{}_{1,2}$ (or one of them) must be out of thermal equilibrium.
That is to say, the rates of $N^{}_i \to \ell^{}_\alpha + H$ decays must
be lower than the Hubble expansion rate of the Universe at temperature
$T \simeq M^{}_i$. In this case one may define the lepton-antilepton asymmetry
$Y^{}_{\rm L} \equiv (n^{}_{\rm L} - n^{}_{\overline{\rm L}})/s$ with
$s$ being the entropy density of the Universe, and then link it linearly to the
CP-violating asymmetries $\varepsilon^{}_i$ (or $\varepsilon^{}_{i \alpha}$).
Note that an exact description of $Y^{}_{\rm L}$ resorts to solving a full set
of Boltzmann equations for the time evolution of relevant particle number densities
\cite{Kolb:1990vq,Barbieri:1999ma,Blanchet:2006be}, but sometimes it is good
enough to make reasonable analytical approximations for the relationship between
$Y^{}_{\rm L}$ and $\varepsilon^{}_i$ (or $\varepsilon^{}_{i \alpha}$)
\cite{Buchmuller:2004nz,Davidson:2008bu,Luty:1992un,Plumacher:1996kc}.
If $M^{}_1$ is much smaller than $M^{}_2$, however, the lepton-number-violating
interactions of $N^{}_1$ can be rapid enough to wash out the lepton-antilepton
asymmetry stemming from $\varepsilon^{}_2$, and hence only $\varepsilon^{}_1$
can contribute to thermal leptogenesis.

At this point it is worth clarifying the meaning of ``unflavored" and ``flavored"
leptogenesis scenarios, which are associated respectively with
$\varepsilon^{}_i$ and $\varepsilon^{}_{i\alpha}$. In the so-called
``unflavored" case, the equilibrium temperature $T$ of $N^{}_i$ is so high
($\gtrsim 10^{12}$ GeV) that the Yukawa interactions of charged leptons
are unable to distinguish one lepton flavor from another. That is to say,
all the relevant Yukawa interactions are blind to lepton flavors. If the
equilibrium temperature $T$ lies in the range $10^9 ~{\rm GeV} \lesssim T
\lesssim 10^{12} ~{\rm GeV}$, then the Yukawa interactions of $\tau$-leptons become
faster than the (inverse) decays of $N^{}_i$ (or equivalently comparable to the
expansion rate of the Universe), and hence
the $\tau$-flavor effects must be taken into account, leading us to
the {\it $\tau$-flavored} leptogenesis which is dependent on the CP-violating
asymmetries $\varepsilon^{}_{1\tau}$ and $\varepsilon^{}_{2\tau}$
\cite{Davidson:2008bu,Barbieri:1999ma,Blanchet:2006be,
Endoh:2003mz,Pilaftsis:2004xx,Abada:2006ea,Nardi:2006fx}. When
$10^5 ~{\rm GeV} \lesssim T \lesssim 10^{9} ~{\rm GeV}$ holds, both
$\mu$ and $\tau$ flavors take effect in thermal leptogenesis; and given
$T \lesssim 10^5$ GeV, all the three flavor-dependent CP-violating asymmetries
$\varepsilon^{}_{i e}$, $\varepsilon^{}_{i \mu}$ and $\varepsilon^{}_{i \tau}$
contribute to leptogenesis.

(c) To convert a net lepton-antilepton asymmetry $Y^{}_{\rm L}$
associated closely with the CP-violating asymmetries $\varepsilon^{}_i$ (or
$\varepsilon^{}_{i \alpha}$) to a net baryon-antibaryon asymmetry
$Y^{}_{\rm B} \equiv (n^{}_{\rm B} - n^{}_{\overline{\rm B}})/s$,
the leptogenesis mechanism should take effect in the temperature range
$10^2 ~{\rm GeV} \lesssim T \lesssim 10^{12} ~{\rm GeV}$
in which the non-perturbative $(B-L)$-conserving sphaleron interactions may
stay in thermal equilibrium and can therefore be very efficient
\cite{Kuzmin:1985mm,Klinkhamer:1984di}. To be explicit, such a conversion
can be expressed as \cite{Kolb:1990vq,Harvey:1990qw}
\begin{eqnarray}
\left. Y^{}_{\rm B}\right|^{}_{\rm equilibrium} =
- c \left. Y^{}_{\rm L}\right|^{}_{\rm initial} \; , \quad
c = \left\{ \begin{array}{ll}
28/79 & ({\rm SM}) \; , \\
8/23 & ({\rm MSSM}) \; .
\end{array} \right.
\label{eq:2.24}
\end{eqnarray}
It is obvious that $Y^{}_{\rm L}$ must be negative to yield a positive $Y^{}_{\rm B}$,
so as to account for the observed value of $\eta$ given in Eq.~(\ref{eq:2.19})
with the help of the relation $\eta = s Y^{}_{\rm B} /n^{}_\gamma \simeq 7.04 Y^{}_{\rm B}$ \cite{Xing:2011zza}. As mentioned above, the evolution of $Y^{}_{\rm L}$ with
temperature $T$ can be computed by solving the relevant Boltzmann equations
\cite{Buchmuller:2004nz,Davidson:2008bu,Luty:1992un,Plumacher:1996kc,
Kolb:1990vq,Barbieri:1999ma,Blanchet:2006be}.

\subsection{Some other minimal seesaw scenarios}
\label{section 2.3}

Motivated by the principle of Occam's razor, one may also consider the simplified
versions of some other seesaw mechanisms. For the sake of illustration, here we
take three examples of this kind: the minimal type-III seesaw, the minimal
type-(I+II) seesaw and the minimal inverse seesaw \cite{Xing:2019vks}.

{\it Example (1): the minimal type-III seesaw scenario}. Similar to the minimal
type-I seesaw discussed above, the minimal type-III seesaw involves only two
$\rm SU(2)^{}_{\rm L}$ fermion triplets \cite{Goswami:2018jar}. As one can see from Eq.~(\ref{eq:1.15}), the rank of $M^{}_\Sigma$ is equal to two in this case. The
rank of $M^{}_\nu$ is therefore equal to two, as determined by the seesaw formula
given in Eq.~(\ref{eq:1.18}). One of the three active neutrinos turns out to be
massless at the tree and one-loop levels, allowing one of the two Majorana phases
to be removed from the PMNS matrix $U$. Some phenomenological consequences of
such a minimal type-III seesaw scenario on various lepton-flavor-violating
and lepton-number-violating processes have been explored in detail in
Refs.~\cite{Goswami:2018jar} and \cite{Biggio:2019eeo}.

{\it Example (2): the minimal type-(I+II) seesaw scenario}. This particular seesaw
scenario is a simplified version of the type-(I+II) seesaw mechanism --- a
straightforward combination of the type-I and type-II seesaws as described by
Eqs.~(\ref{eq:1.11}) and (\ref{eq:1.13}), respectively. Such a ``hybrid" seesaw
mechanism can naturally be embedded into a left-right symmetric model or an SO(10)
GUT \cite{Akhmedov:2006de,Borah:2016iqd,Ohlsson:2019sja}, but it consists of
many more free parameters than the type-I or type-II seesaw itself. A simple way to
reduce the number of free parameters is to introduce only a single heavy Majorana
neutrino state $N$ with mass $M^{}_N$ besides the $\rm SU(2)^{}_{\rm L}$ Higgs triplet
state with a high mass scale $M^{}_\Delta$ \cite{Gu:2006wj,Chan:2007ng,Chao:2008mq,
Ren:2008yi}, and the resulting seesaw is just the minimal type-(I+II) seesaw with
\begin{eqnarray}
M^{}_\nu \simeq 2 v^2 \frac{\lambda^{}_\Delta Y^{}_\Delta}{M^{}_\Delta} -
\frac{M^{}_{\rm D} M^T_{\rm D}}{M^{}_N} \;
\label{eq:2.25}
\end{eqnarray}
in the leading-order approximation, where the relevant notations are self-explanatory
as one can understand from Eq.~(\ref{eq:1.18}). It is obvious that $M^{}_{\rm D}$ is
a $3\times 1$ mass matrix, and thus the second term on the right-hand side of
Eqs.~(\ref{eq:2.25}) is actually a rank-one $3\times 3$ mass matrix. In this case
one may properly adjust the flavor structures of $Y^{}_\Delta$ and $M^{}_{\rm D}$
such that the effective Majorana neutrino mass matrix $M^{}_\nu$ can fit current
neutrino oscillation data. Such a special seesaw scenario is also able to account
for the observed baryon-antibaryon asymmetry of the Universe via the corresponding
minimal type-(I+II) leptogenesis (see, e.g., Refs.~\cite{Gu:2006wj,Chan:2007ng}).

{\it Example (3): The minimal inverse seesaw scenario}. This seesaw scenario is
a simplified version of the inverse (or double) seesaw mechanism, which in general
contains three heavy Majorana neutrino states, three SM gauge-singlet neutrino
states and one scalar singlet state $\Phi$ besides the SM particle content \cite{Wyler:1982dd,Mohapatra:1986bd}. The
most remarkable advantage of the inverse seesaw picture is that it can naturally
lower the seesaw scale to the TeV scale, which is experimentally accessible at
the LHC \cite{Xing:2009in}. But its disadvantage is also obvious: too many new
particles and too many free parameters. That is why we are motivated to simplify the
conventional inverse seesaw by allowing for only two heavy Majorana neutrino
states and only two SM gauge-singlet neutrino states. In such a minimal inverse
seesaw scenario the effective light Majorana neutrino mass matrix is a rank-two
matrix of the approximate form \cite{Malinsky:2009df}
\begin{eqnarray}
M^{}_\nu \simeq M^{}_{\rm D} \hspace{0.05cm} (M^T_S)^{-1} \mu \hspace{0.05cm}
(M^{}_S)^{-1} M^T_{\rm D} \; ,
\label{eq:2.26}
\end{eqnarray}
where $\mu$ is a $2\times 2$ mass matrix which is purely related to the SM
gauge-singlet neutrinos, $M^{}_{\rm D} = Y^{}_\nu v$ is a $3\times 2$ mass
matrix with $Y^{}_\nu$ being the standard Yukawa coupling matrix,
and $M^{}_S = Y^{}_S \langle \Phi\rangle$ is a $2\times 2$ mass matrix
with $Y^{}_S$ being a Yukawa-like coupling matrix associated with
the SM gauge-singlet neutrinos. Note that the mass scale of $\mu$ is expected
to be naturally small (e.g., at the keV level), as dictated by 't Hooft's
naturalness criterion \cite{tHooft:1979rat}. So the tiny mass scale of
$M^{}_\nu$ is attributed to both the small mass scale of $\mu$ and a
strongly suppressed ratio of the mass scale of $M^{}_{\rm D}$ to that of
$M^{}_S$, which actually satisfies the spirit of the multiple seesaw
mechanism \cite{Xing:2009hx,Bonnet:2009ej,Liao:2010cc}. It is clear that
the phenomenological consequences of $M^{}_\nu$ in Eqs.~(\ref{eq:2.26})
are the same as those given in the minimal type-I seesaw scenario. Possible
collider signatures and some other low-energy consequences of this minimal
inverse seesaw scenario, such as lepton-flavor-violating processes,
lepton-number-violating processes and dark-matter candidates, have been
discussed in depth in the literature (see, e.g., Refs.~\cite{Malinsky:2009df,
Hirsch:2009ra,Mondal:2012jv,Abada:2014vea,Abada:2014zra,CarcamoHernandez:2019eme}).

Of course, the predictability and testability of a given minimal seesaw
scenario can be further enhanced if its flavor structures are further determined or
constrained by some proper flavor symmetries. We are going to discuss this
interesting aspect of the minimal type-I seesaw in sections \ref{section 5} and \ref{section 6}.

\setcounter{equation}{0}
\section{Parametrizations of the minimal seesaw texture}
\label{section 3}

As already pointed out in section \ref{section 2}, the neutrino sector of the minimal seesaw texture generally contains eleven physically-relevant parameters, but there are only seven low-energy  observables. A burning question then arises as how to parameterize this texture (i.e., the Dirac neutrino mass matrix $M^{}_{\rm D}$ in the mass basis of two right-handed neutrinos) and in which parametrization the model parameters can be connected to the low-energy flavor parameters in a transparent way and the treatments of related physical processes (e.g., leptogenesis) can be simplified to some extent. In this section we introduce five theoretically well-motivated and practically useful parametrizations of $M^{}_{\rm D}$, which may find respective applications in some specific minimal seesaw scenarios.

\subsection{An exact Euler-like parametrization}
\label{section 3.1}

Since the overall neutrino mass matrix in the minimal seesaw mechanism is a symmetric
$5\times 5$ matrix, it can be diagonalized by the following transformation:
\begin{eqnarray}
{\cal U}^\dagger
\pmatrix{ 0 & M^{}_{\rm D} \cr M^{T}_{\rm D} & M^{}_{\rm R} \cr}
{\cal U}^* =
\pmatrix{ D^{}_\nu & 0 \cr 0 & D^{}_N \cr} \; ,
\label{eq:3.1.1}
\end{eqnarray}
where $D^{}_\nu \equiv {\rm Diag}\{m^{}_1, m^{}_2, m^{}_3\}$ with either $m^{}_1 =0$
or $m^{}_3 =0$, $D^{}_N \equiv {\rm Diag}\{M^{}_1, M^{}_2\}$, and
the $5\times 5$ unitary matrix $\cal U$ can be decomposed into a product
of three more specific $5\times 5$ unitary matrices of the form \cite{Xing:2007zj,Xing:2011ur}
\begin{eqnarray}
{\cal U} =
\pmatrix{ I & 0 \cr 0 & U^{\prime}_0 \cr}
\pmatrix{ A & R \cr S & B \cr}
\pmatrix{ U^{}_0 & 0 \cr 0 & I \cr}
\label{eq:3.1.2}
\end{eqnarray}
with $U^{}_0$ being a $3\times 3$ unitary matrix, $U^\prime_0$ being a
$2\times 2$ unitary matrix and $I$ being the identity matrix of either
rank three or rank two. The advantage of this decomposition is that
$U^{}_0$ and $U^{\prime}_0$ are responsible respectively for flavor mixing in the
light and heavy sectors, while $A$, $B$, $R$ and $S$ describe the interplay
between these two sectors. The unitarity of $\cal U$ allows us to obtain
\begin{eqnarray}
A A^\dagger + R R^\dagger = B B^\dagger + S S^\dagger = I \; ,
\nonumber \\
A S^\dagger + R B^\dagger = A^\dagger R + S^\dagger B = 0 \; ,
\nonumber \\
A^\dagger A + S^\dagger S = B^\dagger B + R^\dagger R = I \; .
\label{eq:3.1.3}
\end{eqnarray}
Switching off $R$ and $S$ (i.e., $R = S = 0$), one will be left with
no correlation between the light and heavy neutrino sectors. To
explicitly parameterize $\cal U$ in terms of the rotation and phase
angles, one may introduce ten two-dimensional complex rotation
matrices $O^{}_{ij}$ (for $1\leq i < j \leq 5$) in a five-dimensional
flavor space,
\begin{eqnarray}
O^{}_{12} = \pmatrix{ c^{}_{12} & \hat{s}^*_{12} & 0 & 0 &
0 \cr -\hat{s}^{}_{12} & c^{}_{12} & 0 & 0 & 0 \cr 0 & 0 & 1
& 0 & 0 \cr 0 & 0 & 0 & 1 & 0 \cr 0 & 0 & 0 & 0 & 1 \cr} \; ,
\quad
O^{}_{13} = \pmatrix{ c^{}_{13} & 0 & \hat{s}^*_{13}
& 0 & 0 \cr 0 & 1 & 0 & 0 & 0 \cr -\hat{s}^{}_{13} & 0 &
c^{}_{13} & 0 & 0 \cr 0 & 0 & 0 & 1 & 0 \cr 0 & 0 & 0 & 0
& 1 \cr} \; ,
\nonumber \\
O^{}_{23} = \pmatrix{ 1 & 0 & 0 & 0 & 0 \cr 0 &
c^{}_{23} & \hat{s}^*_{23} & 0 & 0 \cr 0 & -\hat{s}^{}_{23} &
c^{}_{23} & 0 & 0 \cr 0 & 0 & 0 & 1 & 0 \cr 0 & 0 & 0 & 0 & 1 \cr} \; ,
\quad
O^{}_{14} = \pmatrix{ c^{}_{14} & 0 & 0 & \hat{s}^*_{14} &
0 \cr 0 & 1 & 0 & 0 & 0 \cr 0 & 0 & 1 & 0 & 0 \cr
-\hat{s}^{}_{14} & 0 & 0 & c^{}_{14} & 0 \cr 0 & 0 & 0 & 0 & 1 \cr} \; ,
\nonumber \\
O^{}_{24} = \pmatrix{ 1 & 0 & 0 & 0 & 0 \cr 0 &
c^{}_{24} & 0 & \hat{s}^*_{24} & 0 \cr 0 & 0 & 1 & 0 & 0 \cr
0 & -\hat{s}^{}_{24} & 0 & c^{}_{24} & 0 \cr 0 & 0 & 0 & 0 & 1 \cr} \; ,
\quad
O^{}_{34} = \pmatrix{ 1 & 0 & 0 & 0 & 0 \cr 0 & 1 & 0 &
0 & 0 \cr 0 & 0 & c^{}_{34} & \hat{s}^*_{34} & 0 \cr 0 & 0 &
-\hat{s}^{}_{34} & c^{}_{34} & 0 \cr 0 & 0 & 0 & 0 & 1 \cr} \; ,
\nonumber \\
O^{}_{15} = \pmatrix{ c^{}_{15} & 0 & 0 & 0 &
\hat{s}^*_{15} \cr 0 & 1 & 0 & 0 & 0 \cr 0 & 0 & 1 & 0 & 0 \cr
0 & 0 & 0 & 1 & 0 \cr -\hat{s}^{}_{15} & 0 & 0 & 0 & c^{}_{15} \cr} \; ,
\quad
O^{}_{25} = \pmatrix{ 1 & 0 & 0 & 0 & 0 \cr 0 &
c^{}_{25} & 0 & 0 & \hat{s}^*_{25} \cr 0 & 0 & 1 & 0 & 0 \cr
0 & 0 & 0 & 1 & 0 \cr 0 & -\hat{s}^{}_{25} & 0 & 0 & c^{}_{25} \cr} \; ,
\nonumber \\
O^{}_{35} = \pmatrix{ 1 & 0 & 0 & 0 & 0 \cr 0 & 1 & 0 &
0 & 0 \cr 0 & 0 & c^{}_{35} & 0 & \hat{s}^*_{35} \cr 0 & 0 &
0 & 1 & 0 \cr 0 & 0 & -\hat{s}^{}_{35} & 0 & c^{}_{35} \cr} \; ,
\quad
O^{}_{45} = \pmatrix{ 1 & 0 & 0 & 0 & 0 \cr 0 & 1 & 0 &
0 & 0 \cr 0 & 0 & 1 & 0 & 0 \cr 0 & 0 & 0 & c^{}_{45} &
\hat{s}^*_{45} \cr 0 & 0 & 0 & -\hat{s}^{}_{45} & c^{}_{45} \cr} \;
\label{eq:3.1.4}
\end{eqnarray}
with $c^{}_{ij} \equiv \cos\theta^{}_{ij}$ and
$\hat{s}^{}_{ij} \equiv e^{{\rm i} \delta^{}_{ij}} \sin\theta^{}_{ij}$
(for $1 \leq i < j \leq 5$); and then assign them as
\begin{eqnarray}
\pmatrix{ U^{}_0 & 0 \cr 0 & I \cr}
= O^{}_{23} O^{}_{13} O^{}_{12} \; ,
\nonumber \\
\pmatrix{ A & R \cr S & B \cr}
= O^{}_{35} O^{}_{25} O^{}_{15} O^{}_{34} O^{}_{24} O^{}_{14} \; ,
\nonumber \\
\pmatrix{ I & 0 \cr 0 & U^{\prime}_0 \cr} = O^{}_{45} \; .
\label{eq:3.1.5}
\end{eqnarray}
Among the ten flavor mixing angles and ten CP-violating phases,
eight of them appear in the light ($U^{}_0$) and heavy ($U^\prime_0$) sectors:
\begin{eqnarray}
U^{}_0 = \pmatrix{ c^{}_{12} c^{}_{13} & \hat{s}^*_{12}
c^{}_{13} & \hat{s}^*_{13} \cr \vspace{-0.45cm} \cr
-\hat{s}^{}_{12} c^{}_{23} -
c^{}_{12} \hat{s}^{}_{13} \hat{s}^*_{23} & c^{}_{12} c^{}_{23} -
\hat{s}^*_{12} \hat{s}^{}_{13} \hat{s}^*_{23} & c^{}_{13}
\hat{s}^*_{23} \cr \vspace{-0.45cm} \cr
\hat{s}^{}_{12} \hat{s}^{}_{23} - c^{}_{12}
\hat{s}^{}_{13} c^{}_{23} & -c^{}_{12} \hat{s}^{}_{23} -
\hat{s}^*_{12} \hat{s}^{}_{13} c^{}_{23} & c^{}_{13} c^{}_{23} \cr} \; ,
\nonumber \\
U^{\prime}_0 = \pmatrix{ c^{}_{45} & \hat{s}^*_{45} \cr \vspace{-0.45cm} \cr
-\hat{s}^{}_{45} & c^{}_{45} \cr} \; ;
\label{eq:3.1.6}
\end{eqnarray}
and all the others appear in $A$, $B$, $R$ and $S$:
\begin{eqnarray}
\hspace{-2cm} A = \pmatrix{ c^{}_{14} c^{}_{15} & 0 & 0 \cr \vspace{-0.35cm} \cr
-c^{}_{14} \hat{s}^{}_{15} \hat{s}^*_{25}
-\hat{s}^{}_{14} \hat{s}^*_{24} c^{}_{25} &
c^{}_{24} c^{}_{25} & 0 \cr \vspace{-0.35cm} \cr
- c^{}_{14} \hat{s}^{}_{15} c^{}_{25} \hat{s}^*_{35}
+ \hat{s}^{}_{14} \hat{s}^*_{24} \hat{s}^{}_{25} \hat{s}^*_{35}
- \hat{s}^{}_{14} c^{}_{24} \hat{s}^*_{34} c^{}_{35} &
-c^{}_{24} \hat{s}^{}_{25} \hat{s}^*_{35}
-\hat{s}^{}_{24} \hat{s}^*_{34} c^{}_{35} &
c^{}_{34} c^{}_{35} \cr} \; ,
\nonumber \\
\hspace{-2cm} R = \pmatrix{ \hat{s}^*_{14} c^{}_{15} &
\hat{s}^*_{15} \cr \vspace{-0.4cm} \cr
- \hat{s}^*_{14} \hat{s}^{}_{15} \hat{s}^*_{25}
+ c^{}_{14} \hat{s}^*_{24} c^{}_{25} &
c^{}_{15} \hat{s}^*_{25} \cr \vspace{-0.4cm} \cr
- \hat{s}^*_{14} \hat{s}^{}_{15}
c^{}_{25} \hat{s}^*_{35} - c^{}_{14} \hat{s}^*_{24} \hat{s}^{}_{25}
\hat{s}^*_{35} + c^{}_{14} c^{}_{24} \hat{s}^*_{34} c^{}_{35} &
c^{}_{15} c^{}_{25} \hat{s}^*_{35} \cr} \; ;
\label{eq:3.1.7}
\end{eqnarray}
and
\begin{eqnarray}
\hspace{-2cm} S = \pmatrix{ -\hat{s}^{}_{14} c^{}_{24} c^{}_{34} &
-\hat{s}^{}_{24} c^{}_{34} & -\hat{s}^{}_{34} \cr \vspace{-0.2cm} \cr
\hat{s}^{}_{14} c^{}_{24} \hat{s}^{*}_{34} \hat{s}^{}_{35}
+ \hat{s}^{}_{14} \hat{s}^{*}_{24} \hat{s}^{}_{25} c^{}_{35}
- c^{}_{14} \hat{s}^{}_{15} c^{}_{25} c^{}_{35} &
\hat{s}^{}_{24} \hat{s}^{*}_{34} \hat{s}^{}_{35} - c^{}_{24}
\hat{s}^{}_{25} c^{}_{35} & -c^{}_{34} \hat{s}^{}_{35} \cr} \; ,
\nonumber \\
\hspace{-2cm} B = \pmatrix{ c^{}_{14} c^{}_{24} c^{}_{34} & 0 \cr \vspace{-0.35cm} \cr
-c^{}_{14} c^{}_{24} \hat{s}^{*}_{34} \hat{s}^{}_{35} -
c^{}_{14} \hat{s}^{*}_{24} \hat{s}^{}_{25} c^{}_{35}
-\hat{s}^{*}_{14} \hat{s}^{}_{15} c^{}_{25} c^{}_{35} &
c^{}_{15} c^{}_{25} c^{}_{35} \cr} \; .
\label{eq:3.1.8}
\end{eqnarray}
Note that only $U \equiv A U^{}_0$ and $R$ are relevant for
the standard weak charged-current interactions of five neutrinos
with three charged leptons:
\begin{eqnarray}
-{\cal L}^{}_{\rm cc} = \frac{g}{\sqrt{2}} \ \overline{\left(e~~
\mu~~ \tau\right)^{}_{\rm L}} \ \gamma^\mu \left[ U
\pmatrix{ \nu^{}_1 \cr \nu^{}_2 \cr \nu^{}_3 \cr}^{}_{\rm L}
+ R \pmatrix{ N^{}_1 \cr N^{}_2 \cr}^{}_{\rm L} \right] W^-_\mu
+ {\rm h.c.} \; ,
\label{eq:3.1.9}
\end{eqnarray}
where $U$ is just the $3\times 3$ PMNS matrix responsible for the three
active neutrino mixing in the seesaw mechanism, and $R$
measures the strength of charged-current interactions between two
heavy Majorana neutrinos and three charged leptons. The correlation
between $U$ and $R$ is given by $U U^\dagger = A A^\dagger = I - R R^\dagger$.

Possible deviation of $U$ from exact unitarity (i.e., from the unitary matrix $U^{}_0$)
is described by $A \neq I$ or equivalently $R \neq 0$, but this has been strongly
constrained by current neutrino oscillation experiments, lepton-flavor-violating
processes and precision electroweak data
\cite{Antusch:2006vwa,Antusch:2014woa,Blennow:2016jkn}.
As a result, the angles $\theta^{}_{ij}$ (for $i=1,2,3$ and $j=4,5$)
are expected to be at most at the ${\cal O}(0.1)$ level.
The smallness of these six light-heavy flavor mixing angles allows us to make
an excellent approximation for Eq.~(\ref{eq:3.1.7}):
\begin{eqnarray}
A \simeq I - \frac{1}{2} \pmatrix{ s^2_{14} +
s^2_{15} & 0 & 0 \cr \vspace{-0.4cm} \cr
2\hat{s}^{}_{14} \hat{s}^*_{24} + 2\hat{s}^{}_{15} \hat{s}^*_{25} &
s^2_{24} + s^2_{25} & 0 \cr \vspace{-0.4cm} \cr
2\hat{s}^{}_{14} \hat{s}^*_{34} + 2\hat{s}^{}_{15} \hat{s}^*_{35} &
2\hat{s}^{}_{24} \hat{s}^*_{34} + 2\hat{s}^{}_{25} \hat{s}^*_{35} &
s^2_{34} + s^2_{35} \cr} \; ,
\nonumber \\
R \simeq \pmatrix{ \hat{s}^*_{14} & \hat{s}^*_{15} \cr \vspace{-0.4cm} \cr
\hat{s}^*_{24} & \hat{s}^*_{25} \cr \vspace{-0.4cm} \cr
\hat{s}^*_{34} & \hat{s}^*_{35} \cr} \; .
\label{eq:3.1.10}
\end{eqnarray}
It is clear that possible unitarity-violating effects in $U$ are at or below the
${\cal O}(10^{-2})$ level.

It is worth stressing that Eq.~(\ref{eq:3.1.1}) can actually lead us to the
{\it exact} seesaw relation between light and heavy Majorana neutrinos,
\begin{eqnarray}
U D^{}_\nu U^T + R D^{}_N R^T = 0 \; ,
\label{eq:3.1.11}
\end{eqnarray}
from which one may define the effective light Majorana neutrino mass
matrix
\begin{eqnarray}
M^{}_\nu \equiv U D^{}_\nu U^T = -R D^{}_N R^T \; .
\label{eq:3.1.12}
\end{eqnarray}
In this case the popular but approximate seesaw formula $M^{}_\nu \simeq -M^{}_{\rm D}
M^{-1}_{\rm R} M^T_{\rm D}$ can be easily derived from Eq.~ (\ref{eq:3.1.12})
with the help of the approximations
$M^{}_{\rm D} \simeq R D^{}_N U^{\prime T}_0$ and
$M^{}_{\rm R} \simeq U^\prime_0 D^{}_N U^{\prime T}_0$ as assured by
Eqs. (\ref{eq:3.1.1}) and (\ref{eq:3.1.2}).

The exact seesaw relation in Eq.~(\ref{eq:3.1.11}) may help determine
the masses of two heavy Majorana neutrinos in terms of the light Majorana
neutrino masses $m^{}_i$ (for $i=1,2,3$) and the flavor mixing parameters
of $U$ and $R$ \cite{Xing:2009ce}. To illustrate this interesting point,
here we focus only on the relation
\begin{eqnarray}
\left(U D^{}_\nu U^T\right)^{}_{ee} = \sum^3_{i=1}
U^2_{ei} m^{}_i = - \left(R D^{}_N R^T\right)^{}_{ee}
= - \sum^2_{k=1} R^2_{ek} M^{}_k \; .
\label{eq:3.1.13}
\end{eqnarray}
Considering both the real and imaginary parts of Eq.~(\ref{eq:3.1.13}),
we can therefore arrive at
\begin{eqnarray}
M^{}_1 = +\frac{\displaystyle {\rm Re}R^2_{e2}
\sum^3_{i=1} \left(m^{}_i {\rm Im} U^2_{ei}\right) - {\rm Im} R^2_{e2}
\sum^3_{i=1} \left(m^{}_i {\rm Re} U^2_{ei}\right)}{\displaystyle {\rm
Re}R^2_{e1} {\rm Im}R^2_{e2} - {\rm Im}R^2_{e1} {\rm Re}R^2_{e2}} \; ,
\nonumber \\
M^{}_2 = -\frac{\displaystyle {\rm Re}R^2_{e1}
\sum^3_{i=1} \left(m^{}_i {\rm Im} U^2_{ei}\right) - {\rm Im} R^2_{e1}
\sum^3_{i=1} \left(m^{}_i {\rm Re} U^2_{ei}\right)}{\displaystyle {\rm
Re}R^2_{e1} {\rm Im}R^2_{e2} - {\rm Im}R^2_{e1} {\rm Re}R^2_{e2}} \; .
\label{eq:3.1.14}
\end{eqnarray}
In view of Eq.~(\ref{eq:3.1.7}), we have
\begin{eqnarray}
U^{}_{e1} = c^{}_{12} c^{}_{13} c^{}_{14} c^{}_{15} \; ,
\quad
U^{}_{e2} = \hat{s}^{*}_{12} c^{}_{13} c^{}_{14} c^{}_{15} \; ,
\quad
U^{}_{e3} = \hat{s}^{*}_{13} c^{}_{14} c^{}_{15} \; ;
\nonumber \\
R^{}_{e1} = \hat{s}^{*}_{14} c^{}_{15} \; ,
\quad
R^{}_{e2} = \hat{s}^{*}_{15} \; .
\label{eq:3.1.15}
\end{eqnarray}
Substituting Eq.~(\ref{eq:3.1.15}) into Eq.~(\ref{eq:3.1.14}), we obtain
\begin{eqnarray}
M^{}_1 = - \frac{m^{}_2 s^2_{12} c^2_{13} \sin
\left(\phi^{}_2 + \phi\right) + m^{}_3 s^2_{13} \sin
\left(\phi^{}_2 - \phi\right)}{\sin \left(\phi^{}_2 -
\phi^{}_1\right)} \cdot \frac{c^2_{14}}{s^2_{14}} \; ,
\nonumber \\
M^{}_2 = + \frac{m^{}_2 s^2_{12} c^2_{13} \sin
\left(\phi^{}_1 + \phi\right) + m^{}_3 s^2_{13} \sin
\left(\phi^{}_1 - \phi\right)}{\sin \left(\phi^{}_2 -
\phi^{}_1\right)} \cdot \frac{c^2_{14}c^2_{15}}{s^2_{15}} \;
\label{eq:3.1.16}
\end{eqnarray}
with $\phi \equiv \delta^{}_{13} - \delta^{}_{12}$, $\phi^{}_1
\equiv 2\delta^{}_{14} - \left(\delta^{}_{12} +
\delta^{}_{13}\right)$ and $\phi^{}_2 \equiv 2\delta^{}_{15} -
\left(\delta^{}_{12} + \delta^{}_{13}\right)$ in the normal
neutrino mass ording case (i.e., $m^{}_1 =0$); or
\begin{eqnarray}
M^{}_1 = - \frac{m^{}_1 c^2_{12} \sin \left(\phi^\prime_2
- \phi^\prime\right) + m^{}_2 s^2_{12} \sin \left(\phi^\prime_2 +
\phi^\prime\right)}{\sin \left(\phi^\prime_2 -
\phi^\prime_1\right)} \cdot \frac{c^2_{13}c^2_{14}}{s^2_{14}} \; ,
\nonumber \\
M^{}_2 = + \frac{m^{}_1 c^2_{12} \sin \left(\phi^\prime_1
- \phi^\prime\right) + m^{}_2 s^2_{12} \sin \left(\phi^\prime_1 +
\phi^\prime\right)}{\sin \left(\phi^\prime_2 -
\phi^\prime_1\right)} \cdot
\frac{c^2_{13}c^2_{14}c^2_{15}}{s^2_{15}} \;
\label{eq:3.1.17}
\end{eqnarray}
with $\phi^\prime \equiv - \delta^{}_{12}$, $\phi^\prime_1 \equiv
2\delta^{}_{14} - \delta^{}_{12}$ and $\phi^\prime_2 \equiv
2\delta^{}_{15} - \delta^{}_{12}$ in the inverted mass
ordering case (i.e., $m^{}_3 =0$). The positiveness of $M^{}_1$ and
$M^{}_2$ require that the CP-violating phases in Eq.~(\ref{eq:3.1.16})
or Eq.~(\ref{eq:3.1.17}) should not all be vanishing. Instead, they
must take proper and nontrivial values. Of course, such results will
not be useful unless precision measurements of the fine
unitarity violation of $U$ at low energies become possible.

\subsection{A crude parametrization}
\label{section 3.2}

In this subsection we discuss a crude parametrization of $M^{}_{\rm D}$ given in Eq.~(\ref{eq:2.2}).
In this parametrization, the CP-violating asymmetry $\varepsilon^{}_1$ defined in Eq.~(\ref{eq:2.22})
is proportional to the following combination of the parameters in $M^{}_{\rm D}$:
\begin{eqnarray}
\overline \varepsilon^{}_1 \equiv \frac{{\rm Im} \left[(M^\dagger_{\rm D} M^{}_{\rm D})^{2}_{12}\right]} {(M^\dagger_{\rm D} M^{}_{\rm D})^{}_{11}} = \frac{ {\rm Im}[ (a^*_1 b^{}_1 + a^*_2 b^{}_2 + a^*_3 b^{}_3 )^2 ]} {(|a^{}_1|^2 + |a^{}_2|^2 + |a^{}_3|^2) } \;.
\label{eq:3.2.1}
\end{eqnarray}
All the parameters in $M^{}_{\rm D}$ are apparently involved in this expression.
With the help of  the seesaw formula, one can partially reconstruct $M^{}_{\rm D}$ in terms of
the low-energy flavor parameters. On the one hand, the seesaw formula gives the
effective Majorana mass matrix $M^{}_\nu$ for three light neutrinos as in Eq.~(\ref{eq:2.3}).
On the other hand, $M^{}_\nu$ can be reconstructed in terms of the low-energy
flavor parameters via the relations in Eq.~(\ref{eq:2.4}), by which its entries are
explicitly expressed as
\begin{eqnarray}
M^{}_{ee} & = &
\overline{m^{}_2} s^2_{12} c^2_{13} + m^{}_3 \tilde s^{*2}_{13} \; ,
\nonumber \\
M^{}_{\mu\mu} & = &  \overline m^{}_2 \left(c^{}_{12} c^{}_{23} - s^{}_{12} \tilde
s^{}_{13} s^{}_{23} \right)^2 + m^{}_3 c^2_{13} s^2_{23} \; ,
\nonumber \\
M^{}_{\tau\tau} & = & \overline m^{}_2 \left(c^{}_{12} s^{}_{23} + s^{}_{12} \tilde
s^{}_{13} c^{}_{23} \right)^2 + m^{}_3 c_{13}^2 c_{23}^2 \; ,
\nonumber \\
M^{}_{e\mu} & = & \overline m^{}_2 s^{}_{12} c^{}_{13} \left( c^{}_{12} c^{}_{23} -
s^{}_{12} \tilde s^{}_{13} s^{}_{23} \right)
 + m^{}_3 c^{}_{13} \tilde s^{*}_{13} s^{}_{23} \; ,
\nonumber \\
M^{}_{e\tau} & = &  -\overline m^{}_2 s^{}_{12} c^{}_{13} \left(c^{}_{12} s^{}_{23} +
s^{}_{12}\tilde s^{}_{13} c^{}_{23} \right)
 + m^{}_3 c^{}_{13} \tilde s^{*}_{13}c^{}_{23} \; ,
\nonumber \\
M^{}_{\mu\tau} & = & -\overline m_2^{} \left(c^{}_{12} s^{}_{23} +
s^{}_{12}\tilde  s^{}_{13} c^{}_{23} \right) \left(c^{}_{12}
c^{}_{23} - s^{}_{12} \tilde s^{}_{13} s^{}_{23} \right)
+ m^{}_3 c^2_{13} c^{}_{23} s^{}_{23} \;
\label{eq:3.2.2}
\end{eqnarray}
in the $m^{}_1 =0$ case; or
\begin{eqnarray}
M^{}_{ee} & = & m^{}_1 c^2_{12} c^2_{13}+
\overline m^{}_2 s^2_{12} c^2_{13} \;,
\nonumber \\
M^{}_{\mu\mu} & = & m^{}_1 \left(s^{}_{12}
c^{}_{23} + c^{}_{12} \tilde s^{}_{13} s^{}_{23} \right)^2
 + \overline m^{}_2 \left(c^{}_{12} c^{}_{23} - s^{}_{12} \tilde
s^{}_{13} s^{}_{23} \right)^2  \; ,
\nonumber \\
M^{}_{\tau\tau} & = & m^{}_1
\left(s^{}_{12} s^{}_{23} - c^{}_{12} \tilde s^{}_{13}
c^{}_{23}\right)^2
 + \overline m^{}_2 \left(c^{}_{12} s^{}_{23} + s^{}_{12} \tilde
s^{}_{13} c^{}_{23} \right)^2  \; ,
\nonumber \\
M^{}_{e\mu} & = & -m^{}_1 c^{}_{12}
c^{}_{13} \left(s^{}_{12} c^{}_{23} + c^{}_{12} \tilde s^{}_{13}
s^{}_{23} \right)
 + \overline m^{}_2 s^{}_{12} c^{}_{13} \left( c^{}_{12} c^{}_{23} -
s^{}_{12} \tilde s^{}_{13} s^{}_{23} \right) \; ,
\nonumber \\
M^{}_{e\tau} & = & m^{}_1 c^{}_{12} c^{}_{13}
\left(s^{}_{12} s^{}_{23} - c^{}_{12} \tilde s^{}_{13} c^{}_{23}\right)
 -\overline m^{}_2 s^{}_{12} c^{}_{13} \left(c^{}_{12} s^{}_{23} +
s^{}_{12}\tilde s^{}_{13} c^{}_{23} \right) \; ,
\nonumber \\
M^{}_{\mu\tau} & = & -m^{}_1 \left(s^{}_{12}
s^{}_{23} - c^{}_{12} \tilde s^{}_{13} c^{}_{23} \right) \left(
s^{}_{12} c^{}_{23}+ c^{}_{12} \tilde s^{}_{13} s^{} _{23} \right)
\nonumber \\
&& -\overline m_2^{} \left(c^{}_{12} s^{}_{23} +
s^{}_{12}\tilde  s^{}_{13} c^{}_{23} \right) \left(c^{}_{12}
c^{}_{23} - s^{}_{12} \tilde s^{}_{13} s^{}_{23} \right) \;
\label{eq:3.2.3}
\end{eqnarray}
in the $m^{}_3 =0$ case, where $\overline m^{}_2 \equiv m^{}_2 e^{2{\rm i}\sigma}$
and $\tilde s^{}_{13} \equiv s^{}_{13} e^{{\rm i}\delta}$ have been defined for the
sake of notational simplicity.

Identifying the effective Majorana neutrino mass matrix $M^{}_\nu$ in Eq.~(\ref{eq:2.3})
with that in Eq.~(\ref{eq:2.4}) yields the following connections between the model
parameters and the low-energy flavor parameters contained in $M^{}_{\alpha \beta}$:
\begin{eqnarray}
\hspace{-2cm}  -\pmatrix{ \vspace{0.2cm}
\displaystyle \frac{a^2_1}{M^{}_1} + \frac{b^2_1}{M^{}_2}
& ~ \displaystyle \frac{a^{}_1 a^{}_2}{M^{}_1} + \frac{b^{}_1 b^{}_2}{M^{}_2} ~
& \displaystyle \frac{a^{}_1 a^{}_3}{M^{}_1} + \frac{b^{}_1 b^{}_3}{M^{}_2} \cr
\vspace{0.2cm}
\displaystyle \frac{a^{}_1 a^{}_2}{M^{}_1} + \frac{b^{}_1 b^{}_2}{M^{}_2}
& \displaystyle \frac{a^{2}_2 }{M^{}_1} + \frac{b^{2}_2 }{M^{}_2}
& \displaystyle \frac{a^{}_2 a^{}_3}{M^{}_1} + \frac{b^{}_2 b^{}_3}{M^{}_2} \cr
\displaystyle \frac{a^{}_1 a^{}_3}{M^{}_1} + \frac{b^{}_1 b^{}_3}{M^{}_2}
& \displaystyle \frac{a^{}_2 a^{}_3}{M^{}_1} + \frac{b^{}_2 b^{}_3}{M^{}_2}
& \displaystyle \frac{a^{2}_3}{M^{}_1} + \frac{ b^{2}_3}{M^{}_2} \cr }
= \pmatrix{
M^{}_{ee} & M^{}_{e\mu} & M^{}_{e\tau} \cr
M^{}_{e\mu} & M^{}_{\mu\mu} & M^{}_{\mu\tau} \cr
M^{}_{e\tau} & M^{}_{\mu\tau} & M^{}_{\tau\tau} \cr } \;.
\label{eq:3.2.4}
\end{eqnarray}
Given the complex and symmetric natures of $M^{}_\nu$, it seems at first sight that this equation provides us with twelve constraint equations (with each independent entry contributing two). However, two of them are actually redundant due to the fact of $\det(M^{}_\nu) =0$. This means that only ten model parameters can be determined from the low-energy flavor parameters by solving the independent constraint equations. Before proceeding, let us clarify a subtle issue:
the rephasing of left-handed neutrinos for removing the unphysical phases in the PMNS matrix $U$ does not necessarily coincide with that for removing the unphysical phases in $M^{}_{\rm D}$. In the basis where all the unphysical phases are absent in $U$, all the six phases in $M^{}_{\rm D}$ are generally non-vanishing. So there are actually fourteen model parameters to be determined through the reconstruction. Taking $M^{}_{1,2}$ and two other model parameters as the inputs, one can solve the remaining model parameters from Eq.~(\ref{eq:3.2.4}). For instance, in terms of $M^{}_{1}$ and
$M^{}_2$, the modulus and phase of $a^{}_1$ (or $b^{}_1$) and the low-energy flavor parameters contained in $M^{}_{\alpha \beta}$, the remaining model parameters are explicitly solved as (for $M^{}_{ee} \neq 0$) \cite{Barger:2003gt}
\begin{eqnarray}
\hspace{-1cm} b^{}_1 = \eta^{}_{b^{}_1} \sqrt{M^{}_2 \left( - M^{}_{ee} - \frac{a^2_1}{M^{}_1} \right)}  \hspace{0.3cm}
\left[ {\rm or} \ a^{}_1 = \eta^{}_{a^{}_1} \sqrt{M^{}_1 \left( - M^{}_{ee} - \frac{b^2_1}{M^{}_2} \right)} \right] \;, \nonumber \\
\hspace{-1cm} a^{}_2 = \frac{1}{M^{}_{ee}} \left[ a^{}_1 M^{}_{e\mu} + \eta^{}_{b^{}_2} b^{}_1 \sqrt{\frac{M^{}_1}{M^{}_2} \left( M^{}_{ee} M^{}_{\mu\mu}- M^2_{e\mu} \right) }  \right] \;, \nonumber \\
\hspace{-1cm} b^{}_2 = \frac{1}{M^{}_{ee}} \left[ b^{}_1 M^{}_{e\mu} - \eta^{}_{b^{}_2} a^{}_1 \sqrt{\frac{M^{}_2}{M^{}_1} \left( M^{}_{ee} M^{}_{\mu\mu}- M^2_{e\mu} \right) }  \right] \;, \nonumber \\
\hspace{-1cm} a^{}_3 = \frac{1}{M^{}_{ee}} \left[ a^{}_1 M^{}_{e\tau} + \eta^{}_{b^{}_3} b^{}_1 \sqrt{\frac{M^{}_1}{M^{}_2} \left( M^{}_{ee} M^{}_{\tau\tau}- M^2_{e\tau} \right) }  \right] \;, \nonumber \\
\hspace{-1cm} b^{}_3 = \frac{1}{M^{}_{ee}} \left[ b^{}_1 M^{}_{e\tau} - \eta^{}_{b^{}_3} a^{}_1 \sqrt{\frac{M^{}_2}{M^{}_1} \left( M^{}_{ee} M^{}_{\tau\tau}- M^2_{e\tau} \right) }  \right] \;, \nonumber \\
\hspace{-1cm} M^{}_{\mu \tau} = - \frac{1}{M^{}_{ee}} \left[ M^{}_{e\mu} M^{}_{e\tau} + \eta^{}_{b^{}_2} \eta^{}_{b^{}_3}  \sqrt{ \left( M^{}_{ee} M^{}_{\mu\mu}- M^2_{e\mu} \right) \left( M^{}_{ee} M^{}_{\tau \tau}- M^2_{e\tau} \right) }  \right] \;,
\label{eq:3.2.5}
\end{eqnarray}
where $\eta^{}_{a^{}_1}, \eta^{}_{b^{}_1}, \eta^{}_{b^{}_2}, \eta^{}_{b^{}_3} = \pm 1$. Note that the last equation, which is a consistency condition for $\det(M^{}_\nu) =0$, helps us specify the sign convention of $\eta^{}_{b^{}_2} \eta^{}_{b^{}_3}$.
Thanks to the invariance of Eq.~(\ref{eq:3.2.4}) under the simultaneous interchanges $a^{}_1 \leftrightarrow a^{}_2$, $b^{}_1 \leftrightarrow b^{}_2$, $M^{}_{ee} \leftrightarrow M^{}_{\mu \mu}$ and $M^{}_{e\tau} \leftrightarrow M^{}_{\mu \tau}$, one can readily derive the reconstruction in terms of $a^{}_2$ or $b^{}_2$ (for $M^{}_{\mu\mu} \neq 0$) by making the corresponding interchanges in Eq.~(\ref{eq:3.2.5}). Similarly, the reconstruction in terms of $a^{}_3$ or $b^{}_3$ (for $M^{}_{\tau\tau} \neq 0$) can be derived by taking advantage of the invariance of Eq.~(\ref{eq:3.2.4}) under the simultaneous interchanges $a^{}_1 \leftrightarrow a^{}_3$, $b^{}_1 \leftrightarrow b^{}_3$, $M^{}_{ee} \leftrightarrow M^{}_{\tau \tau}$ and $M^{}_{e\mu} \leftrightarrow M^{}_{\mu \tau}$.

A complete reconstruction of $M^{}_{\rm D}$ in terms of the low-energy flavor parameters (up to $M^{}_1$ and $M^{}_2$) will become possible if the model parameters are reduced by two. The most natural way of doing so is to have one texture zero or one equality between two entries of $M^{}_{\rm D}$. For the one-zero scheme, we take the case of $b^{}_1 =0$ as example, while the results for other one-zero cases can be analogously derived with the help of the observations made below Eq.~(\ref{eq:3.2.5}). To be explicit, the complete reconstruction in the $b^{}_1 =0$ case is given by
\begin{eqnarray}
\hspace{-1cm} a^{}_1 = \eta^{}_{a^{}_1} \sqrt{-M^{}_1  M^{}_{ee} } \;,
\hspace{0.5cm} a^{}_2 = - \eta^{}_{a^{}_1} \sqrt{- \frac{  M^{}_1 } { M^{}_{ee} } } M^{}_{e\mu}\;, \hspace{0.5cm} a^{}_3 = - \eta^{}_{a^{}_1} \sqrt{ - \frac{  M^{}_1 } { M^{}_{ee} } } M^{}_{e\tau} \;, \nonumber \\
\hspace{-1cm} b^{}_2 = \eta^{}_{a^{}_1} \eta^{}_{b^{}_2} \sqrt{\frac{M^{}_2}{M^{}_{ee}} \left( M^2_{e\mu} - M^{}_{ee} M^{}_{\mu\mu} \right) }  \;, \nonumber \\
\hspace{-1cm} b^{}_3 =\eta^{}_{a^{}_1}  \eta^{}_{b^{}_3} \sqrt{\frac{M^{}_2}{M^{}_{ee}} \left( M^2_{e\tau} - M^{}_{ee} M^{}_{\tau\tau} \right) }   \;.
\label{eq:3.2.6}
\end{eqnarray}
By inputting these results into $\overline \varepsilon^{}_1$ defined in Eq.~(\ref{eq:3.2.1}), one arrives at
\begin{eqnarray}
\hspace{-1cm} \frac{\overline \varepsilon^{}_1}{M^{}_2} =  \frac{{\rm Im} \left[ \left(  \eta^{}_{b^{}_2} M^*_{e\mu} \sqrt{M^{}_{ee} M^{}_{\mu\mu}- M^2_{e\mu}} +  \eta^{}_{b^{}_3} M^*_{e\tau} \sqrt{M^{}_{ee} M^{}_{\tau\tau}- M^2_{e\tau}} \right)^2 \right] } {|M^{}_{ee}| \left( |M^{}_{ee}|^2 +  |M^{}_{e\mu}|^2 +  |M^{}_{e\tau}|^2 \right) }  \;,
\label{eq:3.2.7}
\end{eqnarray}
which is completely expressed in terms of the low-energy flavor parameters. This result establishes a direct link between the high-energy and low-energy flavor parameters.
As for the one-equality scheme, let us take the case of $a^{}_1 =b^{}_1$ as example.
Then the complete reconstruction in this case turns out to be
\begin{eqnarray}
a^{}_1 = \eta^{}_{a^{}_1} \sqrt{ - \frac{M^{}_1 M^{}_2 M^{}_{ee}}{M^{}_1+M^{}_2} } \;, \nonumber \\
a^{}_2 = - \eta^{}_{a^{}_1} \sqrt{ - \frac{M^{}_1 M^{}_2 }{M^{}_{ee} \left(M^{}_1+M^{}_2\right) } } \left[ M^{}_{e\mu} + \eta^{}_{b^{}_2} \sqrt{ \frac{M^{}_1}{M^{}_2} \left( M^{}_{ee} M^{}_{\mu\mu}- M^2_{e\mu} \right) } \right]  \;, \nonumber \\
b^{}_2 = - \eta^{}_{a^{}_1} \sqrt{ - \frac{M^{}_1 M^{}_2 }{M^{}_{ee} \left(M^{}_1+M^{}_2\right) } } \left[ M^{}_{e\mu} - \eta^{}_{b^{}_2} \sqrt{ \frac{M^{}_2}{M^{}_1} \left( M^{}_{ee} M^{}_{\mu\mu}- M^2_{e\mu} \right) } \right]  \;, \nonumber \\
a^{}_3 = - \eta^{}_{a^{}_1} \sqrt{ - \frac{M^{}_1 M^{}_2 }{M^{}_{ee} \left(M^{}_1+M^{}_2\right) } } \left[ M^{}_{e\tau} + \eta^{}_{b^{}_3} \sqrt{ \frac{M^{}_1}{M^{}_2} \left( M^{}_{ee} M^{}_{\tau\tau}- M^2_{e\tau} \right) } \right]  \;, \nonumber \\
b^{}_3 = - \eta^{}_{a^{}_1} \sqrt{ - \frac{M^{}_1 M^{}_2 }{M^{}_{ee} \left(M^{}_1+M^{}_2\right) } } \left[ M^{}_{e\tau} - \eta^{}_{b^{}_3} \sqrt{ \frac{M^{}_2}{M^{}_1} \left( M^{}_{ee} M^{}_{\tau\tau}- M^2_{e\tau} \right) } \right]  \;.
\label{eq:3.2.8}
\end{eqnarray}
A detailed study on the reconstructions of all the one-zero and one-equality cases can be found in Ref.~\cite{Barger:2003gt}.

\subsection{The bi-unitary parametrization}
\label{section 3.3}

As we have seen, in the crude parametrization the model parameters relevant for the low-energy flavor parameters and leptogenesis are completely jumbled together. In order to disentangle these two sets of parameters, some deliberate parametrizations of $M^{}_{\rm D}$ must be invoked.
In the light of the expression of $\varepsilon^{}_i$ in Eq.~(\ref{eq:2.22}), one may wish to decompose $M^{}_{\rm D}$ into the product of a unitary matrix $U^{}_{\rm L}$ and some other matrices. As will be seen, such a decomposition pattern has the following two merits: (1) it is immediate to see that $U^{}_{\rm L}$ will cancel out in the expression of $\varepsilon^{}_i$, thereby reducing the parameters involved in the leptogenesis calculations; (2) the resulting PMNS matrix can also be expressed as the product of the same $U^{}_{\rm L}$ and some other matrices, which provides a fresh insight for realizing some particular lepton flavor mixing patterns (see the discussions at the end of section \ref{section 3.4}). From the mathematical point of view, there are two commonly-used approaches to realize this kind of decomposition: the singular value decomposition where a generic matrix is decomposed into the successive product of a unitary matrix, a diagonal matrix and another unitary matrix; and the QR decomposition where a generic matrix is decomposed into the product of a unitary matrix and a triangular matrix. The singular value and QR parametrizations of $M^{}_{\rm D}$, which are referred to respectively as the bi-unitary and triangular parametrizations from now on, will be discussed respectively in this and next subsections.

In the bi-unitary parametrization \cite{Endoh:2002wm}, $M^{}_{\rm D}$ is expressed as $M^{}_{\rm D} = U^{}_{\rm L} \Sigma U^{}_{\rm R}$. Here $U^{}_{\rm L}$ is a $3 \times 3$ unitary matrix and can be parameterized in a similar way as the PMNS lepton flavor mixing matrix:
\begin{eqnarray}
U^{}_{\rm L} = O^{}_{23} (\theta^{}_{23 \rm L}) \hspace{0.05cm} U^{}_{13} (\theta^{}_{13 \rm L}, \delta^{}_{\rm L}) \hspace{0.05cm} O^{}_{12} (\theta^{}_{12 \rm L}) \hspace{0.05cm} P^{}_{\rm L} \;.
\label{eq:3.3.1}
\end{eqnarray}
However, unlike $P^{}_\nu$ in Eq.~(\ref{eq:1.3}), $P^{}_{\rm L} = {\rm Diag}\{ 1, e^{-{\rm i} \sigma^{}_{\rm L} },  1 \}$ only contains one effective phase. This is because the other would-be phase will be rendered ineffective when $U^{}_{\rm L}$ is multiplied by the following $\Sigma$ matrix from the right-hand side:
\begin{eqnarray}
m^{}_1 = 0: \hspace{1cm} \Sigma = \pmatrix{ 0 & 0 \cr n^{}_2 & 0 \cr 0 & n^{}_3 } \;;
\nonumber \\
m^{}_3 = 0: \hspace{1cm} \Sigma = \pmatrix{ n^{}_1 & 0 \cr 0 & n^{}_2 \cr 0 & 0 } \;,
\label{eq:3.3.2}
\end{eqnarray}
with $n^{}_i$ being real and positive.
On the other hand, $U^{}_{\rm R}$ is a $2 \times 2$ unitary matrix and can be parameterized as
\begin{eqnarray}
U^{}_{\rm R} =  \pmatrix{ \cos{\theta^{}_{\rm R}} & \sin{\theta^{}_{\rm R}} \cr
- \sin{\theta^{}_{\rm R}} & \cos{\theta^{}_{\rm R}} }
\pmatrix{ \displaystyle e^{-{\rm i} \gamma^{}_{\rm R}/2 } & 0 \cr
0 & e^{{\rm i} \gamma^{}_{\rm R}/2 } } \;.
\label{eq:3.3.3}
\end{eqnarray}
A simple parameter counting indicates that there are eleven model parameters in total: five in $U^{}_{\rm L}$, two in $\Sigma$, two in $U^{}_{\rm R}$ plus two heavy Majorana neutrino masses.
Finally, it is worth mentioning that $U^{}_{\rm L}$ and $U^{}_{\rm R}$ can be interpreted as the mixing contributions from the left-handed charged-lepton sector and the right-handed neutrino sector, respectively. In the case of $U^{}_{\rm L}$ (or $U^{}_{\rm R}$) being trivially the unity matrix, one can go into a basis where both $M^{}_l$ (or $M^{}_{\rm R}$) and $M^{}_{\rm D}$ are diagonal but $M^{}_{\rm R}$ (or $M^{}_l$) is not.

In the present parametrization, one obtains $\overline \varepsilon^{}_1$ as
\begin{eqnarray}
\overline \varepsilon^{}_1 = \frac{ \left(n^2_i-n^2_j\right)^2 \sin^2{2\theta^{}_{\rm R}} \sin{2\gamma^{}_{\rm R}} } { 4\left(n^2_i \cos^2{\theta^{}_{\rm R}} + n^2_j \sin^2{\theta^{}_{\rm R}}\right) } \; ,
\label{eq:3.3.4}
\end{eqnarray}
with $i =2$ or 1 and $j =3$ or 2 in the $m^{}_1 =0$ or $m^{}_3 =0$ case.
As promised, $U^{}_{\rm L}$ has been cancelled in the expression of $\varepsilon^{}_1$, leaving $\gamma^{}_{\rm R}$ as the only source of CP violation for leptogenesis. Furthermore, $\theta^{}_{\rm R}$ should also be finite so as to yield a non-zero $\varepsilon^{}_1$, implying that the success of leptogenesis crucially relies on the mixing effect of right-handed neutrinos. This can be easily understood from the fact that the generation of $\varepsilon^{}_i$ in the decays of $N^{}_i$ owes to the Feynman diagrams with $N^{}_j$ (for $j \neq i$) running in the loops (see Fig.~\ref{Fig:4}). On the other hand, the seesaw formula yields an effective Majorana mass matrix for three light neutrinos as $M^{}_\nu = - U^{}_{\rm L} \Sigma U^{}_{\rm R} D^{-1}_{\rm N} U^{T}_{\rm R} \Sigma^{T}  U^{T}_{\rm L}$.
It is easy to verify that the resulting PMNS matrix can be expressed as $U = {\rm i} U^{}_{\rm L} V^{}_{\rm R}$ with $V^{}_{\rm R}$ being the unitary matrix for diagonalizing the intermediate matrix $M = - U^\dagger_{\rm L} M^{}_\nu U^*_{\rm L} = \Sigma U^{}_{\rm R} D^{-1}_{\rm N} U^{T}_{\rm R} \Sigma^{T}$; namely, $V^\dagger_{\rm R} M V^{*}_{\rm R} = {\rm Diag}\{0, m^{}_2, m^{}_3\}$  in the $m^{}_1=0$ case or $V^\dagger_{\rm R} M V^{*}_{\rm R} = {\rm Diag}\{m^{}_1, m^{}_2, 0\}$ in the $m^{}_3=0$ case. Explicitly, $M$ is found to take a form as
\begin{eqnarray}
\hspace{-2.5cm}
M = \pmatrix{ 0 & 0 & 0 \cr
0 & \displaystyle  n^2_2 \left( \frac{ \cos^2{\theta^{}_{\rm R}} e^{-{\rm i} \gamma^{}_{\rm R} } } {M^{}_1} + \frac{ \sin^2{\theta^{}_{\rm R}} e^{{\rm i}\gamma^{}_{\rm R}}} {M^{}_2}  \right) &  \displaystyle \frac{1}{2} n^{}_2 n^{}_3 \sin{2\theta^{}_{\rm R}} \left( \frac{ e^{{\rm i} \gamma^{}_{\rm R} } }{M^{}_2} - \frac{ e^{-{\rm i}\gamma^{}_{\rm R}}}{M^{}_1}  \right)  \cr
0 & \displaystyle \frac{1}{2} n^{}_2 n^{}_3 \sin{2\theta^{}_{\rm R}} \left( \frac{ e^{{\rm i} \gamma^{}_{\rm R} } }{M^{}_2} - \frac{ e^{-{\rm i}\gamma^{}_{\rm R}}}{M^{}_1}  \right) & \displaystyle n^2_3 \left( \frac{ \sin^2{\theta^{}_{\rm R}} e^{-{\rm i} \gamma^{}_{\rm R} } }{M^{}_1} + \frac{ \cos^2{\theta^{}_{\rm R}} e^{{\rm i}\gamma^{}_{\rm R}}}{M^{}_2}  \right) } \;
\label{eq:3.3.5}
\end{eqnarray}
in the $m^{}_1=0$ case, or
\begin{eqnarray}
\hspace{-2.5cm}
M = \pmatrix{ \displaystyle  n^2_1 \left( \frac{ \cos^2{\theta^{}_{\rm R}} e^{-{\rm i} \gamma^{}_{\rm R} } } {M^{}_1} + \frac{ \sin^2{\theta^{}_{\rm R}} e^{{\rm i}\gamma^{}_{\rm R}}} {M^{}_2}  \right) & \displaystyle \frac{1}{2} n^{}_1 n^{}_2 \sin{2\theta^{}_{\rm R}} \left( \frac{ e^{{\rm i} \gamma^{}_{\rm R} } }{M^{}_2} - \frac{ e^{-{\rm i}\gamma^{}_{\rm R}}}{M^{}_1}  \right) & 0 \cr
\displaystyle \frac{1}{2} n^{}_1 n^{}_2 \sin{2\theta^{}_{\rm R}} \left( \frac{ e^{{\rm i} \gamma^{}_{\rm R} } }{M^{}_2} - \frac{ e^{-{\rm i}\gamma^{}_{\rm R}}}{M^{}_1}  \right) & \displaystyle n^2_2 \left( \frac{ \sin^2{\theta^{}_{\rm R}} e^{-{\rm i} \gamma^{}_{\rm R} } }{M^{}_1} + \frac{ \cos^2{\theta^{}_{\rm R}} e^{{\rm i}\gamma^{}_{\rm R}}}{M^{}_2}  \right) &  0  \cr
0 & 0 & 0 } \;
\label{eq:3.3.6}
\end{eqnarray}
in the $m^{}_3=0$ case. Correspondingly, $V^{}_{\rm R}$ is expected to take the form
\begin{eqnarray}
\hspace{-1cm} m^{}_1 = 0: \hspace{1cm}  V^{}_{\rm R} = \pmatrix{ 1 & 0 & 0 \cr 0 & \cos{\theta} & \sin{\theta} \hspace{0.05cm} e^{-{\rm i} \phi} \cr
0 & - \sin{\theta} \hspace{0.05cm} e^{{\rm i} \phi} & \cos{\theta} }
\pmatrix{ 1 & & \cr & e^{{\rm i} \alpha } & \cr  &  & e^{{\rm i} \beta } } \;;
\nonumber \\
\hspace{-1cm} m^{}_3 = 0: \hspace{1cm}  V^{}_{\rm R} = \pmatrix{ \cos{\theta} & \sin{\theta} \hspace{0.05cm} e^{-{\rm i} \phi} & 0 \cr  - \sin{\theta} \hspace{0.05cm} e^{{\rm i} \phi} & \cos{\theta} & 0 \cr
0 & 0 & 1 }
\pmatrix{  e^{{\rm i} \alpha } & & \cr   & e^{{\rm i} \beta } & \cr & & 1 } \;,
\label{eq:3.3.7}
\end{eqnarray}
where the first part of $V^{}_{\rm R}$ is dedicated to diagonalizing $M$, and its second part is to make the obtained neutrino mass eigenvalues real and positive. After a straightforward calculation, we arrive at
\begin{eqnarray}
\tan{2\theta} = \frac{ 2\left|M^*_{ii} M^{}_{ij} + M^*_{ij} M^{}_{jj} \right| } {\left| M^{}_{jj}|^2 - | M^{}_{ii}\right|^2 } \;, \nonumber \\
\phi = \arg \left( M^*_{ii} M^{}_{ij} + M^*_{ij} M^{}_{jj} \right) \;, \nonumber \\
m^{}_i e^{ 2 {\rm i} \alpha } = M^{}_{ii} \cos^2{\theta} + M^{}_{jj} \sin^2{\theta} \hspace{0.05cm} e^{-2{\rm i} \phi} - M^{}_{ij} \sin{2\theta} \hspace{0.05cm} e^{- {\rm i} \phi} \;, \nonumber \\
m^{}_j e^{ 2 {\rm i} \beta }  = M^{}_{jj} \cos^2{\theta} + M^{}_{ii} \sin^2{\theta} \hspace{0.05cm} e^{ 2{\rm i} \phi} + M^{}_{ij} \sin{2\theta} \hspace{0.05cm} e^{ {\rm i} \phi} \;,
\label{eq:3.3.8}
\end{eqnarray}
with $i =2$ or 1 and $j =3$ or 2 in the $m^{}_1 =0$ or $m^{}_3 =0$ case, where $M^{}_{ij}$ denotes the $(i, j)$ element of $M$.

Note that the PMNS matrix thus obtained (i.e., $U = {\rm i} U^{}_{\rm L} V^{}_{\rm R}$) is not of the standard-parametrization form as shown in Eq.~(\ref{eq:1.3}).
To confront the predictions of $U$ with the experimental data, one should better extract the resulting standard-parametrization parameters according to the formulas
\begin{eqnarray}
s^{}_{13} = |U^{}_{e3}| \;, \hspace{1cm} s^{}_{12} = \frac{|U^{}_{e2}|}{\sqrt{1-|U^{}_{e3}|^2}} \;,
\hspace{1cm} s^{}_{23} = \frac{|U^{}_{\mu 3}|}{\sqrt{1-|U^{}_{e3}|^2}} \;, \nonumber \\
\delta = {\arg}\left( \displaystyle \frac{ \displaystyle U^{}_{e 2} U^{}_{\mu 3} U^{*}_{e 3} U^{*}_{\mu 2} + s^{2}_{12} c^2_{13} s^{2}_{13} s^{2}_{23}  }{ c^{}_{12} s^{}_{12} c^2_{13} s^{}_{13} c^{}_{23} s^{}_{23}} \right) \;,
\nonumber \\
\rho = {\arg}\left(U^{}_{e1} U^{*}_{e3}\right) - \delta \;, \hspace{1cm} \sigma =  {\arg}\left(U^{}_{e2} U^{*}_{e3}\right) - \delta \;.
\label{eq:3.3.9}
\end{eqnarray}
Of course, only a single Majorana CP phase (i.e., $\sigma$ in the $m^{}_1=0$ case or
$\sigma-\rho$ in the $m^{}_3 = 0$ case) is physically relevant.

\subsection{The triangular parametrization}
\label{section 3.4}

This subsection is devoted to the triangular parametrization \cite{Branco:2002xf, Fujihara:2005pv}, in which $M^{}_{\rm D}$ is expressed as $M^{}_{\rm D} = U^{}_{\rm L} \Delta$ with $U^{}_{\rm L}$ retaining the form in Eq.~(\ref{eq:3.3.1}) and $\Delta$ being a $3 \times 2$ triangular matrix parameterized as
\begin{eqnarray}
m^{}_1 = 0: \hspace{1cm} \Delta = \pmatrix{ 0 & 0 \cr \tau^{}_{21} & 0 \cr \tau^{}_{31} e^{ {\rm i} \gamma } & \tau^{}_{32} } \;;  \nonumber \\
m^{}_3 = 0: \hspace{1cm} \Delta = \pmatrix{ \tau^{}_{11}  &  0 \cr \tau^{}_{21} e^{ {\rm i} \gamma}  & \tau^{}_{22} \cr 0 & 0 } \;,
\label{eq:3.4.1}
\end{eqnarray}
where $\tau^{}_{ij}$ and $\gamma$ are real free parameters. Here there are also eleven model parameters in total: five in $U^{}_{\rm L}$, four in $\Delta$ plus two heavy Majorana neutrino masses.
Such a decomposition can simply be accomplished by employing the Schmidt orthogonalization method.
The triangular parametrization of $M^{}_{\rm D}$ in Eq.~(\ref{eq:2.2}) yields a form of $U^{}_{\rm L}$ as
\begin{eqnarray}
U^{}_{\rm L} = \pmatrix{
\vspace{0.2cm}
\displaystyle \frac{ a^{*}_2 b^{*}_3 - a^{*}_3 b^{*}_2 } { \sqrt{ a^2 b^2 - | a^{}_i b^*_i|^2 } }  & \displaystyle \frac{ b^2 a^{}_1 - a^{}_i b^*_i b^{}_1 } { b \sqrt{ a^2 b^2 - | a^{}_i b^*_i|^2 } } & \displaystyle \frac{b^{}_1}{ b } \cr
\vspace{0.2cm}
\displaystyle \frac{ a^{*}_3 b^{*}_1 - a^{*}_1 b^{*}_3 } { \sqrt{ a^2 b^2 - | a^{}_i b^*_i|^2 } } & \displaystyle \frac{ b^2 a^{}_2 - a^{}_i b^*_i b^{}_2 } { b \sqrt{ a^2 b^2 - | a^{}_i b^*_i|^2 } } & \displaystyle \frac{b^{}_2}{ b } \cr
\displaystyle \frac{ a^{*}_1 b^{*}_2 - a^{*}_2 b^{*}_1 } { \sqrt{ a^2 b^2 - | a^{}_i b^*_i|^2 } } & \displaystyle \frac{ b^2 a^{}_3 - a^{}_i b^*_i b^{}_3 } { b \sqrt{ a^2 b^2 - | a^{}_i b^*_i|^2 } } & \displaystyle \frac{b^{}_3}{ b }
} \;,
\label{eq:3.4.2}
\end{eqnarray}
in the $m^{}_1 =0$ case; or
\begin{eqnarray}
U^{}_{\rm L} = \pmatrix{
\vspace{0.2cm}
\displaystyle \frac{ b^2 a^{}_1 - a^{}_i b^*_i b^{}_1 } { b \sqrt{ a^2 b^2 - | a^{}_i b^*_i|^2 } } & \displaystyle \frac{b^{}_1}{ b } & \displaystyle \frac{ a^{*}_2 b^{*}_3 - a^{*}_3 b^{*}_2 } { \sqrt{ a^2 b^2 - | a^{}_i b^*_i|^2 } } \cr
\vspace{0.2cm}
\displaystyle \frac{ b^2 a^{}_2 - a^{}_i b^*_i b^{}_2 } { b \sqrt{ a^2 b^2 - | a^{}_i b^*_i|^2 } } & \displaystyle \frac{b^{}_2}{ b }  & \displaystyle \frac{ a^{*}_3 b^{*}_1 - a^{*}_1 b^{*}_3 } { \sqrt{ a^2 b^2 - | a^{}_i b^*_i|^2 } } \cr
\displaystyle \frac{ b^2 a^{}_3 - a^{}_i b^*_i b^{}_3 } { b \sqrt{ a^2 b^2 - | a^{}_i b^*_i|^2 } } & \displaystyle \frac{b^{}_3}{ b } & \displaystyle \frac{ a^{*}_1 b^{*}_2 - a^{*}_2 b^{*}_1 } { \sqrt{ a^2 b^2 - | a^{}_i b^*_i|^2 } }
} \;,
\label{eq:3.4.3}
\end{eqnarray}
in the $m^{}_3 =0$ case, where $a \equiv \sqrt{a^{}_i a^{*}_i}$ and $b \equiv \sqrt{b^{}_i b^*_i}$ are defined, and the summation over $i$ is implied. On the other hand, $\Delta$ can be obtained via the relation $\Delta = U^{\dagger}_{\rm L} M^{}_{\rm D}$ as
\begin{eqnarray}
\tau^{}_{21} = \frac{1}{b} \sqrt{ a^2 b^2 - | a^{}_i b^*_i|^2 } \;, \hspace{1cm} \tau^{}_{31} e^{ {\rm i} \gamma } =  \frac{1}{b} a^{}_i b^*_i \;, \hspace{1cm} \tau^{}_{32} = b \;,
\label{eq:3.4.4}
\end{eqnarray}
in the $m^{}_1 =0$ case; or
\begin{eqnarray}
\tau^{}_{11} = \frac{1}{b} \sqrt{ a^2 b^2 - | a^{}_i b^*_i|^2 } \;, \hspace{1cm} \tau^{}_{21} e^{ {\rm i} \gamma } =  \frac{1}{b} a^{}_i b^*_i \;, \hspace{1cm} \tau^{}_{22} = b \;,
\label{eq:3.4.5}
\end{eqnarray}
in the $m^{}_3 =0$ case.

In the present parametrization, $\overline \varepsilon^{}_1$ turns out to be
\begin{eqnarray}
m^{}_1 = 0: \hspace{1cm} \overline \varepsilon^{}_1 = - \frac{ \tau^2_{31} \tau^2_{32} \sin{2\gamma } } { \tau^2_{21} + \tau^2_{31} } \; ;
\nonumber \\
m^{}_3 = 0: \hspace{1cm} \overline \varepsilon^{}_1 = - \frac{ \tau^2_{21} \tau^2_{22} \sin{2\gamma } } { \tau^2_{11} + \tau^2_{21} } \; .
\label{eq:3.4.6}
\end{eqnarray}
Like in the bi-unitary parametrization, here only four parameters in $M^{}_{\rm D}$ (one of which is the CP phase $\gamma$) enter in the expression of $\varepsilon^{}_1$. Furthermore, the effective Majorana mass matrix $M^{}_\nu  = - U^{}_{\rm L} \Delta D^{-1}_{\rm N} \Delta^{T} U^{T}_{\rm L}$ for three light neutrinos will also lead to a lepton flavor mixing matrix of the form $U = {\rm i} U^{}_{\rm L} V^{}_{\rm R}$ with $V^{}_{\rm R}$ being the unitary matrix for diagonalizing the intermediate matrix $M = - U^\dagger_{\rm L} M^{}_\nu U^*_{\rm L} = \Delta D^{-1}_{\rm N} \Delta^{T}$:
\begin{eqnarray}
m^{}_1 = 0: \hspace{1cm} M = \pmatrix{ \vspace{0.2cm}
0 & 0 & 0 \cr
\vspace{0.2cm}
0 & \displaystyle \frac{\tau^2_{21}}{M^{}_1} &  \displaystyle \frac{\tau^{}_{21} \tau^{}_{31} e^{ {\rm i} \gamma } }{M^{}_1} \cr
0 & \displaystyle \frac{\tau^{}_{21} \tau^{}_{31} e^{ {\rm i} \gamma } }{M^{}_1} & \displaystyle \frac{\tau^2_{32}}{M^{}_2} } \;; \nonumber \\
m^{}_3 = 0: \hspace{1cm} M = \pmatrix{ \vspace{0.2cm}
\displaystyle \frac{\tau^2_{11}}{M^{}_1} & \displaystyle \frac{\tau^{}_{11} \tau^{}_{21} e^{ {\rm i} \gamma } }{M^{}_1} & 0 \cr
\vspace{0.2cm}
\displaystyle \frac{\tau^{}_{11} \tau^{}_{21} e^{ {\rm i} \gamma } }{M^{}_1} & \displaystyle \frac{\tau^2_{22}}{M^{}_2}  &  0 \cr
0 & 0 & 0 } \;.
\label{eq:3.4.7}
\end{eqnarray}
One may calculate $V^{}_{\rm R}$ in a similar way as formulated in section \ref{section 3.3}, such as  Eqs.~(\ref{eq:3.3.7})---(\ref{eq:3.3.8}). Without going into the details, the following observations which provide a fresh insight for realizing some particular lepton flavor mixing patterns can be made. In the particular case of $\tau^{}_{31}$ (for $m^{}_1 =0$) or $\tau^{}_{21}$ (for $m^{}_3 =0$) being vanishing (i.e., two columns of $M^{}_{\rm D}$ being complex orthogonal to each other), $V^{}_{\rm R}$ will trivially be the unity matrix, leaving us with $U= U^{}_{\rm L}$. In this case two columns of $U$ are proportional respectively to two columns of $M^{}_{\rm D}$, and the remaining one is specified by the unitarity of $U$. This means that, to obtain the PMNS matrix $U$ of a particular form, one may require two columns of $M^{}_{\rm D}$ to be proportional respectively to two columns of $U$.
This is just the so-called {\it form dominance} scenario, which can be realized with the help of some flavor symmetries \cite{King:2013eh,King:2003jb}. But one should keep in mind that leptogenesis does not work in such a scenario, as can be easily seen from Eq.~(\ref{eq:3.4.6}). In the generic case of $\tau^{}_{31}$ (for $m^{}_1 =0$) or $\tau^{}_{21}$ (for $m^{}_3 =0$) being nonzero, it is easy to see that the first or third column of $U^{}_{\rm L}$ remains unaffected by $V^{}_{\rm R}$. This means that, to obtain the PMNS matrix $U$ with its first column (in the $m^{}_1 =0$ case) or third column (in the $m^{}_3 =0$ case) being of a particular form, one may require both columns of $M^{}_{\rm D}$ to be complex orthogonal to the form under consideration. As one will see, these observations can provide some intuitive explanations for the interesting results derived in section~\ref{section 6.2}.

\subsection{The Casas-Ibarra parametrization}
\label{section 3.5}

The most popular and useful parametrization of $M^{}_{\rm D}$ is the one put forward by Casas and Ibarra \cite{Casas:2001sr, Ibarra:2003up}, because its parameters relevant for the low-energy flavor parameters and those associated with leptogenesis are disentangled to the max. In this parametrization $M^{}_{\rm D}$ is also expressed as the product of a unitary matrix $U^{}_{\rm L}$ and some other matrices, but now $U^{}_{\rm L}$ is completely identical with the PMNS matrix $U$. To rationalize such a parametrization, one may start from the following relation between the model parameters and low-energy flavor parameters bridged by the effective Majorana mass matrix for three light neutrinos:
\begin{eqnarray}
-M^{}_{\rm D} D^{-1/2}_{N} D^{-1/2}_{N} M^{T}_{\rm D} = M^{}_{\nu}
= U D^{1/2}_\nu D^{1/2}_\nu U^{T} \;,
\label{eq:3.5.1}
\end{eqnarray}
where $D^{}_N$ and $D^{}_\nu$ have been defined below Eq.~(\ref{eq:3.1.1}), and
the definitions of $D^{-1/2}_{N}$ and $D^{1/2}_\nu$ are self explanatory. This relation tells us that $M^{}_{\rm D} D^{-1/2}_{N}$ is equivalent to $U D^{1/2}_\nu$ up to some uncertainties described by a $3 \times 2$ complex matrix $O$:
\begin{eqnarray}
M^{}_{\rm D} D^{-1/2}_{N} = {\rm i} U D^{1/2}_\nu O \;,
\label{eq:3.5.2}
\end{eqnarray}
where $O$ satisfies the normalization relation
\begin{eqnarray}
m^{}_1 = 0: \hspace{1cm} O O^{T} = {\rm Diag}\{0, 1, 1\} \; ;
\nonumber \\
m^{}_3 = 0: \hspace{1cm} O O^{T} = {\rm Diag}\{1, 1, 0\} \; .
\label{eq:3.5.3}
\end{eqnarray}
In this way we obtain the Casas-Ibarra parametrization of $M^{}_{\rm D}$ as follows:
\begin{eqnarray}
M^{}_{\rm D} = {\rm i} U D^{1/2}_\nu O D^{1/2}_{N} \;.
\label{eq:3.5.4}
\end{eqnarray}
An explicit form of $O$ is given by
\begin{eqnarray}
m^{}_1 = 0: \hspace{1cm} O = \pmatrix{ 0 & 0 \cr \cos{z} & -  \zeta \sin{z} \cr \sin{z} &  \zeta  \cos{z} } \;;  \nonumber \\
m^{}_3 = 0: \hspace{1cm} O = \pmatrix{ \cos{z} & -  \zeta  \sin{z} \cr \sin{z} &  \zeta  \cos{z} \cr 0 & 0 } \;,
\label{eq:3.5.5}
\end{eqnarray}
where $z$ is a complex parameter (for which the relation $\cos^2 z + \sin^2 z =1$ always holds) and $\zeta = \pm 1$ accounts for the possibility of two discrete choices.
For simplicity and definiteness, in the subsequent discussions we just fix $\zeta = + 1$ without loss of generality. By substituting the explicit form of $O$ into Eq.~(\ref{eq:3.5.4}), we obtain the entries of $M^{}_{\rm D}$ as
\begin{eqnarray}
\left(M^{}_{\rm D}\right)^{}_{\alpha 1} = +{\rm i}\sqrt{M^{}_1} \left( U^{}_{\alpha i} \sqrt{m^{}_i} \cos z +  U^{}_{\alpha j} \sqrt{m^{}_j} \sin z \right) \;, \nonumber \\
\left(M^{}_{\rm D}\right)^{}_{\alpha 2} = - {\rm i} \sqrt{M^{}_2} \left( U^{}_{\alpha i} \sqrt{m^{}_i} \sin z -  U^{}_{\alpha j} \sqrt{m^{}_j} \cos z \right) \;,
\label{eq:3.5.6}
\end{eqnarray}
with $i =2$ (or $i=1$) and $j =3$ (or $j=2$) in the $m^{}_1 =0$ (or $m^{}_3 =0$) case.

In the present parametrization, all the model parameters bear clear meanings: seven of them are simply the low-energy flavor parameters, two of them are the heavy Majorana neutrino masses, and the implication of $O$ can be interpreted as follows. First of all,
the low-energy flavor parameters are independent of $O$ at all. However, $O$ plays a crucial role in determining the CP-violating effects in leptogenesis:
\begin{eqnarray}
\frac{\overline \varepsilon^{}_1}{M^{}_2} = \frac{(m^2_i - m^2_j) {\rm Im} \left( \sin^2 z \right) } { m^{}_i \left|\cos z\right|^2+ m^{}_j \left|\sin z\right|^2 } \; ,
\label{eq:3.5.7}
\end{eqnarray}
with $i =2$ or ($i=1$) and $j =3$ or ($j=2$) in the $m^{}_1 =0$ (or $m^{}_3 =0$) case.
Like in the bi-unitary and triangular parametrizations, $U$ has been canceled out in the expression of $\varepsilon^{}_1$. To generate a nonzero $\varepsilon^{}_1$, both the modulus and phase of $z$ should be finite. Given the above facts, one can conclude that the low-energy flavor parameters and unflavored thermal leptogenesis are dependent on two distinct sets of CP phases (i.e., those contained in $U$ and $O$, respectively). It is therefore impossible to establish a direct link between the low-energy CP violation and unflavored leptogenesis. This means that an observation (or absence) of leptonic CP violation at low energies will not necessarily imply a non-vanishing (or vanishing) baryon-antibaryon asymmetry through unflavored thermal leptogenesis. However, it has been pointed out that there exist some physical effects which can invalidate such a conclusion. The most notable example of this kind is the flavor effects which will become relevant if leptogenesis takes place at a temperature below $T\sim 10^{12}$ GeV. This will be the subject of section \ref{section 4.2}. Another obvious example is the non-unitarity effects which can lead to $U^\dagger U \neq I$, implying that  $U$ will not be fully cancelled out in the expression of $\varepsilon^{}_1$ \cite{Xing:2009vb,Rodejohann:2009cq,Antusch2010Non}. In addition, the renormalization-group running effect is also found to be potentially capable of inducing a non-unitarity-like effect, which will be discussed in section \ref{section 4.5}.

Note that $O$ can be regarded as a dominance matrix characterizing the weights of the contributions of each heavy Majorana neutrino to each light Majorana neutrino mass. To
see this point clearly, let us use the orthogonality of $O$ to express $m^{}_i$ as
follows:
\begin{eqnarray}
m^{}_i = \sum^{}_j \left(m^{}_i O^2_{ij}\right) = \sum^{}_j \left(m^{}_i \frac{O^2_{ij} M^{}_j}{M^{}_j}\right) \;.
\label{eq:3.5.8}
\end{eqnarray}
For example, in the case of $O^{2}_{21} \gg O^{2}_{22}$, the generation of $m^{}_2$ can be mainly attributed to $N^{}_1$. Such an interpretation in terms of weights is straightforward for the rotational part of $O$. But one should take care of the boost part of $O$ (controlled by the phase of $z$), for which $O^2_{ij}$ are not necessarily positive or smaller than one.
Although the situation becomes more subtle in this case, the weights argument still holds \cite{Masina:2002qh}.

\setcounter{equation}{0}
\setcounter{figure}{0}

\section{Leptogenesis in the minimal seesaw model}
\label{section 4}

This section deals with the implementations of several typical leptogenesis scenarios in the minimal seesaw mechanism. In section 4.1 we consider the classical vanilla leptogenesis scenario where the heavy Majorana neutrino mass spectrum is assumed to be hierarchical (i.e., $M^{}_2 \gtrsim 3 M^{}_1$). In this case only the contribution of the lighter heavy neutrino  $N^{}_1$ to the final baryon-antibaryon asymmetry is taken into account, and that of the heavier neutrino $N^{}_2$ is assumed to be negligible in consideration of the fact that it may be erased away by the $N^{}_1$-related interactions or the reheating temperature $T^{}_{\rm RH}$ of the Universe after inflation may be too low to guarantee the production of a sufficient amount of $N^{}_2$'s. Furthermore, the flavor compositions (i.e., the so-called flavor effects) of the lepton doublet state coupling with $N^{}_1$ are not considered. In spite of these simplifications, the vanilla leptogenesis scenario grasps most of the main features of leptogenesis and provides a good starting point for incorporating more potentially important effects one by one. We shall relax the above constraint conditions in the following three subsections. In section 4.2 the striking consequences of flavor effects on leptogenesis are explored. In section 4.3 the possible contribution of $N^{}_2$ to the final baryon number asymmetry is studied. In section 4.4 the implications of a particular scenario with $N^{}_1$ and $N^{}_2$ being nearly degenerate in their masses are investigated. Given a huge gap between the seesaw and electroweak scales, we look at the impacts of small but important quantum corrections on leptogenesis in section 4.5.

\subsection{The vanilla leptogenesis}
\label{section 4.1}

In the vanilla leptogenesis scenario, the final baryon number asymmetry can be expressed as a product of several suppression factors:
\begin{eqnarray}
Y^{}_{\rm B} = - c r \varepsilon^{}_1 \kappa^{}_1  \;,
\label{eq:4.1.1}
\end{eqnarray}
where $c$ describes the conversion from the lepton-antilepton asymmetry to the baryon-antibaryon asymmetry via the sphaleron processes as shown in Eq.~(\ref{eq:2.24}), $\varepsilon^{}_1$ denotes the CP-violating asymmetry in the decays of $N^{}_1$ as given in Eq.~(\ref{eq:2.22}), $r \equiv n^{\rm eq}_{N^{}_1}/s |^{}_{T \gg M^{}_1}= 135 \zeta(3)/(4 \pi^4 g^{}_*) \simeq 3.9 \times 10^{-3}$ (with $g^{}_* =106.75$ in the SM or 228.75 in the MSSM) quantifies the ratio of the equilibrium $N^{}_1$ number density to the entropy density at temperature $T \gg M^{}_1$, and $\kappa^{}_1$ is an efficiency factor accounting for the washout effect due to various inverse decays and scattering processes. A calculation of the value of $\kappa^{}_1$ relies on numerically solving a full set of Boltzmann equations, and this constitutes the most difficult part of evaluating $Y^{}_{\rm B}$.

In the neglect of the $N^{}_i$-mediated $\Delta L =2$ scattering processes ($LH \to \bar L \bar H$ and $L L \to \bar H \bar H$ as well as their inverse processes), which is justified for $M^{}_1 \ll 10^{14}$ GeV --- a condition that holds in most of the viable minimal seesaw models, $\kappa^{}_1$ depends only upon the washout mass parameter \cite{Buchmuller:1996pa}
\begin{eqnarray}
\widetilde m^{}_1 \equiv 8 \pi \frac{v^2}{M^2_1} \Gamma^{}_{1} = \frac{\left( M^\dagger_{\rm D} M^{}_{\rm D} \right)^{}_{11}}{M^{}_1} \;,
\label{eq:4.1.2}
\end{eqnarray}
where $\Gamma^{}_{1} = (Y^\dagger_\nu Y^{}_\nu)^{}_{11} M^{}_1/(8\pi) $ is the decay rate of $N^{}_1$. The ratio of $\widetilde m^{}_1$ to the so-called equilibrium neutrino mass
\begin{eqnarray}
m^{}_* \equiv  8 \pi \frac{v^2}{M^2_1} H(M^{}_1) = \frac{16 \pi^{5/2}}{3\sqrt{5}} \sqrt{g^{}_*} \frac{v^2}{M^{}_{\rm Pl}} \simeq 1.1 \times 10^{-3} \ {\rm eV} \; ,
\label{eq:4.1.3}
\end{eqnarray}
where $H(T) \simeq 1.66 \sqrt{g^{}_*} T^2/M^{}_{\rm Pl}$ is the expansion rate of the Universe at temperature $T$, characterizes the strength of the washout effect. The $\widetilde m^{}_1 < m^{}_*$ and $\widetilde m^{}_1 > m^{}_*$ regimes, or equivalently the $\Gamma^{}_{1} < H (M^{}_1)$ and $\Gamma^{}_{1} > H (M^{}_1)$ regimes, will be referred to respectively as the {\it weak} and {\it strong} washout regimes. A detailed study shows that in the weak washout regime the value of $\kappa^{}_1$ has a strong dependence on the unknown initial conditions (e.g., the initial abundance of $N^{}_1$ and pre-existing asymmetries), rendering the picture not self-contained, while in the strong washout regime this dependence evaporates. Therefore, an optimal situation is expected to be that the washout effect is strong enough to erase any memory of the initial conditions but it is not too strong to allow leptogenesis to work successfully.
For the minimal seesaw model under discussion, one may use the Casas-Ibarra parametrization to explicitly express $\widetilde m^{}_1$ as
\begin{eqnarray}
\widetilde m^{}_1 = m^{}_i \left|\cos z\right|^2+ m^{}_j \left|\sin z\right|^2 \;,
\label{eq:4.1.4}
\end{eqnarray}
where $i =2$ (or $i=1$) and $j =3$ (or $j=2$) in the $m^{}_1 =0$ (or $m^{}_3 =0$) case. One finds that $\widetilde m^{}_1$ is actually independent of $M^{}_1$, and thus $\kappa^{}_1$ is also
independent of $M^{}_1$. Furthermore, it is easy to see \cite{Fujii:2002jw}
\begin{eqnarray}
\widetilde m^{}_1 > m^{}_{\rm min} \left(\left|\cos z\right|^2+ \left|\sin z\right|^2\right) \ge m^{}_{\rm min} \left|\cos^2 z+ \sin^2 z\right| = m^{}_{\rm min} \;,
\label{eq:4.1.5}
\end{eqnarray}
where $m^{}_{\rm min}$ denotes $m^{}_2 \simeq 8.6 \times 10^{-3}$ eV (or $m^{}_1 \simeq 4.9 \times 10^{-2}$ eV) in the $m^{}_1 =0$ (or $m^{}_3 =0$) case, implying that one will be restricted to the strong washout regime. In comparison, there exists no upper bound on $\widetilde m^{}_1$. But one typically has $\widetilde m^{}_1 \lesssim m^{}_{\rm max} \simeq 0.05$ eV (with $m^{}_{\rm max}$ denoting the maximal neutrino mass) if strong fine tunings of the parameters are barred. It is quite impressive that the allowed range of $\widetilde m^{}_1$ as indicated by current experimental data has just the right value for realizing the aforementioned optimal situation of thermal leptogenesis.

For the strong washout regime, the value of $\kappa^{}_1$ can be estimated in an intuitive way as follows. As we know from thermodynamics, the baryon number asymmetry produced in the decays of $N^{}_1$ will be substantially erased by the inverse decays (ID) unless the following out-of-equilibrium condition is satisfied:
\begin{eqnarray}
\Gamma^{}_{1 \rm ID}(T) \simeq \frac{1}{2} \Gamma^{}_{1} e^{-M^{}_1/T} < H(T) \;,
\label{eq:4.1.6}
\end{eqnarray}
where $\Gamma^{}_{1 \rm ID}(T)$ denotes the inverse-decay rate of $N^{}_1$ at temperature $T$.
The critical temperature $T^{}_f$ at which this condition is initially fulfilled (i.e., $\Gamma^{}_{1 \rm ID}(T^{}_f) \simeq H(T^{}_f)$ holds) is found to be $M^{}_1/6 \gtrsim T^{}_f \gtrsim M^{}_1/8$ for the experimentally favored region $8.6 \times 10^{-3} ~{\rm eV} \leq \widetilde m^{}_1 \lesssim 0.05 ~{\rm eV}$ \cite{Buchmuller:2004nz}.
Once the temperature of the Universe drops below $T^{}_f$, most of the produced baryon number asymmetry can survive the washout effect and contribute to today's value of $Y^{}_{\rm B}$. Now that the remaining number density of $N^{}_1$  at $T^{}_f$ is Boltzmann-suppressed by $e^{-M^{}_1/T^{}_f}$, one may approximately obtain
\cite{article}
\begin{eqnarray}
\kappa^{}_1 \simeq \frac{n^{}_{N^{}_1}(T^{}_f)}{n^{}_{N^{}_1}(T \gg M^{}_1)} \simeq e^{-M^{}_1/T^{}_f} \simeq  2 \frac{m^{}_*}{\widetilde m^{}_1} \cdot \frac{T^2_f}{M^2_1} \;,
\label{eq:4.1.7}
\end{eqnarray}
with the help of Eqs.~(\ref{eq:4.1.2}) and (\ref{eq:4.1.3}).
An analytical approximation with a better degree of accuracy, $\kappa^{}_1 \simeq 2 m^{}_* T^{}_f /(\widetilde m^{}_1 M^{}_1)$, has been carefully derived in Ref.~\cite{Buchmuller:2004nz}. We observe that in the strong washout regime $\kappa^{}_1$ is roughly inversely proportional to ${\widetilde m^{}_1}$. In our numerical calculations we are going to make use of the following simpler empirical fit formula for the efficiency factor \cite{Giudice:2003jh}:
\begin{eqnarray}
\frac{1}{\kappa(\widetilde m^{})} \simeq \frac{3.3 \times 10^{-3} ~{\rm eV}}{\widetilde m} + \left( \frac{\widetilde m} {5.5 \times 10^{-4} ~{\rm eV}} \right)^{1.16} \;.
\label{eq:4.1.8}
\end{eqnarray}
This expression is applicable in both strong and weak washout regimes assuming the vanishing initial heavy Majorana neutrino abundances. It is useful to notice that $\kappa(\widetilde m)$ has a maximal value $\sim 0.2$ which is reached at $\widetilde m \simeq m^{}_*$. As a final note, one should bear in mind that the above results are obtained under the prerequisite of $M^{}_1 \ll 10^{14}$ GeV. Otherwise, the $\Delta L =2$ scattering processes would attain equilibrium and greatly suppress the value of $\kappa^{}_1$.

To proceed, let us turn to the CP-violating asymmetry $\varepsilon^{}_1$. In the Casas-Ibarra parametrization $\varepsilon^{}_1$ is expressed as
\begin{eqnarray}
\varepsilon^{}_1 \simeq \frac{3 M^{}_1}{16\pi v^2} \cdot
\frac{ \left(m^2_j - m^2_i\right) {\rm Im} \left(\sin^2 z\right) } { m^{}_i \left|\cos z\right|^2+ m^{}_j \left|\sin z\right|^2 } \;,
\label{eq:4.1.9}
\end{eqnarray}
where $i =2$ (or $i=1$) and $j =3$ (or $j=2$) in the $m^{}_1 =0$ (or $m^{}_3 =0$) case. In obtaining this result from Eq.~(\ref{eq:2.22}), the approximation ${\cal F}(M^2_2/M^2_1) \simeq - 3 M^{}_1/(2M^{}_2)$ has been taken for $M^{}_2 \gg M^{}_1$. We see that in this approximation  $\varepsilon^{}_1$ is proportional to $M^{}_1$ and independent of $M^{}_2$.
It has been pointed out that there exists an upper bound on $|\varepsilon^{}_1|$
\cite{Chankowski:2003rr}. With the help of the reparametrization $\cos^2 z = x+ {\rm i}y$ and $\sin^2 z =1-x -{\rm i} y$ with $x$ and $y$ being real \cite{article}, one finds
\begin{eqnarray}
\left|\varepsilon^{}_1\right| & \simeq & \frac{3 M^{}_1}{16\pi v^2} \cdot
\frac{ \left(m^2_j- m^2_i\right) \left|y\right| } { m^{}_i \sqrt{x^2+y^2}+ m^{}_j \sqrt{\left(1-x\right)^2+y^2} }
\nonumber \\
& \lesssim & \frac{3 M^{}_1}{16\pi v^2} \cdot \frac{ \left(m^2_j- m^2_i\right) \left|y\right| } { m^{}_i \sqrt{1+y^2}+ m^{}_j \left|y\right| }
\nonumber \\
& = & \frac{3 M^{}_1}{16\pi v^2} \cdot \frac{m^2_j- m^2_i} { m^{}_i \sqrt{1+1/y^2}+ m^{}_j  }
\nonumber \\
& \leq &
\frac{3 M^{}_1}{16\pi v^2} \left(m^{}_j- m^{}_i\right)
\equiv |\varepsilon^{}_1|^{}_{\rm max} \;.
\label{eq:4.1.10}
\end{eqnarray}
Three immediate comments on this result are in order.
(1) One can see that $|\varepsilon^{}_1|^{}_{\rm max}$ is reached for $y \to \infty$, which brings about undesired non-perturbative Yukawa couplings. But, in fact, one just needs to have $x\simeq 1$ and $|y|\gtrsim 1$ in order to get a value of $|\varepsilon^{}_1|$ which is not far from $|\varepsilon^{}_1|^{}_{\rm max}$. For example, $x=1$ and $|y|=1$ lead us to $|\varepsilon^{}_1|/|\varepsilon^{}_1|^{}_{\rm max} \simeq 0.94$ (or 0.83) in the $m^{}_1 =0$ (or $m^{}_3 =0$) case. (2) In the $m^{}_3 =0$ case, due to a large cancellation between $m^{}_1$ and $m^{}_2$, the size of $|\varepsilon^{}_1|^{}_{\rm max}$ is suppressed by a factor of about 56 as compared with its size in the $m^{}_1 =0$ case. (3) The bound $|\varepsilon^{}_1|^{}_{\rm max}$ is lower than its counterpart in a generic seesaw model, known as the Davidson-Ibarra (DI) bound \cite{Davidson:2002qv}
\footnote{A more stringent bound has been derived in Ref.~\cite{Hambye:2003rt}.}
\begin{eqnarray}
|\varepsilon^{}_1|^{\rm DI}_{\rm max} \simeq \frac{3 M^{}_1}{16\pi v^2} m^{}_{\rm max} \;,
\label{eq:4.1.11}
\end{eqnarray}
by about $17\%$ (or $99\%$) in the $m^{}_1 =0$ (or $m^{}_3 =0$) case.

Now let us take a look at the upper bound of $|Y^{}_{\rm B}|$ and correspondingly the lower bound of $M^{}_1$ from the requirement of successful thermal leptogenesis. Employing the aforementioned result $\kappa^{}_1 \simeq 2 m^{}_* T^{}_f /(\tilde m^{}_1 M^{}_1)$ and the reparametrization $\cos^2 z = x+ {\rm i}y$ and $\sin^2 z =1-x -{\rm i} y$, we arrive at
\begin{eqnarray}
\left|Y^{}_{\rm B}\right| & \simeq & c r \frac{3 m^{}_* T^{}_f}{8\pi v^2} \cdot
\frac{ \left(m^2_j- m^2_i\right) \left|y\right| } { \left[ m^{}_i \sqrt{x^2+y^2}+ m^{}_j \sqrt{\left(1-x\right)^2+y^2} \right]^2 }  \;.
\label{eq:4.1.12}
\end{eqnarray}
A numerical calculation shows that $|Y^{}_{\rm B}|$ has a maximum value $|Y^{}_{\rm B}|^{}_{\rm max} \simeq 1.4 \times 10^{-21} M^{}_1$ GeV$^{-1}$ at $x\simeq 0.97$ and $y \simeq 0.16$ in the $m^{}_1 =0$ case, or $|Y^{}_{\rm B}|^{}_{\rm max} \simeq 4.4 \times 10^{-24} M^{}_1$ GeV$^{-1}$ at $x\simeq 0.5$ and $y \simeq 0.5$ in the $m^{}_3 =0$ case. Accordingly, requiring $|Y^{}_{\rm B}|^{}_{\rm max}$ to be larger than the observed value of $|Y^{}_{\rm B}|$ yields a lower bound $M^{\rm min}_{1} \simeq 6 \times 10^{10}$ GeV (or $2 \times 10^{13}$ GeV) in the $m^{}_1 =0$ (or $m^{}_3 =0$) case \cite{Chankowski:2003rr}. Note that such a bound is consistent with the neglect of those $\Delta L=2$ scattering processes.

A consensus has nowadays been reached in the Big Bang cosmology: the very early Universe once underwent an inflationary phase driven by the inflaton field. At the end of inflation the inflaton decays into lighter particles, reheating the Universe to a temperature $T^{}_{\rm RH}$. The requirement that a sufficient amount of $N^{}_1$'s (for leptogenesis to work successfully) can be produced places a lower bound on $T^{}_{\rm RH}$ \cite{Croon:2019dfw}. Naively, one may expect that such a bound $T^{\rm min}_{\rm RH}$ roughly coincides with $M^{\rm min}_1$. In the strong washout regime which is the case for the minimal seesaw model, however, $T^{\rm min}_{\rm RH}$ can be about one order of magnitude smaller than $M^{\rm min}_1$. The point is that the baryon number asymmetry that can survive the washout effect is dominantly produced around $T^{}_f $ where the inverse decays eventually departure from equilibrium. Consequently, $T^{\rm min}_{\rm RH}$ just needs to be somewhat larger than $T^{}_f$. A detailed analysis gives $T^{\rm min}_{\rm RH} \simeq M^{\rm min}_1 T^{}_f/(M^{\rm min}_1-2T^{}_f)$ \cite{Buchmuller:2004nz}. For the experimentally favored region $8.6 \times 10^{-3} ~{\rm eV} < \tilde m^{}_1 \lesssim 0.05 ~{\rm eV}$, one has $T^{\rm min}_{\rm RH} \simeq M^{\rm min}_1/5$. In those supersymmetric extensions of the SM, the bound $T^{\rm min}_{\rm RH}$ will be problematic since it is incompatible with the upper bound $T^{\rm max}_{\rm RH} \simeq 10^9$ GeV from the requirement of avoiding the overproduction of gravitinos which might spoil the success of the Big Bang nucleosynthesis \cite{Khlopov1984Is,Ellis:1984eq}.

Finally, we emphasize again that the results in this subsection are obtained under the assumptions of the heavy Majorana neutrino mass spectrum being hierarchical, the contribution of $N^{}_2$ to the final baryon number asymmetry being negligible and the flavor effects playing no role in leptogenesis. When such constraint conditions are relaxed, the situation may change dramatically as will be shown in the next three subsections.

\subsection{Flavored leptogenesis}
\label{section 4.2}

The inclusion of flavor effects on leptogenesis provides a very essential modification for the calculation of the final baryon number asymmetry, as compared with the calculation in the vanilla leptogenesis scenario. The importance of flavor effects follows from that of the washout effect. In the washout processes the lepton doublets participate as the initial states, so one needs to know which flavors are distinguishable before calculating the washout interaction rates. As is known, different lepton flavors are distinguished by their Yukawa couplings $y^{}_\alpha$ (for $\alpha =e, \mu$ and $\tau$). When the $y^{}_\alpha$-related interactions enter into equilibrium, the corresponding lepton flavor $\alpha$ will become distinguishable. By comparing the $y^{}_\alpha$-related interaction rates with the Hubble expansion rate of the Universe, we learn that the $y^{}_\tau$, $y^{}_\mu$ and $y^{}_e$-related interactions are in equilibrium below about $10^{12}$ GeV, $10^9$ GeV and $10^{6}$ GeV, respectively \cite{Campbell:1992jd,Cline:1993bd}. This observation means that the unflavored regime only applies in the temperature range above $10^{12}$ GeV. In the $10^9 ~{\rm GeV} \lesssim T \lesssim 10^{12} ~{\rm GeV}$ range, the $\tau$ flavor becomes distinguishable but $\mu$ and $e$ flavors are still not, leading us to an effective two-flavor regime. When the temperature of the Universe drops below $10^9$ GeV, the indistinguishability between $\mu$ and $e$ flavors will be resolved by the $y^{}_\mu$-related interactions, leading us to a full three-flavor regime.

Flavor effects on leptogenesis may bring about several striking consequences. (1) Contrary to unflavored leptogenesis, flavored leptogenesis allows the PMNS neutrino mixing parameters to directly enter into calculations of the baryon number asymmetry via the Yukawa coupling matrix as shown in Eq.~(\ref{eq:2.21}), making it possible to establish a direct link between the low-energy CP violation and leptogenesis \cite{Blanchet:2006be,Abada:2006ea,Nardi:2006fx,Pascoli:2006ci}. (2) Flavored leptogenesis may successfully take effect in spite of $\varepsilon^{}_1 =0$, a case  which definitely forbids unflavored leptogenesis to work \cite{Nardi:2006fx}. (3) Due to the presence of more CP-violating phases and a potential reduction of the washout effect on a specific flavor in the flavored regime, it is usually easier (with less fine tuning of the model parameters) to account for the observed baryon-antibaryon asymmetry through flavored leptogenesis rather than unflavored leptogenesis. Such qualitatively interesting and quantitatively significant flavor effects should therefore be taken into account in implementing the baryogenesis-via-leptogenesis idea, if the seesaw scale is below $T \sim 10^{12}$ GeV.

Here we explore the implications of flavor effects on leptogenesis in the minimal seesaw model by assuming its heavy Majorana neutrinos to have a hierarchical mass spectrum \cite{Pascoli:2006ci}.
As will be seen later, even after the inclusion of flavor effects, the lower bound on $M^{}_1$ obtained from the requirement of successful leptogenesis remains far above $10^9$ GeV. Hence one just needs to work in the two-flavor regime and realize the so-called $\tau$-flavored leptogenesis. In this regime the lepton state produced in the decays of $N^{}_1$ collapses into a mixture of the $\tau$ component and another component orthogonal to it (denoted by $o$) which is a coherent superposition of $\mu$ and $e$ flavors. Since the CP-violating asymmetries stored in the $\tau$ and $o$ components are subject to different washout effects, they have to be treated separately. In this way the final baryon number asymmetry receives two kinds of contributions \cite{Abada:2006ea}:
\begin{eqnarray}
Y^{}_{\rm B} & = & - c r \left( \varepsilon^{}_{1 \tau} \kappa^{}_{1\tau}  + \varepsilon^{}_{1 o} \kappa^{}_{1o} \right)
\nonumber \\
& = & - c r \left[ \varepsilon^{}_{1 \tau} \kappa \left(\frac{390}{589} \widetilde m^{}_{1\tau} \right) + \varepsilon^{}_{1 o} \kappa \left(\frac{417}{589} \widetilde m^{}_{1o} \right) \right] \;,
\label{eq:4.2.1}
\end{eqnarray}
in which the flavored CP-violating asymmetries are weighted by the corresponding efficiency factors. Note that the coefficients multiplying $\tilde m^{}_{1\tau}$ and $\tilde m^{}_{1o}$ arise from the fact that the lepton doublet asymmetry upon which the washout processes act only constitutes a (major) part of the lepton number asymmetry $Y^{}_{\rm L}$ due to the redistribution of $Y^{}_{\rm L}$ to the singlet leptons via the so-called spectator processes \cite{Buchmuller:2001sr,Nardi:2005hs}. On the one hand, the expressions of flavored CP-violating asymmetries $\varepsilon^{}_{1\alpha}$ have been given in Eq.~(\ref{eq:2.21}). One may define
$\varepsilon^{}_{1 o} \equiv \varepsilon^{}_{1 \mu} + \varepsilon^{}_{1 e}$ in the $\tau$-flavored regime. In the Casas-Ibarra parametrization, $\varepsilon^{}_{1\alpha}$ can be recast as
\cite{Abada:2006ea}
\begin{eqnarray}
\hspace{-0.5cm}
\varepsilon^{}_{1\alpha} & \simeq & \frac{3 M^{}_1}{16 \pi v^2 \left(m^{}_i \left|\cos z\right|^2+ m^{}_j \left|\sin z\right|^2\right)} \left\{ \left( m^2_j \left|U^{}_{\alpha j}\right|^2 - m^2_i \left|U^{}_{\alpha i}\right|^2 \right) {\rm Im}\left(\sin^2 z\right) \right.
\nonumber \\
&& + \sqrt{m^{}_j m^{}_i} \left[ \left(m^{}_j - m^{}_i\right) {\rm Im}\left(U^{}_{\alpha i} U^*_{\alpha j}\right) {\rm Re}\left(\cos z \sin z\right) \right.
\nonumber \\
&& + \left. \left. \left(m^{}_j + m^{}_i\right) {\rm Re}\left(U^{}_{\alpha i} U^*_{\alpha j}\right) {\rm Im}\left(\cos z \sin z\right) \right] \right\} \;,
\label{eq:4.2.2}
\end{eqnarray}
where $i =2$ (or $i=1$) and $j =3$ (or $j=2$) in the $m^{}_1 =0$ (or $m^{}_3 =0$) case. Here we have neglected the term proportional to ${\cal G}(x)$ because it is suppressed by a factor of $M^{}_1/M^{}_2$ compared to the term proportional to ${\cal F}(x)$. On the other hand, the efficiency factors are determined by the flavored washout mass parameters
\begin{eqnarray}
\widetilde m^{}_{1\alpha} = \frac{\left|(M^{}_{\rm D})^{}_{\alpha 1}\right|^2}{M^{}_1}  \;, \hspace{1cm}
\widetilde m^{}_{1o} \equiv \widetilde m^{}_{1\mu} + \widetilde m^{}_{1e} \; .
\label{eq:4.2.3}
\end{eqnarray}
The former can explicitly be expressed as
\begin{eqnarray}
\widetilde m^{}_{1\alpha} = \left| U^{}_{\alpha i} \sqrt{m^{}_i} \cos z +  U^{}_{\alpha j} \sqrt{m^{}_j} \sin z \right|^2 \;,
\label{eq:4.2.4}
\end{eqnarray}
where $i =2$ (or $i=1$) and $j =3$ (or $j=2$) in the $m^{}_1 =0$ (or $m^{}_3 =0$) case.
It is immediate to see that $\widetilde m^{}_{1o} = \widetilde m^{}_1 - \widetilde m^{}_{1\tau}$ holds.

Now let us pay attention to a simple but interesting case in which $\sin z$ is either real or purely imaginary. In this case we are left with $\varepsilon^{}_1 =0$ as indicated by Eq.~(\ref{eq:4.1.9}), and hence flavor effects become crucial to achieve successful leptogenesis. For real (or purely imaginary) $\sin z$, one may relabel $\sin z$ and $\cos z$ as $\sin \theta$ and $\cos \theta$ (or ${\rm i}\sinh y$ and $\cosh y$) with $\theta$ ($y$) being a real parameter. Then Eq.~(\ref{eq:4.2.2}) is accordingly simplified to
\begin{eqnarray}
\hspace{-0.5cm}
\varepsilon^{}_{1\alpha} = \frac{3 M^{}_1}{8 \pi v^2 } \cdot \frac{ \sin 2 \theta}{ m^{}_i \cos^2 \theta + m^{}_j \sin^2 \theta} \sqrt{m^{}_j m^{}_i} \left(m^{}_j - m^{}_i\right) {\rm Im}\left(U^{}_{\alpha i} U^*_{\alpha j}\right) \; ,
\label{eq:4.2.5a}
\end{eqnarray}
or
\begin{eqnarray}
\hspace{-0.5cm}
\varepsilon^{}_{1\alpha} = \frac{3 M^{}_1}{8 \pi v^2 } \cdot \frac{ \sinh 2y}{m^{}_i \cosh^2 y + m^{}_j \sinh^2 y } \sqrt{m^{}_j m^{}_i} \left(m^{}_j + m^{}_i\right) {\rm Re}\left(U^{}_{\alpha i} U^*_{\alpha j}\right) \;.
\label{eq:4.2.5b}
\end{eqnarray}
Note that $\varepsilon^{}_1 =0$ leads to $\varepsilon^{}_{1o} = -\varepsilon^{}_{1\tau}$ and thus we arrive at
\begin{eqnarray}
Y^{}_{\rm B} = - c r \varepsilon^{}_{1 \tau} \left[ \kappa \left(0.66 \widetilde m^{}_{1\tau} \right) - \kappa \left(0.71 \widetilde m^{}_{1o} \right) \right]  \;.
\label{eq:4.2.6}
\end{eqnarray}
Apparently, the total efficiency factor $\kappa^{}_1 = \kappa(0.66 \widetilde m^{}_{1\tau})-\kappa(0.71 \widetilde m^{}_{1o})$ would suffer a large cancellation if $\widetilde m^{}_{1\tau}$ and $\widetilde m^{}_{1o}$ were very close to each other.
Given that the function $\kappa(\widetilde m)$ takes its maximal value at $\widetilde m \simeq m^{}_*$ and the sum of $\widetilde m^{}_{1 \tau}$ and $\widetilde m^{}_{1o}$ (i.e., $\widetilde m^{}_1$) has a lower bound equal to $m^{}_2 \simeq 8 m^{}_*$, it is conceivable that the maximal value of $|\kappa^{}_1|$ will be realized by having either $\tilde m^{}_{1\tau}$ or $\tilde m^{}_{1o}$ around $m^{}_*$. The reason is simply that such a parameter setup will enable one component of $\kappa^{}_1$ to achieve its maximal value and the other to be simultaneously suppressed.
\begin{figure*}[t]
\centering
\includegraphics[width=6in]{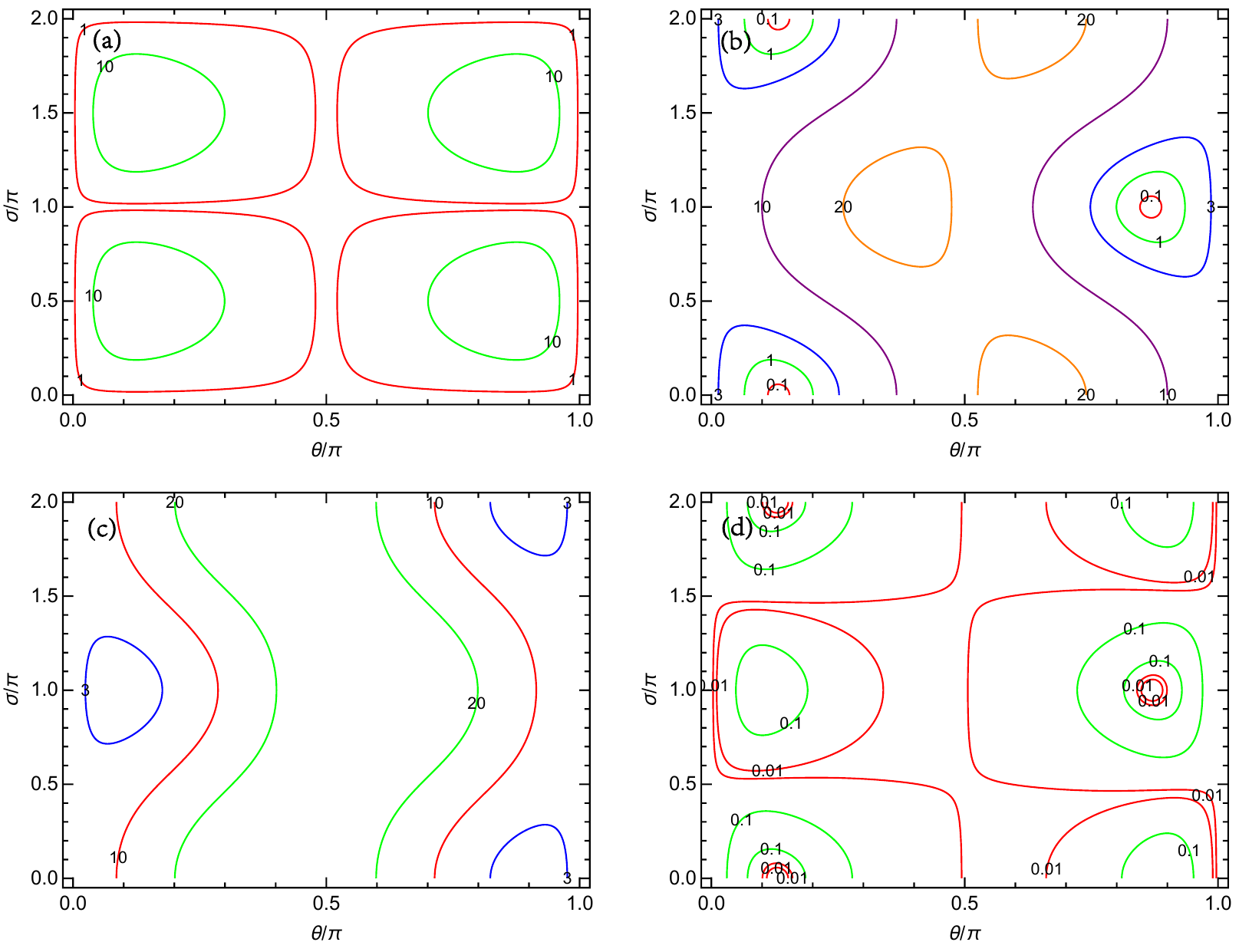}
\caption{ In the $m^{}_1=0$ case with $z$ being real (i.e., $z = \theta$) and $\delta =0$, the contour lines of (a) $|\varepsilon^{}_{1\tau}|/10^{-8}$ for $M^{}_1 =10^{10}$ GeV, (b) $\widetilde m^{}_{1\tau}/m^{}_*$, (c) $\widetilde m^{}_{1o}/m^{}_*$ and (d) $|\kappa^{}_1|$ are shown in the $\theta$-$\sigma$ plane.}
\label{Fig:4-1}
\end{figure*}

\begin{figure*}[t]
\centering
\includegraphics[width=6in]{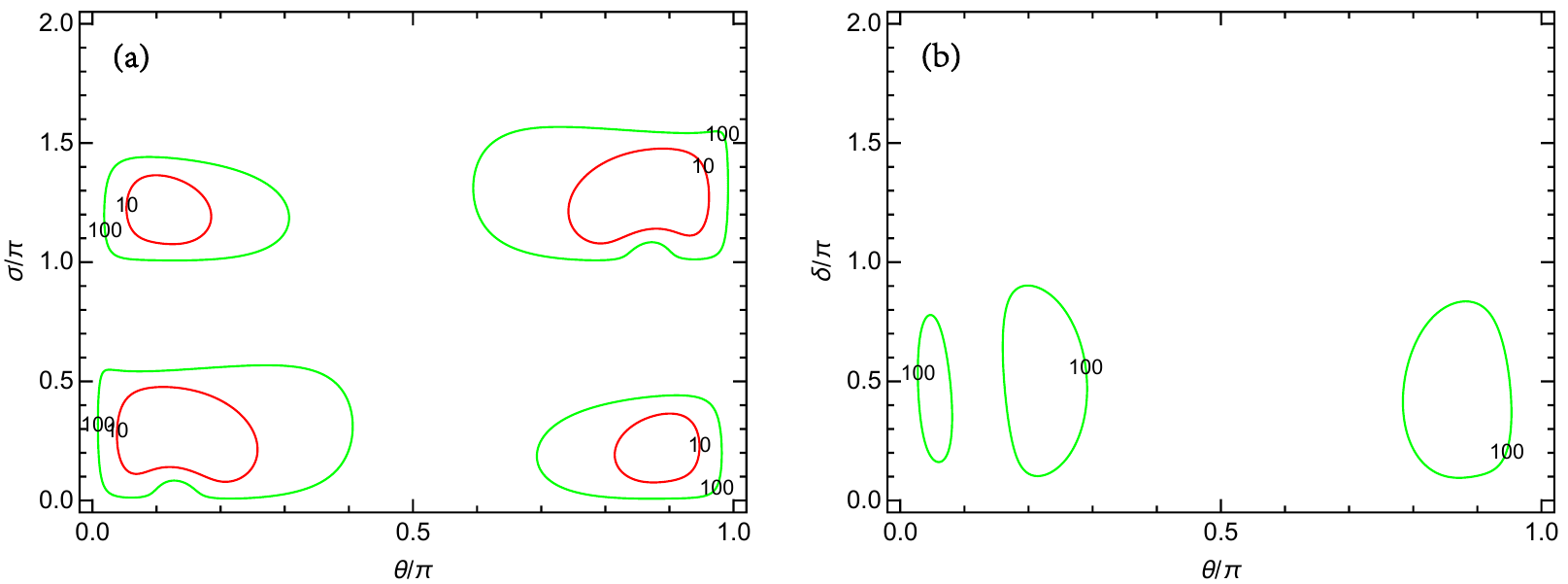}
\caption{In the $m^{}_1=0$ case with $z$ being real (i.e., $z = \theta$), the contour lines of $M^{}_1/(10^{10} \ \rm GeV)$ for successful flavored leptogenesis are shown (a) in the $\theta$-$\sigma$ plane by assuming $\delta=0$ and (b) in the $\theta$-$\delta$ plane by assuming $\sigma=0$.}
\label{Fig:4-2}
\end{figure*}

We first consider the $m^{}_1 =0$ case. For real $\sin z$, $\varepsilon^{}_{1\tau}$ is proportional to
\begin{eqnarray}
{\rm Im}\left(U^{}_{\tau 2} U^*_{\tau 3}\right) = - c^{}_{13} \left[ c^{}_{12} c^{}_{23} s^{}_{23} \sin \sigma + s^{}_{12} s^{}_{13} c^2_{23} \sin \left(\delta+\sigma\right) \right] \;
\label{eq:4.2.7}
\end{eqnarray}
in the standard parametrization of $U$, and $\widetilde m^{}_{1\tau}$ can be expressed as
\begin{eqnarray}
\hspace{-1cm} \widetilde m^{}_{1\tau} \simeq m^{}_2 c^2_{12} s^2_{23} \cos^2 \theta + m^{}_3 c^2_{23} \sin^2 \theta - 2 \sqrt{m^{}_2 m^{}_3} \ c^{}_{12} c^{}_{23} s^{}_{23} \cos\theta \sin\theta \cos\sigma \;
\label{eq:4.2.8}
\end{eqnarray}
by taking $\theta^{}_{13} =0$ as a reasonable approximation. Let us first examine the possibility that $\sigma$ serves as the only source of CP violation by assuming $\delta =0$. In this case it is easy to see that $|\varepsilon^{}_{1\tau}|$ has a maximal value of $1.8 \times 10^{-7} (M^{}_1/10^{10} \ {\rm GeV})$ at $\sigma = \pm \pi/2$ and $\theta = \arctan (\pm \sqrt{m^{}_2/m^{}_3}) \simeq 0.12 \pi$ or $0.88\pi$. For illustration, in Fig.~\ref{Fig:4-1}(a)---Fig.~\ref{Fig:4-1}(d) the contour lines of (a) $|\varepsilon^{}_{1\tau}|/10^{-8}$ for $M^{}_1 =10^{10}$ GeV, (b) $\widetilde m^{}_{1\tau}/m^{}_*$, (c) $\widetilde m^{}_{1o}/m^{}_*$ and (d) $|\kappa^{}_1|$ are shown in the $\theta$-$\sigma$ plane, respectively. In obtaining these (and the following) numerical results, the best-fit values of relevant neutrino oscillation parameters have been used as the typical inputs. It is apparent that the results shown in Fig.~\ref{Fig:4-1} exhibit a symmetry under the joint transformations $\sigma \to \sigma+\pi$ and $\theta \to \pi-\theta$. We see that $\widetilde m^{}_{1\tau}$ may vanish at $\sigma = 0$ or $\pi$ and $\theta \simeq \arctan (\pm \sqrt{m^{}_2/m^{}_3} \ c^{}_{12} s^{}_{23}/c^{}_{23}) \simeq  0.12 \pi$ or $0.88\pi$, and its value is larger than that of $m^{}_*$ in most of the parameter space. And $\widetilde m^{}_{1o}$ is always larger than $m^{}_*$ and can be relatively small in the vicinity of $[\theta, \sigma] \sim [0.1 \pi, \pi]$ and $[0.9 \pi, 0]$. As explained below Eq.~(\ref{eq:4.2.6}), $|\kappa^{}_1|$ is expected to take the maximal value when either $\widetilde m^{}_{1\tau}$ or $\widetilde m^{}_{1o}$ is around $m^{}_*$. We find that $|\kappa^{}_1|$ assumes its local maximal value 0.17 (or 0.15) at $[\theta, \sigma] \simeq [0.22 \pi, 0]$ and $[0.78\pi, \pi]$ (or $[\theta, \sigma] \simeq [0.13 \pi, \pi]$ and $[0.87 \pi, 0]$).
Finally, Fig.~\ref{Fig:4-2}(a) illustrates the contour lines of $M^{}_1/(10^{10} \ \rm GeV)$ for successful leptogenesis in the $\theta$-$\sigma$ plane. Given the fact that flavor effects will disappear for $M^{}_1 \gtrsim 10^{12}$ GeV, only in the parameter regions enclosed by the contour lines 100 of $M^{}_1/(10^{10} \ \rm GeV)$ (equivalent to $M^{}_1 \simeq 10^{12}$ GeV) can successful flavored leptogenesis be achieved. In the special case under discussion the lower bound of $M^{}_1$ is found to be $3.2 \times 10^{10}$ GeV at $[\theta, \sigma] \simeq [0.13 \pi, 0.27 \pi]$ and $[0.87 \pi, 1.27 \pi]$.

\begin{figure*}[t]
\centering
\includegraphics[width=6in]{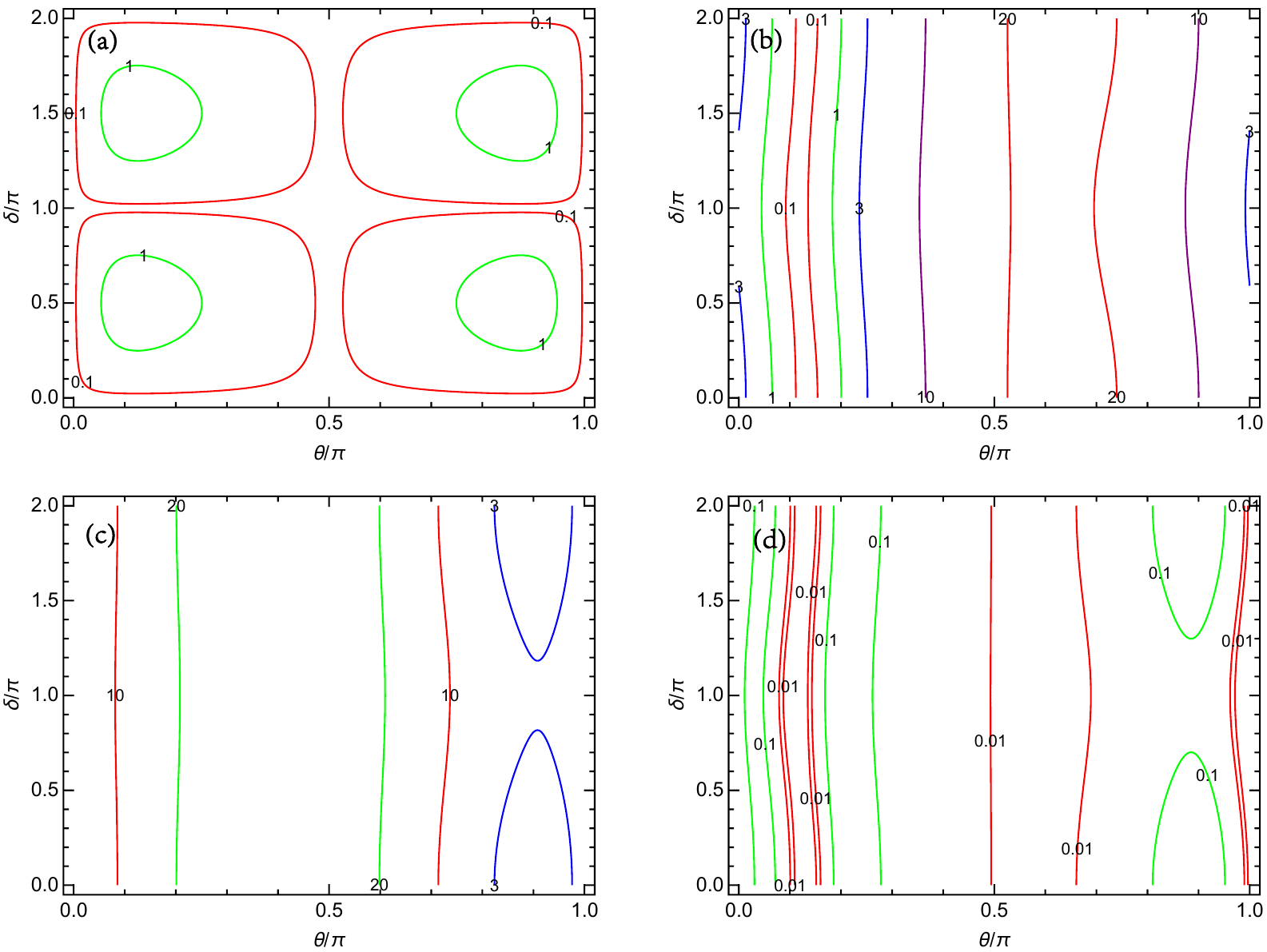}
\caption{In the $m^{}_1=0$ case with $z$ being real (i.e., $z = \theta$) and $\sigma =0$, the contour lines of (a) $|\varepsilon^{}_{1\tau}|/10^{-8}$ for $M^{}_1 =10^{10}$ GeV, (b) $\widetilde m^{}_{1\tau}/m^{}_*$, (c) $\widetilde m^{}_{1o}/m^{}_*$ and (d) $|\kappa^{}_1|$ are illustrated in the $\theta$-$\delta$ plane.}
\label{Fig:4-3}
\end{figure*}

Next, we look at the possibility that $\delta$ serves as the only source of CP violation by assuming $\sigma =0$. In this special case $|\varepsilon^{}_{1\tau}|$ is suppressed by a factor of $s^{}_{13}$ compared to in the former case, as one can see from Eq.~(\ref{eq:4.2.7}); and it has a maximal value of $1.4 \times 10^{-8} (M^{}_1/10^{10} \ {\rm GeV})$ at $\delta = \pm \pi/2$ and $\theta \simeq 0.12 \pi$ or $0.88\pi$. To illustrate, in Figs.~\ref{Fig:4-3}(a)---\ref{Fig:4-3}(d) the contour lines of (a) $|\varepsilon^{}_{1\tau}|/10^{-8}$ for $M^{}_1 =10^{10}$ GeV, (b) $\widetilde m^{}_{1\tau}/m^{}_*$, (c) $\widetilde m^{}_{1o}/m^{}_*$ and (d) $|\kappa^{}_1|$ are shown in the $\theta$-$\delta$ plane, respectively. It turns out that $\widetilde m^{}_{1\tau}$, $\widetilde m^{}_{1o}$ and $|\kappa^{}_1|$ depend weakly on $\delta$ due to its association with the suppression factor $s^{}_{13}$, and their values are mainly determined by $\theta$. We find that $\widetilde m^{}_{1\tau}$ can be around or smaller than $m^{}_*$ for $\theta \sim 0.1 \pi$ and larger than $m^{}_*$ in most of the remaining parameter space, and $\widetilde m^{}_{1o}$ is always larger than $m^{}_*$ and can be relatively small for $\theta \sim 0.9 \pi$. Consequently, $|\kappa^{}_1|$ can have a relatively large value for $\theta \sim 0.1 \pi$ or $0.9 \pi$. Finally, Fig.~\ref{Fig:4-2}(b) shows the contour lines of $M^{}_1/(10^{10} \ \rm GeV)$ for successful flavored leptogenesis in the $\theta$-$\delta$ plane. We see that in this case the allowed parameter space is rather small. And the lower bound of $M^{}_1$ is found to be $3.1 \times 10^{11}$ GeV at $[\theta, \delta] \simeq [0.21 \pi, 0.5 \pi]$.

One may perform a similar study on the possibility that $z$ serves as the only source of CP violation by assuming $\sigma =\delta =0$. However, given the fact that both $\sigma$ and $\delta$ are competent to play such a role in generating CP violation and $z$ is actually inaccessible in any low-energy measurements, we shall not pay much attention to this possibility.

\begin{figure*}[t]
\centering
\includegraphics[width=6in]{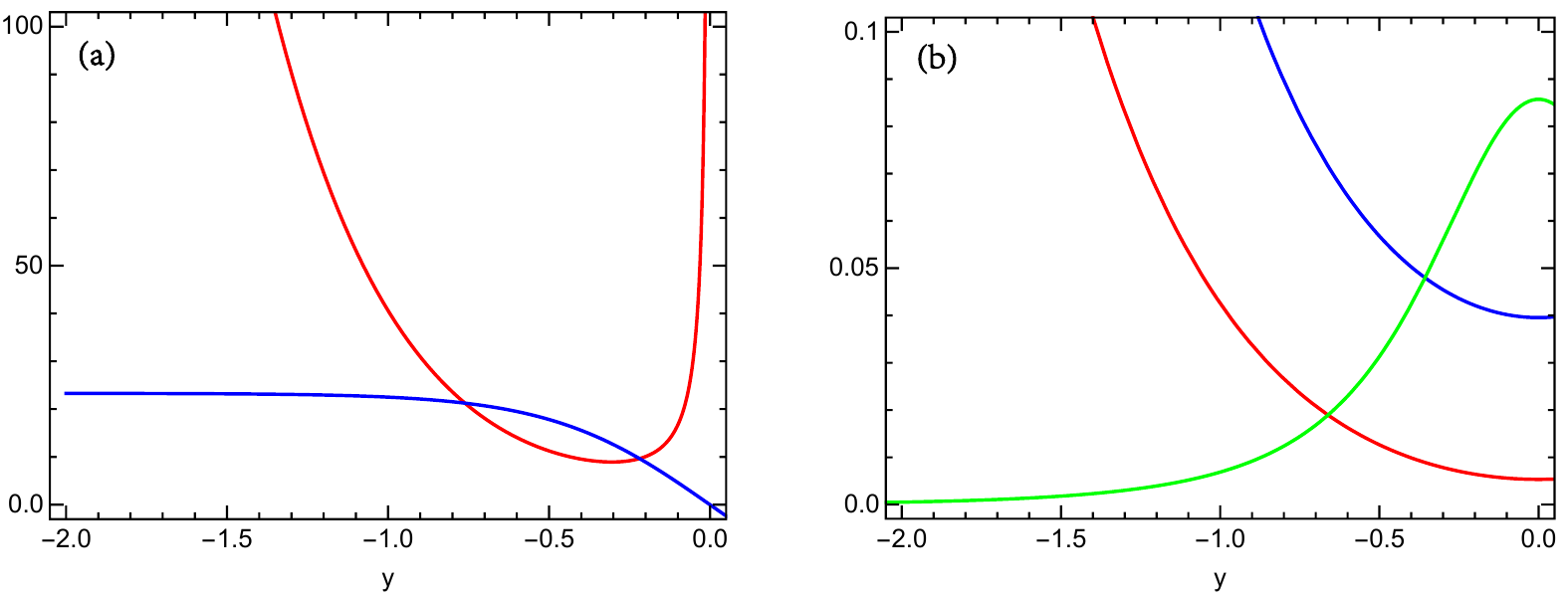}
\caption{In the $m^{}_3 =0$ case with $\delta =\sigma =0$ and $\sin z$ being purely imaginary (i.e., $\sin z = {\rm i} \sinh y$), a numerical illustration of (a) $|\varepsilon^{}_{1\tau}|/10^{-8}$ for $M^{}_1 =10^{10}$ GeV (blue) and $M^{}_1/(10^{10}$ GeV) (red) for successful flavored leptogenesis, (b) $10^{-3}\widetilde m^{}_{1\tau}/m^{}_*$ (red), $10^{-3} \widetilde m^{}_{1o}/m^{}_*$ (blue) and $\kappa^{}_1$ (green) as functions of $y$.}
\label{Fig:4-4}
\end{figure*}

Instead, let us turn to the $m^{}_3 =0$ case where CP violation originating from $z$ is indispensable for successful leptogenesis. This point can be seen from the fact that, for real $\sin z$, $\varepsilon^{}_{1\tau}$ is strongly suppressed by the small mass difference $m^{}_2- m^{}_1$ as indicated by Eq.~(\ref{eq:4.2.5a}) and Eq.~(\ref{eq:2.7}) in the present case.
A numerical exercise shows that $|\varepsilon^{}_{1\tau}|$ has a maximal value of $2 \times 10^{-7}$ for $M^{}_1  \sim  10^{12}$ GeV at $[\theta, \delta, \sigma] \simeq [0.25 \pi, \pi, 0.5 \pi]$. On the other hand, given that $\kappa(\widetilde m)$ has a maximal value of 0.2 as indicated by Eq.~(\ref{eq:4.1.8}), one necessarily has $|\kappa \left(0.66 \widetilde m^{}_{1\tau} \right) - \kappa \left(0.71 \widetilde m^{}_{1o} \right)| \leq 0.2$. It then follows that $|Y^{}_{\rm B}|$ is bounded by $5.5 \times 10^{-11}$ from above, and hence there is no chance to account for the observed value of $Y^{}_{\rm B}$. So we resort to the possibility that $\sin z$ is purely imaginary and serves as the only source of CP violation by assuming $\sigma = \delta =0$. In this special case the expression of $\varepsilon^{}_{1\tau}$ can be readily read off from Eq.~(\ref{eq:4.2.5b}) by taking $\alpha =\tau$,  $i=1$ and $j=2$, and it is further simplified to
\begin{eqnarray}
\varepsilon^{}_{1\tau} \simeq \frac{3 M^{}_1 m^{}_0}{16 \pi v^2 } \cdot \frac{\sinh 2y}{ \cosh^2 y + \sinh^2 y } {\rm Re}\left(U^{}_{\tau 1} U^*_{\tau 2}\right) \;
\label{eq:4.2.9}
\end{eqnarray}
by taking into account $m^{}_1 \simeq m^{}_2 \simeq m^{}_0 \equiv (m^{}_1+m^{}_2)/2$ in the $m^{}_3 =0$ case. Now $\varepsilon^{}_{1\tau}$ does not suffer the suppression from $m^{}_2 - m^{}_1$ any more. As can be seen from Fig.~\ref{Fig:4-4}(a), $|\varepsilon^{}_{1\tau}|$ quickly converges towards its maximal value $2.4 \times 10^{-7} M^{}_1/(10^{10} \ \rm GeV)$ for $|y| \gtrsim 0.5$. In addition, $\widetilde m^{}_{1\tau}/m^{}_*$, $\widetilde m^{}_{1o}/m^{}_*$ and $\kappa^{}_1$ are shown as functions of $y$ in Fig.~\ref{Fig:4-4}(b). It is clear that both $\widetilde m^{}_{1\tau}$ and $\widetilde m^{}_{1o}$ are much larger than $m^{}_*$, and they increase quickly along with the increase of $|y|$. Hence $\kappa^{}_1$ takes its maximal value 0.1 at $y =0$ and decreases quickly along with the increase of $|y|$. Finally, the values of $M^{}_1/(10^{10} \ \rm GeV)$ for successful leptogenesis are shown as a function of $y$ in Fig.~\ref{Fig:4-4}(a). Given the behaviors of $|\varepsilon^{}_{1\tau}|$ and $\kappa^{}_1$ changing with $y$, it is easy to understand that the lower bound of $M^{}_1$ (i.e., $9 \times 10^{10}$ GeV) appears at $y \simeq -0.3$. Note that such a lower limit on $M^{}_1$ is about two orders of magnitude smaller than that obtained in the unflavored leptogenesis regime \cite{Petcov:2005jh}.

Our final remarks are as follows. For the sake of simplicity in the above discussions, we have chosen to switch off two of the three CP-violating parameters (namely, $\delta$, $\sigma$ and ${\rm Im}z$) to highlight the effect of one source of CP violation on leptogenesis. Such a treatment is just for the purpose of illustration, of course. To get a ball-park feeling of the nontrivial interplay among three different possible sources of CP violation in flavored thermal leptogenesis, we refer the reader to Ref.~\cite{Molinaro:2008rg}.

\subsection{Contribution of $N^{}_2$ to leptogenesis}
\label{section 4.3}

In the conventional leptogenesis scenarios one usually assumes $M^{}_3 > M^{}_2 \gg M^{}_1$ to hold. This assumption assures that the $N^{}_2$-generated lepton number asymmetry will in general be substantially washed out by the $N^{}_1$-related interactions and thus can be safely neglected. However, it is not the case in the following several (actually rather generic) circumstances. (1) When the mass spectrum of three heavy Majorana neutrinos is not very hierarchical, the situation will change. (2) Provided the Yukawa couplings of $N^{}_1$ are just too weak (i.e., $\widetilde m^{}_1 \ll m^{}_*$), then the $N^{}_1$-related interactions will be inefficient in facilitating the $N^{}_1$-generated lepton number asymmetry on the one hand and erasing the $N^{}_{2}$-generated asymmetry on the other hand, and thus enhancing the significance of the contribution of $N^{}_2$ to leptogenesis \cite{DiBari:2005st,Vives:2005ra,Blanchet:2006dq}. (3) Even in the case that the Yukawa couplings of $N^{}_1$ are very strong (i.e., $\widetilde m^{}_1 \gg m^{}_*$), the $N^{}_{2}$-generated lepton number asymmetry stored in the flavor direction orthogonal to that coupled with $N^{}_1$ does not suffer the $N^{}_1$-associated washout effect and thus can survive \cite{Barbieri:1999ma,Strumia:2006qk,Engelhard:2006yg,Nielsen:2001fy}. (4) When flavor effects are relevant, it may happen that the $N^{}_1$-associated washout effect in a specific flavor direction is negligible, sparing the $N^{}_{2}$-generated lepton number asymmetry stored in this very flavor direction from being strongly washed out \cite{Vives:2005ra}. To sum up, the $N^{}_{2}$-generated lepton number asymmetry may constitute a non-negligible or even dominant part of the final baryon number asymmetry. It is therefore problematic to assume, {\it a priori}, that this kind of contribution can be safely neglected, unless the reheating temperature of the Universe is simply too low to guarantee the production of a sufficient amount of $N^{}_2$'s.

\begin{figure*}[t]
\centering
\includegraphics[width=4.5in]{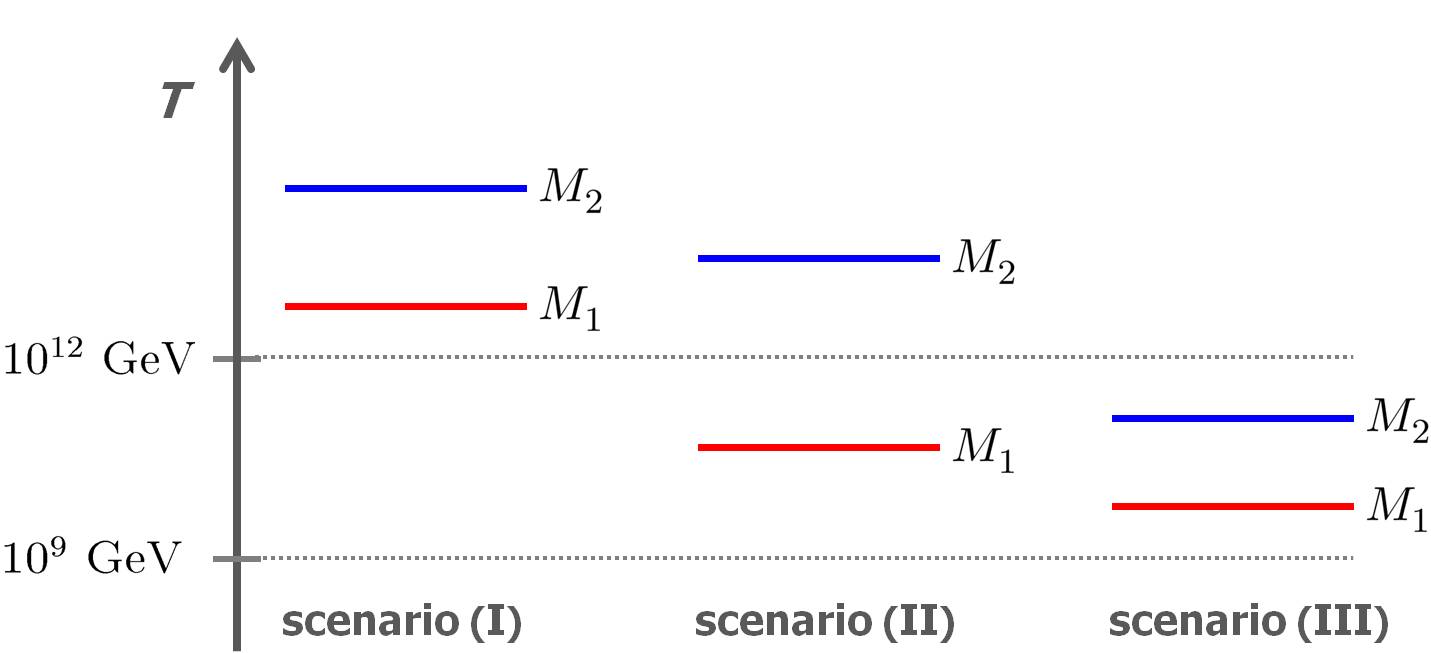}
\caption{A schematic illustration of three possible patterns of the heavy Majorana neutrino mass spectrum corresponding to three different leptogenesis scenarios in the minimal seesaw model. }
\label{Fig:4-5}
\end{figure*}

Here we study the possible contribution of $N^{}_2$ to leptogenesis in the minimal seesaw model, where the mass spectrum of two heavy Majorana neutrinos is still assumed to be hierarchical such that the $N^{}_2$- and $N^{}_1$-associated leptogenesis phases can well be separated. The lepton number asymmetry is produced and washed out by the $N^{}_2$-related interactions in the first stage and by the $N^{}_1$-related interactions in the second stage. A careful analysis shows that, even after the inclusion of the contribution of $N^{}_2$ and flavor effects, the lower bound on $M^{}_1$ from the requirement of successful leptogenesis remains above $10^{9}$ GeV \cite{Antusch:2011nz}. Therefore, as illustrated in Fig.~\ref{Fig:4-5}, there are totally three possible patterns for the heavy Majorana  neutrino mass spectrum corresponding to three different leptogenesis scenarios \cite{Bertuzzo:2010et}: (I) both $M^{}_1$ and $M^{}_2$ are larger than $10^{12}$ GeV, for which unflavored leptogenesis takes effect; (II) $M^{}_1$ lies in the temperature range of $\tau$-flavored leptogenesis (i.e., $10^9 ~{\rm GeV} < T < 10^{12} ~{\rm GeV}$), but $M^{}_2$ remains above $10^{12}$ GeV; (III) both $M^{}_1$ and $M^{}_2$ are in the temperature range of $\tau$-flavored leptogenesis.

In scenario (I), flavor effects have not come into play, so the lepton doublet states produced by the decays of $N^{}_i$ are some coherent superpositions of all the three flavors
\begin{eqnarray}
\ket{\ell^{}_i} = \frac{1}{ \sqrt{(Y^\dagger_{\nu} Y^{}_{\nu})^{}_{ii}} } \sum^{}_\alpha (Y^{}_{\nu})^*_{\alpha i} \ket{\ell^{}_\alpha} \;.
\label{eq:4.3.1}
\end{eqnarray}
After the $N^{}_2$-associated leptogenesis phase, the generated baryon number asymmetry can be expressed as $Y^0_{2\rm B} = -c r \varepsilon^{}_2 \kappa^{}_2$. During the $N^{}_1$-associated leptogenesis phase, the part of $Y^0_{2\rm B}$ stored in the flavor direction parallel (or orthogonal) to $\ket{\ell^{}_1}$ undergoes (or escapes) the $N^{}_1$-related washout effect. After the $N^{}_1$-associated leptogenesis phase, the surviving amount of $Y^0_{2\rm B}$ can be calculated according to
\begin{eqnarray}
Y^{}_{2\rm B} = \left[\left(1-p^{}_{21}\right) + p^{}_{21} \exp \left(-\frac{3\pi \widetilde m^{}_1}{8 m^{}_*}\right)\right] Y^0_{2\rm B}  \;,
\label{eq:4.3.2}
\end{eqnarray}
where
\begin{eqnarray}
p^{}_{21} = \left| \braket{\ell^{}_1 | \ell^{}_2} \right|^2 = \frac{\left|(Y^\dagger_{\nu} Y^{}_{\nu})^{}_{12} \right|^2}{(Y^\dagger_{\nu} Y^{}_{\nu})^{}_{11} (Y^\dagger_{\nu} Y^{}_{\nu})^{}_{22}}  \;
\label{eq:4.3.3}
\end{eqnarray}
quantifies the overlap between $\ket{\ell^{}_2}$ and $\ket{\ell^{}_1}$. Apparently, the smaller $p^{}_{21}$ is the more likely the $N^{}_2$-generated baryon number asymmetry can survive through the $N^{}_1$-associated leptogeneis phase.
Taking into account the $N^{}_1$-generated asymmetry $Y^{}_{1\rm B} = -c r \varepsilon^{}_1 \kappa^{}_1$, the final baryon number asymmetry is simply given by $Y^{}_{\rm B} = Y^{}_{2\rm B} + Y^{}_{1\rm B}$.

In scenario (II), during the $N^{}_1$-associated leptogenesis phase the $\tau$ flavor should be treated separately from the other two flavors which form a coherent superposition state of the form
\begin{eqnarray}
\ket{l^{}_{1o}} = \frac{1}{ \sqrt{\left|(Y^{}_{\nu})^{}_{e 1}\right|^2 + \left|(Y^{}_{\nu})^{}_{\mu 1}\right|^2} } \left[(Y^{}_{\nu})^*_{e 1} \ket{l^{}_e} + (Y^{}_{\nu})^*_{\mu 1} \ket{l^{}_\mu}\right] \;.
\label{eq:4.3.4}
\end{eqnarray}
In this case the parts of $Y^0_{2\rm B}$ stored in the $\tau$ and $1o$ flavor directions are subject to the $N^{}_1$-related washout effect, but the remaining part stays unaffected. After the $N^{}_1$-associated leptogenesis phase, the surviving amount of $Y^0_{2\rm B}$ is expressed as
\begin{eqnarray}
\hspace{-1.3cm} Y^{}_{2\rm B} = \left[ p^{}_{2\tau} \exp\left(-\frac{3\pi \widetilde m^{}_{1\tau}}{8 m^{}_*}\right) + p^{}_{21o} \exp\left(-\frac{3\pi \widetilde m^{}_{1o}}{8 m^{}_*}\right) + 1- p^{}_{2\tau} - p^{}_{21o} \right] Y^0_{2\rm B} \;,
\label{eq:4.3.5}
\end{eqnarray}
where
\begin{eqnarray}
p^{}_{2\tau} = \left|\braket{ l^{}_{\tau} | l^{}_2}\right|^2 = \frac{1}{ (Y^\dagger_{\nu} Y^{}_{\nu})^{}_{22} } \left|(Y^{}_{\nu})^{}_{\tau 2}\right|^2 \;,
\nonumber \\
p^{}_{2 1o} = \left|\braket{ l^{}_{1o} | l^{}_2}\right|^2 = \frac{\left|(Y^{}_{\nu})^{}_{e 1} (Y^{}_{\nu})^{*}_{e 2} + (Y^{}_{\nu})^{}_{\mu 1} (Y^{}_{\nu})^{*}_{\mu 2} \right|^2 }{ (Y^\dagger_{\nu} Y^{}_{\nu})^{}_{22} \left[\left|(Y^{}_{\nu})^*_{e 1}\right|^2 + \left|(Y^{}_{\nu})^*_{\mu 1}\right|^2\right]} \;.
\label{eq:4.3.6}
\end{eqnarray}
The $N^{}_1$-generated baryon number asymmetry turns out to be $Y^{}_{1\rm B} =- c r(\varepsilon^{}_{1\tau} \kappa^{}_{1\tau} + \varepsilon^{}_{1o} \kappa^{}_{1o} )$.

In scenario (III), during the $N^{}_2$-associated leptogenesis phase one should also treat the $\tau$ flavor separately from the other two flavors which form a coherent superposition state $\ket{l^{}_{2o}}$ analogous to $\ket{l^{}_{1o}}$. In this case the baryon number asymmetry generated after the $N^{}_2$-associated leptogenesis phase becomes $Y^0_{2\rm B} = - c r (\varepsilon^{}_{2\tau} \kappa^{}_{2\tau} + \varepsilon^{}_{2o} \kappa^{}_{2o} )$ with $\varepsilon^{}_{2o} \equiv \varepsilon^{}_{2e}+ \varepsilon^{}_{2\mu}$, of which the amount surviving through the $N^{}_1$-associated leptogenesis phase is given by
\begin{eqnarray}
Y^{}_{2\rm B} & = & -c r \left\{ \varepsilon^{}_{2\tau} \kappa^{}_{2\tau} \exp\left(-\frac{3\pi \widetilde m^{}_{1\tau}}{8 m^{}_*}\right) \right.
\nonumber \\
&& + \left. \varepsilon^{}_{2o} \kappa^{}_{2o} \left[ \left(1-p^{}_{2o1o}\right)
+ p^{}_{2o1o} \exp\left(-\frac{3\pi \widetilde m^{}_{1o}}{8 m^{}_*}\right) \right] \right\} \;,
\label{eq:4.3.7}
\end{eqnarray}
where
\begin{eqnarray}
p^{}_{2o 1o} \equiv \left|\braket{ l^{}_{1o} | l^{}_{2o}}\right|^2 = \frac{  \left| (Y^{}_{\nu})^{}_{e 1} (Y^{}_{\nu})^{*}_{e 2} + (Y^{}_{\nu})^{}_{\mu 1} (Y^{}_{\nu})^{*}_{\mu 2}\right|^2 }{ \left[\left|(Y^{}_{\nu})^{}_{e 1}\right|^2 + \left|(Y^{}_{\nu})^{}_{\mu 1}\right|^2 \right] \left[\left|(Y^{}_{\nu})^{}_{e 2}\right|^2 + \left|(Y^{}_{\nu})^{}_{\mu 2}\right|^2 \right]} \;
\label{eq:4.3.8}
\end{eqnarray}
has been defined in a way similar to the definition of $p^{}_{2 1o}$.

Note that the unflavored and flavored CP-violating asymmetries $\varepsilon^{}_2$ and $\varepsilon^{}_{2\alpha}$ for the decays of $N^{}_2$ have been given in Eqs.~(\ref{eq:2.21}) and (\ref{eq:2.22}), respectively. The corresponding efficiency factors $\kappa^{}_2$ and $\kappa^{}_{2\alpha}$ are determined by the unflavored and flavored washout mass parameters $\widetilde m^{}_2$ and $\widetilde m^{}_{2\alpha}$ (with $\widetilde m^{}_{2o} \equiv \widetilde m^{}_{2e}+ \widetilde m^{}_{2\mu}$), respectively. Here $\widetilde m^{}_2$ and $\widetilde m^{}_{2\alpha}$ are defined in the same way as $\widetilde m^{}_1$ and $\widetilde m^{}_{1\alpha}$. In the Casas-Ibarra parametrization the expressions of $\widetilde m^{}_2$ and $\widetilde m^{}_{2\alpha}$ are given by
\begin{eqnarray}
\widetilde m^{}_2 = m^{}_i \left|\sin z\right|^2 + m^{}_j \left|\cos z\right|^2 \;,
\nonumber \\
\widetilde m^{}_{2\alpha} = \left| U^{}_{\alpha i} \sqrt{m^{}_i} \sin z -  U^{}_{\alpha j} \sqrt{m^{}_j} \cos z \right|^2 \;,
\label{eq:4.3.9}
\end{eqnarray}
where $i =2$ (or $i=1$) and $j =3$ (or $j=2$) in the $m^{}_1 =0$ (or $m^{}_3 =0$) case.
Given a hierarchical heavy Majorana neutrino mass spectrum with $M^{}_2 \gtrsim 3 M^{}_1$, the CP-violating asymmetries $\varepsilon^{}_2$ and $\varepsilon^{}_{2\alpha}$ can approximate to \cite{Antusch:2011nz}
\begin{eqnarray}
\hspace{-1.8cm} \varepsilon^{}_2 \simeq \frac{ M^{}_1 \left(m^2_i - m^2_j\right){\rm Im} \left( \sin^2 z \right) } { 4 \pi v^2 \widetilde m^{}_2 } \cdot \frac{M^{}_1}{M^{}_2} \left[ \ln \left( \frac{M^{}_2}{M^{}_1} \right) - 1  \right] \;,
\nonumber \\
\hspace{-1.8cm} \varepsilon^{}_{2\alpha} \simeq \frac{M^{}_1}{8\pi v^2 \widetilde m^{}_2} \left[ m^{}_i m^{}_j \left( |U^{}_{\alpha i}|^2 - |U^{}_{\alpha j}|^2 \right) {\rm Im}\left(\sin^2 z\right) \right.
\nonumber \\
\hspace{-0.63cm} + \sqrt{m^{}_i m^{}_j} \left(m^{}_i -m^{}_j\right) {\rm Im}\left(U^*_{\alpha i} U^{}_{\alpha j}\right) {\rm Re}\left(\sin z \cos^* z\right) \left(\left|\cos z\right|^2 + \left|\sin z\right|^2\right)
\nonumber \\
\hspace{-0.63cm} - \sqrt{m^{}_i m^{}_j} \left(m^{}_i + m^{}_j\right) {\rm Re}\left(U^*_{\alpha i} U^{}_{\alpha j}\right) {\rm Im}\left(\sin z \cos^* z\right) \left(\left|\cos z\right|^2 - \left|\sin z\right|^2\right) \;.
\label{eq:4.3.10}
\end{eqnarray}
Here we have neglected the term proportional to ${\cal F}(z)$ because it is suppressed by the factor $M^{}_1/M^{}_2$ as compared with the term proportional to ${\cal G}(z)$.

\begin{figure*}[t]
\centering
\includegraphics[width=6in]{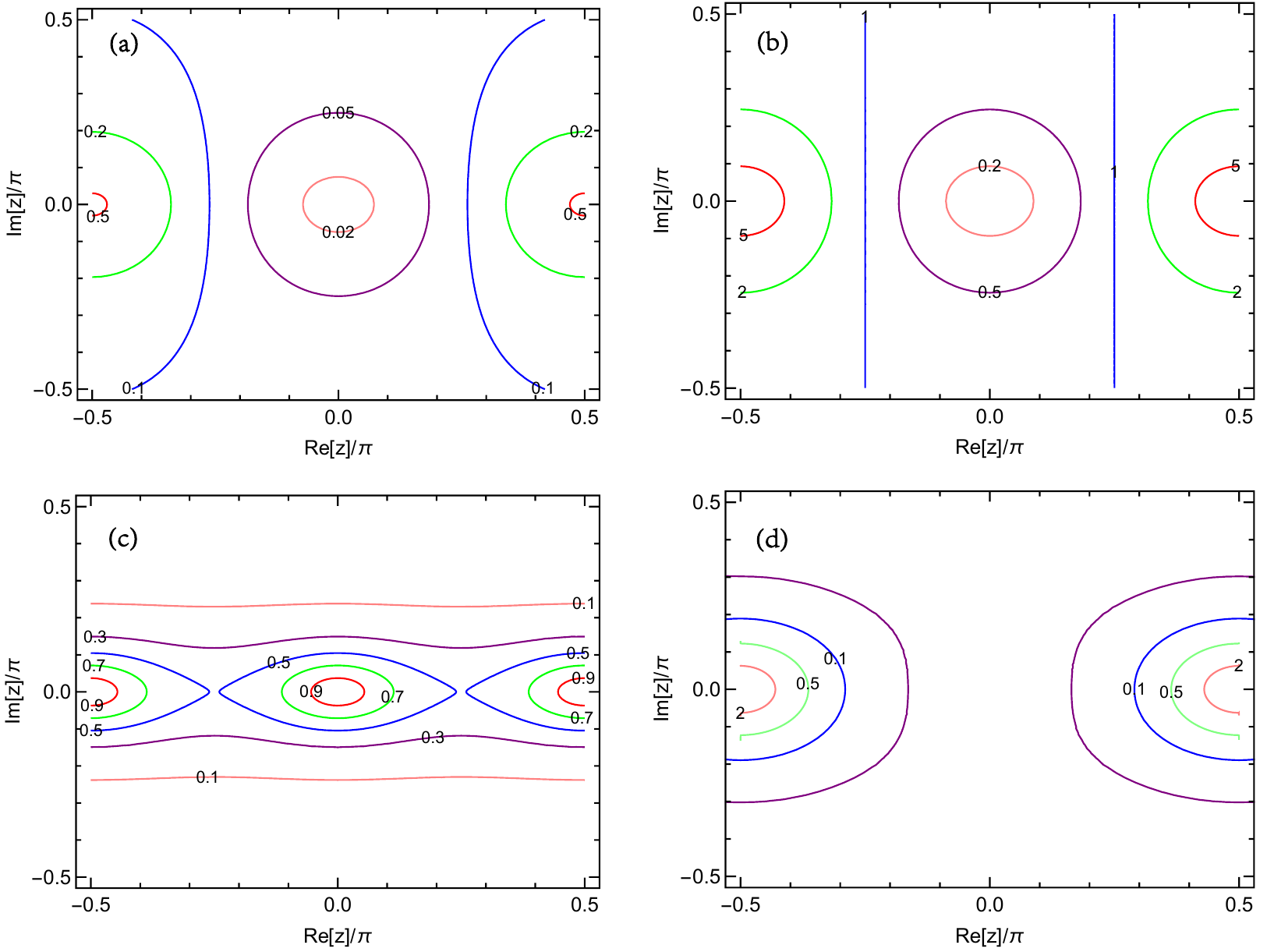}
\caption{A numerical illustration of scenario (I): the contour lines of (a) $|\varepsilon^{}_2/\varepsilon^{}_1|$, (b) $\kappa^{}_2/\kappa^{}_1$, (c) $1-p^{}_{21}$ and (d) $Y^{}_{2 \rm B}/Y^{}_{1 \rm B}$ in the ${\rm Re}(z)$-${\rm Im}(z)$ plane for $m^{}_1 =0$ and $M^{}_2 = 3 M^{}_1$. Note that $p^{}_{21}$ is dependent upon $\sigma$ and $\delta$, and these two phases have been marginalized to maximize $1-p^{}_{21}$ in obtaining the results in Fig.~\ref{Fig:4-6}(c).}
\label{Fig:4-6}
\end{figure*}

Now we are ready to examine how important the contribution of $N^{}_2$ to leptogenesis can be. In scenario (I), the ratio $\varepsilon^{}_2/\varepsilon^{}_1$ is suppressed by the ratio $M^{}_1 {\cal F}(z)/[M^{}_2 {\cal F}(x)]$, which lies in the range of 0.1 to 0.2 for $M^{}_2 \gtrsim 3 M^{}_1$. In the $m^{}_1 =0$ case, given the fact that $m^{}_3 \simeq 6 m^{}_2$ and $\varepsilon^{}_2/\varepsilon^{}_1 \propto \widetilde m^{}_1/\widetilde m^{}_2$, the above suppression can be offset for $|\sin z|^2 \gg |\cos z|^2$ (corresponding to $z \simeq \pm \pi/2$) which leads to $\widetilde m^{}_1/\widetilde m^{}_2 \simeq m^{}_3/m^{}_2$. It is easy to see that $|\sin z|^2 \gg |\cos z|^2$ is also advantageous to the enhancement of $\kappa^{}_2/\kappa^{}_1$ which approximates to $\widetilde m^{}_1/\widetilde m^{}_2$ in the strong washout regime. These qualitative analyses are supported by our numerical results in Fig.~\ref{Fig:4-6}(a) and Fig.~\ref{Fig:4-6}(b), which show the contour lines of $|\varepsilon^{}_2/\varepsilon^{}_1|$ and $\kappa^{}_2/\kappa^{}_1$ in the ${\rm Re}[z]$-${\rm Im}[z]$ plane. Here and hereafter, $M^{}_2 = 3 M^{}_1$ is taken as a benchmark value.
Furthermore, as can be seen from Fig.~\ref{Fig:4-6}(c), the parameter regions for enhancing $1-p^{}_{21}$ are also around $z \simeq \pm \pi/2$. These observations allow us to conclude that the contribution of $N^{}_2$ to leptogenesis may become important in the regions $z \simeq \pm \pi/2$. As shown in Fig.~\ref{Fig:4-6}(d), $Y^{}_{2\rm B}$ can be comparable with $Y^{}_{1\rm B}$ in such parameter regions. It is obvious that the results in Fig.~\ref{Fig:4-6} are periodic in $\pi$ along the ${\rm Re}[z]$ axis.
In the $m^{}_3 =0$ case, one always has $\widetilde m^{}_1 \simeq \widetilde m^{}_2$ because of $m^{}_1 \simeq m^{}_2$. So a similar way of enhancing $Y^{}_{2\rm B}/Y^{}_{1\rm B}$ as described above does not work anymore, leaving us with a negligible ($\lesssim 10\%$) contribution of $N^{}_2$ to leptogenesis.
\begin{figure*}[t]
\centering
\includegraphics[width=6in]{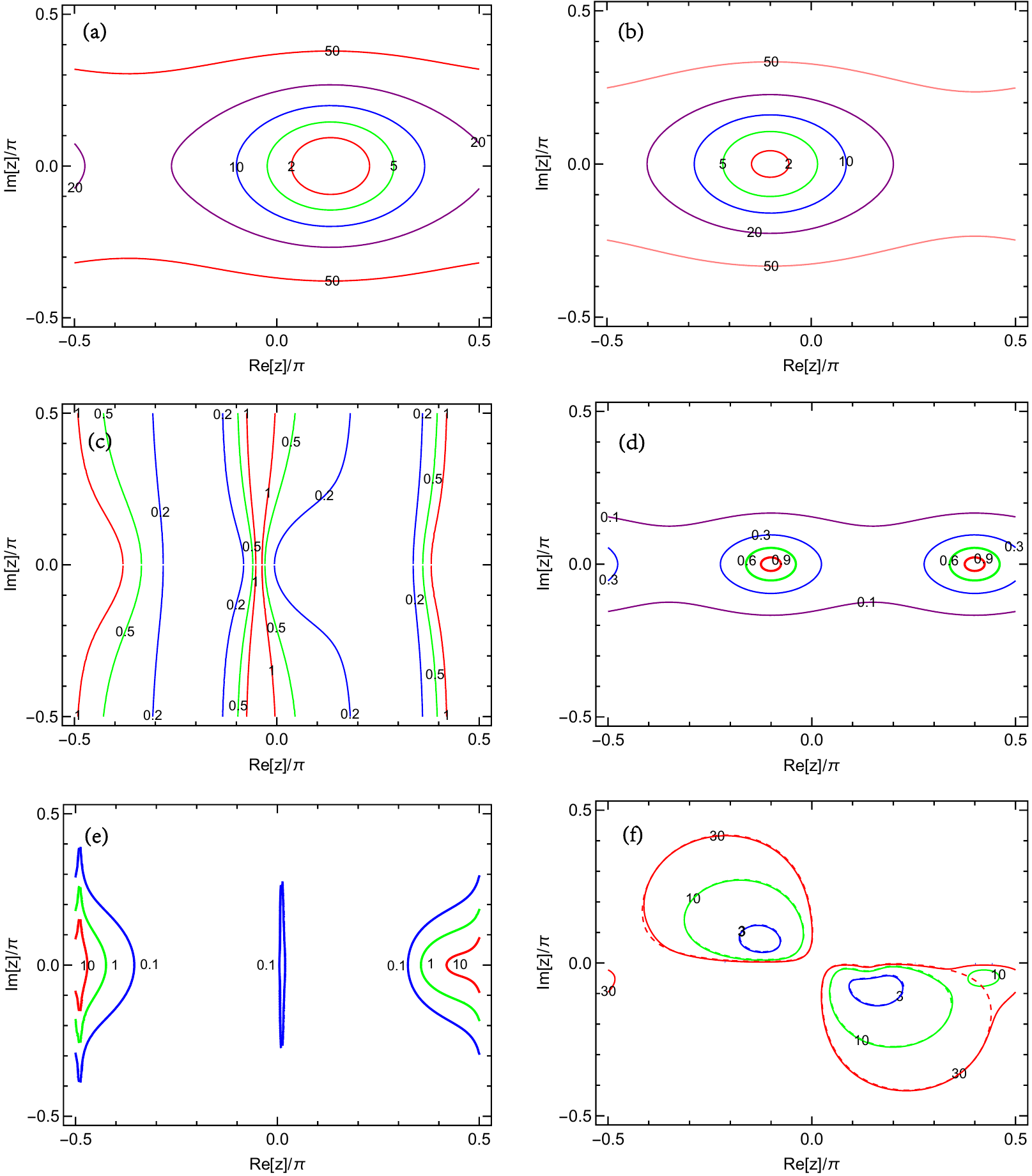}
\caption{A numerical illustration of scenario (III): the contour lines of (a) $\widetilde m^{}_{1\tau}/m^{}_*$, (b) $\widetilde m^{}_{1o}/m^{}_*$, (c) $|\varepsilon^{}_{2o}/\varepsilon^{}_{1o}|$, (d) $1-p^{}_{2o1o}$, (e) $|Y^{}_{2\rm B}/Y^{}_{1\rm B}|$ and (f) $M^{}_1/(10^{10} \ {\rm GeV})$ for successful leptogenesis in the ${\rm Re}(z)$-${\rm Im}(z)$ plane for $m^{}_1 =0$, $M^{}_2 = 3 M^{}_1$ and $\delta =\sigma =0$. In Fig.~\ref{Fig:4-7}(f) the solid (dashed) lines are obtained by including (excluding) the contribution of $N^{}_2$ to leptogenesis.}
\label{Fig:4-7}
\end{figure*}

Given that scenario (II) is qualitatively a scenario between scenario (I) and scenario (III), we directly proceed to consider scenario (III), where the contribution of $N^{}_2$ to leptogenesis can be dominant. In Fig.~\ref{Fig:4-7}(a) and Fig.~\ref{Fig:4-7}(b) we illustrate the contour lines of $\widetilde m^{}_{1\tau}/m^{}_*$ and $\widetilde m^{}_{1o}/m^{}_*$ in the ${\rm Re}[z]$-${\rm Im}[z]$ plane. In obtaining such results, we have taken $\sigma =\delta =0$ as the benchmark values. It turns out that the $N^{}_1$-related washout effects on the $\tau$ and $1o$ flavor directions are both strong, implying that only an amount of $-c r \varepsilon^{}_{2o} \kappa^{}_{2o}(1-p^{}_{2o1o})$ in $Y^0_{2\rm B}$ can escape the strong $N^{}_1$-associated washout effect. This point is easily seeable in Eq.~(\ref{eq:4.3.7}). Thanks to the term proportional to ${\cal G}(z)$, the $N^{}_2$-related CP-violating asymmetries $\varepsilon^{}_{2\alpha}$ are not suppressed as compared with their $N^{}_1$-related counterparts. For illustration, in Fig.~\ref{Fig:4-7}(c) we show the contour lines of $|\varepsilon^{}_{2o}/\varepsilon^{}_{1o}|$ in the ${\rm Re}[z]$-${\rm Im}[z]$ plane. One can see that $|\varepsilon^{}_{2o}|$ is comparable to or even larger than $|\varepsilon^{}_{1o}|$ for $z \simeq 0$ or $\pm \pi/2$. With the help of the observation that the values of $\widetilde m^{}_{2o}/m^{}_*$ can be obtained from those of $\widetilde m^{}_{1o}/m^{}_*$ by making the replacement $z \to z + \pi/2$ as shown in Eqs.~(\ref{eq:4.2.4}) and (\ref{eq:4.3.9}), from Fig.~\ref{Fig:4-7}(b) it can be inferred that $\kappa^{}_{2o}$ takes its maximal value (corresponding to the minimal value of $\widetilde m^{}_{2o}$) at $z \simeq \pm \pi/2$. Furthermore, Fig.~\ref{Fig:4-7}(d) tells us that $1-p^{}_{2o1o}$ can be close to unity for $z \simeq 0$ or $\pm \pi/2$.
Given these results, it appears that the parameter regions for enhancing $\varepsilon^{}_{2o}$, $\kappa^{}_{2o}$ and $1-p^{}_{2o1o}$ overlap at $z \simeq \pm \pi/2$. On the other hand, it is easy to see that $z \simeq \pm \pi/2$ also tend to suppress the contribution of $N^{}_1$ to leptogenesis. Hence $Y^{}_{2\rm B}$ can be much larger than $Y^{}_{1\rm B}$ for $z \simeq \pm \pi/2$, as shown in Fig.~\ref{Fig:4-7}(e). Finally, Fig.~\ref{Fig:4-7}(f) shows the contour lines of $M^{}_1/(10^{10} \ {\rm GeV})$ for successful leptogenesis in the ${\rm Re}[z]$-${\rm Im}[z]$ plane. We see that including the contribution of $N^{}_2$ to leptogenesis enlarges the parameter region for successful letogenesis at $z \simeq \pm \pi/2$.

It is worth mentioning that the parameter regions $z \simeq \pm \pi/2$ correspond to the so-called sequential dominance where $m^{}_3$ (or $m^{}_2$) is dominantly produced by $N^{}_1$ (or $N^{}_2$) \cite{King:2003jb}. This point can be easily inferred from the extreme case of $z$ being exactly $\pi/2$, for which $m^{}_3$ (or $m^{}_2$) is completely produced by $N^{}_1$ (or $N^{}_2$). But it should be noted that leptogenesis will definitely not work in this extreme case because the crucial ingredient $(Y^\dagger_\nu Y^{}_\nu)^{}_{12}$ in the CP-violating asymmetries for leptogenesis will be proportional to $U^*_{e 3} U^{}_{e 2} + U^*_{\mu 3} U^{}_{\mu 2} + U^*_{\tau 3} U^{}_{\tau 2} = 0$ due to the unitarity of $U$.

\subsection{Resonant leptogenesis}
\label{section 4.4}

In the above discussions, we have been assuming the heavy Majorana neutrino mass spectrum to be hierarchical. If this restriction is loosened, a dramatic effect may arise: the CP-violating asymmetries $\varepsilon^{}_i$ for the decays of $N^{}_i$ will receive strong enhancements if $N^{}_1$ and $N^{}_2$ are nearly degenerate in their masses \cite{Pilaftsis:1997jf,Pilaftsis:2003gt}. This effect comes from the interference between tree-level and one-loop self-energy decay amplitudes \cite{Flanz:1994yx,Covi:1996wh,Buchmuller:1997yu}, and its contribution to $\varepsilon^{}_i$ in the quasi-degeneracy regime is described by Eq.~(\ref{eq:2.23}). For the extreme case of $M^{}_2 - M^{}_1 \simeq \Gamma^{}_i/2$, $\varepsilon^{}_i$ can be resonantly enhanced as follows:
\begin{eqnarray}
\varepsilon^{}_i \simeq - \frac{1}{2} \cdot \frac{{\rm Im}\left[(Y^\dagger_\nu
Y^{}_\nu)^2_{12}\right]}{(Y^\dagger_\nu Y^{}_\nu)^{}_{11}
(Y^\dagger_\nu Y^{}_\nu)^{}_{22}}  \; ,
\label{eq:4.4.1}
\end{eqnarray}
which can be as large as $50\%$ in magnitude.
Such an interesting observation implies that {\it resonant} leptogenesis may work at an energy scale (e.g., TeV) much lower than the conventional seesaw scale, and thus can help evade the aforementioned tension between the lower bound of $T^{}_{\rm RH}$ required by a successful leptogenesis scenario in the $M^{}_2 \gg M^{}_1$ case and the upper bound of $T^{}_{\rm RH}$ required by avoiding the overproduction of gravitinos in a supersymmetric extension of the SM. Moreover, the possibilities of $M^{}_i \sim {\cal O}(1)$ TeV to ${\cal O}(10)$ TeV are also interesting in the sense that $N^{}_i$ may potentially manifest themselves in some high-energy frontier experiments, such as the on-going experiment based on the Large Hadron Collider (LHC) and the upcoming experiment based on a much higher energy collider. These distinctive features make the resonant leptogensis scenario quite appealing in phenomenology of particle physics and cosmology.

As already argued by 't Hooft \cite{tHooft:1979rat}, a dimensionless parameter can be regarded as a ``naturally small" parameter only in the case that switching off this parameter will enhance the system's degrees of symmetry. Following this guiding principle, we invoke some softly broken symmetries (which otherwise would protect the exact degeneracy between $M^{}_1$ and $M^{}_2$) to realize the small mass splitting of $M^{}_1$ and $M^{}_2$ that is typically required in the resonant leptogensis scenario. Here let us introduce two examples along this line of thought. The first example is the minimal inverse seesaw model \cite{Malinsky:2009df}, where the softly broken lepton number symmetry is utilized \cite{Branco:1988ex,Shaposhnikov:2006nn,Kersten:2007vk}. In this model $N^{}_1$ and $N^{}_2$ are assigned the lepton numbers $+1$ and $-1$, respectively. If the lepton number conservation holds exactly, the full $5 \times 5$ neutrino mass matrix will take a form like
\begin{eqnarray}
M^{}_{\nu +N} = \pmatrix{
0 & y^{}_1 v & 0 \cr
y^{T}_1 v & 0 & M \cr
0 & M & 0
}  \; ,
\label{eq:4.4.2}
\end{eqnarray}
where $y^{}_1$ is a column vector containing the ${\cal O}(1)$ Yukawa couplings $y^{}_{\alpha 1}$ for $l^{}_{\alpha \rm L}$ and $N^{}_1$. Transforming the right-handed neutrino fields from their flavor basis to their mass basis, one arrives at
\begin{eqnarray}
M^{\prime}_{\nu +N} = \frac{1}{\sqrt{2}} \pmatrix{
0 & {\rm i} y^{}_{1} v & y^{}_{ 1} v \cr
{\rm i} y^{T}_{ 1} v & \sqrt{2} M & 0 \cr
y^{T}_{ 1} v & 0 & \sqrt{2} M
}  \; .
\label{eq:4.4.3}
\end{eqnarray}
As a result, the two heavy neutrino masses are exactly degenerate, and the three light Majorana neutrinos remain massless. Hence the lepton number symmetry must be broken to a small extent, so as to generate finite but tiny masses for three light neutrinos and a small mass splitting between $M^{}_1$ and $M^{}_2$. To be specific, this can be done by adding a Majorana mass term $\mu$ for $N^{}_2$ in Eq.~(\ref{eq:4.4.2}) which explicitly breaks the lepton number symmetry and thus should be naturally small with respect to $M$. In this case the light Majorana neutrino masses and the heavy Majorana neutrino mass splitting are $\mu y^{}_1 y^T_1 v^2/M^2$ and $\mu$, respectively.

The second example of this kind \cite{Pilaftsis:2003gt,Rink:2016knw} is based on the so-called Froggatt-Nielsen (FN) mechanism \cite{Froggatt1979Hierarchy}. In this model $N^{}_1$ and $N^{}_2$ are assigned the $\rm U(1)^{}_{FN}$ charges $+1$ and $-1$, respectively; and the SM fields do not carry any $\rm U(1)^{}_{FN}$ charge. In addition, two scalar fields $\phi^{}_1$ and $\phi^{}_2$ with the respective $\rm U(1)^{}_{FN}$ charges $-1$ and $+1$ but no SM quantum numbers are introduced, and they will be responsible for the FN symmetry breaking through their acquiring the vacuum expectation values. Under such a setting, some neutrino mass terms can only arise from the higher-dimension operators which are suppressed by the FN symmetry scale $\Lambda$ (i.e., a common mass scale of the FN fields). To the leading order, the Lagrangian relevant for the neutrino masses appears as
\begin{eqnarray}
- {\cal L}^{\rm mass}_{\nu +N} = & \frac{y^{}_1}{\Lambda} \overline{l^{}_{\rm L }} \widetilde H N^{}_1 \phi^{}_1  + \frac{y^{}_{2}}{\Lambda} \overline{l^{}_{\rm L}} \widetilde H N^{}_2 \phi^{}_2 + M \overline {N^{c}_1} N^{}_2 \nonumber \\
& + \frac{y^{}_{11}}{2 \Lambda} \phi^2_1 \overline {N^{c}_1} N^{}_1 + \frac{y^{}_{22}}{2 \Lambda} \phi^2_2 \overline {N^{c}_2} N^{}_2 + {\rm h.c.} \; ,
\label{eq:4.4.4}
\end{eqnarray}
where $y^{}_i$ are defined in the same way as $y^{}_1$ in Eq.~(\ref{eq:4.4.2}) and $y^{}_{ii}$ are the ${\cal O}(1)$ coefficients. After the FN and electroweak symmetry breaking, the full $5 \times 5$ neutrino mass matrix turns out to be
\begin{eqnarray}
M^{}_{\nu +N} = \pmatrix{
0 & \epsilon^{}_1 y^{}_{ 1} v  & \epsilon^{}_2  y^{}_{2} v \cr
\epsilon^{}_1 y^{T}_{ 1} v & \epsilon^2_1 y^{}_{11} \Lambda & M \cr
\epsilon^{}_2 y^{T}_{ 2} v & M & \epsilon^2_2 y^{}_{22} \Lambda
}  \; ,
\label{eq:4.4.5}
\end{eqnarray}
with $\epsilon^{}_i \equiv \langle \phi^{}_i \rangle/\Lambda$. Under the natural assumptions of $\Lambda \sim M$ and $\epsilon^{}_i \ll 1$, one obtains the light Majorana neutrino masses of ${\cal O}(\epsilon^2_i v^2/M)$ and two nearly degenerate heavy Majorana neutrinos with a mass splitting of ${\cal O}(\epsilon^{2}_i M)$.

Given $M^{}_1 \simeq M^{}_2 < T \simeq 10^5$ GeV, it is obvious that both $N^{}_1$ and $N^{}_2$ will contribute to resonant leptogenesis, in which all the three flavors participate. In this case the final baryon number asymmetry can be calculated by means of
\begin{eqnarray}
Y^{}_{\rm B} = - c r \sum^{}_{\alpha} \kappa \left( \widetilde m^{}_\alpha\right) \sum^{}_i \varepsilon^{}_{i\alpha}  \;,
\label{eq:4.4.6}
\end{eqnarray}
where $\widetilde m^{}_\alpha = \widetilde m^{}_{1 \alpha} + \widetilde m^{}_{2 \alpha}$ and
\begin{eqnarray}
\hspace{-1.5cm} \varepsilon^{}_{i\alpha} = \frac{{\rm Im}\left\{ (Y^*_\nu)^{}_{\alpha i} (Y^{}_\nu)^{}_{\alpha j}
\left[ M^{}_j (Y^\dagger_\nu Y^{}_\nu)^{}_{ij} + M^{}_i (Y^\dagger_\nu Y^{}_\nu)^{}_{ji} \right] \right\} }{8\pi (Y^\dagger_\nu Y^{}_\nu)^{}_{ii}} \cdot \frac{M^{}_i \Delta M^2_{ij}}{(\Delta M^2_{ij})^2 + M^2_i \Gamma^2_j} \;,
\label{eq:4.4.7}
\end{eqnarray}
where $\Delta M^2_{ij} \equiv M^2_i - M^2_j$ has been defined, and $i$ and $j$ (for $i \neq j$) run over 1 and 2. Note that the efficiency factor for each flavor $\alpha$ is determined by the sum of the corresponding flavored washout mass parameters of two heavy Majorana neutrinos.
In the Casas-Ibarra parametrization, $\varepsilon^{}_{i \alpha}$ and $\widetilde m^{}_\alpha$ can be recast as
\begin{eqnarray}
\varepsilon^{}_{i \alpha} & = & \frac{ 1 }{2\pi v^2 \widetilde m^{}_i } \cdot \frac{ \xi M^3_0 }{ 4 \xi^2 M^2_0 + \Gamma^2_j} \left(m^{}_2 - m^{}_3\right) {\rm Re}\left(\cos z \sin^* z\right)
\nonumber \\
&& \times \left[ \left(m^{}_2 \left|U^{}_{\alpha 2}\right|^2 + m^{}_3 \left|U^{}_{\alpha 3}\right|^2\right) {\rm Im}\left(\cos z \sin^* z\right) \right.
\nonumber \\
&& \left. + \sqrt{m^{}_2 m^{}_3} \left(\left|\cos z\right|^2 + \left|\sin z\right|^2\right) {\rm Im}\left(U^{*}_{\alpha 2} U^{}_{\alpha 3}\right) \right] \;,
\nonumber \\
\widetilde m^{}_\alpha & = & \left(m^{}_2 \left|U^{}_{\alpha 2}\right|^2 + m^{}_3 \left|U^{}_{\alpha 3}\right|^2\right) \left(\left|\cos z\right|^2 + \left|\sin z\right|^2\right)
\nonumber \\
&& + 4 \sqrt{m^{}_2 m^{}_3} \ {\rm Im}\left(\cos z \sin^* z\right) {\rm Im}\left(U^{*}_{\alpha 2} U^{}_{\alpha 3}\right) \;,
\label{eq:4.4.8}
\end{eqnarray}
in the $m^{}_1= 0$ case; and the similar results in the $m^{}_3 =0$ case can be obtained from Eq.~(\ref{eq:4.4.8}) by simply replacing the subscripts 2 and 3 with 1 and 2. Here we have taken advantage of $M^{}_2 \simeq M^{}_1 $ and defined $\xi \equiv (M^{}_2 -M^{}_1)/M^{}_1$ and $M^{}_0 \equiv (M^{}_1 + M^{}_2)/2$. Note that the dependence of $\varepsilon^{}_{i \alpha}$ on $M^{}_0$ and $\xi$ can be factorized out as $f(M^{}_0, \xi) \equiv \xi M^3_0/(4 \xi^2 M^2_0 + \Gamma^2_j)$. In the range of $\xi \gg \Gamma^{}_j/M^{}_0$ or $\xi \ll \Gamma^{}_j/M^{}_0$, one may simplify $f(M^{}_0, \xi)$ to $M^{}_0/(4 \xi)$ or $\xi M^3_0/\Gamma^2_j = 64 \pi^2 \xi v^4/(\widetilde m^2_j M^{}_0)$. We see that $f(M^{}_0, \xi)$ is inversely proportional and proportional to $\xi$ in these two interesting ranges, respectively. And the values of $\xi$ needed for producing certain values of $f(M^{}_0, \xi)$ are proportional to $M^{}_0$ in both cases. There always exist two values of $\xi$ (denoted by $\xi^{}_1$ and $\xi^{}_2$), which reside respectively in the above two ranges and satisfy the condition $\xi^{}_1 \xi^{}_2 \simeq \Gamma^2_j/(4M^2_0)$, can give the same value of $f(M^{}_0, \xi)$.

\begin{figure*}[t]
\centering
\includegraphics[width=6in]{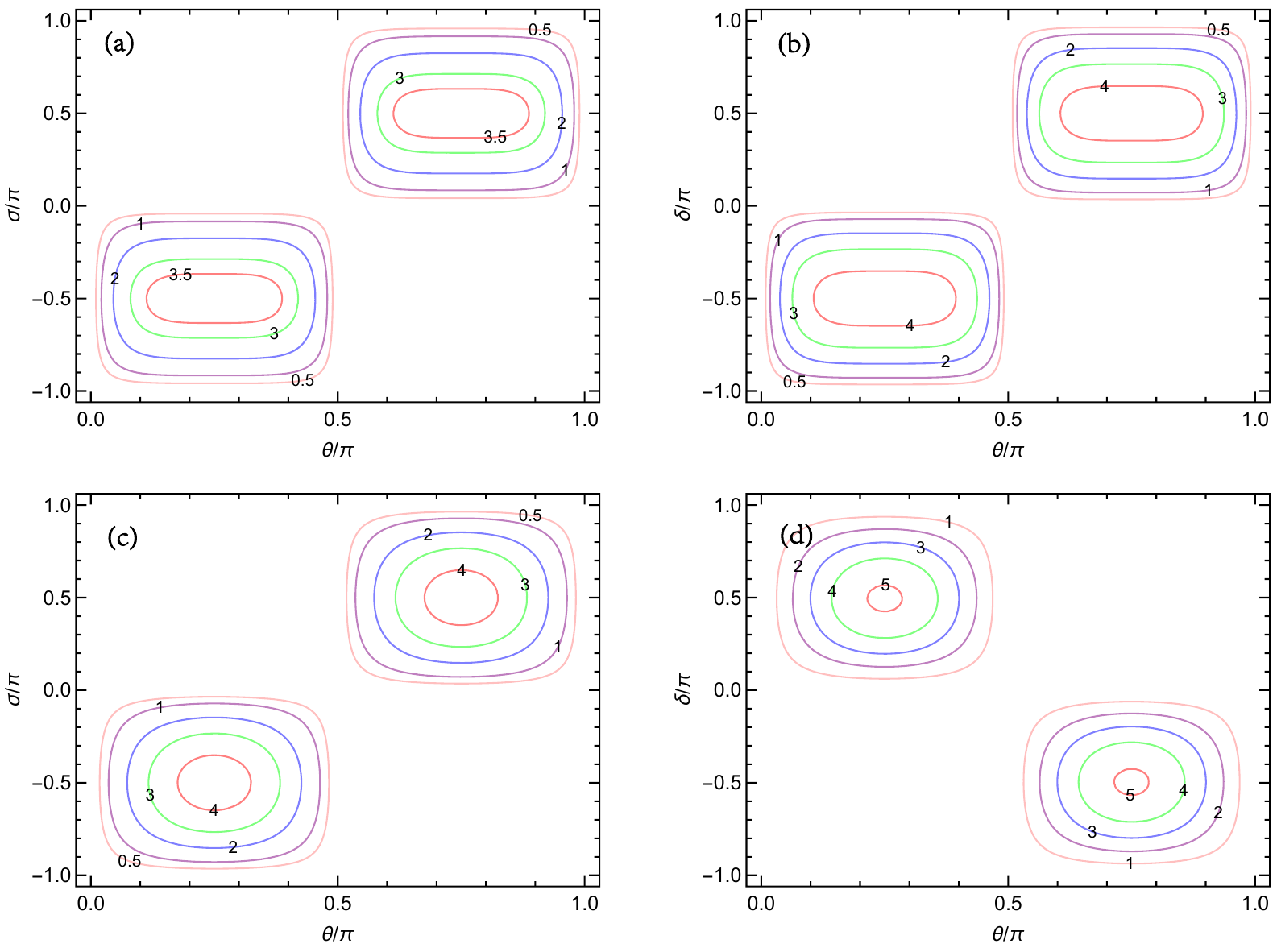}
\caption{In the scenario of assuming $z$ to be real (i.e., $z = \theta$) and taking $M^{}_0 = 1$ TeV, the contour lines of $\xi$ for successful resonant leptogenesis are shown in the $\theta$-$\sigma$ plane with $\delta=0$ for (a) $m^{}_1=0$ and (c) $m^{}_3 =0$; and in the $\theta$-$\delta$ plane with $\sigma=0$ for (b) $m^{}_1=0$ and (d) $m^{}_3 =0$. The units in these four cases are (a) $10^{-9}$, (b) $10^{-9}$, (c) $10^{-11}$ and (d) $10^{-12}$, respectively. }
\label{Fig:4-8}
\end{figure*}

Now we restrict our analysis to the possibility of $z$ being real (renamed as $\theta$). In this case the expressions of $\varepsilon^{}_{i \alpha}$ and $\widetilde m^{}_\alpha$ in Eq.~(\ref{eq:4.4.8}) can be simplified to
\begin{eqnarray}
\varepsilon^{}_{i \alpha} & = & \frac{ \sqrt{m^{}_2 m^{}_3} }{4\pi v^2 \widetilde m^{}_i } \cdot \frac{ \xi M^3_0 }{ 4 \xi^2 M^2_0 + \Gamma^2_j} \left(m^{}_2 - m^{}_3\right) \sin 2\theta \ {\rm Im}\left(U^{*}_{\alpha 2} U^{}_{\alpha 3}\right) \;,
\nonumber \\
\widetilde m^{}_\alpha & = & m^{}_2 \left|U^{}_{\alpha 2}\right|^2 + m^{}_3 \left|U^{}_{\alpha 3}\right|^2  \;,
\label{eq:4.4.9}
\end{eqnarray}
in the $m^{}_1=0$ case; and the results in the $m^{}_3 =0$ case can be obtained from Eq.~(\ref{eq:4.4.9}) by replacing the subscripts 2 and 3 with 1 and 2. Let us first consider the $m^{}_1 =0$ case. In Fig.~\ref{Fig:4-8}(a) we show the contour lines of $\xi/10^{-9}$ for successful leptogenesis in the $\theta$-$\sigma$ plane by assuming $\delta =0$ (i.e., $\sigma$ is the only source of CP violation). Here and hereafter, $M^{}_0 =1$ TeV is taken as a benchmark value. It is obvious that the numerical results exhibit a symmetry under the transformation $\theta \to \pi/2 -\theta$ or $\delta \to \pi-\delta$, or the joint transformations $\theta \to \pi -\theta$ and $\delta \to - \delta$.
We find that $\xi$ has a maximal value of $3.5 \times 10^{-9}$ at $[\theta, \delta] = [\pi/4, -\pi/2]$ and $[3 \pi/4, \pi/2]$. Note that we have only shown the results in the range $\xi \gg \Gamma^{}_j/M^{}_0 \sim  {\cal O}(10^{-14})$. The corresponding results in the range $\xi \ll \Gamma^{}_j/M^{}_0$, which are of ${\cal O}(10^{-19})$, can be obtained with the help of the observation made below Eq.~(\ref{eq:4.4.8}). For the case of $\sigma =0$, similar results are obtained as shown in Fig.~\ref{Fig:4-8}(b). The fact that the values of $\xi$ obtained in the present case are comparable to those in the previous case seems to be in conflict with a naive expectation that they should be smaller by about a factor of $s^{}_{13}$ such that the suppression on the effect of $\delta$ from the accompanying $s^{}_{13}$ factor can be offset. This issue can be understood as follows. Due to the smallness of $|U^{}_{e3}| =s^{}_{13}$, $\widetilde m^{}_e \simeq 3.4 m^{}_*$ is much smaller than $\widetilde m^{}_{\mu} \simeq 28 m^{}_*$ and $\widetilde m^{}_\tau \simeq 22 m^{}_*$, making the $e$-flavor contribution to the final baryon number asymmetry dominant. And one has $\varepsilon^{}_{i e} \propto {\rm Im}(U^{*}_{e 2} U^{}_{e 3}) = - s^{}_{12} c^{}_{13} s^{}_{13} \sin \sigma$ (or $- s^{}_{12} c^{}_{13} s^{}_{13} \sin \delta$) in the $\delta =0$ (or $\sigma =0$) case. Namely, in the present case the effect of $\sigma$ is subject to the equal suppression as $\delta$.
In the $m^{}_3 =0$ case, the magnitudes of $\varepsilon^{}_{i \alpha}$ are commonly suppressed by $m^{}_2 -m^{}_1$ as compared with those in the $m^{}_1 =0$ case. Consequently, the values of $\xi$ for successful leptogenesis are further reduced so as to offset the suppression from $m^{}_2 -m^{}_1$, as can be seen from Fig.~\ref{Fig:4-8}(c) and Fig.~\ref{Fig:4-8}(d).
Now the results of $\xi$ obtained in the case of $\sigma =0$ are indeed smaller by about a factor of $s^{}_{13}$ than those in the case of $\delta =0$.
Finally, it is worth pointing out that for other values of $M^{}_0$ the results of $\xi$ can be obtained from those for $M^{}_0 =1$ TeV by a simple rescaling based on the aforementioned fact that the values of $\xi$ needed for producing certain values of $f(M^{}_0, \xi)$ and thus $\varepsilon^{}_{i \alpha}$ are proportional to $M^{}_0$ in both the $\xi \gg \Gamma^{}_j/M^{}_0$ and $\xi \ll \Gamma^{}_j/M^{}_0$ ranges.

\subsection{RGE effects on leptogenesis}
\label{section 4.5}

Given that leptogenesis is expected to take effect at the seesaw scale, which is much higher than the electroweak scale, it is in general necessary to bridge the gap between leptogenesis and low-energy flavor parameters by means of the RGEs. Furthermore, the light particles involved in leptogenesis will acquire their effective thermal masses proportional to the temperature due to their fast interactions with the hot bath in the early Universe, and this may induce appreciable corrections to several ingredients in the calculations of leptogenesis. A comprehensive study of the renormalization and thermal corrections to leptogenesis has been carried out in Ref.~\cite{Giudice:2003jh}. For the strong washout regime in which we are particularly interested, the following effects are found to be most relevant \cite{Giudice:2003jh}: (1) above all, one needs to renormalize neutrino masses and flavor mixing parameters from low energies (usually taken as the electroweak scale $\Lambda^{}_{\rm EW}$) up to the seesaw scale (usually taken as the mass scale $M^{}_1$ of the lightest heavy Majorana neutrino) where they are used to reconstruct $M^{}_{\rm D}$ --- a key ingredient for the calculation of leptogenesis; (2) the top-quark Yukawa coupling and electroweak gauge coupling constants, which control the rates of the dominant $\Delta L=1$ washout scattering processes, should also be renormalized to the leptogenesis scale; (3) the thermal corrections to the Higgs mass, which may greatly reduce the rates of the $\Delta L =1$ washout scattering processes involving the top quark, should be taken into account.

There are two different but complementary approaches for renormalizing the flavor parameters of three light Majorana neutrinos. In the so-called ``diagonalize and run" approach \cite{Casas:1999tg,Antusch:2003kp}, one simply focuses on the running behaviors of neutrino masses and flavor mixing parameters themselves between $\Lambda^{}_{\rm EW}$ and $M^{}_1$. But in the so-called ``run and diagonalize" approach, one first runs the effective neutrino mass matrix $M^{}_\nu$ from one energy scale to another, and then diagonalizes it to obtain neutrino masses, flavor mixing angles and CP-violating phases. Here we adopt the latter approach, because it is more convenient and transparent in describing the connection between leptogenesis and low-energy flavor parameters via the seesaw formula. At the one-loop level, the RGE for $M^{}_{\nu}$ is given by \cite{Chankowski:1993tx,Babu:1993qv,Antusch:2001ck,Antusch:2001vn,Ohlsson:2013xva}
\begin{eqnarray}
16 \pi^2 \frac{ {\rm d} M^{}_{\nu} }{{\rm d} t} = \gamma \left( Y^{\dagger}_{l} Y^{}_{l} \right)^{T} M^{}_{\nu}
+ \gamma M^{}_{\nu} \left( Y^{\dagger}_{l} Y^{}_{l} \right) + \alpha M^{}_{\nu} \; ,
\label{eq:4.5.1}
\end{eqnarray}
where
\begin{eqnarray}
&& \gamma^{}_{\rm SM}= - \frac{3}{2} \;,  \hspace{1cm} \alpha^{}_{\rm SM} \simeq -3 g^2_2 + 6 y^2_t + \lambda  \;; \nonumber \\
&& \gamma^{}_{\rm MSSM}= 1 \;, \hspace{1cm} \alpha^{}_{\rm MSSM} \simeq -\frac{6}{5} g^2_1 - 6 g^2_2 + 6 y^2_t  \;.
\label{eq:4.5.2}
\end{eqnarray}
The meanings of the symbols in these equations have been explained below Eq.~(\ref{eq:2.13}).
An integration of Eq.~(\ref{eq:4.5.1}) allows us to obtain $M^{}_{\nu}(M^{}_1)$ from $M^{}_{\nu}(\Lambda^{}_{\rm EW})$ or vice versa, according to \cite{Ellis:1999my,Chankowski:1999xc}
\begin{eqnarray}
M^{}_{\nu} (M^{}_1) = I^{-1}_{0} T^{-1}_l
M^{}_{\nu} (\Lambda^{}_{\rm EW}) T^{-1}_l  \;,
\label{eq:4.5.3}
\end{eqnarray}
where $T^{}_l \simeq {\rm Diag} \{1, 1, I^{}_\tau\}$ is an excellent approximation thanks to $m^{}_e \ll m^{}_\mu \ll m^{}_\tau$, and the definitions of $I^{}_0$ and $I^{}_\tau$ can be found in Eq.~(\ref{eq:2.16}).

In the SM with $M^{}_1 \sim 10^{10}$ GeV, the deviation of $I^{}_\tau$ from unity is described by $\Delta^{}_\tau \equiv I^{}_\tau- 1 \simeq - \gamma^{}_{\rm SM} y^2_\tau \ln(M^{}_1/\Lambda^{}_{\rm EW})/(16\pi^2) \simeq 2 \times 10^{-5}$, which appears negligibly small. As a result, the RGE running effects just lead us to an appreciable overall rescaling factor $I^{-1}_0 \sim 1.2$, which is relevant for the absolute neutrino mass scale; but they do not modify the structure of $M^{}_\nu$, which is relevant for the neutrino mass hierarchy and flavor mixing parameters. In the MSSM, however, $y^{2}_\tau = (1+ \tan^2{\beta}) m^2_\tau/v^2$ holds and thus the magnitude of $\Delta^{}_\tau$ can be significantly enhanced by large values of $\tan \beta$. To be specific, one has $\Delta^{}_\tau \simeq - 0.01 (\tan \beta/30)^2$ for $M^{}_1 \sim 10^{10}$ GeV. An appreciable value of $\Delta^{}_\tau$ will modify the structure of $M^{}_\nu$, and thus correct the neutrino mixing angles and CP-violating phases in a nontrivial way. To illustrate, we are going to discuss two interesting scenarios in which successful leptogenesis is triggered by the RGE running effects characterized by the $\tan \beta$-enhanced $\Delta^{}_\tau$. But let us first of all point out two different aspects of the MSSM as compared with the SM in dealing with thermal leptogenesis \cite{Plumacher1998Baryon,Bruce1993Inflation}. (1) Due to the existence of two Higgs doublets, the neutrino and charged-lepton Yukawa coupling matrices are $Y^{}_\nu = M^{}_{\rm D}\sqrt{1+\tan^2 \beta}/(v \tan \beta)$ and $Y^{}_l = M^{}_l \sqrt{1+\tan^2 \beta}/v$ in the MSSM, respectively. In addition to the aforementioned enhancement of $\Delta^{}_\tau$, the temperature values dividing different flavor regimes also receive an enhancement factor of $1+\tan^2 \beta$. For example, the temperature value dividing the unflavored and flavored leptogenesis regimes becomes $\sim (1+\tan^2 \beta)10^{12}$ GeV.
(2) The doubling of the particle spectrum in the MSSM matters. For given values of $M^{}_i$, $Y^{}_\nu$ and $Y^{}_l$, the total effect of supersymmetry on the final baryon number asymmetry can simply be summarized as a constant factor:
\begin{eqnarray}
\left. \frac{Y^{\rm MSSM}_{\rm B}}{Y^{\rm SM}_{\rm B}} \right|^{}_{M^{}_i, Y^{}_\nu, Y^{}_l} \simeq \left\{ \begin{array}{l} \sqrt{2} \hspace{0.5cm} ({\rm strong \ washout} ) \; ; \\ 2 \sqrt{2} \hspace{0.5cm} ({\rm weak \ washout}) \; . \end{array} \right.
\label{eq:4.5.4}
\end{eqnarray}
Of course, a specific seesaw-plus-leptogenesis model usually works in either the MSSM
or the SM extended with heavy Majorana neutrinos, instead of both of them.

As mentioned in section \ref{section 3.5}, in the original Casas-Ibarra parametrization the PMNS neutrino mixing matrix $U$ is cancelled out in the expression of the unflavored CP-violating asymmetries $\varepsilon^{}_i$. But this observation will not be true anymore if the RGE-induced quantum correction is taken into consideration. On the one hand, substituting $M^{}_\nu (\Lambda^{}_{\rm EW})  = U D^{}_\nu U^T$ at the electroweak scale into Eq.~(\ref{eq:4.5.3}) yields the expression of $M^{}_\nu (M^{}_1)$ at the seesaw scale:
\begin{eqnarray}
M^{}_{\nu} (M^{}_1) = I^{-1}_{0} T^{-1}_l U D^{}_\nu U^T T^{-1}_l  \;.
\label{eq:4.5.5}
\end{eqnarray}
On the other hand, the seesaw formula goes as
\begin{eqnarray}
M^{}_\nu(M^{}_1) = - M^{}_{\rm D} D^{-1}_{N} M^T_{\rm D} \;,
\label{eq:4.5.6}
\end{eqnarray}
where $M^{}_{\rm D}$ is defined at the seesaw scale.
A comparison between Eq.~(\ref{eq:4.5.5}) and Eq.~(\ref{eq:4.5.6}) immediately leads us to the RGE-corrected Casas-Ibarra parametrization \cite{Xing:2020erm,Xing:2020ghj}:
\begin{eqnarray}
M^{}_{\rm D}(M^{}_1) = {\rm i}I^{-1/2}_0 \hspace{0.06cm} T^{-1}_l U D^{1/2}_\nu \hspace{0.06cm} O D^{1/2}_{N}  \;,
\label{eq:4.5.7}
\end{eqnarray}
in which $U$ and $D^{}_\nu$ take their values at the electroweak scale. It is easy to check that $U$ will not be cancelled out in the expression of $\varepsilon^{}_i$ due to $T^{}_l \neq I$ as a result of the $\Delta^{}_\tau$-induced RGE effect \cite{Cooper:2011rh}.
We subsequently study the implications of this striking effect for unflavored leptogenesis. Note that in the SM and MSSM frameworks unflavored leptogenesis takes effect when $M^{}_1$ lies in the temperature regimes $T > 10^{12}$ GeV and $T > 10^{12} \hspace{0.06cm} (1+\tan^2 \beta)$ GeV, respectively.

\begin{figure*}[t]
\centering
\includegraphics[width=6in]{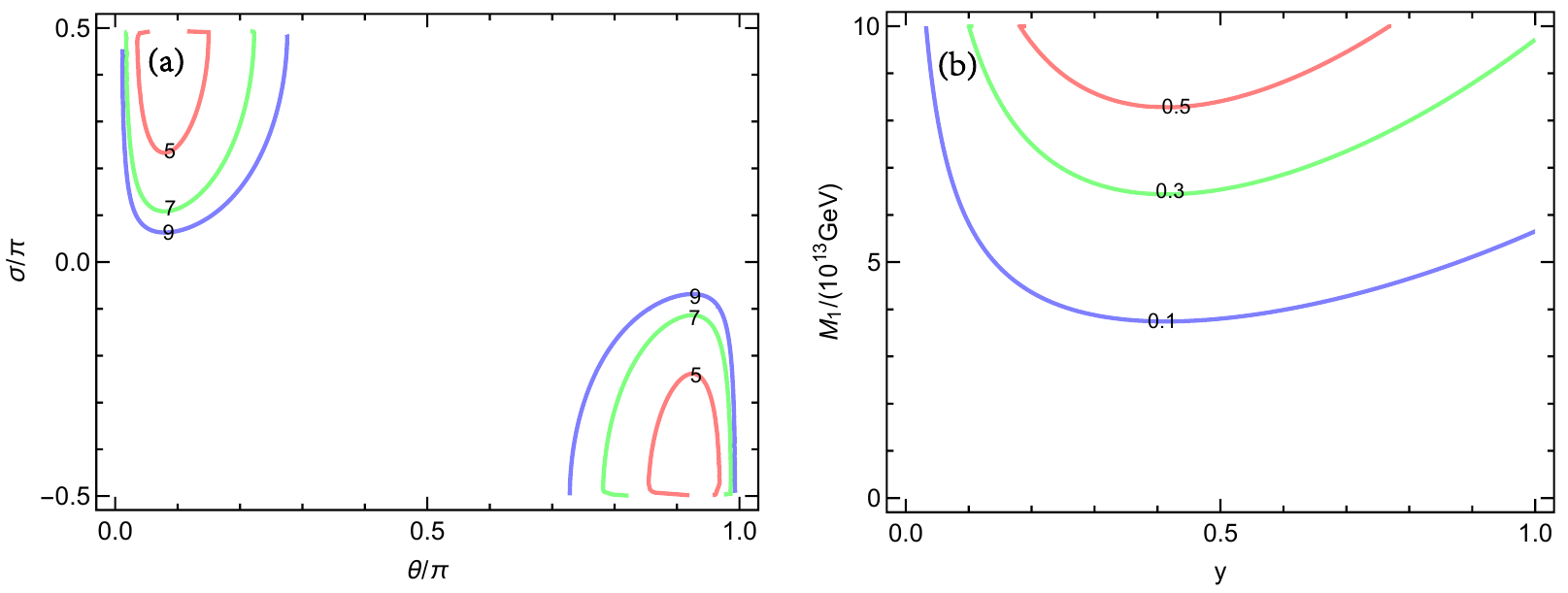}
\caption{(a) In the case of $z$ being real (i.e., $z = \theta$) together with $m^{}_1 =0$ and $\delta =0$, the contour lines of $M^{}_1/(10^{13} \ {\rm GeV})$ for successful unflavored leptogenesis in the $\theta$-$\sigma$ plane. (b) In the case of $\sin z$ being purely imaginary (i.e., $\sin z = {\rm i} \sinh y$ with $y$ being real) together with $m^{}_3 =0$ and $\sigma = \delta =0$, the contour lines of the ratio of $Y^{}_{\rm B}$ to the observed value of $Y^{}_{\rm B}$ in the $y$-$M^{}_1$ plane. }
\label{Fig:4-9}
\end{figure*}

The arbitrary orthogonal matrix $O$ in Eq.~(\ref{eq:4.5.7}) can be explicitly parameterized in terms of a complex parameter $z$ in the minimal seesaw model, as shown in Eq.~(\ref{eq:3.5.5}). Here we focus on the particular case of $z$ being real (renamed as $\theta$), in which leptogenesis can only originate from the nontrivial CP-violating phases of $U$ with the aid of the aforementioned RGE running effects.
In this scenario one has
\begin{eqnarray}
\varepsilon^{}_1 & \simeq & - \frac{3 \Delta^{}_\tau M^{}_1}{8 \pi v^2 I^{}_0} \cdot
\frac{ \sqrt{m^{}_i m^{}_j} \left( m^{}_j - m^{}_i \right) \sin 2\theta }{ m^{}_i \cos^2 \theta + m^{}_j \sin^2 \theta } {\rm Im}   \left( U^{}_{\tau i} U^*_{\tau j} \right) \;,
\nonumber \\
\widetilde m^{}_1 & \simeq & I^{-1}_0 \left( m^{}_i \cos^2 \theta + m^{}_j \sin^2 \theta \right)  \;,
\label{eq:4.5.8}
\end{eqnarray}
where $i =2$ (or $i=1$) and $j =3$ (or $j=2$) in the $m^{}_1 =0$ (or $m^{}_3 =0$) case.
As expected, the size of $\varepsilon^{}_1$ is directly controlled by $\Delta^{}_\tau$ and the imaginary part of $U$. Let us first consider the $m^{}_1 =0$ case.
In Fig.~\ref{Fig:4-9}(a) we illustrate the contour lines of $M^{}_1$ for successful unflavored leptogenesis in the $\theta$-$\sigma$ plane by assuming $\delta =0$.
In our numerical calculations we have required $\tan^2 \beta$ to take its maximally-allowed values for the  unflavored leptogenesis regime to hold (i.e., $\tan^2 \beta \simeq M^{}_1/(10^{12} \ {\rm GeV})$). It is obvious that the results exhibit a symmetry under the joint transformations $\theta \to \pi -\theta$ and $\sigma \to -\sigma$. One can see that there only exist some narrow parameter regions for unflavored leptogenesis to work successfully. Recall that for $M^{}_1 \gtrsim 10^{14}$ GeV the $\Delta L =2$ scattering processes would enter into equilibrium and greatly suppress the efficiency factor. The minimal value of $M^{}_1$ allowing for successful leptogenesis is found to be $4.1 \times 10^{13}$ GeV which is reached at $[\theta, \sigma] \simeq [0.08\pi , 0.5\pi]$ or $[0.92\pi, -0.5\pi]$. If $\sigma =0$ is assumed to ascribe leptogenesis purely to $\delta$, one finds that $\varepsilon^{}_1$ is suppressed by a factor of $s^{}_{13}$ as compared with its size in the previous case.
Hence there exists no parameter space for unflavored leptogenesis to work successfully. Now let us consider the $m^{}_3 =0$ case, in which $\varepsilon^{}_1$ suffers a strong suppression from $m^{}_2 - m^{}_1$ compared to its size in the $m^{}_1=0$ case. This suppression makes leptogenesis quite impotent. We therefore resort to the scenario of $\sin z$ being purely imaginary, in which unflavored leptogenesis still cannot work unless the RGE-induced quantum corrections are included. In terms of the reparametrization $\cos z = \cosh y$ and $\sin z = {\rm i} \sinh y$ with $y$ being a real parameter, $\varepsilon^{}_1$ becomes
\begin{eqnarray}
\varepsilon^{}_1 & \simeq & - \frac{3 \Delta^{}_\tau M^{}_1}{8 \pi v^2 I^{}_0} \cdot
\frac{ \sqrt{m^{}_1 m^{}_2} \left( m^{}_2 + m^{}_1 \right) \sinh 2y }{ m^{}_1 \cosh^2 y + m^{}_2 \sinh^2 y } {\rm Re} \left( U^{}_{\tau 1} U^*_{\tau 2} \right)
\;,
\label{eq:4.5.9}
\end{eqnarray}
which is not suppressed by $m^{}_2-m^{}_1$ anymore.
Unfortunately, Fig.~\ref{Fig:4-9}(b) shows that it is still difficult for the produced baryon number asymmetry to reach the observed value.

We proceed to discuss an appealing scenario that the extremely small mass splitting between $M^{}_1$ and $M^{}_2$ needed for realizing the low-scale resonant leptogenesis scenario is due to the RGE running effects \cite{GonzalezFelipe:2003fi,Turzynski:2004xy,Joaquim:2005zv,Babu:2008kp,Achelashvili:2016trx,Achelashvili:2017nqp}. At a superhigh scale $\Lambda$ (such as the GUT scale) where the heavy Majorana neutrino masses are generated, $N^{}_1$ and $N^{}_2$ are assumed to possess the exactly degenerate masses; namely, $M^{}_1 = M^{}_2 \equiv M^{}_0 < \Lambda$. Then, the RGE evolution of $M^{}_1$ and $M^{}_2$ down to the leptogenesis scale $\sim M^{}_0$ will eventually break their degeneracy. The one-loop RGE of $M^{}_i$ reads \cite{Casas:1999tp}
\begin{eqnarray}
16 \pi^2 \frac{ {\rm d} M^{}_i}{ {\rm d} t} = 2 h \left( Y^\dagger_\nu Y^{}_\nu \right)^{}_{ii} M^{}_i  \;,
\label{eq:4.5.10}
\end{eqnarray}
with $h^{}_{\rm SM}= 1$ and $h^{}_{\rm MSSM}= 2$.
For a reason to be explained below, to achieve successful leptogenesis, the renormalization corrections to $Y^{}_\nu$ should also be taken into account. In the mass bases of both right-handed neutrinos and charged leptons, the one-loop RGE for $Y^{}_\nu$ is \cite{Casas:1999tp,Chankowski:2001mx,Antusch2002Neutrino}
\begin{eqnarray}
16 \pi^2 \frac{ {\rm d} Y^{}_\nu}{{\rm d} t} = f Y^{}_\nu - \left[ a Y^{}_l Y^\dagger_l + b Y^{}_\nu Y^\dagger_\nu  \right] Y^{}_\nu + Y^{}_\nu A   \;,
\label{eq:4.5.11}
\end{eqnarray}
where
\begin{eqnarray}
f^{}_{\rm SM} =3 {\rm Tr}(Y^{}_u Y^\dagger_u) + 3 {\rm Tr}(Y^{}_d Y^\dagger_d) + {\rm Tr}(Y^{}_l Y^\dagger_l) + {\rm Tr}(Y^{}_\nu Y^\dagger_\nu) - \frac{9}{20} g^2_1 - \frac{9}{4} g^2_2  \;,
\nonumber \\
f^{}_{\rm MSSM} =3 {\rm Tr}(Y^{}_u Y^\dagger_u) + {\rm Tr}(Y^{}_\nu Y^\dagger_\nu) - \frac{3}{5} g^2_1 - 3 g^2_2 \;,
\nonumber \\
a^{}_{\rm SM} = -b^{}_{\rm SM} = \frac{3}{2} \;, \hspace{1cm} a^{}_{\rm MSSM} = \frac{1}{3} b^{}_{\rm MSSM} = -1 \;,
\label{eq:4.5.12}
\end{eqnarray}
and $A$ is a $2\times2$ anti-Hermitian matrix. The nonzero entries of $A$ are given by
\begin{eqnarray}
A^{}_{12} = \frac{M^{}_1+M^{}_2}{M^{}_2-M^{}_1} {\rm Re}\left[(Y^\dagger_\nu Y^{}_\nu)^{}_{12}\right] + {\rm i} \frac{M^{}_2-M^{}_1}{M^{}_1+M^{}_2} {\rm Im}\left[(Y^\dagger_\nu Y^{}_\nu)^{}_{12}\right] = - A^{*}_{21} \;.
\label{eq:4.5.13}
\end{eqnarray}
Apparently, $A^{}_{12}$ and $A^{}_{21}$ are singular at the scale $\Lambda$  where $M^{}_1 = M^{}_2$ holds exactly, unless one imposes the condition ${\rm Re}[(Y^\dagger_\nu Y^{}_\nu)^{}_{12}] =0$. However, a purely imaginary $(Y^\dagger_\nu Y^{}_\nu)^{}_{12}$ will lead to $\varepsilon^{}_i \propto {\rm Im}[(Y^\dagger_\nu Y^{}_\nu)^2_{12}] =0$ and thus prohibit successful leptogenesis. That is why one needs to include the renormalization corrections to $Y^{}_\nu$, which provide a unique source for the generation of finite ${\rm Re}[(Y^\dagger_\nu Y^{}_\nu)^{}_{12}]$ and thus finite $\varepsilon^{}_i$ that is proportional to both ${\rm Im}[(Y^\dagger_\nu Y^{}_\nu)^{}_{12}]$ and ${\rm Re}[(Y^\dagger_\nu Y^{}_\nu)^{}_{12}]$. Note that the condition ${\rm Re}[(Y^\dagger_\nu Y^{}_\nu)^{}_{12}] =0$ at the scale $\Lambda$  can always be fulfilled by taking advantage of the orthogonal-rotation freedom of two right-handed neutrinos due to their exact mass degeneracy. To be specific, an orthogonal rotation matrix $R$ of the right-handed neutrinos with the angle
\begin{eqnarray}
\tan 2 \theta = \frac{2 {\rm Re}[(Y^\dagger_\nu Y^{}_\nu)^{}_{12}]}{(Y^\dagger_\nu Y^{}_\nu)^{}_{22} - (Y^\dagger_\nu Y^{}_\nu)^{}_{11}}  \;
\label{eq:4.5.14}
\end{eqnarray}
does not affect the right-handed neutrino mass matrix $M^{}_{\rm R}$ which is proportional to the unity matrix $I$, but it will transform $Y^{}_\nu$ to $Y^{\prime}_\nu \equiv Y^{}_\nu R$ and $Y^\dagger_\nu Y^{}_\nu$ to the following desired form (i.e., ${\rm Re}[(Y^{\dagger\prime}_\nu Y^{\prime}_\nu)^{}_{12}] =0$):
\begin{eqnarray}
Y^{\prime\dagger}_\nu Y^{\prime}_\nu & = & \pmatrix{
(Y^\dagger_\nu Y^{}_\nu)^{}_{11} & {\rm i} {\rm Im}\left[(Y^\dagger_\nu Y^{}_\nu)^{}_{12}\right] \cr
- {\rm i} {\rm Im}\left[(Y^\dagger_\nu Y^{}_\nu)^{}_{12}\right] & (Y^\dagger_\nu Y^{}_\nu)^{}_{22}}
\nonumber \\
&& - \tan\theta \hspace{0.08cm} {\rm Re}\left[(Y^\dagger_\nu Y^{}_\nu)^{}_{12}\right] \pmatrix{1 & 0 \cr
0 & -1}  \;.
\label{eq:4.5.15}
\end{eqnarray}
In other words, the renormalization corrections help us single out a proper right-handed neutrino mass basis, which would otherwise be subject to the orthogonal-rotation arbitrariness. With the aid of Eq.~(\ref{eq:4.5.15}), it is straightforward to show that $O$ in the Casas-Ibarra parametrization of $Y^{\prime}_\nu$ can be conveniently reparameterized as $\cos z = \cosh y $ and $\sin z = {\rm i} \sinh y$ with $y$ being a real parameter.

Given the definition $\xi \equiv (M^{}_2-M^{}_1)/M^{}_1$ and Eq.~(\ref{eq:4.5.10}), one obtains
\begin{eqnarray}
16 \pi^2 \frac{{\rm d} \xi }{{\rm d} t} = 2 h \left(\xi +1\right) \left[ \left( Y^{\prime \dagger}_\nu Y^\prime_\nu \right)^{}_{22} -  \left( Y^{\prime \dagger}_\nu Y^\prime_\nu \right)^{}_{11} \right]  \;,
\label{eq:4.5.16}
\end{eqnarray}
which yields a finite $\xi$ at the leptogenesis scale:
\begin{eqnarray}
\xi & \simeq  & \frac{h}{8 \pi^2} \left[ \left( Y^{\prime \dagger}_\nu Y^\prime_\nu \right)^{}_{22} -  \left( Y^{\prime \dagger}_\nu Y^\prime_\nu \right)^{}_{11} \right] \ln \left(\frac{M^{}_0}{\Lambda} \right) \nonumber \\
& =  & \frac{h M^{}_0 \left(m^{}_j -m^{}_i\right)}{8 \pi^2 v^2}  \ln \left(\frac{M^{}_0}{\Lambda} \right)
\;,
\label{eq:4.5.17}
\end{eqnarray}
where $i =2$ or ($i=1$) and $j =3$ or ($j=2$) in the $m^{}_1 =0$ (or $m^{}_3 =0$) case.
On the other hand, Eq.~(\ref{eq:4.5.11}) allows us to derive the RGE for $Y^{\prime \dagger}_\nu Y^\prime_\nu$ \cite{Joaquim:2005zv}:
\begin{eqnarray}
16 \pi^2 \frac{{\rm d} Y^{\prime \dagger}_\nu Y^\prime_\nu }{{\rm d} t} = 2 f Y^{\prime \dagger}_\nu Y^\prime_\nu - 2 b \left(Y^{\prime \dagger}_\nu Y^\prime_\nu\right)^2 - 2 a Y^{\prime \dagger}_\nu Y^{}_l Y^\dagger_l Y^\prime_\nu + \left[Y^{\prime \dagger}_\nu Y^\prime_\nu, A\right] \;
\label{eq:4.5.18}
\end{eqnarray}
with $[Y^{\prime \dagger}_\nu Y^\prime_\nu, A] = Y^{\prime \dagger}_\nu Y^\prime_\nu A - A Y^{\prime \dagger}_\nu Y^\prime_\nu$, which in turn yields the RGE for ${\rm Re}[(Y^{\prime \dagger}_\nu Y^\prime_\nu)^{}_{12}]$ as
\begin{eqnarray}
16 \pi^2 \frac{{\rm d} {\rm Re}\left[(Y^{\prime \dagger}_\nu Y^\prime_\nu)^{}_{12}\right] }{{\rm d} t} & \simeq & \frac{2 h}{\xi}  \left[ \left( Y^{\prime \dagger}_\nu Y^\prime_\nu \right)^{}_{11} -  \left( Y^{\prime \dagger}_\nu Y^\prime_\nu \right)^{}_{22} \right] {\rm Re}\left[\left(Y^{\prime \dagger}_\nu Y^\prime_\nu\right)^{}_{12}\right] \nonumber \\
&& -2 a {\rm Re}\left[\left(Y^{\prime \dagger}_\nu Y^{}_l Y^\dagger_l Y^\prime_\nu\right)^{}_{12}\right]   \;.
\label{eq:4.5.19}
\end{eqnarray}
Taking into account that ${\rm Re}[(Y^{\prime \dagger}_\nu Y^\prime_\nu)^{}_{12}]$ vanishes at the scale $\Lambda$, one obtains
\begin{eqnarray}
{\rm Re}\left[\left(Y^{\prime \dagger}_\nu Y^\prime_\nu\right)^{}_{12}\right] \simeq - \frac{a y^2_\tau}{16 \pi^2} {\rm Re}\left[\left(Y^{\prime}_\nu\right)^{*}_{\tau 1} \left(Y^{\prime}_\nu\right)^{}_{\tau 2} \right]  \ln \left(\frac{M^{}_0}{\Lambda} \right) \;,
\label{eq:4.5.20}
\end{eqnarray}
at the leptogenesis scale.
Substituting the above results into Eq.~(\ref{eq:2.23}) and making use of the Casas-Ibarra parametrization of $Y^{\prime}_\nu$, we arrive at
\begin{eqnarray}
\hspace{-1.5cm} \left(m^{}_2 \cosh^2 y + m^{}_3 \sinh^2 y\right) \varepsilon^{}_1 & \simeq & \left(m^{}_2 \sinh^2 y + m^{}_3 \cosh^2 y\right) \varepsilon^{}_2
\nonumber \\
& \simeq & \frac{a y^2_\tau \sqrt{m^{}_2 m^{}_3}\left(m^{}_2 + m^{}_3\right) \sinh 2y}{32 \pi h \left(m^{}_2 - m^{}_3\right)} {\rm Re}\left(U^{}_{\tau 2} U^*_{\tau 3}\right) \;
\label{eq:4.5.21}
\end{eqnarray}
in the $m^{}_1 =0$ case; or
\begin{eqnarray}
\varepsilon^{}_1 \simeq  \varepsilon^{}_2
\simeq  - \frac{a h y^2_\tau \left(m^2_2 -m^2_1\right) \sinh 2y}{8 \pi^3 m^2_0 \left(\cosh^2 y + \sinh^2 y\right)^3 } {\rm Re}\left(U^{}_{\tau 1} U^*_{\tau 2}\right)\ln^2 \left(\frac{M^{}_0}{\Lambda} \right) \;
\label{eq:4.5.22}
\end{eqnarray}
in the $m^{}_3 =0$ case. In obtaining these results, we have used $\xi  \gg \Gamma^{}_j/M^{}_0 = \widetilde m^{}_j M^{}_0/(8 \pi v^2) $ in the $m^{}_1=0$ case or $\xi \ll \Gamma^{}_j/M^{}_0$ and $m^{}_1 \simeq m^{}_2 \simeq m^{}_0 \equiv (m^{}_1 +m^{}_2)/2$ in the $m^{}_3=0$ case to simplify the expressions. It is interesting to notice that $\varepsilon^{}_1$ and $\varepsilon^{}_2$ are not explicitly dependent upon $M^{}_0$ in the $m^{}_1 =0$ case.
The final baryon number asymmetry can be calculated from $Y^{}_{\rm B} \simeq - c r(\varepsilon^{}_1 + \varepsilon^{}_2) \kappa(\widetilde m^{}_1 + \widetilde m^{}_2)$.
Unfortunately, the obtained value of $Y^{}_{\rm B}$ is smaller than the observed value of $Y^{}_{\rm B}$ by about one order of magnitude. But in the MSSM where $y^2_\tau$ can be enhanced by a factor of $1+\tan^2 \beta$, it will be possible to achieve successful leptogenesis for large values of $\tan \beta$ (e.g., $\tan \beta > 3$) \cite{GonzalezFelipe:2003fi,Turzynski:2004xy,Joaquim:2005zv}.

\setcounter{equation}{0}
\setcounter{figure}{0}
\section{Simplified textures of the minimal seesaw model}
\label{section 5}

As we have discussed, even in the minimal seesaw model the model parameters are more than the low-energy flavor parameters by two in number. In order to achieve some experimentally testable predictions, one needs to further reduce the number of the model parameters (at least by three). In the literature there are two widely-used approaches to simplify the flavor textures of the minimal seesaw model. The first approach is to start from some phenomenological or empirical points of view and take some entries of the neutrino mass matrices $M^{}_{\rm D}$ and $M^{}_{\rm R}$ either to be vanishing or to have some linear relations (e.g., equalities).
In the second approach one employs some kinds of flavor symmetries to determine or constrain the textures of $M^{}_{\rm D}$ and $M^{}_{\rm R}$, which may typically predict a part of the flavor mixing parameters to be the constants (such as $\theta^{}_{23} = \pi/4$ and $\delta = \pm \pi/2$). These two approaches, which will be discussed respectively in sections 5 and 6, are complementary to each other in some sense. On the one hand, some particular textures of the neutrino mass matrices to be studied in the first approach may provide some useful clues to uncover underlying lepton flavor symmetries. On the other hand, some enlightening flavor symmetries may help us to naturally realize some particular textures of the neutrino mass matrices in the minimal seesaw framework.

\subsection{Two-zero textures of $M^{}_{\rm D}$}
\label{section 5.1}

In this subsection we study possible zero textures of the Dirac neutrino mass matrix $M^{}_{\rm D}$ in the basis where both the charged-lepton mass matrix $M^{}_l$ and the right-handed Majorana neutrino mass matrix $M^{}_{\rm R}$ are diagonal. A texture zero of $M^{}_{\rm D}$ means that the corresponding entry is either exactly vanishing or vanishingly small as compared with its neighboring entries. It is worth pointing out that this approach has been implemented in the quark sector to successfully establish some testable relationships between small quark flavor mixing angles and strong quark mass hierarchies \cite{Weinberg:1977hb, Fritzsch:1977vd, Fritzsch:1999ee}.

Following the spirit of Occam's razor, let us first ascertain the maximal number of texture zeros that can be imposed on $M^{}_{\rm D}$ without bringing about obvious inconsistencies with current experimental data. It is immediate to see that $M^{}_{\rm D}$ with more than three texture zeros can never be phenomenologically viable. In this case the entries in at least one row of $M^{}_{\rm D}$ have to vanish, rendering the corresponding left-handed neutrino field to be completely decoupled from the right-handed neutrino fields and thus from the model itself. Such a consequence is definitely incompatible with the well established three-flavor neutrino mixing picture. So we turn to consider the three-zero textures of $M^{}_{\rm D}$, which can be classified into three categories based on the positions of three zero entries. The first category includes those textures whose two entries in one row of $M^{}_{\rm D}$ are both vanishing, and hence it should be abandoned for the same reason as given above. In the second category all the three entries in one column of $M^{}_{\rm D}$ are vanishing (i.e., the rank of $M^{}_{\rm D}$ is one), leaving us with only a single massive light Majorana neutrino --- a result inconsistent with current neutrino oscillation data either. An example for the zero textures of $M^{}_{\rm D}$ in the third category can be expressed as
\begin{eqnarray}
M^{}_{\rm D} = \pmatrix{
0 & \times \cr
\times & 0 \cr
\times & 0 } \; ,
\label{eq:5.1.1}
\end{eqnarray}
in which ``$\times$" denotes a nonzero entry. Permutating the rows and columns of $M^{}_{\rm D}$ in Eq.~(\ref{eq:5.1.1}) allows us to arrive at five other three-zero textures of $M^{}_{\rm D}$. But it is easy to check that all of them dictate one left-handed neutrino field to be completely decoupled from the other two, and hence this category should also be abandoned. To conclude, all the possible textures of $M^{}_{\rm D}$ with three or more zero entries have been ruled out in the chosen basis for a minimal seesaw model.

Let us proceed to consider all the possible two-zero textures of $M^{}_{\rm D}$ \cite{Frampton:2002qc}, which can also be classified into three categories based on the positions of two zero entries. The first category is those textures with two vanishing entries in the same row of $M^{}_{\rm D}$, and it should be abandoned for the same phenomenological reason as mentioned above. The second category (named as category A) contains six two-zero textures of $M^{}_{\rm D}$, whose vanishing entries are located in different rows and columns of $M^{}_{\rm D}$ as follows:
\begin{eqnarray}
{\rm A}^{}_1: \pmatrix{
0 & \times \cr
\times & 0 \cr
\times & \times } \; , \hspace{1cm}
{\rm A}^{}_2: \pmatrix{
0 & \times \cr
\times & \times \cr
\times & 0 } \; , \hspace{1cm}
{\rm A}^{}_3: \pmatrix{
\times & \times \cr
0 & \times \cr
\times & 0 } \; ,
\nonumber \\
{\rm A}^{}_4: \pmatrix{
\times & 0   \cr
0 & \times   \cr
\times & \times } \; , \hspace{1cm}
{\rm A}^{}_5: \pmatrix{
\times & 0 \cr
\times & \times \cr
0 & \times } \; , \hspace{1cm}
{\rm A}^{}_6: \pmatrix{
\times & \times \cr
\times & 0   \cr
0 & \times } \; .
\label{eq:5.1.2}
\end{eqnarray}
The third category (named as category B) also includes six two-zero textures of $M^{}_{\rm D}$, whose vanishing entries are located in the same column of $M^{}_{\rm D}$ as follows:
\begin{eqnarray}
{\rm B}^{}_1: \pmatrix{
0 & \times \cr
0 & \times \cr
\times & \times } \; , \hspace{1cm}
{\rm B}^{}_2: \pmatrix{
0 & \times \cr
\times & \times \cr
0 & \times } \; , \hspace{1cm}
{\rm B}^{}_3: \pmatrix{
\times & \times \cr
0 & \times \cr
0 & \times } \; ,
\nonumber \\
{\rm B}^{}_4: \pmatrix{
\times & 0  \cr
\times & 0  \cr
\times & \times } \; , \hspace{1cm}
{\rm B}^{}_5: \pmatrix{
\times & 0 \cr
\times & \times \cr
\times & 0 } \; , \hspace{1cm}
{\rm B}^{}_6: \pmatrix{
\times & \times \cr
\times & 0 \cr
\times & 0 } \; .
\label{eq:5.1.3}
\end{eqnarray}
Apparently, patterns $\rm A^{}_{4, 5, 6}$ (or $\rm B^{}_{4, 5, 6}$) can be obtained from patterns $\rm A^{}_{1, 2, 3}$ (or $\rm B^{}_{1, 2, 3}$) by interchanging the two columns of $M^{}_{\rm D}$. Among patterns $\rm A^{}_1$, $\rm A^{}_2$ and $\rm A^{}_3$ (or $\rm B^{}_1$, $\rm B^{}_{2}$ and $\rm B^{}_{3}$), one of them can be achieved from another by permuting the rows of $M^{}_{\rm D}$.
None of these two-zero textures of $M^{}_{\rm D}$ can be immediately excluded without taking into account current neutrino oscillation data, and thus some of them are expected to be potentially viable in neutrino phenomenology. It is worth pointing out that only a single phase parameter in such a two-zero texture of $M^{}_{\rm D}$ can survive the rephasing of relevant left-handed neutrino fields and thus it has a physical meaning. This observation makes it possible to establish a direct connection between leptogenesis and CP violation at low energies in a concrete minimal seesaw model.

Given that the number of free parameters in a generic minimal seesaw model is more than the number of low-energy flavor parameters by two, imposing one texture zero on $M^{}_{\rm D}$ in the chosen basis will make a balance between them. In the Casas-Ibarra parametrization of $M^{}_{\rm D}$, for example, the free parameter $z$ can be determined from the texture zero $(M^{}_{\rm D})^{}_{\alpha 1} =0$ or $(M^{}_{\rm D})^{}_{\alpha 2} =0$ through Eq.~(\ref{eq:3.5.6}) as follows:
\begin{eqnarray}
(M^{}_{\rm D})^{}_{\alpha 1} =0: \hspace{1cm}
\tan z = - \frac{U^{}_{\alpha i} \sqrt{m^{}_i}} {U^{}_{\alpha j} \sqrt{m^{}_j}} \; ,
\nonumber \\
(M^{}_{\rm D})^{}_{\alpha 2} =0: \hspace{1cm}
\tan z = +\frac{ U^{}_{\alpha j} \sqrt{m^{}_j}} {U^{}_{\alpha i} \sqrt{m^{}_i} } \; ,
\label{eq:5.1.4}
\end{eqnarray}
in which $i =2$ (or $i=1$) and $j =3$ (or $j=2$) in the $m^{}_1 =0$ (or $m^{}_3 =0$) case. If another texture zero is imposed on $M^{}_{\rm D}$, the number of model parameters will be fewer than the number of low-energy flavor parameters by two. Then we are left with two predictions from such a more concrete minimal seesaw scenario.
For patterns ${\rm A}^{}_1$---${\rm A}^{}_6$ with two texture zeros $(M^{}_{\rm D})^{}_{\alpha 1} = (M^{}_{\rm D})^{}_{\beta 2} =0$, one may use Eq.~(\ref{eq:5.1.4}) to obtain the following correlations between neutrino masses and flavor mixing parameters at low energies:
\begin{eqnarray}
m^{}_i U^{}_{\alpha i} U^{}_{\beta i} + m^{}_j U^{}_{\alpha j} U^{}_{\beta j} =0 \; .
\label{eq:5.1.5}
\end{eqnarray}
Note that Eq.~(\ref{eq:5.1.5}) is actually equivalent to $\langle m\rangle^{}_{\alpha\beta} \equiv (M^{}_\nu)^{}_{\alpha \beta} =0$ for the light Majorana neutrino mass matrix $M^{}_\nu$ in the $m^{}_1 =0$ or $m^{}_3 =0$ case, and this result can be easily understood by using the seesaw formula $M^{}_\nu = -M^{}_{\rm D} D^{-1}_N M^T_{\rm D}$ in the chosen $M^{}_{\rm R} = D^{}_N$ basis \cite{Guo:2003cc}. As for patterns ${\rm B}^{}_1$---${\rm B}^{}_6$ with $(M^{}_{\rm D})^{}_{\alpha k} = (M^{}_{\rm D})^{}_{\beta k} =0$ (for $k=1$ or 2), one finds that Eq.~(\ref{eq:5.1.4}) may lead to the predictions
\begin{eqnarray}
U^{}_{\alpha i} U^{}_{\beta j} = U^{}_{\alpha j} U^{}_{\beta i} \; .
\label{eq:5.1.6}
\end{eqnarray}
If two of the two-zero textures of $M^{}_{\rm D}$ differ only by the interchange of two columns of $M^{}_{\rm D}$ (e.g., patterns ${\rm A}^{}_1$ and ${\rm A}^{}_4$), it is easy to verify that they will lead us to the same result as obtained in Eq.~(\ref{eq:5.1.5}) or (\ref{eq:5.1.6}).

\begin{figure*}[t]
\centering
\includegraphics[width=6.33in]{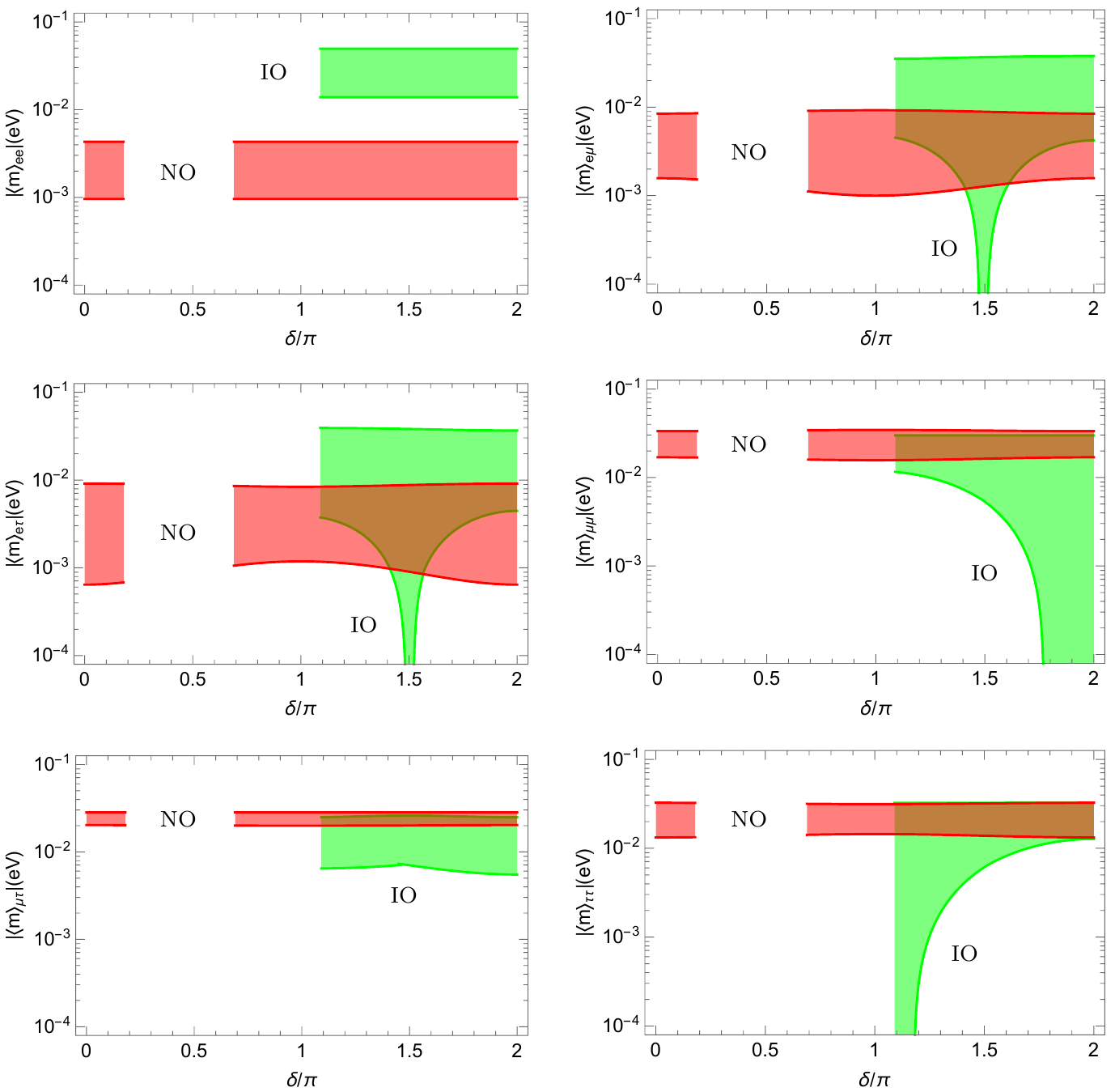}
\caption{In the minimal seesaw model with a choice of the
$M^{}_l = D^{}_l$ basis, the $3\sigma$ ranges of six independent elements
$|\langle m\rangle^{}_{\alpha \beta}|$ (for $\alpha, \beta = e, \mu, \tau$)
of the effective Majorana neutrino mass matrix $M^{}_\nu$ as functions of the
Dirac CP phase $\delta$, where both the normal neutrino mass ordering (NO, or $m^{}_1 =0$) and the inverted ordering (IO, or $m^{}_3 =0$) have been taken into account \cite{Xing:2019vks}.}
\label{Fig:5-1}
\end{figure*}

With the help of the above expressions, we are ready to confront the potentially viable two-zero textures of $M^{}_{\rm D}$ with current neutrino oscillation data \cite{Harigaya:2012bw}. Let us first consider the $m^{}_1 =0$ case. For patterns ${\rm A}^{}_{1}$ and ${\rm A}^{}_{4}$ which predict $(M^{}_\nu)^{}_{e \mu} =0$, Eq.~(\ref{eq:3.2.2}) tells us that Eq.~(\ref{eq:5.1.5}) can be explicitly expressed as
\begin{eqnarray}
m^{}_2 s^{}_{12} \left(c^{}_{12} c^{}_{23} - s^{}_{12} s^{}_{13} s^{}_{23} e^{{\rm i} \delta}\right) + m^{}_3  s^{}_{13} s^{}_{23} e^{-{\rm i} \left(\delta + 2\sigma\right) } =0 \; .
\label{eq:5.1.7}
\end{eqnarray}
But a simple numerical exercise shows that the maximal magnitude (corresponding to $\delta =\pi$) of the first term in Eq.~(\ref{eq:5.1.7}) is actually smaller than the magnitude of the second term in Eq.~(\ref{eq:5.1.7}); namely,
\begin{eqnarray}
m^{}_2 s^{}_{12} (c^{}_{12} c^{}_{23} + s^{}_{12} s^{}_{13} s^{}_{23}) < m^{}_3  s^{}_{13} s^{}_{23} \; .
\label{eq:5.1.8}
\end{eqnarray}
Hence these two terms have no chance to completely cancel each other to allow Eq.~(\ref{eq:5.1.7}) to hold \cite{Xing:2014yka}. This point can also be seen from Fig.~\ref{Fig:5-1}, in which the allowed ranges of $|\langle m\rangle^{}_{\alpha \beta}| = |(M^{}_\nu)^{}_{\alpha \beta}|$ (for $\alpha, \beta = e, \mu, \tau$) are obtained by inputting current neutrino oscillation data at the $3\sigma$ level and permitting $\sigma$ to take arbitrary values. Fig.~\ref{Fig:5-1} tells us that patterns ${\rm A}^{}_{2, 5}$ and ${\rm A}^{}_{3, 6}$, which lead respectively to $(M^{}_\nu)^{}_{e\tau} =0$ and $(M^{}_\nu)^{}_{\mu \tau} =0$, are also disfavored by current  experimental data. When patterns ${\rm B}^{}_{1}$ and ${\rm B}^{}_{4}$ are considered, Eq.~(\ref{eq:5.1.6}) is explicitly expressed as
\begin{eqnarray}
s^{}_{12} s^{}_{23} - c^{}_{12} s^{}_{13} c^{}_{23} e^{- {\rm i} \delta} =0  \; ,
\label{eq:5.1.9}
\end{eqnarray}
which yields a too large value of $s^{}_{13} = t^{}_{12} t^{}_{23} \simeq 0.79$. So these two patterns have already been ruled out. A similar analysis shows that the same is true for patterns ${\rm B}^{}_{2, 5}$ and ${\rm B}^{}_{3, 6}$ \cite{Harigaya:2012bw}.
To summarize, in the $m^{}_1 =0$ case and in the chosen basis all the two-zero textures of $M^{}_{\rm D}$ are disfavored by current neutrino oscillation data. Ref.~\cite{Zhang:2015tea} shows that this conclusion does not change even if the RGE running effects on relevant neutrino flavor parameters are taken into account.

Let us continue with the $m^{}_3 =0$ case. Fig.~\ref{Fig:5-1} shows that in this case $(M^{}_\nu)^{}_{e\mu} =0$ is allowed and thus patterns ${\rm A}^{}_{1}$ and ${\rm A}^{}_{4}$ can be viable for $\delta \simeq \pm \pi/2$. Such values of $\delta$ coincide with the preliminary experimental results of $\delta$. This observation is supported by the analytical analysis made in Ref.~\cite{Harigaya:2012bw} and the numerical calculations done in Ref.~\cite{Zhang:2015tea}. Here we make a reanalysis of patterns ${\rm A}^{}_{1}$ and ${\rm A}^{}_{4}$ in the $m^{}_3 =0$ case with the help of Eq.~(\ref{eq:3.2.3}). Then Eq.~(\ref{eq:5.1.5}) can be explicitly expressed as
\begin{eqnarray}
\hspace{-1cm}
m^{}_1 c^{}_{12} \left(s^{}_{12} + c^{}_{12} s^{}_{13}
\tan \theta^{}_{23} e^{ {\rm i} \delta} \right) - m^{}_2 s^{}_{12} \left( c^{}_{12}  -
s^{}_{12} s^{}_{13} \tan \theta^{}_{23} e^{ {\rm i} \delta} \right) e^{2 {\rm i} \sigma} =0 \; ,
\label{eq:5.1.10}
\end{eqnarray}
from which we obtain
\begin{eqnarray}
s^{}_{13} e^{{\rm i}\delta} = \frac{\left(m^{}_2 e^{2{\rm i}\sigma}- m^{}_1\right) \sin 2 \theta^{}_{12}  } {2 \left(m^{}_1 c^2_{12} + m^{}_2 s^2_{12} e^{2{\rm i}\sigma} \right) \tan \theta^{}_{23} }
\simeq \frac{ \left( e^{2{\rm i}\sigma}- 1\right) \sin 2 \theta^{}_{12}  } { 2 \left( c^2_{12} + s^2_{12} e^{2{\rm i}\sigma} \right) \tan \theta^{}_{23} }  \; ,
\label{eq:5.1.11}
\end{eqnarray}
where the approximate equality between $m^{}_1$ and $m^{}_2$ in the $m^{}_3 =0$ case has been utilized to simplify the expression. In order to get a small value of $s^{}_{13}$ as observed, a significant cancellation between $e^{2{\rm i}\sigma}$ and $1$ is required, which can be achieved for $\sigma$ to be close to $0$. In this situation Eq.~(\ref{eq:5.1.11}) can approximate to
\begin{eqnarray}
s^{}_{13} e^{{\rm i}\delta}
\simeq {\rm i} \frac{\sin 2 \theta^{}_{12}  } {\tan \theta^{}_{23} } \sigma \; ,
\label{eq:5.1.12}
\end{eqnarray}
which subsequently points to $\delta \simeq \pm \pi/2$ and $\sigma \simeq s^{}_{13} \tan \theta^{}_{23}/\sin 2 \theta^{}_{12}$. In another way, $\delta$ and $\sigma$ are directly solved from Eq.~(\ref{eq:5.1.10}) as
\begin{eqnarray}
\cos \delta & \simeq & \frac{ \left(m^2_2 - m^2_1\right) \sin 2 \theta^{}_{12} }{4 m^2_2 s^{}_{13} \tan \theta^{}_{23} } - \frac{s^{}_{13} \tan \theta^{}_{23}  }{\tan 2 \theta^{}_{12}} \; ,
\nonumber \\
\cos 2 \sigma & \simeq & 1 - \frac{2 s^2_{13} \tan^2 \theta^{}_{23}}{\sin^2 2\theta^{}_{12}} \; ,
\label{eq:5.1.13}
\end{eqnarray}
which numerically give $\delta \simeq \pm 0.49 \pi$ and $\sigma \simeq \pm 0.06 \pi$, in good agreement with the above analytical analysis. We notice that such values of $\sigma$ render $|(M^{}_\nu)^{}_{ee}|$ to take some values which are close to its maximally-allowed value $m^{}_1 c^2_{12} c^2_{13}+ m^{}_2 s^2_{12} c^2_{13} \simeq 0.05$ eV (corresponding to $\sigma =0$), a good news for the $0\nu 2\beta$-decay experiments. Because of the approximate $\mu$-$\tau$ symmetry, patterns ${\rm A}^{}_{2}$ and ${\rm A}^{}_{5}$ with the prediction $(M^{}_\nu)^{}_{e\tau} =0$
are also viable (see Fig.~\ref{Fig:5-1} and Ref.~\cite{Xing:2017cwb}). The explicit expression of $(M^{}_\nu)^{}_{e\tau} =0$ and the resulting predictions for $\delta$ and $\sigma$ can be obtained from Eqs.~(\ref{eq:5.1.10}) and (\ref{eq:5.1.13}) by making the replacement $\tan \theta^{}_{23} \to - \cot \theta^{}_{23}$. Numerically, we obtain $\delta \simeq \pm 0.5 \pi$ and $\sigma \simeq \mp 0.04 \pi$ for patterns ${\rm A}^{}_{2}$ and ${\rm A}^{}_{5}$.
In comparison, $(M^{}_\nu)^{}_{\mu\tau} =0$ and thus patterns ${\rm A}^{}_{3, 6}$ are disfavored by current experimental data. Moreover, patterns ${\rm B}^{}_1$---${\rm B}^{}_6$ are also found to be phenomenologically inviable. To summarize, in the $m^{}_3 =0$ case only patterns ${\rm A}^{}_{1, 4}$ and ${\rm A}^{}_{2, 5}$ of $M^{}_{\rm D}$ are consistent with the present experimental results.

It is worthwhile to investigate whether there exist some simple relations among the nonzero entries of four viable two-zero textures of $M^{}_{\rm D}$ so that they can be further simplified.
For this purpose, we present the numerical reconstructions of $M^{}_{\rm D}$ for patterns ${\rm A}^{}_{1}$ and ${\rm A}^{}_{2}$ by taking the best-fit values of neutrino oscillation parameters listed in Table~\ref{Table:1} (for the $m^{}_3 =0$ case) as the typical inputs:
\begin{eqnarray}
{\rm A}^{}_1: \hspace{0.5cm} M^{}_{\rm D} = \pmatrix{
0 & 0.22 \hspace{0.05cm} e^{-1.53{\rm i}} \sqrt{M^{}_2} \cr
0.15 \sqrt{M^{}_1} & 0 \cr
0.17 \sqrt{M^{}_1} & 0.05 \hspace{0.05cm} e^{-1.53{\rm i}} \sqrt{M^{}_2} } \; ;
\nonumber \\
{\rm A}^{}_2: \hspace{0.5cm} M^{}_{\rm D} = \pmatrix{
0 & 0.22 \hspace{0.05cm} e^{+1.63{\rm i}} \sqrt{M^{}_2} \cr
0.14 \sqrt{M^{}_1} & 0.04 \hspace{0.05cm} e^{+1.63{\rm i}} \sqrt{M^{}_2} \cr
0.17 \sqrt{M^{}_1} & 0 } \; ,
\label{eq:5.1.14}
\end{eqnarray}
while the results for patterns ${\rm A}^{}_{4}$ and ${\rm A}^{}_{5}$ just differ by an interchange of two columns of $M^{}_{\rm D}$ in Eq.~(\ref{eq:5.1.14}). Here we have redefined the phases of three left-handed neutrino fields to make the entries in the first column of $M^{}_{\rm D}$ real and those in the second column of $M^{}_{\rm D}$ share a common phase. A short inspection of Eq.~(\ref{eq:5.1.14}) shows the approximate equality between $(M^{}_{\rm D})^{}_{\mu 1}$ and $(M^{}_{\rm D})^{}_{\tau 1}$. A burning question turns out to be: can they be exactly equal to each other within the error bars of neutrino oscillation parameters? A direct numerical calculation tells us that this is the case at the $3\sigma$ level (i.e., the $3\sigma$ ranges of $\Delta m^2_{21}$, $\Delta m^2_{31}$, $\sin^2\theta^{}_{12}$, $\sin^2\theta^{}_{13}$, $\sin^2\theta^{}_{23}$ and $\delta$ are input). Integrated with $(M^{}_{\rm D})^{}_{\mu 1}= (M^{}_{\rm D})^{}_{\tau 1}$, pattern ${\rm A}^{}_{1}$ will lead to $(M^{}_\nu)^{}_{e\mu} =0$ and $(M^{}_\nu)^{}_{\mu\mu} = (M^{}_\nu)^{}_{\mu \tau}$, and pattern ${\rm A}^{}_{2}$ will result in $(M^{}_\nu)^{}_{e\tau} =0$ and $(M^{}_\nu)^{}_{\mu \tau} = (M^{}_\nu)^{}_{\tau\tau}$. To quantify the compatibility of current experimental data with these relations, we introduce a $\chi^2$ function of the form
\begin{eqnarray}
\chi^2 = \sum_i \left( \frac{ {\mathcal O}^{}_{i} - \overline {\mathcal O}^{}_{i} }{ \sigma^{}_{i } } \right)^2 \; ,
\label{eq:5.1.15}
\end{eqnarray}
where the sum is over the quantities $\Delta m^2_{21}$, $\Delta m^2_{31}$, $\sin^2\theta_{12}$, $\sin^2\theta_{13}$, $\sin^2\theta_{23}$ and $\delta$, and ${\mathcal O}^{}_i$, $\overline {\mathcal O}^{}_i$ and $\sigma^{}_{i }$ stand respectively for their predicted values, best-fit values and $1\sigma$ errors. In the case where the data are best fitted (corresponding to the minimization of $\chi^2$), our predictions for the low-energy flavor parameters are shown in Table~\ref{Table:2}. We point out that $\theta^{}_{23}$ is restricted to $\pi/4$, a special value which can be analytically derived from the relations $(M^{}_\nu)^{}_{e\mu} =0$ and $(M^{}_\nu)^{}_{\mu\mu} = (M^{}_\nu)^{}_{\mu \tau}$ of pattern ${\rm A}^{}_{1}$ or from the relations $(M^{}_\nu)^{}_{e\tau} =0$ and $(M^{}_\nu)^{}_{\mu \tau} = (M^{}_\nu)^{}_{\tau\tau }$ of pattern ${\rm A}^{}_{2}$ \cite{Barreiros2018Minimal}.
\begin{table}[t]
\caption{In the $m^{}_3 =0$ case, the predictions for low-energy flavor parameters from a combination of pattern ${\rm A}^{}_{1}$ or ${\rm A}^{}_{2}$ with the equality $(M^{}_{\rm D})^{}_{\mu 1}= (M^{}_{\rm D})^{}_{\tau 1}$. }
\label{Table:2}
\vspace{0.1cm}
\centering
\begin{tabular}{ccccccccc} \br
& $\chi^2_{\rm min}$ & $\displaystyle \frac{\Delta m^2_{21} } {10^{-5} \ {\rm eV}^2 }$ & $\displaystyle \frac{-\Delta m^2_{31} } {10^{-3} \ {\rm eV}^2 }$ & $\displaystyle \frac{\sin^2\theta_{12}} {10^{-1} }$ & $\displaystyle \frac{ \sin^2\theta_{13} } {10^{-2} }$ & $\displaystyle \frac{\sin^2\theta_{23} } {10^{-1} }$ & $\displaystyle \frac{\delta} {\pi}$ & $\displaystyle \frac{ \sigma} {\pi}$  \\
\mr
${\rm A}^{}_1$ &  21 & 7.33 & 2.45 & 3.07 & 2.23 & 5.00 & $-$0.50 & $-$0.05   \\
${\rm A}^{}_2$  & 20 & 7.55 & 2.44 & 3.08 & 2.28 & 5.00 & 0.50 & $-$0.05   \\
\br
\end{tabular}
\end{table}

We proceed to study the implications of four viable two-zero textures of $M^{}_{\rm D}$ for leptogenesis.
Given the Casas-Ibarra parametrization of these textures, Eq.~(\ref{eq:5.1.4}) allows us to determine the free parameter $z$ in terms of the neutrino masses and flavor mixing quantities as follows:
\begin{eqnarray}
{\rm A}^{}_{1, 2}: \hspace{0.5cm} \tan z =  - \sqrt{ \frac{m^{}_1}{m^{}_2} } \cot \theta^{}_{12} \hspace{0.05cm} e^{- {\rm i} \sigma}  \;;
\nonumber \\
{\rm A}^{}_{4, 5}: \hspace{0.5cm} \tan z =  +\sqrt{ \frac{m^{}_2}{m^{}_1} } \tan \theta^{}_{12} \hspace{0.05cm} e^{{+\rm i} \sigma} \; .
\label{eq:5.1.16}
\end{eqnarray}
Then we arrive at a direct connection between the CP-violating asymmetry $\varepsilon^{}_1$ associated with unflavored leptogenesis and the Dirac CP phase $\delta$ of $U$ \cite{Frampton:2002qc,Harigaya:2012bw}:
\begin{eqnarray}
{\rm A}^{}_{1, 4}: \hspace{0.5cm} \varepsilon^{}_1 \propto {\rm Im}\left(\sin^2 z\right) = \mp \frac{1}{2} \sin 2\theta^{}_{12} \tan \theta^{}_{23} \sin \theta^{}_{13} \sin \delta \;; \nonumber \\
{\rm A}^{}_{2, 5}: \hspace{0.5cm} \varepsilon^{}_1 \propto {\rm Im}\left(\sin^2 z\right) = \pm \frac{1}{2} \sin 2\theta^{}_{12} \cot \theta^{}_{23} \sin \theta^{}_{13} \sin \delta \; ,
\label{eq:5.1.17}
\end{eqnarray}
in which ``$-$" corresponds to patterns ${\rm A}^{}_{1,5}$, and ``$+$ is related to patterns ${\rm A}^{}_{2,4}$. In the vanilla leptogenesis scenario described in section~\ref{section 4.1}, pattern ${\rm A}^{}_1$ leads us to
\begin{eqnarray}
\varepsilon^{}_1 \simeq \mp 2.4 \times 10^{-6} \left( \frac{M^{}_1}{10^{13} \ {\rm GeV}} \right) \; ,
\nonumber \\
Y^{}_{\rm B} \simeq \pm 1.7 \times 10^{-11} \left( \frac{M^{}_1}{10^{13} \ {\rm GeV}} \right) \; ,
\label{eq:5.1.18}
\end{eqnarray}
where ``$\mp$" of $\varepsilon^{}_1$ and ``$\pm$" of $Y^{}_{\rm B}$ correspond to ${\rm sign}(\delta) = \pm$. Therefore, successful unflavored leptogenesis can be achieved for $M^{}_1 \simeq 5 \times 10^{13}$ GeV and ${\rm sign}(\delta) = +$. As for pattern ${\rm A}^{}_4$, the correlation between the signs of $\delta$ and $Y^{}_{\rm B}$ is found to be opposite. Namely, ${\rm sign}(\delta) = -$ is needed for successful leptogenesis. A similar calculation shows that leptogenesis can work successfully in pattern ${\rm A}^{}_2$ for $M^{}_1 \simeq 7 \times 10^{13}$ GeV and ${\rm sign}(\delta) = -$; and in pattern  ${\rm A}^{}_5$ the correlation between the signs of $\delta$ and $Y^{}_{\rm B}$ is also found to be opposite \cite{Harigaya:2012bw}. A further study along this line of thought has been carried out in Ref.~\cite{Zhang:2015tea} by taking into account the flavored and resonant leptogenesis scenarios. It turns out that flavored leptogenesis fails to reproduce the observed baryon number asymmetry, as opposed to the result obtained for a general minimal seesaw model (see section \ref{section 4.2}). The reason is that the flavored CP-violating asymmetries are highly suppressed as compared with the unflavored ones in the four two-zero textures of $M^{}_{\rm D}$ under discussion. For example, one has $\varepsilon^{}_{1 e} = \varepsilon^{}_{1\mu} =0$ as a result of $(Y^{}_\nu)^{}_{e1} = (Y^{}_\nu)^{}_{\mu 2} =0$ in pattern ${\rm A}^{}_1$, and thus $\varepsilon^{}_{1\tau} = \varepsilon^{}_1 \propto M^{}_1$ holds. Since flavored leptogenesis takes effect only when $M^{}_1 \lesssim 10^{12}$ GeV, which is far smaller than $M^{}_1 \simeq 5 \times 10^{13}$ obtained above, the resultant value of $Y^{}_{\rm B}$ is not big enough to match the observation.

Finally, let us briefly comment on how to realize the zero textures of $M^{}_{\rm D}$ in the minimal seesaw model. The Abelian flavor symmetries have commonly been invoked for this purpose \cite{Grimus:2004hf}
\footnote{ Note that there are also some particular models where the non-Abelian flavor symmetries \cite{Raby:2003ay, Kuchimanchi:2002yu, Kuchimanchi:2002fi, Dutta:2003ps} or extra dimensions \cite{Harigaya:2012bw,Raidal:2002xf} have been invoked to realize the two-zero textures of $M^{}_{\rm D}$. }.
Here we take the $Z^{}_7 \times Z^{}_4$ symmetries as an illustration. The $Z^{}_7 \times Z^{}_4$ charges of the lepton and Higgs fields are assigned in a way as follows:
\begin{eqnarray}
l^{}_{e \rm L}, e^{}_{\rm R} \sim (\omega^6,1) \; , \hspace{1cm} l^{}_{\mu \rm L}, \mu^{}_{\rm R} \sim (\omega^5,1) \; , \hspace{1cm} l^{}_{\tau \rm L}, \tau^{}_{\rm R} \sim (\omega^2,1) \; , \nonumber \\
 N^{}_{1 \rm R} \sim (\omega,{\rm i}) \; , \hspace{1cm} N^{}_{2 \rm R} \sim (\omega^3,{\rm i})  \; , \hspace{1cm} H \sim (1, 1) \; ,
\label{eq:5.1.19}
\end{eqnarray}
where $\omega \equiv \exp{\left({\rm i}2\pi/7\right)}$. One can see that the $Z^{}_4$ (or $Z^{}_7$) symmetry serves to distinguish different types (flavors) of the lepton fields.
Under such an assignment, the trilinears $\overline{l^{}_{\alpha \rm L}} H \alpha^{}_{\rm R}$ carry no $Z^{}_7 \times Z^{}_4$ charges and will produce the flavor-diagonal charged-lepton mass terms as usual by means of the Higgs mechanism. In comparison, the trilinears $\overline{l^{}_{\alpha \rm L}} H \beta^{}_{\rm R}$ (for $\alpha \neq \beta$) carry nontrivial $Z^{}_7 \times Z^{}_4$ charges, prohibiting the flavor-off-diagonal charged-lepton mass terms. In this way the desired diagonal charged-lepton mass matrix (i.e., $M^{}_l = D^{}_l$) can be realized. On the other hand, the bilinears $\overline{N^{c}_{ i \rm R}} N^{}_{j \rm R}$ carry the following $Z^{}_7 \times Z^{}_4$ charges:
\begin{eqnarray}
\pmatrix{
(\omega^2, -1) & (\omega^4, -1)  \cr
(\omega^4, -1) & (\omega^6, -1)
} \; ,
\label{eq:5.1.20}
\end{eqnarray}
prohibiting the right-handed neutrino mass terms when the $Z^{}_7 \times Z^{}_4$ symmetries maintain. But it is possible to generate the right-handed neutrino mass terms by introducing some $\rm SU(2)^{}_{\rm L}$-singlet scalars with appropriate $Z^{}_7 \times Z^{}_4$ charges, which will spontaneously break the $Z^{}_7 \times Z^{}_4$ symmetries by acquiring nonzero vacuum expectation values. To be specific, two $\rm SU(2)^{}_{\rm L}$-singlet scalars $\phi^{}_1$ and $\phi^{}_2$ with the $Z^{}_7 \times Z^{}_4$ charges $(\omega^5, -1)$ and $(\omega, -1)$ will lead to the right-handed neutrino mass terms
\begin{eqnarray}
y^{}_{11} \langle \phi^{}_1 \rangle \overline{N^{c}_{ 1 \rm R}} N^{}_{1 \rm R} + y^{}_{22} \langle \phi^{}_2 \rangle \overline{N^{c}_{ 2 \rm R}} N^{}_{2 \rm R} \; ,
\label{eq:5.1.21}
\end{eqnarray}
where $y^{}_{ii}$ are the Yukawa-like coefficients and $\langle \phi^{}_i \rangle$ denote the vacuum expectation values of $\phi^{}_i$ (for $i=1,2$). Note that a mass term relating $N^{}_{1 \rm R}$ to $N^{}_{2 \rm R}$ remains absent due to the lack of an $\rm SU(2)^{}_{\rm L}$-singlet scalar with the $Z^{}_7 \times Z^{}_4$ charges $(\omega^3, -1)$. In this way the desired diagonal right-handed neutrino mass matrix (i.e., $M^{}_{\rm R} = D^{}_N$) can also be realized.
Finally, the trilinears $\overline{l^{}_{\alpha \rm L}} \tilde H N^{}_{i \rm R}$ carry the $Z^{}_7 \times Z^{}_4$ charges as
\begin{eqnarray}
\pmatrix{
(\omega^2, {\rm i}) & (\omega^4, {\rm i})  \cr
(\omega^3, {\rm i}) & (\omega^5, {\rm i}) \cr
(\omega^6, {\rm i}) & (\omega, {\rm i})
} \; .
\label{eq:5.1.22}
\end{eqnarray}
The above exercises tell us that one can actually make any given entries of $M^{}_{\rm D}$ either vanishing or nonzero with the help of suitable Abelian flavor symmetries.
For instance, pattern $\rm A^{}_1$ can be realized by introducing some $\rm SU(2)^{}_{\rm L}$-singlet scalars with the $Z^{}_7 \times Z^{}_4$ charges $(\omega^3, -{\rm i})$, $(\omega^4, -{\rm i})$, $(\omega, -{\rm i})$ and $(\omega^6, -{\rm i})$, which are responsible for generating the corresponding nonzero $(e, 2)$, $(\mu, 1)$, $(\tau, 1)$ and $(\tau, 2)$ entries of $M^{}_{\rm D}$; while the $(e, 1)$ and $(\mu, 2)$ entries can remain vanishing if one does not introduce any $\rm SU(2)^{}_{\rm L}$-singlet scalars with the $Z^{}_7 \times Z^{}_4$ charges $(\omega^5, -{\rm i})$ and $(\omega^2, -{\rm i})$. Now that the powers of $\omega$ in six entries of $M^{}_{\rm D}$ shown in Eq.~(\ref{eq:5.1.22}) are all different, all the possible zero textures of $M^{}_{\rm D}$ can be achieved in an analogous way.

\subsection{When $M^{}_{\rm R}$ is not diagonal}
\label{section 5.2}

Now we turn to the zero textures of $M^{}_{\rm D}$ in the basis where the right-handed neutrino mass matrix $M^{}_{\rm R}$ is not diagonal anymore \cite{Goswami:2008rt,Barreiros2018Minimal,Barreiros:2018bju} (see Ref.~\cite{Barreiros:2020mnr} for a study on the case of the charged-lepton mass matrix $M^{}_l$ being non-diagonal). To be specific,
the following three textures of symmetric $M^{}_{\rm R}$ will be considered:
\begin{eqnarray}
{\rm T}^{}_1: \pmatrix{ 0 & \times \cr \times & 0 } \; , \hspace{1cm}
{\rm T}^{}_2: \pmatrix{ 0 & \times \cr \times & \times } \; , \hspace{1cm}
{\rm T}^{}_3: \pmatrix{ \times & \times \cr \times & 0 } \; .
\label{eq:5.2.1}
\end{eqnarray}
Note that any of textures $\rm T^{}_{1, 2, 3}$ of $M^{}_{\rm R}$ can always be made real via a proper redefinition of the phases of two right-handed neutrino fields. Therefore, we simply take $M^{}_{\rm R}$ to be real without loss of generality.

In Eq.~(\ref{eq:3.5.4}) the Casas-Ibarra parametrization of $M^{}_{\rm D}$ has been formulated by taking the diagonal basis $M^{}_{\rm R} = D^{}_N$. To incorporate textures $\rm T^{}_{1, 2, 3}$ of $M^{}_{\rm R}$ given by Eq.~(\ref{eq:5.2.1}) into this parametrization, one needs to diagonalize each of them by means of a unitary transformation as follows:
\begin{eqnarray}
O^T_{\rm R} M^{}_{\rm R} O^{}_{\rm R} = D^{}_N \; .
\label{eq:5.2.2}
\end{eqnarray}
Meanwhile, $M^{}_{\rm D}$ is transformed into $M^{\prime}_{\rm D} = M^{}_{\rm D} O^{}_{\rm R}$.
Now that $M^{\prime}_{\rm D}$ can be parameterized in the Casas-Ibarra form as $M^{\prime}_{\rm D} = {\rm i} U D^{1/2}_\nu O D^{1/2}_{N}$, one immediately arrives at a modified Casas-Ibarra parametrization of $M^{}_{\rm D}$ as
\begin{eqnarray}
M^{}_{\rm D} = {\rm i} U D^{1/2}_\nu O D^{1/2}_{N} O^\dagger_{\rm R} \; .
\label{eq:5.2.3}
\end{eqnarray}
For texture $\rm T^{}_1$, the two heavy Majorana neutrinos are exactly degenerate in mass (i.e., $M^{}_1 = M^{}_2$) and thus
\begin{eqnarray}
O^{}_{\rm R} = \frac{1}{\sqrt 2} \pmatrix{ 1 & 1 \cr - 1  & 1 } P^{}_N R \; ,
\label{eq:5.2.4}
\end{eqnarray}
where $P^{}_N = {\rm Diag} \{ {\rm i}, 1 \}$ ensures the positivity of $M^{}_{1, 2}$ and $R$ is an arbitrary orthogonal matrix arising due to the degeneracy between $M^{}_1$ and $M^{}_2$.
It is easy to see that $R$ can be absorbed in Eq.~(\ref{eq:5.2.3}) via a redefinition of $O$ because of $D^{}_N \propto I$, so we may simply neglect $R$ but keep in mind that now $O$ inherits its arbitrariness. As for texture $\rm T^{}_2$ (or $\rm T^{}_3$), the corresponding $O^{}_{\rm R}$ is given by
\begin{eqnarray}
O^{}_{\rm R} = \pmatrix{ \cos \theta & \sin \theta \cr - \sin \theta & \cos \theta } P^{}_N \; ,
\label{eq:5.2.5}
\end{eqnarray}
where $\tan \theta = \sqrt{M^{}_1/M^{}_2}$ (or $\sqrt{M^{}_2/M^{}_1}$).

With the help of the above expressions, we are ready to study possible implications of the aforementioned texture zeros imposed on $M^{}_{\rm D}$. Let us first consider texture $\rm T^{}_1$ \cite{Achelashvili:2016nkr}. From Eqs.~(\ref{eq:5.2.3}) and (\ref{eq:5.2.4}), we get
the following explicit expression:
\begin{eqnarray}
(M^{}_{\rm D})^{}_{\alpha 1} =0: \hspace{0.5cm}
\left(\sqrt{m^{}_i} U^{}_{\alpha i} + {\rm i} \sqrt{m^{}_j} U^{}_{\alpha j} \right) \left(\cos z - {\rm i} \sin z\right) = 0 \; ,
\nonumber \\
(M^{}_{\rm D})^{}_{\alpha 2} =0: \hspace{0.5cm}
\left(\sqrt{m^{}_i} U^{}_{\alpha i} - {\rm i} \sqrt{m^{}_j} U^{}_{\alpha j} \right) \left(\cos z + {\rm i} \sin z\right) = 0 \; ,
\label{eq:5.2.6}
\end{eqnarray}
where $i =2$ (or $i=1$) and $j =3$ (or $j=2$) in the $m^{}_1 =0$ (or $m^{}_3 =0$) case.
In order for Eq.~(\ref{eq:5.2.6}) to hold for arbitrary values of $z$, one must have
\begin{eqnarray}
\sqrt{m^{}_i} U^{}_{\alpha i} \pm {\rm i} \sqrt{m^{}_j} U^{}_{\alpha j} =0
\hspace{0.5cm} \Longrightarrow \hspace{0.5cm}
m^{}_i U^2_{\alpha i} + m^{}_j U^{2}_{\alpha j} = 0 \; ,
\label{eq:5.2.7}
\end{eqnarray}
which is equivalent to $(M^{}_\nu)^{}_{\alpha \alpha} =0$. In fact, this result can be directly obtained from the seesaw formula: a combination of $(M^{}_{\rm D})^{}_{\alpha 1} =0$ or $(M^{}_{\rm D})^{}_{\alpha 2} =0$ with texture $\rm T^{}_1$ of $M^{}_{\rm R}$ simply leads to  $(M^{}_\nu)^{}_{\alpha \alpha} =0$ via $M^{}_\nu = -M^{}_{\rm D} M^{-1}_{\rm R} M^T_{\rm D}$. Fig.~\ref{Fig:5-1} tells us that in the $m^{}_1= 0$ case none of $(M^{}_\nu)^{}_{\alpha \alpha}$ is allowed to vanish within the $3\sigma$ level. But in the $m^{}_3= 0$ case $(M^{}_\nu)^{}_{\mu\mu} =0$ or $(M^{}_\nu)^{}_{\tau\tau} =0$ can hold for $\delta \simeq 0$ or $\pi$, respectively. Hence it is phenomenologically viable to require $(M^{}_{\rm D})^{}_{\mu k} =0$ or $(M^{}_{\rm D})^{}_{\tau k} =0$ (for $k=1$ or 2). However, the possibility of $(M^{}_{\rm D})^{}_{\mu k} = (M^{}_{\rm D})^{}_{\tau k} =0$ for texture $\rm T^{}_1$ of $M^{}_{\rm R}$ has been ruled out by current experimental data. Given $m^{}_3= 0$, $(M^{}_\nu)^{}_{\mu\mu} =0$ is explicitly expressed as
\begin{eqnarray}
& m^{}_1 \left(s^{}_{12}
+ c^{}_{12} s^{}_{13} \tan \theta^{}_{23} e^{{\rm i} \delta} \right)^2
 + m^{}_2 \left(c^{}_{12} - s^{}_{12}
s^{}_{13} \tan \theta^{}_{23} e^{{\rm i} \delta} \right)^2 e^{2{\rm i} \sigma} = 0   \; ,
\label{eq:5.2.8}
\end{eqnarray}
from which $\delta$ and $\sigma$ are approximately obtained as
\begin{eqnarray}
\cos \delta \simeq \frac{1}{2 \tan 2\theta^{}_{12} s^{}_{13} \tan \theta^{}_{23} }  \; ,
\nonumber \\
\sin \sigma \simeq - \frac{1}{\sin 2\theta^{}_{12} } \; .
\label{eq:5.2.9}
\end{eqnarray}
In the same case the explicit expression of $(M^{}_\nu)^{}_{\tau\tau} =0$ and the corresponding predictions for $\delta$ and $\sigma$ can be achieved from Eqs.~(\ref{eq:5.2.8}) and (\ref{eq:5.2.9}) by making the replacement $\tan \theta^{}_{23} \to - \cot \theta^{}_{23}$. Our numerical calculations show that $(M^{}_\nu)^{}_{\mu\mu} =0$ or $(M^{}_\nu)^{}_{\tau\tau} =0$ can hold at the $1\sigma$ or $3\sigma$ level and lead to $\delta \simeq 0$ or $\pi$ together with  $\sigma \simeq \pi/2$, respectively. Of course, leptogenesis has no way to work unless the degeneracy between $M^{}_1$ and $M^{}_2$ is properly lifted (as discussed in section~\ref{section 4.5}) \cite{Zhao:2021tgi}.

For texture $\rm T^{}_{2}$ (or $\rm T^{}_{3}$) of $M^{}_{\rm R}$, Eqs.~(\ref{eq:5.2.3}) and  (\ref{eq:5.2.5}) allow us to obtain $(M^{}_\nu)^{}_{\alpha \alpha} =0$ for
the imposition of $(M^{}_{\rm D})^{}_{\alpha 1} =0$ (or $(M^{}_{\rm D})^{}_{\alpha 2} =0$) as can be checked by using the seesaw formula, and
\begin{eqnarray}
& \tan z = - \frac{ {\rm i} M^{}_1 \sqrt{m^{}_i} U^{}_{\alpha i} + M^{}_2 \sqrt{m^{}_j} U^{}_{\alpha j} } { {\rm i} M^{}_1 \sqrt{m^{}_j} U^{}_{\alpha j} - M^{}_2 \sqrt{m^{}_i} U^{}_{\alpha i} }
\nonumber \\
( {\rm or} \hspace{0.3cm} & \left. \tan z = - \frac{ {\rm i} M^{}_1 \sqrt{m^{}_i} U^{}_{\alpha i} - M^{}_2 \sqrt{m^{}_j} U^{}_{\alpha j} } { {\rm i} M^{}_1 \sqrt{m^{}_j} U^{}_{\alpha j} + M^{}_2 \sqrt{m^{}_i} U^{}_{\alpha i}  } \right) \; ,
\label{eq:5.2.10}
\end{eqnarray}
for the imposition of $(M^{}_{\rm D})^{}_{\alpha 2} =0$ (or $(M^{}_{\rm D})^{}_{\alpha 1} =0$).
Next, let us consider the possibility of simultaneously imposing two texture zeros on $M^{}_{\rm D}$. Case (1): a combination of $(M^{}_{\rm D})^{}_{\alpha 1} = (M^{}_{\rm D})^{}_{\beta 1} =0$ (or $(M^{}_{\rm D})^{}_{\alpha 2} = (M^{}_{\rm D})^{}_{\beta 2} =0$) with texture $\rm T^{}_2$ (or $\rm T^{}_3$)  of $M^{}_{\rm R}$ will lead us to $(M^{}_\nu)^{}_{\alpha \alpha} = (M^{}_\nu)^{}_{\beta \beta} = (M^{}_\nu)^{}_{\alpha \beta} =0$ via the seesaw relation, a result which is disfavored by current experimental data (see Fig.~\ref{Fig:5-1}). Case (2): with the help of Eq.~(\ref{eq:5.2.10}), we find that a combination of $(M^{}_{\rm D})^{}_{\alpha 2} = (M^{}_{\rm D})^{}_{\beta 2} =0$ (or $(M^{}_{\rm D})^{}_{\alpha 1} = (M^{}_{\rm D})^{}_{\beta 1} =0$) with texture $\rm T^{}_2$ (or $\rm T^{}_3$) of $M^{}_{\rm R}$ will lead us to a constraint equation on the lepton flavor mixing parameters as Eq.~(\ref{eq:5.1.6}), which is not viable either (see section~\ref{section 5.1} again). Case (3): considering the possibilities of two texture zeros located in two different columns of $M^{}_{\rm D}$ as patterns ${\rm A}^{}_1$---${\rm A}^{}_6$ listed in Eq.~(\ref{eq:5.1.2}), one can easily see that a combination of $(M^{}_{\rm D})^{}_{\alpha 1} = (M^{}_{\rm D})^{}_{\beta 2} =0$ (or $(M^{}_{\rm D})^{}_{\alpha 2} = (M^{}_{\rm D})^{}_{\beta 1} =0$) with texture $\rm T^{}_2$ (or $\rm T^{}_3$) of $M^{}_{\rm R}$ will result in $(M^{}_\nu)^{}_{\alpha \alpha} =0$ and a determination of $z$ as in Eq.~(\ref{eq:5.2.10}). The numerical calculations in Ref.~\cite{Barreiros2018Minimal} show that a combination of patterns ${\rm A}^{}_{3, 4}$ (or ${\rm A}^{}_{1,6}$) or ${\rm A}^{}_{5, 6}$ (or ${\rm A}^{}_{2, 3}$) of $M^{}_{\rm D}$ with texture $\rm T^{}_2$ (or $\rm T^{}_3$) of $M^{}_{\rm R}$ can be phenomenologically viable at the $1\sigma$ or $3\sigma$ level in the $m^{}_3 =0$ case, which lead to $\delta \simeq 0$ or $\pi$ and $\sigma \simeq \pi/2$. For a viable combination as described above, one can of course investigate whether there exist some simple relations among the nonzero entries of $M^{}_{\rm D}$ so that its texture can be further simplified as we have done in section~\ref{section 5.1} \cite{Barreiros2018Minimal}. As for the study of possible implications of such zero textures of $M^{}_{\rm D}$ and $M^{}_{\rm R}$ on leptogenesis, we refer the reader to Ref.~\cite{Barreiros2018Minimal}, where it is found that successful leptogenesis can be achieved for $M^{}_1 \sim 10^{14}$ GeV.

\subsection{Hybrid textures of $M^{}_{\rm D}$}
\label{section 5.3}

Now we examine some hybrid textures of $M^{}_{\rm D}$ in the basis where both the charged-lepton mass matrix $M^{}_l$ and the right-handed neutrino mass matrix $M^{}_{\rm R}$ are diagonal (i.e., $M^{}_l = D^{}_l$ and $M^{}_{\rm R} = D^{}_N$). In the literature a hybrid texture usually refers to the coexistence of texture zeros and equalities for the entries of a given mass matrix \cite{Kaneko:2005yz}. Here we consider not only equalities \cite{Goswami2010Hybrid} but also simple ratios for the entries of $M^{}_{\rm D}$. For the sake of convenience, we define
\begin{eqnarray}
a^{}_{12} = \frac{(M^{}_{\rm D})^{}_{e 1}}{(M^{}_{\rm D})^{}_{\mu 1}} \; , \hspace{1cm}
a^{}_{13} = \frac{(M^{}_{\rm D})^{}_{e 1}}{(M^{}_{\rm D})^{}_{\tau 1}} \; , \hspace{1cm}
a^{}_{23} = \frac{(M^{}_{\rm D})^{}_{\mu 1}}{(M^{}_{\rm D})^{}_{\tau 1}} \; , \nonumber \\
b^{}_{12} = \frac{(M^{}_{\rm D})^{}_{e 2}}{(M^{}_{\rm D})^{}_{\mu 2}} \; , \hspace{1cm}
b^{}_{13} = \frac{(M^{}_{\rm D})^{}_{e 2}}{(M^{}_{\rm D})^{}_{\tau 2}} \; , \hspace{1cm}
b^{}_{23} = \frac{(M^{}_{\rm D})^{}_{\mu 2}}{(M^{}_{\rm D})^{}_{\tau 2}} \; .
\label{eq:5.3.1}
\end{eqnarray}
In the following we are going to check whether such ratios can take some special values after one or two texture zeros are imposed on $M^{}_{\rm D}$. To be more explicit, only the simple numbers 1/3, 1/2, 1, 2 and 3 will be taken into account for $|a^{}_{ij}|$ and $|b^{}_{ij}|$.
Given that in the $m^{}_1 =0$ case the two-zero textures of $M^{}_{\rm D}$ are disfavored by current neutrino oscillation data and in the $m^{}_3=0$ case some viable two-zero textures of $M^{}_{\rm D}$ with simple values of $a^{}_{ij}$ and $b^{}_{ij}$ have been studied in section~\ref{section 5.1}, here we simply focus on the one-zero texture of $M^{}_{\rm D}$ \cite{King:2002nf,Brahmachari:2006es}. Moreover, we just consider the possibilities of $(M^{}_{\rm D})^{}_{\alpha 1} =0$, since  $(M^{}_{\rm D})^{}_{\alpha 2} =0$ will lead to the same low-energy consequences.

Following a top-down approach, one may first assume some special values of $a^{}_{ij}$ and $b^{}_{ij}$ and then check their consistencies with current experimental data. Motivated by Eq.~(\ref{eq:5.1.4}), however, we choose to adopt a bottom-up approach as follows. Imposing a texture zero on $M^{}_{\rm D}$ leads to the determination of $z$ in terms of other flavor parameters. The remaining entries of $M^{}_{\rm D}$ can be subsequently expressed in terms of the low-energy flavor parameters and the heavy Majorana neutrino masses $M^{}_i$. It turns out that the dependence of $a^{}_{ij}$ and $b^{}_{ij}$ on $M^{}_i$ is cancelled out, so we can infer what special values they may take by simply taking into account current neutrino oscillation data.
In comparison, this is not the case for the ratios of two entries located in different columns of $M^{}_{\rm D}$, and hence such a case will not be considered here.

We first take a look at the $m^{}_1 =0$ case. Let us begin with $(M^{}_{\rm D})^{}_{e1} =0$. Above all, the allowed ranges of $|a^{}_{23}|$ and $|b^{}_{ij}|$ are numerically obtained in Table~\ref{Table:3} by varying the neutrino oscillation parameters in their $3\sigma$ intervals and treating $\sigma$ as a free phase parameter.
These results verify the conclusion that patterns $\rm A^{}_{1, 2, 4, 5}$ of $M^{}_{\rm D}$ in Eq.~(\ref{eq:5.1.2}) are phenomenologically disfavored. Nevertheless, the fact that $|b^{}_{23}|$ can be much smaller (or larger) than one indicates that such zero textures can approximately hold at an empirically acceptable level \cite{Rink:2016vvl}. It should be noted that when $|a^{}_{23}|$ and $|b^{}_{ij}|$ can individually take some special values, $a^{}_{23}$ and $b^{}_{ij}$ can also do so thanks to the freedom of rephasing the left-handed neutrino fields. So one may directly infer what special values $a^{}_{23}$ and $b^{}_{ij}$ can individually take with the help of the results listed in Table~\ref{Table:3}. In the following, we explore possible textures of $M^{}_{\rm D}$ where its one (or even both) column is of a particular alignment.
\begin{table}[t]
\caption{In the $a^{}_\alpha \equiv (M^{}_{\rm D})^{}_{\alpha 1} =0$ cases (for $\alpha = e, \mu, \tau$), the allowed ranges of $|a^{}_{ij}|$ and $|b^{}_{ij}|$ as constrained by current experimental data at the $3\sigma$ level. }
\label{Table:3}
\vspace{0.1cm}
\centering
\begin{tabular}{cccccccc} \br
&  & $|a^{}_{12}|$ & $|a^{}_{13}|$ & $|a^{}_{23}|$ & $|b^{}_{12}|$ & $|b^{}_{13}|$ & $|b^{}_{23}|$ \\
\mr
$m^{}_1 =0$ & $a^{}_e =0$  & 0 &  0 & 0.49$-$2.3  & 0.12$-$3.6 & 0.12$-$5.1 & 0.13$-$10.3 \\
 & $a^{}_\mu =0$  & $-$ &  0.29$-$0.68 & 0 & 0.05$-$0.51 & 0.05$-$0.40 & 0.59$-$1.6 \\
 & $a^{}_\tau =0$  & 0.26$-$0.66 &  $-$ & $-$  & 0.04$-$0.38 & 0.04$-$0.59 & 0.66$-$1.94 \\
$m^{}_3 =0$ & $a^{}_e =0$  & 0 &  0 & 0.77$-$1.2  & 0.39$-\infty$ & 0.38$-\infty$ & 0$-\infty$ \\
 & $a^{}_\mu =0$  & $-$ &  3.8$-$5.2 & 0  & 0$-\infty$ & 0$-$5.3 & 0$-$2.4 \\
 & $a^{}_\tau =0$  & 4.1$-$5.5 &  $-$ & $-$  & 0$-$6.3 & 0$-\infty$ & 0.37$-\infty$ \\
\br
\end{tabular}
\end{table}

Now we explore the textures of $M^{}_{\rm D}$ where all of $b^{}_{ij}$ can simultaneously take some special values. Also due to the freedom of rephasing the left-handed neutrino fields, one just needs to require the moduli $|b^{}_{ij}|$ to simultaneously take some special values.
For such textures the second column of $M^{}_{\rm D}$, denoted as $(M^{}_{\rm D})^{}_{2}$, can be fixed up to an overall factor. For simplicity, here we only consider those textures in which  $(M^{}_{\rm D})^{}_2$ is proportional to $(1, 1, 1)^{T}$, $(1, 1, 2)^{T}$, $(1, 2, 2)^{T}$, $(1, 1, 3)^{T}$, $(1, 3, 3)^{T}$ or a version obtained from permutating the three entries of $(M^{}_{\rm D})^{}_2$, such as $(1, 2, 1)^{T}$ derived from $(1, 1, 2)^{T}$. Table~\ref{Table:4} enumerates all the particular patterns of $(M^{}_{\rm D})^{}_2$ consistent with current  experimental data at the $3\sigma$ level, together with their phenomenological consequences. Our numerical results are obtained in such a way that the data are best fitted, corresponding to the minimizations of the $\chi^2$ function defined in Eq.~(\ref{eq:5.1.15}).
\begin{table}[t]
\caption{In the $m^{}_1 =0$ case with $(M^{}_{\rm D})^{}_{\alpha 1} =0$, the phenomenological consequences obtained from some particular patterns of $(M^{}_{\rm D})^{}_1$ and $(M^{}_{\rm D})^{}_2$ defined to be the first and second columns of $M^{}_{\rm D}$. The units of $\Delta m^2_{21}$ and $\Delta m^2_{31}$ are $10^{-5}$ eV$^2$ and $10^{-3}$ eV$^2$, respectively; and the symbol ``$\times$" denotes an unspecified number.}
\label{Table:4}
\vspace{0.1cm}
\centering
\begin{tabular}{cccccccccc} \br
$(M^{}_{\rm D})^{T}_1$ & $(M^{}_{\rm D})^{T}_2$ & $\chi^2_{\rm min}$ & $\Delta m^2_{21}$ & $\Delta m^2_{31}$ & $s^2_{12}$ & $s^2_{13}$ & $s^2_{23}$ & $\delta/\pi$ & $\sigma/\pi$  \\
\mr
$(0, \times, \times)$ & $(1, 1, 2)$  &  0 & 7.39 & 2.53 & 0.310 & 0.0224 & 0.580 &  $\pm 0.63$ & $\pm 0.29$  \\
&  &  &  &  &  &  &  & $\pm 0.98$ & $\pm 0.10$  \\
& $(1, 2, 1)$ &  0 & 7.39 & 2.53 & 0.310 & 0.0224 & 0.580 & $\pm 0.12$ & $\pm 0.80$  \\
&  &  &  &  &  &  &  & $\pm 0.46$ & $\pm 0.62$  \\
& $(1, 2, 2)$ &  0 & 7.39 & 2.53 & 0.310 & 0.0224 & 0.580 & $\pm 0.06$ & $\pm 0.71$  \\
&  &  &  &  &  &  &  & $\pm 0.99$ & $\pm 0.78$  \\
& $(1, 1, 3)$ & 10 & 7.64 & 2.51 & 0.318 & 0.0216 & 0.535 &  $\pm 0.40$ & $\pm 0.31$  \\
& $(1, 3, 1)$ &  0 & 7.39 & 2.53 & 0.310 & 0.0224 & 0.580 & $\pm 0.40$ & $\pm 0.87$  \\
&  &  &  &  &  &  &  & $\pm 0.73$ & $\pm 0.55$  \\
& $(1, 3, 3)$ &  0 & 7.39 & 2.53 & 0.310 & 0.0224 & 0.580 & $\pm 0.21$ & $\pm 0.16$  \\
&  &  &  &  &  &  &  & $\pm 0.72$ & $\pm 0.92$  \\
$(0, 1, 1)$ & $(1, 1, 3)$ & 24 & 7.68 & 2.52 & 0.319 & 0.0211 & 0.509 & $\pm 0.49$ & $\pm 0.22$ \\
& $(1, 3, 1)$ & 12 & 7.17 & 2.54 & 0.318 & 0.0231 & 0.528 & $\pm 0.46$ & $\pm 0.81$ \\
$(\times, 0, \times)$ & $(1, 3, 3)$ & 0.7 & 7.42 & 2.52 & 0.306 & 0.0229 & 0.576 & $\pm 0.76$ & $\pm 0.65$  \\
$(\times, \times, 0)$ & $(1, 3, 3)$ & 0 & 7.39 & 2.53 & 0.310 & 0.0224 & 0.580 & $\pm 0.49$ & $\pm 0.15$ \\ &  & &  &  &  & & & $\pm 0.82$ & $\pm 0.14$  \\
\br
\end{tabular}
\end{table}

We proceed to explore possible textures of $M^{}_{\rm D}$ where $a^{}_{23}$ and all of $b^{}_{ij}$  can simultaneously take some special values. In this case both columns of $M^{}_{\rm D}$ are fully determined up to some overall factors.
Note that when $|a^{}_{23}|$ and $|b^{}_{23}|$ simultaneously take some special values, it does not necessarily mean that $a^{}_{23}$ and $b^{}_{23}$ can also do so unless the phase relationship $\arg{[(M^{}_{\rm D})^{}_{\mu 2}]} - \arg{[(M^{}_{\rm D})^{}_{\mu 1}]} = \arg{[(M^{}_{\rm D})^{}_{\tau 2}]} - \arg{[(M^{}_{\rm D})^{}_{\tau 1}]}$ holds. We point out that $a^{}_{23} =1$ and $b^{}_{23} =1$ can never hold simultaneously because otherwise $M^{}_{\rm D}$ would acquire a $\mu$-$\tau$ permutation symmetry, which will lead to $\theta^{}_{13} =0$ (see section~\ref{section 6.1}).
Our numerical calculations show that only the combinations of $(M^{}_{\rm D})^{}_1 \propto (0, 1, 1)^{T}$ with $(M^{}_{\rm D})^{}_2 \propto (1, 1, 3)^{T}$ and $(1, 3, 1)^{T}$ can be consistent with current experimental data at the $3\sigma$ level (see Table~\ref{Table:4}). Remarkably, the constant neutrino flavor mixing pattern resulting from these two special cases is the so-called TM1 mixing pattern, which is phenomenologically interesting and will be discussed in section~\ref{section 6.2}.

A study on the possibility of $(M^{}_{\rm D})^{}_{\mu 1} =0$ or $(M^{}_{\rm D})^{}_{\tau 1} =0$ can be carried out in a similar way. For these two cases, the allowed ranges of $|a^{}_{ij}|$ and $|b^{}_{ij}|$ are also listed in Table~\ref{Table:3}; and only the texture $(M^{}_{\rm D})^{}_2 \propto (1, 3, 3)^{T}$ can be consistent with current experimental data at the $3\sigma$ level (see Table~\ref{Table:4}).

A similar study in the $m^{}_3 =0$ case is also done. The allowed ranges of $|a^{}_{ij}|$ and $|b^{}_{ij}|$ in the assumption of $(M^{}_{\rm D})^{}_{\alpha 1} =0$ are also listed in Table~\ref{Table:3}. For $(M^{}_{\rm D})^{}_{e 1} =0$, the fact that $|b^{}_{ij}|$ (or $|b^{}_{23}|$) may approach infinity (or zero) verifies the consistencies of patterns ${\rm A}^{}_{1, 2, 4, 5}$ in Eq.~(\ref{eq:5.1.2}) with current experimental data. Similarly, $|b^{}_{12}|$ and $|b^{}_{13}|$ may also approach zero in the assumption of $(M^{}_{\rm D})^{}_{\mu 1} =0$ or $(M^{}_{\rm D})^{}_{\tau 1} =0$.
All the textures consistent with the low-energy measurements at the $3\sigma$ level are enumerated in Table~\ref{Table:5}, so are their phenomenological consequences.
\begin{table}[h]
\caption{In the $m^{}_3 =0$ case the phenomenological consequences obtained from some particular forms of $(M^{}_{\rm D})^{}_1$ and $(M^{}_{\rm D})^{}_2$, where the units of $\Delta m^2_{21}$ and $\Delta m^2_{31}$ are $10^{-5}$ eV$^2$ and $10^{-3}$ eV$^2$, respectively. }
\label{Table:5}
\vspace{0.1cm}
\centering
\begin{tabular}{cccccccccc} \br
$(M^{}_{\rm D})^{T}_1$ & $(M^{}_{\rm D})^{T}_2$ & $\chi^2_{\rm min}$ & $\Delta m^2_{21}$ & $-\Delta m^2_{31}$ & $s^2_{12}$ & $s^2_{13}$ & $s^2_{23}$ & $\delta/\pi$ & $\sigma/\pi$  \\
\mr
$(0, \times, \times)$ & $(1, 1, 1)$ & 0 & 7.39 & 2.44 & 0.310 & 0.0226 & 0.584 & $\pm 0.10/0.53$ & $\pm 0.66$ \\
& $(2, 1, 1)$ & 0 & 7.39 & 2.44 & 0.310 & 0.0226 & 0.584 & $\pm 0.03/0.79$ & $\pm 0.79$ \\
& $(1, 2, 2)$ & 3.0 & 7.43 & 2.45 & 0.331 & 0.0225 & 0.566 & $\pm 0.05$ & $\pm 0.52$ \\
& $(3, 1, 1)$ & 0 & 7.39 & 2.44 & 0.310 & 0.0226 & 0.584 & $\pm 0.02/0.88$ & $\pm 0.86$ \\
$(\times, 0, \times)$ & $(1, 1, 1)$ & 0 & 7.39 & 2.44 & 0.310 & 0.0226 & 0.584 & $\pm 0.31/0.94$ & $\pm 0.25$ \\
& $(2, 1, 1)$ & 0 & 7.39 & 2.44 & 0.310 & 0.0226 & 0.584 & $\pm 0.22$ & $\pm 0.36$ \\
&  &  &  &  &  &  &  & $\pm 0.78$ & $\pm 0.64$ \\
& $(1, 2, 2)$ & 1.3 & 7.41 & 2.42 & 0.311 & 0.0224 & 0.568 & $\pm 0.59$ & $\pm 0.18$ \\
& $(3, 1, 1)$ & 0 & 7.39 & 2.44 & 0.310 & 0.0226 & 0.584 & $\pm 0.29$ & $\pm 0.45$ \\
&  &  &  &  &  &  &  & $\pm 0.56$ & $\pm 0.56$ \\
& $(1, 3, 3)$ & 2.5 & 7.46 & 2.44 & 0.308 & 0.0223 & 0.558 & $\pm 0.58$ & $\pm 0.14$ \\
$(\times, \times, 0)$ & $(1, 1, 1)$ & 0 & 7.39 & 2.44 & 0.310 & 0.0226 & 0.584 & $\pm 0.41/0.75$ & $\pm 0.18$ \\
& $(2, 1, 1)$ & 0 & 7.39 & 2.44 & 0.310 & 0.0226 & 0.584 & $\pm 0.37/0.99$ & $\pm 0.32$ \\
& $(1, 2, 2)$ & 1.1 & 7.42 & 2.44 & 0.308 & 0.0226 & 0.566 & $\pm 0.53$ & $\pm 0.07$ \\
& $(3, 1, 1)$ & 0 & 7.39 & 2.44 & 0.310 & 0.0226 & 0.584 & $\pm 0.48$ & $\pm 0.42$ \\
&  &  &  &  &  &  &  & $\pm 0.85$ & $\pm 0.58$ \\
& $(1, 3, 3)$ & 2.1 & 7.30 & 2.46 & 0.301 & 0.0228 & 0.564 & $\pm 0.52$ & $\pm 0.04$ \\
\br
\end{tabular}
\end{table}

\setcounter{equation}{0}
\setcounter{figure}{0}
\section{Flavor symmetries embedded in the minimal seesaw}
\label{section 6}

As pointed out in section~\ref{section 1.1}, the structure and parameters of the PMNS lepton flavor mixing matrix $U$ bear several remarkable features. (1) There is an approximate equality between the magnitudes of each entry of $U$ in the second row and its counterpart in the third row (i.e., $|U^{}_{\mu i}| \sim |U^{}_{\tau i}|$ for $i=1,2,3$) as demonstrated by Eq.~(\ref{eq:1.7}). Such an observation points towards an underlying symmetry between $\mu$ and $\tau$ flavors (i.e., the $\mu$-$\tau$ flavor symmetry; or more exactly, the $\nu^{}_\mu$-$\nu^{}_\tau$ symmetry \cite{Xing:2015fdg}). (2) The lepton flavor mixing angles are close to some special values, such as $\sin \theta^{}_{12} \sim 1/\sqrt{3}$ and $\sin \theta^{}_{23} \sim 1/\sqrt{2}$ (and $\theta^{}_{13} \sim 0$ used to be a popular conjecture). Therefore, to a good approximation, $U$ can be described by a few simple numbers (i.e., 0, 1, 2 and 3) and their square roots. A striking example of this kind is the tribimaximal (TBM) mixing pattern \cite{Harrison:2002er, Xing:2002sw, Harrison:2002kp}
\begin{eqnarray}
U^{}_{\rm TBM}= \displaystyle \frac{1}{\sqrt 6} \pmatrix{
2 & \sqrt{2} & 0 \cr
-1 & \sqrt{2} & \sqrt{3} \cr
1 & - \sqrt{2} & \sqrt{3} \cr
} \; .
\label{eq:6.1.1}
\end{eqnarray}
In the literature some extensive studies have shown that a particular mixing pattern such as $U^{}_{\rm TBM}$ can be naturally realized with the help of some discrete non-Abelian flavor symmetries \cite{Altarelli:2010gt, King:2013eh}.
(3) A preliminary but exciting measurement of the Dirac CP phase $\delta$ also hints at a special value \cite{Abe:2019vii}: $\delta \sim - \pi/2$. A convincing explanation of such a suggestive result may require the use of proper flavor symmetries to predict or constrain the CP phases. A realistic way of model building is expected to combine the minimal seesaw model with a kind of flavor symmetry. In this section we are going to embed the $\mu$-$\tau$ reflection symmetry (section~\ref{section 6.1}), trimaximal mixing (section~\ref{section 6.2}) or CP symmetries \cite{Li:2017zmk} (section~\ref{section 6.3}) in the minimal seesaw mechanism
\footnote{See Refs.~\cite{Zhang:2009ac,Park:2011zt,Yang2011Minimal,Zhao:2011pv,Yasue:2012zv,Ding:2017hdv,Wang:2019ovr, Wang:2020lxk} for the phenomenological studies of how to embed some other interesting flavor symmetries in the minimal seesaw scheme. } .

\subsection{The $\mu$-$\tau$ reflection symmetry}
\label{section 6.1}

Before its value was experimentally determined, the extreme possibility of $\theta^{}_{13}$ being vanishingly small was popularly conjectured. On the other hand, $\theta^{}_{23} \sim \pi/4$ has been firmly established from both atmospheric and long-baseline neutrino oscillation experiments. If one takes $\theta^{}_{13} =0$ and $\theta^{}_{23} =\pi/4$ as a starting point for model building, then the lepton flavor mixing matrix appears to be
\begin{eqnarray}
U = \displaystyle \frac{1}{\sqrt 2} \pmatrix{
\sqrt{2} \cos \theta^{}_{12} & \sqrt{2} \sin \theta^{}_{12} & 0 \cr
 -\sin \theta^{}_{12} & \cos \theta^{}_{12} & 1 \cr
\sin \theta^{}_{12} & - \cos \theta^{}_{12} & 1 \cr
} \; ,
\label{eq:6.1.2}
\end{eqnarray}
where $\theta^{}_{12}$ is arbitrary. In the basis of $M^{}_l = D^{}_l$ the reconstruction of $M^{}_\nu = U D^{}_\nu U^T$ in terms of such a special pattern of $U$ leads us to a special texture of $M^{}_\nu$ whose entries satisfy the relations $(M^{}_\nu)^{}_{e\mu} = (M^{}_\nu)^{}_{e\tau}$ and $(M^{}_\nu)^{}_{\mu\mu} = (M^{}_\nu)^{}_{\tau\tau}$. Such a special form of $M^{}_\nu$ possesses a simple flavor symmetry with respect to the $\mu$-$\tau$ interchange transformation $\nu^{}_{\mu {\rm L}} \leftrightarrow \nu^{}_{\tau {\rm L}}$ \cite{Fukuyama:1997ky, Ma:2001mr, Lam:2001fb, Balaji:2001ex}:
\begin{eqnarray}
G^\dagger M^{}_{\nu} G^* = M^{}_{\nu} \; , \quad G = \pmatrix{
1 & 0 & 0 \cr 0 & 0 & 1 \cr 0 & 1 & 0
} \; .
\label{eq:6.1.3}
\end{eqnarray}
However, the experimental observation of a relatively large $\theta^{}_{13}$ motivates us to forsake this interesting symmetry \cite{He:2011kn, Ge:2010js}. But its variant --- the so-called $\mu$-$\tau$ reflection symmetry \cite{Harrison:2002et} (see Ref.~\cite{Chen:2015siy} for its generalization) --- offers an appealing possibility of explaining current neutrino oscillation data \cite{Xing:2015fdg}.
This symmetry is defined in the way that $M^{}_\nu$ keeps invariant (i.e., $G^\dagger M^{}_\nu G^* = M^{*}_\nu$) with respect to a combination of the $\mu$-$\tau$ interchange and charge conjugation transformations as follows:
\begin{eqnarray}
\nu^{}_{e {\rm L}} \leftrightarrow \nu^{c}_{e {\rm L}} \; , \quad \nu^{}_{\mu {\rm L}} \leftrightarrow \nu^{c}_{\tau {\rm L}} \; , \quad \nu^{}_{\tau {\rm L}} \leftrightarrow \nu^{c}_{\mu {\rm L}} \; .
\label{eq:6.1.4}
\end{eqnarray}
In this case the entries of $M^{}_\nu$ are subject to the constraints
\begin{eqnarray}
(M^{}_\nu)^{}_{e\mu} = (M^{}_\nu)^*_{e\tau} \; , \quad (M^{}_\nu)^{}_{\mu\mu} = (M^{}_\nu)^*_{\tau\tau}  \; ,
\nonumber \\
(M^{}_\nu)^{}_{ee} = (M^{}_\nu)^{*}_{ee} \; , \quad
(M^{}_\nu)^{}_{\mu\tau} = (M^{}_\nu)^{*}_{\mu\tau} \; ,
\label{eq:6.1.5}
\end{eqnarray}
which amount to six real constraint equations and lead to the same number of predictions for the lepton flavor mixing parameters \cite{Grimus:2003yn}:
\begin{eqnarray}
\theta^{}_{23} = \frac{\pi}{4} \; , \quad \delta = \pm \frac{\pi}{2} \; , \quad \rho = 0 \;\; {\rm or} \;\; \frac{\pi}{2} \; , \quad \sigma = 0 \;\; {\rm or} \;\; \frac{\pi}{2} \; ,
\nonumber \\
\phi^{}_e = \frac{\pi}{2} \; , \quad \phi^{}_\mu = - \phi^{}_\tau \; ,
\label{eq:6.1.6}
\end{eqnarray}
where $\phi^{}_\alpha$ (for $\alpha = e, \mu, \tau$) are the unphysical phases. In comparison, the other two flavor mixing angles $\theta^{}_{12}$ and $\theta^{}_{13}$ are unconstrained.

Now let us embed the $\mu$-$\tau$ reflection symmetry in the minimal seesaw mechanism \cite{Kitabayashi:2016zec, Liu:2017frs, Nath:2018hjx,King:2018kka,King:2019tbt}
\footnote{The so-called $\mu$-$\tau$ reflection {\it antisymmetry} has been combined with the minimal seesaw model in Ref.~\cite{Samanta:2017kce}, where $G^\dagger M^{}_\nu G^* = - M^{*}_\nu$ is required to hold. In this case one is left with $(M^{}_\nu)^{}_{e\mu} = - (M^{}_\nu)^*_{e\tau}$, $(M^{}_\nu)^{}_{\mu\mu} = - (M^{}_\nu)^*_{\tau\tau}$, $(M^{}_\nu)^{}_{ee} =
-(M^{}_\nu)^{*}_{ee}$ and $(M^{}_\nu)^{}_{\mu\tau} = - (M^{}_\nu)^{*}_{\mu\tau}$.
Since $G^\dagger \left({\rm i} M^{}_\nu\right) G^* = ({\rm i} M^{}_\nu)^*$ holds, $M^{}_\nu$ and $M^\prime_\nu \equiv {\rm i} M^{}_\nu$ are expected to have the same phenomenological consequences.}.
Note that one of the six real constraint equations in Eq.~(\ref{eq:6.1.5}) is redundant due to ${\rm Det}(M^{}_\nu) =0$. Accordingly, the $\mu$-$\tau$ symmetry predictions for the lepton flavor mixing parameters is also reduced by one (i.e., $\rho =0$ or $\pi/2$ is irrelevant in physics).
Such an embedding is highly predictive, allowing us to reconstruct $M^{}_\nu$ by taking into account both the symmetry predictions and current neutrino oscillation data.
The results of $|(M^{}_\nu)^{}_{\alpha \beta}|$
for $(\delta, \sigma) = (\pi/2, 0)$ and $(\pi/2, \pi/2)$ are shown in Table~\ref{Table:6}, where the uncertainties correspond to the $3\sigma$ errors of $\Delta m^2_{21}$, $\Delta m^2_{31}$, $\theta^{}_{12}$ and $\theta^{}_{13}$. Because $|(M^{}_\nu)^{}_{\alpha \beta}|$ keep invariant under the transformation $\delta \to -\delta$ in the case of $\sigma =0$ or $\pi/2$, as one can see from Eqs.~(\ref{eq:3.2.2}) and (\ref{eq:3.2.3}), the results for $(\delta, \sigma) = (-\pi/2, 0)$ and $(-\pi/2, \pi/2)$ are the same as those for $(\delta, \sigma) = (\pi/2, 0)$ and $(\pi/2, \pi/2)$.
\begin{table}[t]
\caption{ The results of $|(M^{}_\nu)^{}_{\alpha \beta}|$ for $(\delta, \sigma) = (\pi/2, 0)$ and $(\pi/2, \pi/2)$ in the $m^{}_1 =0$ or $m^{}_3 =0$ case, where the $3\sigma$ ranges of $\Delta m^2_{21}$, $\Delta m^2_{31}$, $\theta^{}_{12}$ and $\theta^{}_{13}$ have been taken into account, and $\theta^{}_{23} = \pi/4$ is fixed by the $\mu$-$\tau$ symmetry. }
\label{Table:6}
\vspace{0.1cm}
\begin{indented}
\item[]
\hspace{-2cm}
\begin{tabular}{ccccccc} \br
 & $\delta$ &  $\sigma$ & $|(M^{}_\nu)^{}_{ee}|$ (meV) & $|(M^{}_\nu)^{}_{e\mu}|$ (meV) &  $|(M^{}_\nu)^{}_{\mu\mu}|$ (meV) & $|(M^{}_\nu)^{}_{\mu\tau}|$ (meV) \\
\mr
\vspace{0.2cm}
$m^{}_1 =0$ & $\pi/2$ & 0 & 3.23---4.31 & 5.76---6.64 & 26.7---28.3 & 20.8---22.4 \\
 & $\pi/2$ & $\pi/2$ & 0.96---2.06 & 5.36---6.03 & 20.8---22.5 & 26.7---28.4  \\
\mr
\vspace{0.2cm}
$m^{}_3 =0$ & $\pi/2$ & 0 & 47.3---49.6 & 4.89---5.55 & 23.8---24.9 & 24.9---26.1 \\
 & $\pi/2$ & $\pi/2$ & 13.2---22.5 & 30.5---33.9 & 9.98---14.0 & 6.93---11.7  \\
\br
\end{tabular}
\end{indented}
\end{table}

When the complete Lagrangian of the minimal seesaw model is concerned, the transformation properties of two right-handed neutrino fields $N^{}_{\alpha {\rm R}}$ (for $\alpha = \mu$ and $\tau$) under the $\mu$-$\tau$ reflection symmetry should also be specified. This can be simply done in a way parallel to that for the left-handed neutrino fields \cite{Nath:2018hjx}:
\begin{eqnarray}
N^{}_{\mu {\rm R}} \leftrightarrow N^{c}_{\tau {\rm R}} \; , \quad N^{}_{\tau {\rm R}} \leftrightarrow N^{c}_{\mu {\rm R}} \; .
\label{eq:6.1.7}
\end{eqnarray}
In this case the symmetry requirement $G^{\prime T} M^{}_{\rm R} G^{\prime} = M^{*}_{\rm R}$ with $G^\prime = \pmatrix{ 0 & 1 \cr 1 & 0}$ for the right-handed neutrino mass matrix $M^{}_{\rm R}$ yields the following constraints on its entries
\begin{eqnarray}
(M^{}_{\rm R})^{}_{11} = (M^{}_{\rm R})^*_{22} \; , \quad (M^{}_{\rm R})^{}_{12} = (M^{}_{\rm R})^{*}_{12} \; .
\label{eq:6.1.8}
\end{eqnarray}
On the other hand, the symmetry requirement $G^\dagger M^{0}_{\rm D} G^{\prime} = (M^{0}_{\rm D})^*$ for the Dirac neutrino mass matrix $M^{0}_{\rm D}$ forces its entries to satisfy the relations
\begin{eqnarray}
(M^{0}_{\rm D})^{}_{e 1} = (M^{0}_{\rm D})^{*}_{e 2} \; , \quad (M^{0}_{\rm D})^{}_{\mu 1} = (M^{0}_{\rm D})^{*}_{\tau 2} \; , \quad (M^{0}_{\rm D})^{}_{\tau 1} = (M^{0}_{\rm D})^{*}_{\mu 2}
 \; .
\label{eq:6.1.9}
\end{eqnarray}
One can easily verify that such textures of $M^{}_{\rm R}$ and $M^{0}_{\rm D}$ {\it do} produce an $M^{}_\nu$ of the form described in Eq.~(\ref{eq:6.1.5}) via the seesaw formula. Then we turn to the mass basis of two right-handed neutrinos via a unitary transformation $U^{}_{\rm R}$ which diagonalizes $M^{}_{\rm R}$. By substituting $M^{}_{\rm R} = U^*_{\rm R} D^{}_N U^\dagger_{\rm R}$ into $G^{\prime T} M^{}_{\rm R} G^{\prime} = M^{*}_{\rm R}$, we see
that $U^{}_{\rm R}$ can be decomposed into $U^{0}_{\rm R} P^{}_N$ where $(G^\prime U^{0}_{\rm R})^* = U^{0}_{\rm R}$ (i.e., $(U^{0}_{\rm R})^{}_{1i} = (U^{0}_{\rm R})^{*}_{2i}$ hold for $i=1,2$) and
$P^{2}_N = {\rm Diag} \{ \eta^{}_1 , \eta^{}_2 \}$ with $\eta^{}_i = \pm 1$.
Under such a basis transformation, $M^{0}_{\rm D}$ is transformed to $M^{}_{\rm D} = M^{0}_{\rm D} U^{}_{\rm R}$,
which can be decomposed into $M^{\prime}_{\rm D} P^{}_N$ with the entries of $M^{\prime}_{\rm D} = M^{0}_{\rm D} U^{0}_{\rm R}$ satisfying the relations
\begin{eqnarray}
(M^{\prime}_{\rm D})^{}_{\mu 1} = (M^{\prime }_{\rm D})^{*}_{\tau 1} \; , \quad (M^{\prime}_{\rm D})^{}_{\mu 2} = (M^{\prime}_{\rm D})^{*}_{\tau 2} \; ,
\nonumber \\
(M^{\prime }_{\rm D})^{}_{e 1} = (M^{\prime }_{\rm D})^{*}_{e 1} \; , \quad
(M^{\prime }_{\rm D})^{}_{e 2} = (M^{\prime }_{\rm D})^{*}_{e 2} \; .
\label{eq:6.1.10}
\end{eqnarray}
It is worth mentioning that there exists an equivalent but simpler way to specify the transformation properties of $N^{}_{\alpha {\rm R}}$; namely,
\begin{eqnarray}
N^{}_{\mu {\rm R}} \leftrightarrow N^{c}_{\mu {\rm R}} \; , \hspace{1cm} N^{}_{\tau {\rm R}} \leftrightarrow N^{c}_{\tau {\rm R}} \; .
\label{eq:6.1.11}
\end{eqnarray}
In this case the symmetry requirements $M^{*}_{\rm R} = M^{}_{\rm R}$ and $(G^\dagger M^{0}_{\rm D})^* = M^{0}_{\rm D}$ render $M^{}_{\rm R}$ and $M^{0}_{\rm D}$ to be real and of the form described in Eq.~(\ref{eq:6.1.10}), respectively.
After transforming the flavor basis of two right-handed neutrinos into their mass basis via $U^{}_{\rm R}=U^{0}_{\rm R} P^{}_N$ with $U^{0}_{\rm R}$ being real now, the resultant Dirac neutrino mass matrix $M^{}_{\rm D} = M^{0}_{\rm D} U^{}_{\rm R}$ acquires the same form as in the previous case. For simplicity but without loss of generality, we choose to work in the mass basis of right-handed neutrinos in the following discussions.

We proceed to study the implications of this specific minimal seesaw model on thermal leptogenesis. Let us first figure out the Casas-Ibarra parametrization of $M^{}_{\rm D}$.
A careful analysis reveals that the $O$ matrix should take a form as $P^\dagger_\nu O^0 P^{}_N$, where $P^{}_\nu = {\rm Diag}\{ 1, e^{{\rm i}\sigma}, e^{-{\rm i}\delta} \}$ and $O^0$ is a real $3\times 2$ matrix whose first (or third) row is vanishing in the $m^{}_1 =0$ (or $m^{}_3 =0$) case \cite{Mohapatra:2015gwa}.
By comparing such a form of $O$ with that in Eq.~(\ref{eq:3.5.5}), we arrive at
\begin{eqnarray}
m^{}_1 =0: \hspace{1cm} \cos z = e^{-{\rm i}\sigma} O^0_{21} \sqrt{\eta^{}_1} \; , \hspace{1cm}
\sin z = e^{{\rm i}\delta} O^0_{31} \sqrt{\eta^{}_1} \; ; \nonumber \\
m^{}_3 =0: \hspace{1cm} \cos z = O^0_{11} \sqrt{\eta^{}_1} \; , \hspace{1cm}
\sin z = e^{-{\rm i}\sigma} O^0_{21} \sqrt{\eta^{}_1} \; ,
\label{eq:6.1.12}
\end{eqnarray}
for which the normalization relation in Eq.~(\ref{eq:3.5.3}) reads
\begin{eqnarray}
m^{}_1 =0: \hspace{1cm} e^{-2{\rm i}\sigma} (O^0_{21})^2 - (O^0_{31})^2 =\eta^{}_1  \; ; \nonumber \\
m^{}_3 =0: \hspace{1cm} (O^0_{11})^2 + e^{-2{\rm i}\sigma} (O^0_{21})^2 = \eta^{}_1 \; ,
\label{eq:6.1.13}
\end{eqnarray}
where $\delta = \pm \pi/2$ has been taken into account.
Depending on the values of $\sigma$ and $\eta^{}_1$, Eq.~(\ref{eq:6.1.13}) acts as a definition of the real orthogonal matrix or a hyperbola:
\begin{eqnarray}
m^{}_1 =0: \hspace{0.5cm}
\left\{ \begin{array}{l}
(O^0_{21})^2 - (O^0_{31})^2 = \pm 1 \; , \hspace{0.5cm} {\rm for } \hspace{0.2cm} (\sigma, \eta^{}_1) = (0 , \pm 1) \; , \\
(O^0_{21})^2 + (O^0_{31})^2 = 1 \; , \hspace{0.5cm} {\rm for } \hspace{0.2cm} (\sigma, \eta^{}_1) = (\pi/2 , -1) \; ;
\end{array}
\right. \nonumber \\
m^{}_3 =0: \hspace{0.5cm}
\left\{ \begin{array}{l}
(O^0_{11})^2 + (O^0_{21})^2 = 1 \; , \hspace{0.5cm} {\rm for } \hspace{0.2cm} (\sigma, \eta^{}_1) = (0 , 1) \; , \\
(O^0_{11})^2 - (O^0_{21})^2 = \pm 1 \; , \hspace{0.5cm} {\rm for } \hspace{0.2cm} (\sigma, \eta^{}_1) = (\pi/2 , \pm 1) \; .
\end{array}
\right.
\label{eq:6.1.14}
\end{eqnarray}
So one may parameterize $(O^0_{21}, O^0_{31})$ as $(\cos \theta, \sin \theta)$, $(\cosh y, \sinh y)$ and $(\sinh y, \cosh y)$ in the $m^{}_1 =0$ case with $(\sigma, \eta^{}_1) = (\pi/2 , -1)$, $(0, 1)$ and $(0, -1)$, respectively; or parameterize
$(O^0_{11}, O^0_{21})$) as $(\cos \theta, \sin \theta)$, $(\cosh y, \sinh y)$ and $(\sinh y, \cosh y)$ in the $m^{}_3 =0$ case with $(\sigma, \eta^{}_1) = (0 , 1)$, $(\pi/2, 1)$ and $(\pi/2, -1)$, respectively.

\begin{figure*}[t]
\centering
\includegraphics[width=6in]{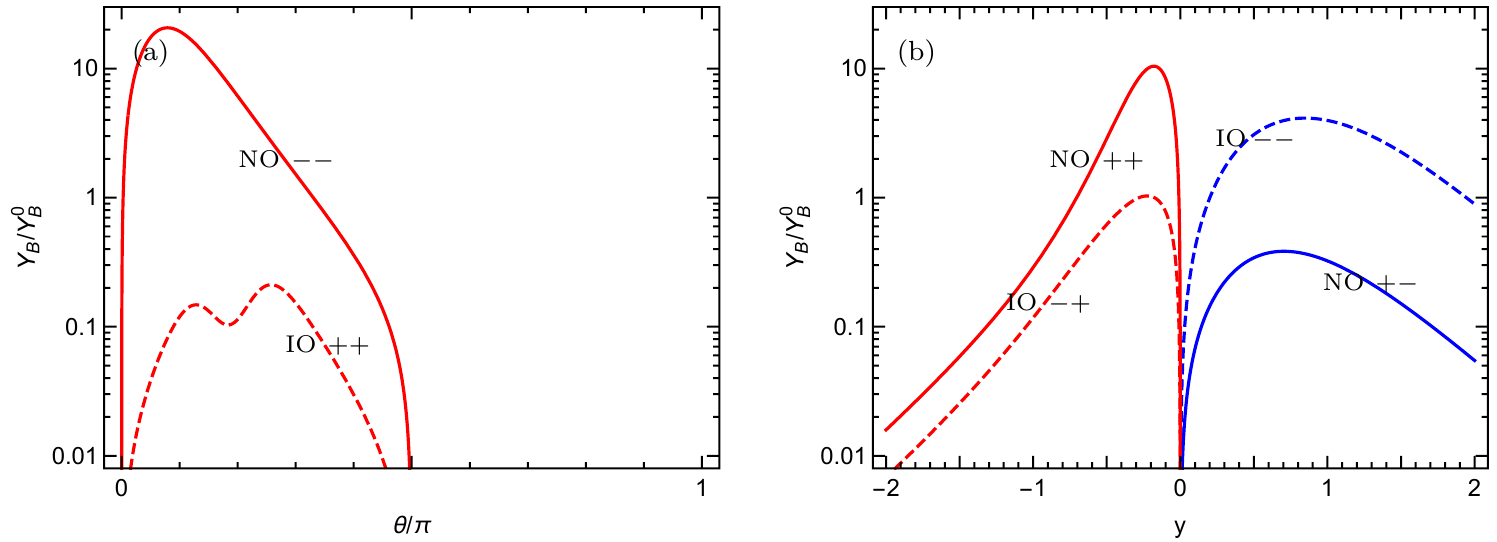}
\caption{(a) $Y^{}_{\rm B}/Y^{0}_{\rm B}$ as functions of $\theta$ in the $m^{}_1 =0$ case with $(\sigma, \eta^{}_1) = (\pi/2, -1)$ (denoted by ``NO $--$"), or in the $m^{}_3 =0$ case with $(\sigma, \eta^{}_1) = (0, 1)$ (labelled by ``IO $++$"); (b) $Y^{}_{\rm B}/Y^{0}_{\rm B}$ as functions of $y$ in the $m^{}_1 =0$ case with $(\sigma, \eta^{}_1) = (0, 1)$ (denoted by ``NO $++$") and $(0, -1)$ (denoted by ``NO $+-$"), or in the $m^{}_3 =0$ case with $(\sigma, \eta^{}_1) = (\pi/2, 1)$ (labelled by ``IO $-+$") and $(\pi/2, -1)$ (labelled by ``IO $--$"). In our calculations $\delta = -\pi/2$ has been fixed.}
\label{Fig:6-1}
\end{figure*}

Due to the special form of $M^{}_{\rm D}$, the CP-violating asymmetry $\varepsilon^{}_i$ for leptogenesis vanishes, as one can see from Eq.~(\ref{eq:2.22}) \cite{Harrison:2002et}. To achieve successful leptogenesis, one may choose to softly break the $\mu$-$\tau$ reflection symmetry \cite{Ahn:2008hy}. Fortunately, there is another way out --- one may retain this symmetry but resort to the flavor effects \cite{Mohapatra:2015gwa}.
As discussed in section~\ref{section 4.2}, we need to work in the two-flavor regime (i.e., the regime of the $\tau$ flavor and a combination of the $\mu$ and $e$ flavors) in the minimal seesaw model with a hierarchical mass spectrum of two heavy Majorana neutrinos. In this regime the final baryon number asymmetry can be calculated according to Eq.~(\ref{eq:4.2.1}). In view of Eq.~(\ref{eq:2.21}), it is easy to check that the special form of $M^{}_{\rm D}$ leads to the following relations for the flavored CP-violating asymmetries:
\begin{eqnarray}
\varepsilon^{}_{ie} =0 \; , \hspace{1cm} \varepsilon^{}_{i\mu} = - \varepsilon^{}_{i\tau} \; ,
\hspace{1cm} \varepsilon^{}_{io} =  - \varepsilon^{}_{i\tau} \; ,
\label{eq:6.1.15}
\end{eqnarray}
which will help us simplify Eq.~(\ref{eq:4.2.1}) to Eq.~(\ref{eq:4.2.6}). We see that there is a {\it partial} cancellation between the contributions of the $\tau$ flavor and its orthogonal combination. But the net result is nonzero because these two contributions are washed out in different ways. Then the issue becomes quantitative.
Fig.~\ref{Fig:6-1} shows the ratio of the generated baryon number asymmetry $Y^{}_{\rm B}$ to the observed one $Y^{0}_{\rm B}$ as functions of the parameter $\theta$ or $y$ for various cases listed in Eq.~(\ref{eq:6.1.14}). Here the maximally allowed value of $M^{}_1$ (i.e., $M^{}_1 \sim 10^{12}$ GeV) for which the flavored leptogenesis scenario remains viable is taken as a benchmark input; and the results for smaller values of $M^{}_1$ can be obtained by scaling down the lines in Fig.~\ref{Fig:6-1} proportionally. We see that successful leptogenesis can be achieved in the $m^{}_1 =0$ case with $(\sigma, \eta^{}_1) = (\pi/2, -1)$ and $(0, 1)$ or in the $m^{}_3 =0$ case with $(\sigma, \eta^{}_1) = (\pi/2, -1)$ and $(\pi/2, 1)$, but the latter is only marginally viable.
Moreover, we find that the minimally allowed value of $M^{}_1$ for flavored leptogenesis to work is around $10^{11}$ GeV \cite{Mohapatra:2015gwa}.

We recall that the scale of thermal leptogenesis can be greatly lowered in a viable {\it resonant} leptogenesis scenario, realizing the three-flavor regime. However, successful leptogenesis is found to be precluded by the $\mu$-$\tau$ reflection symmetry in such a scenario \cite{Mohapatra:2015gwa}. This point can be easily seen from the following expression for the final baryon number asymmetry \cite{Buchmuller:2001sr,Nardi:2005hs}:
\begin{eqnarray}
\hspace{-0.5cm} Y^{}_{i\rm B} & = & - c r \left( \varepsilon^{}_{i \tau} \kappa^{}_{i\tau}  + \varepsilon^{}_{i \mu} \kappa^{}_{i\mu} + \varepsilon^{}_{i e} \kappa^{}_{ie} \right)
\nonumber \\
& = & - c r \left[ \varepsilon^{}_{i \tau} \kappa \left(\frac{344}{537} \widetilde m^{}_{i\tau} \right) + \varepsilon^{}_{i \mu} \kappa \left(\frac{344}{537} \widetilde m^{}_{i\mu} \right) + \varepsilon^{}_{i e} \kappa \left(\frac{453}{537} \widetilde m^{}_{ie} \right) \right] \; ,
\label{eq:6.1.16}
\end{eqnarray}
in combination with the results in Eq.~(\ref{eq:6.1.15}) and the equalities $\widetilde m^{}_{i\mu} = \widetilde m^{}_{i\tau}$ due to the special form of $M^{}_{\rm D}$ as one can check with the help of Eq.~(\ref{eq:4.2.3}).

Let us turn to the issue of how to break the $\mu$-$\tau$ reflection symmetry by considering the following two aspects. On the experimental side, current neutrino oscillation data have shown a mild preference for $\theta^{}_{23} > \pi/4$. But the present measurement of $\delta$ is rather preliminary, although its best-fit value is close to $-\pi/2$. Hence it will be no surprise if the ongoing and upcoming precision neutrino oscillation experiments finally establish some appreciable deviations of $\theta^{}_{23}$ and $\delta$ from $\pi/4$ and $-\pi/2$, respectively. In any case the $\mu$-$\tau$ reflection symmetry can serve as the simplest benchmark flavor symmetry for understanding the origin of the observed lepton flavor mixing pattern, since the aforementioned deviations are not expected to be very large. So one may introduce some proper symmetry-breaking terms to account for the deviations of this kind. On the other side, there are also two good theoretical reasons for us to consider the breaking of this flavor symmetry or any others in this connection. (1) The underlying flavor symmetry responsible for neutrino mass generation and lepton flavor mixing is naturally expected to show up at an energy scale far above the electroweak scale, such as the seesaw scale. Therefore, to confront the flavor-symmetry predictions with the low-energy measurements, one should take into account the RGE running effects. Such quantum effects will inevitably break the $\mu$-$\tau$ reflection symmetry and other kinds of flavor symmetries. (2) In the practice of building specific flavor-symmetry models, it often occurs that the applied flavor symmetries only hold exactly at the leading order and are softly broken by some higher-order contributions.

Here we first give a {\it generic} treatment of the $\mu$-$\tau$ reflection symmetry breaking \cite{Xing:2015fdg,Zhao:2017yvw,Liu:2017frs,Nath:2018hjx}
\footnote{See Refs.~\cite{Grimus:2004cc,Liao:2012xm,Gupta2013How} for a similar study of the $\mu$-$\tau$ interchange symmetry breaking.}
and then consider the {\it specific} symmetry breaking effects as a result of the RGE evolution. From the phenomenological point of view, corresponding to the symmetry conditions in Eq.~(\ref{eq:6.1.5}), the following six real parameters can be used to measure the strength of $\mu$-$\tau$ reflection symmetry breaking:
\begin{eqnarray}
\epsilon^{}_{1} + {\rm i} \epsilon^{}_{2} =\displaystyle \frac{M^{}_{e \mu}- M^{*}_{e\tau}}{M^{}_{e \mu}+M^{*}_{e\tau}} \; , \hspace{1cm}
\epsilon^{}_{3} + {\rm i} \epsilon^{}_{4} =\displaystyle \frac{M^{}_{\mu\mu}-M^{*}_{\tau \tau}}{M^{}_{\mu \mu}+ M^{*}_{\tau\tau}} \; , \nonumber \\
\epsilon^{}_{5}=\displaystyle \frac{{\rm Im}\left(M^{}_{ee}\right)}{{\rm Re}\left(M^{}_{ee}\right)} \; , \hspace{1cm}
\epsilon^{}_{6}=\displaystyle \frac{{\rm Im}\left(M^{}_{\mu \tau}\right)}{{\rm Re}\left(M^{}_{\mu \tau}\right)} \; .
\label{eq:6.1.17}
\end{eqnarray}
In order to keep the $\mu$-$\tau$ reflection symmetry as a good approximation, these six parameters should be small enough (e.g., $\lesssim 0.1$).
Note that $\epsilon^{}_i$ are not all independent but subject to the following consistency equation obtained from the requirement of ${\rm Det}(M^{}_\nu) =0$ in the minimal seesaw framework:
\begin{eqnarray}
c^{}_1 \left(\epsilon^{}_3 - 2 \epsilon^{}_1\right) + c^{}_2 \left(\epsilon^{}_5 - 2 \epsilon^{}_2\right) + c^{}_3 \left(\epsilon^{}_4 - 2 \epsilon^{}_2\right) + c^{}_4 \left(\epsilon^{}_6 -  \epsilon^{}_4\right) =0  \; ,
\label{eq:6.1.18}
\end{eqnarray}
where $c^{}_i$ are some coefficients to be determined from the fit of $M^{}_\nu$ to current experimental data. On the other hand, nonzero $\epsilon^{}_i$ will induce some deviations of the flavor mixing parameters from their values in the symmetry limit as shown in Eq.~(\ref{eq:6.1.6}) with the disposal of $\rho$. Let us define
\begin{eqnarray}
&& \Delta \phi^{}_e \equiv \phi^{}_e - \frac{\pi}{2} \; , \hspace {1cm}
\Delta \phi^{}_{\mu\tau} \equiv \frac{\phi^{}_\mu + \phi^{}_\tau}{2} - 0 \; ,
\hspace {1cm} \Delta \theta^{}_{23} \equiv \theta^{}_{23} - \frac{\pi}{4} \; ,  \nonumber \\
&& \Delta \delta \equiv \delta - \delta^0 \; , \hspace {1cm}
\Delta \sigma \equiv \sigma - \sigma^0 \; ,
\label{eq:6.1.19}
\end{eqnarray}
with $\delta^0 = \pm \pi/2$ and $\sigma^0 = 0$ or $\pi/2$. By doing perturbation expansions for these small quantities in the reconstructions of the conditions in Eq.~(\ref{eq:6.1.17}) via $M^{}_\nu = U D^{}_\nu U^T$, to the leading order, we arrive at the following constraint equations for them:
\begin{eqnarray}
\hspace{-1.5cm} 2 m^{}_3 s^2_{13} \Delta \delta + 2 \overline m^{}_2 s^2 _{12} \Delta \sigma
= \left( m^{}_{11} - m^{}_3 s^2_{13} \right) ( \epsilon^{}_5 - 2 \Delta \phi^{}_e)  \; , \nonumber \\
\hspace{-1.5cm} - 4 m^{}_{12} \bar s^{}_{13} \Delta \theta^{}_{23} - 2 m^{}_{11} s^2_{13} \Delta \delta
- 2 \overline m^{}_2 c^2_{12} \Delta \sigma =(m^{}_3 - m^{}_{22}) (\epsilon^{}_6 - 2 \Delta \phi^{}_{\mu\tau})  \;, \nonumber\\
\hspace{-1.5cm} \left[ m^{}_{12} - {\rm i} \left( m^{}_{11} + m^{}_3 \right) \bar{s}^{}_{13} \right] \Delta \theta^{}_{23} + \left( m^{}_{11} - m^{}_3 \right) \bar{s}^{}_{13} \Delta \delta + 2 \overline m^{}_2 s^{}_{12} \left( {\rm i} c^{}_{12} + s^{}_{12} \bar{s}^{}_{13} \right)  \Delta \sigma \nonumber \\
= - \left[ m^{}_{12} +  {\rm i} (m^{}_{11} + m^{}_3) \bar s^{}_{13}  \right]
(\epsilon^{}_{1} + {\rm i} \epsilon^{}_{2} - {\rm i} \Delta \phi^{}_e - {\rm i} \Delta \phi^{}_{\mu \tau}) \; , \nonumber \\
\hspace{-1.5cm} - 2 \left( m^{}_{22} - m^{}_3 \right) \Delta \theta^{}_{23}
- 2 \left( m^{}_{12} + {\rm i} m^{}_{11} \bar{s}^{}_{13} \right) \bar{s}^{}_{13} \Delta \delta + 2 \overline m^{}_2 c^{}_{12} \left( {\rm i} c^{}_{12} + 2 s^{}_{12} \bar{s}^{}_{13} \right) \Delta \sigma \nonumber \\
= \left( m^{}_{22} + m^{}_3 + 2{\rm i} m^{}_{12} \bar s^{}_{13} \right)
(\epsilon^{}_{3} + {\rm i} \epsilon^{}_{4} - 2{\rm i}\Delta \phi^{}_{\mu\tau}) \; ,
\label{eq:6.1.20}
\end{eqnarray}
with $m^{}_1 = 0$ (or $m^{}_3 = 0$) in the NO (or IO) case, $\overline m^{}_2 \equiv m^{}_2 e^{2{\rm i} \sigma^0}$, $\bar s^{}_{13} \equiv - {\rm i} s^{}_{13} e^{{\rm i} \delta^0}$ and
\begin{eqnarray}
\hspace{-1.5cm} m^{}_{11} = m^{}_1 c^2_{12} + \overline m^{}_2 s^2_{12} \; ,  \quad
m^{}_{12} = ( m^{}_1 - \overline m^{}_2) c^{}_{12} s^{}_{12} \; ,  \quad
m^{}_{22} =  m^{}_1 s^2_{12} + \overline m^{}_2 c^2_{12} \; .
\label{eq:6.1.21}
\end{eqnarray}
Of course, only five of the six real equations in Eq.~(\ref{eq:6.1.20}) are independent. In such a way there is a balance between the number of the quantities to be determined and that of the constraint equations for them. That is why we have taken into account the unphysical phases; otherwise such a balance would not be achieved.
Once a particular set of $\epsilon^{}_i$ is given by a symmetry-breaking mechanism (e.g., the RGE running effects), the resulting $\Delta \theta^{}_{23}$, $\Delta \delta$ and $\Delta \sigma$ can be figured out by solving Eq.~(\ref{eq:6.1.20}).

Now we consider the $\mu$-$\tau$ reflection symmetry breaking induced by the RGE running effects.
The ``run and diagonalize" approach mentioned in section~\ref{section 4.5} can help make clear how the neutrino mass matrix as a whole gets renormalized or how its initial $\mu$-$\tau$ reflection symmetry is gradually broken. Assuming that the effective Majorana neutrino mass matrix $M^{}_\nu$ possesses the exact $\mu$-$\tau$ reflection symmetry at a superhigh energy scale denoted as $\Lambda^{}_{\mu \tau}$, one may obtain its RGE-corrected low-energy counterpart at the one-loop level by reversing the procedure in Eq.~(\ref{eq:4.5.3}):
\begin{eqnarray}
M^{}_{\nu} (\Lambda^{}_{\rm EW}) = I^{}_{0} T^{}_l
M^{}_{\nu} (\Lambda^{}_{\mu \tau}) \hspace{0.03cm} T^{}_l  \; .
\label{eq:6.1.22}
\end{eqnarray}
One can see that $M^{}_{\nu} (\Lambda^{}_{\rm EW})$ receives the $\mu$-$\tau$ reflection symmetry breaking due to the difference between $y^{}_\mu$ and $y^{}_\tau$ appearing in $T^{}_l$, and the latter is effectively measured by $\Delta^{}_\tau$ as defined below Eq.~(\ref{eq:4.5.3}). Furthermore, we find $\epsilon^{}_3 \simeq 2 \epsilon^{}_1 \simeq \Delta^{}_\tau$ and $\epsilon^{}_{2} = \epsilon^{}_{4} = \epsilon^{}_{5} = \epsilon^{}_{6}  =0$, which obviously obey the consistency equation in Eq.~(\ref{eq:6.1.18}). Then, a numerical calculation of Eq.~(\ref{eq:6.1.20}) gives $\Delta \theta^{}_{23}$, $\Delta \delta$ and $\Delta \sigma$ as (for $\delta^0 = -\pi/2$)
\begin{eqnarray}
\sigma^0 = 0: \hspace{0.5cm} \Delta \theta^{}_{23} \simeq 0.64 \Delta^{}_\tau  \; ,  \hspace{0.5cm} \Delta \delta \simeq 0.39 \Delta^{}_\tau \; ,  \hspace{0.5cm} \Delta \sigma \simeq -0.03 \Delta^{}_\tau \; ; \nonumber \\
\sigma^0 = \frac{ \pi}{2} : \hspace{0.5cm} \Delta \theta^{}_{23} \simeq 0.40 \Delta^{}_\tau  \; ,  \hspace{0.5cm} \Delta \delta \simeq -0.65 \Delta^{}_\tau \; ,  \hspace{0.5cm} \Delta \sigma \simeq 0.02 \Delta^{}_\tau \; ,
\label{eq:6.1.23}
\end{eqnarray}
in the $m^{}_1 =0$ case; or
\begin{eqnarray}
\sigma^0 = 0: \hspace{0.5cm} \Delta \theta^{}_{23} \simeq -0.54 \Delta^{}_\tau  \; ,  \hspace{0.5cm} \Delta \delta \simeq 0.05 \Delta^{}_\tau \; ,  \hspace{0.5cm} \Delta \sigma \simeq 0.17 \Delta^{}_\tau \; ; \nonumber \\
\sigma^0 = \frac{ \pi}{2} : \hspace{0.5cm} \Delta \theta^{}_{23} \simeq -0.54 \Delta^{}_\tau  \; ,  \hspace{0.5cm} \Delta \delta \simeq -28 \Delta^{}_\tau \; ,  \hspace{0.5cm} \Delta \sigma \simeq  -10 \Delta^{}_\tau \; ,
\label{eq:6.1.24}
\end{eqnarray}
in the $m^{}_3 =0$ case.
Note that there exists a correlation between the (positive or negative) sign of $\Delta \theta^{}_{23}$ and the (normal or inverted) mass ordering of three light neutrinos \cite{Luo:2014upa,Huan:2018lzd,Zhu:2018dvj,Huang:2020kgt}.
Quantitatively, $\Delta^{}_\tau$ is negligibly small in the SM (see section~\ref{section 4.5}). But in the MSSM $\Delta^{}_\tau$ can be enhanced to ${\cal O}(10^{-2})$ by taking large values of $\tan \beta$ (e.g., $\Delta^{}_\tau \simeq 0.01$ for $\Lambda^{}_{\mu \tau} = 10^{10}$ GeV and $\tan \beta = 30$). However, even with such a significant enhancement, the RGE-induced quantum corrections are practically negligible except that $\Delta \delta$ and $\Delta \sigma$ can be appreciable in the $m^{}_3 =0$ case with $\sigma^0 = \pi/2$.

On the other hand, the ``diagonalize and run" approach allows us to study the RGE running behaviours of the flavor parameters of three massive neutrinos individually. To be complete and explicit, here we present their RGEs in the $m^{}_1 =0$ or $m^{}_3 =0$ limit \cite{Antusch:2003kp,Mei:2004rn}.
The one-loop RGEs of three neutrino masses are given by
\begin{eqnarray}
16 \pi^2 \frac{{\rm d} m^{}_1}{{\rm d} t} \simeq \left(\alpha + 2 \gamma y^2_\tau s^2_{12} s^2_{23}\right) m^{}_1  \; , \nonumber \\
16 \pi^2 \frac{{\rm d} m^{}_2}{{\rm d} t} \simeq \left(\alpha + 2 \gamma y^2_\tau c^2_{12} s^2_{23}\right) m^{}_2  \; , \nonumber \\
16 \pi^2 \frac{{\rm d} m^{}_3}{{\rm d} t} \simeq \left(\alpha + 2 \gamma y^2_\tau c^2_{23}\right) m^{}_3 \;
\label{eq:6.1.25}
\end{eqnarray}
with $m^{}_1 =0$ (or $m^{}_3 =0$) in the NO (or IO) case, where the meanings of $\alpha$ and $\gamma$ can be found in Eq.~(\ref{eq:4.5.2}). One can see that the initially vanishing neutrino mass is stable against the one-loop RGE corrections. However, as shown in section~\ref{section 2.2}, it will become nonzero (although extremely tiny) when the two-loop RGE effects are taken into account. As for the initially nonzero neutrino masses, since the $\alpha$ term is dominant and yields a common rescaling factor $I^{}_0$ as discussed below Eq.~(\ref{eq:4.5.2}), their running behaviors are essentially identical. The one-loop RGEs of three lepton flavor mixing angles and two CP-violating phases read
\begin{eqnarray}
16 \pi^2 \frac{{\rm d} \theta^{}_{12}}{{\rm d} t} & \simeq & - \gamma y^2_\tau c^{}_{12} s^{}_{12}  s^2_{23} \; ,
\nonumber \\
16 \pi^2 \frac{{\rm d} \theta^{}_{13}}{{\rm d} t} & \simeq & - 2 \gamma y^2_\tau \xi c^{}_{12} s^{}_{12} c^{}_{23} s^{}_{23} \cos (2\sigma + \delta) \; ,
\nonumber \\
16 \pi^2 \frac{{\rm d} \theta^{}_{23}}{{\rm d} t} & \simeq & - \gamma y^2_\tau c^{}_{23} s^{}_{23} \left( 1 + 2 \xi c^2_{12} \cos 2\sigma \right) \; ,
\nonumber \\
16 \pi^2 \frac{{\rm d} \sigma }{{\rm d} t} & \simeq & 2  \gamma y^2_\tau \xi c^2_{12} (c^2_{23} - s^2_{23})  \sin 2\sigma \; ,
\nonumber \\
16 \pi^2 \frac{{\rm d} \delta }{{\rm d} t} & \simeq & 2 \gamma y^2_\tau  \xi \left[ c^{}_{12} s^{}_{12} s^{-1}_{13} c^{}_{23} s^{}_{23} \sin (2\sigma+\delta) + s^2_{12} c^2_{23} \sin 2(\sigma+\delta) \right.
\nonumber \\
&& \left.  -c^2_{12} (c^2_{23} - s^2_{23}) \sin 2\sigma \right] \;
\label{eq:6.1.26a}
\end{eqnarray}
in the $m^{}_1 =0$ case; or
\begin{eqnarray}
16 \pi^2 \frac{{\rm d} \theta^{}_{12}}{{\rm d} t} & \simeq & - 2 \gamma y^2_\tau \xi^{-2} c^{}_{12} s^{}_{12}  s^2_{23} (1+\cos 2\sigma) \; ,
\nonumber \\
16 \pi^2 \frac{{\rm d} \theta^{}_{13}}{{\rm d} t} & \simeq & \gamma y^2_\tau c^{2}_{23} s^{}_{13} \; ,
\nonumber \\
16 \pi^2 \frac{{\rm d} \theta^{}_{23}}{{\rm d} t} & \simeq & \gamma y^2_\tau c^{}_{23} s^{}_{23} \; ,
\nonumber \\
16 \pi^2 \frac{{\rm d} \sigma }{{\rm d} t} & \simeq & - 2 \gamma y^2_\tau \xi^{-2} s^2_{12} s^2_{23} \sin 2\sigma \; ,
\nonumber \\
16 \pi^2 \frac{{\rm d} \delta }{{\rm d} t} & \simeq & 2 \gamma y^2_\tau \xi^{-2} s^2_{23} \sin 2\sigma \;
\label{eq:6.1.26b}
\end{eqnarray}
in the $m^{}_3 =0$ case, where $\xi \equiv \sqrt{\Delta m^2_{21}/|\Delta m^2_{31}|} \simeq 0.17$ has been defined. The above RGEs can help us understand the results in Eqs.~(\ref{eq:6.1.23}) and (\ref{eq:6.1.24}). For example, the coefficient $0.54$ for $\Delta \theta^{}_{23}/(-\Delta^{}_\tau)$ obtained in the $m^{}_3 =0$ case can be understood as the factor $c^{}_{23} s^{}_{23}$ which governs the RGE running behaviours of $\theta^{}_{23}$. And in the $m^{}_3 =0$ case the quantum corrections to $\theta^{}_{12}$, $\delta$ and $\sigma$ can be appreciable because of the presence of an enhancement factor $\xi^{-2} \simeq 34$ in their RGEs.

\subsection{The trimaximal mixing patterns}
\label{section 6.2}

\begin{figure*}[h]
\centering
\includegraphics[width=6in]{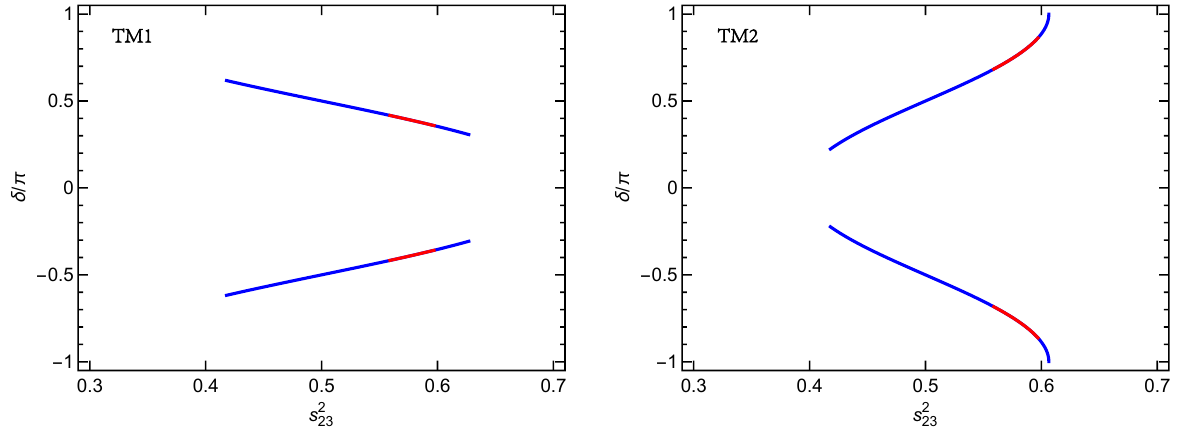}
\caption{The predictions for $\delta$ as functions of $s^2_{23}$ (red: $1 \sigma$ range; blue: $3\sigma$ range) in the TM1 and TM2 cases. }
\label{Fig:6-2}
\end{figure*}

If the $\mu$-$\tau$ interchange symmetry is combined with $\sin \theta^{}_{12} = 1/\sqrt{3}$, we shall arrive at the simple and elegant TBM mixing pattern $U^{}_{\rm TBM}$ in Eq.~(\ref{eq:6.1.1}) \cite{Harrison:2002er, Xing:2002sw, Harrison:2002kp}.
But the experimental observation of a relatively large $\theta^{}_{13}$ motivates us to modify this flavor mixing pattern. An economical and predictive way out is to keep the first or second column of $U^{}_{\rm TBM}$ unchanged but modify its other two columns within the unitarity constraints (see also Ref.~\cite{Perez:2020nqq}). Such a revised version of $U^{}_{\rm TBM}$ can be easily obtained by multiplying $U^{}_{\rm TBM}$ from the right-hand side with a rotation matrix either in the (2,3) plane or in the (1,3) plane \cite{Bjorken:2005rm, Xing:2006ms, He:2006qd, Albright:2008rp, Albright:2010ap}:
\begin{eqnarray}
U^{}_{\rm TM1}= U^{}_{\rm TBM} \pmatrix{
1 & 0  & 0 \cr
0 & \cos \theta & \sin \theta \hspace{0.03cm} e^{ -{\rm i} \phi} \cr
0 & -\sin \theta \hspace{0.03cm} e^{ {\rm i} \phi} & \cos \theta \cr
} \;; \nonumber \\
U^{}_{\rm TM2}= U^{}_{\rm TBM} \pmatrix{
\cos \theta & 0  & \sin \theta \hspace{0.03cm} e^{ -{\rm i} \phi} \cr
0 & 1 & 0 \cr
-\sin \theta \hspace{0.03cm} e^{ {\rm i} \phi} & 0 & \cos \theta \cr
} \; .
\label{eq:6.2.1}
\end{eqnarray}
The flavor mixing patterns $U^{}_{\rm TM1}$ and $U^{}_{\rm TM2}$ are usually referred to as the first and second trimaximal mixing patterns, respectively.
With the help of Eq.~(\ref{eq:3.3.9}), we see that these two flavor mixing patterns can accommodate an arbitrary $\theta^{}_{13}$ and basically retain the TBM value of $\theta^{}_{12}$. Namely,
\begin{eqnarray}
{\rm TM1}: \hspace{1cm} s^{}_{13} = \frac{1}{\sqrt 3} \sin \theta \; , \hspace{1cm}
s^{2}_{12} = \frac{1-3 s^{2}_{13}}{3 - 3s^{2}_{13}} \;; \nonumber \\
{\rm TM2}: \hspace{1cm} s^{}_{13} = \frac{2}{\sqrt 6} \sin \theta \; , \hspace{1cm}
s^{2}_{12} = \frac{1}{3 - 3s^{2}_{13}} \; .
\label{eq:6.2.2}
\end{eqnarray}
Given the $3\sigma$ range of $s^2_{13}$, $s^2_{12}$ is predicted to lie in the range 0.317---0.319 (or in the range 0.340---0.342) in the TM1 (or TM2) case, which is in agreement with current experimental result at the $1\sigma$ (or $3\sigma$) level.
Furthermore, there is a correlation among $\theta^{}_{13}$, $\theta^{}_{23}$ and $\delta$ as follows \cite{Albright:2008rp}:
\begin{eqnarray}
{\rm TM1}: \hspace{1cm} \tan{2\theta^{}_{23}} \cos \delta = - \frac{1-5 s^2_{13}}{2 \sqrt{2} s^{}_{13} \sqrt{1- 3 s^2_{13}} } \;;
\nonumber \\
{\rm TM2}: \hspace{1cm} \tan{2\theta^{}_{23}} \cos \delta = \frac{1-2 s^2_{13}}{s^{}_{13} \sqrt{2- 3 s^2_{13}} } \; .
\label{eq:6.2.3}
\end{eqnarray}
The predictions of these two relations for $\delta$ are shown as functions of $s^{2}_{23}$ in Fig.~\ref{Fig:6-2}, where $\delta= \pm \pi/2$ can be obtained from $s^2_{23} =1/2$.

Now we study a simple realization of the TM1 and TM2 mixing patterns in the minimal seesaw scheme.
We first derive possible textures of the Dirac neutrino mass matrix $M^{}_{\rm D}$ that can lead to these two flavor mixing patterns in the mass bases of both charged leptons and right-handed neutrinos. It is a good starting point to note that the neutrino mass matrix in Eq.~(\ref{eq:2.3}) can be transformed to the following intermediate form by means of the $U^{}_{\rm TBM}$ transformation \cite{Shimizu:2017fgu}:
\begin{eqnarray}
\hspace{-1.5cm} M^\prime_\nu = - \pmatrix{
\vspace{0.1cm}
\displaystyle \frac{A^2_1}{6 M^{}_1} + \frac{B^2_1}{6 M^{}_2} & \displaystyle \frac{A^{}_1 A^{}_2}{3\sqrt{2} M^{}_1} + \frac{B^{}_1 B^{}_2}{3\sqrt{2} M^{}_2}  & \displaystyle \frac{A^{}_1 A^{}_3}{2\sqrt{3} M^{}_1} + \frac{B^{}_1 B^{}_3}{2\sqrt{3} M^{}_2}  \cr
\vspace{0.1cm}
\displaystyle \frac{A^{}_1 A^{}_2}{3\sqrt{2} M^{}_1} + \frac{B^{}_1 B^{}_2}{3\sqrt{2} M^{}_2} & \displaystyle \frac{A^2_2}{3 M^{}_1}+ \frac{B^2_2}{3 M^{}_2} &  \displaystyle \frac{A^{}_2 A^{}_3}{\sqrt{6} M^{}_1}+ \frac{B^{}_2 B^{}_3}{\sqrt{6} M^{}_2} \cr
\displaystyle \frac{A^{}_1 A^{}_3}{2\sqrt{3} M^{}_1} + \frac{B^{}_1 B^{}_3}{2\sqrt{3} M^{}_2} & \displaystyle \frac{A^{}_2 A^{}_3}{\sqrt{6} M^{}_1}+ \frac{B^{}_2 B^{}_3}{\sqrt{6} M^{}_2} &  \displaystyle \frac{A^2_3}{2 M^{}_1}+ \frac{ B^2_3}{2 M^{}_2} \cr
} \;
\label{eq:6.2.4}
\end{eqnarray}
with $M^\prime_\nu \equiv U^{\dagger}_{\rm TBM} M^{}_\nu U^{*}_{\rm TBM}$ and
\begin{eqnarray}
A^{}_1 = 2 a^{}_1 - a^{}_2 + a^{}_3 \; , \hspace{1cm} A^{}_2 = a^{}_1 + a^{}_2 - a^{}_3
\; , \hspace{1cm} A^{}_3 = a^{}_2 + a^{}_3 \; , \nonumber \\
B^{}_1 = 2 b^{}_1 - b^{}_2 + b^{}_3 \; , \hspace{1cm} B^{}_2 = b^{}_1 + b^{}_2 - b^{}_3
\; , \hspace{1cm} B^{}_3 = b^{}_2 + b^{}_3 \; .
\label{eq:6.2.5}
\end{eqnarray}
With the help of this result, let us make some immediate observations.
(1) In the case of $A^{}_1 = B^{}_1 = 0$, one is left with
\begin{eqnarray}
M^{}_{\rm D}= \pmatrix{
\vspace{0.1cm}
\displaystyle \frac{a^{}_2 -a^{}_3}{2} & \displaystyle \frac{b^{}_2 -b^{}_3}{2}  \cr
a^{}_2 & b^{}_2  \cr
a^{}_3 & b^{}_3 \cr
} \; .
\label{eq:6.2.6}
\end{eqnarray}
Then $M^\prime_\nu$ is simplified to
\begin{eqnarray}
\hspace{-1.cm} M^\prime_\nu = - \pmatrix{
\vspace{0.1cm}
0 & 0 & 0 \cr
\vspace{0.1cm}
0 & \displaystyle \frac{3(a^{}_2-a^{}_3)^2}{4 M^{}_1}+ \frac{3(b^{}_2-b^{}_3)^2}{4 M^{}_2} &  \displaystyle \frac{3(a^{2}_2-a^{2}_3)}{2\sqrt{6} M^{}_1}+ \frac{3(b^{2}_2-b^{2}_3)}{2\sqrt{6} M^{}_2} \cr
0 & \displaystyle \frac{3(a^{2}_2-a^{2}_3)}{2\sqrt{6} M^{}_1}+ \frac{3(b^{2}_2-b^{2}_3)}{2\sqrt{6} M^{}_2} &  \displaystyle \frac{(a^{}_2+a^{}_3)^2}{2 M^{}_1}+ \frac{ (b^{}_2+b^{}_3)^2}{2 M^{}_2} \cr
} \; ,
\label{eq:6.2.7}
\end{eqnarray}
which can be diagonalized by a (2,3) rotation matrix and yields $m^{}_1 =0$. Therefore, the resulting lepton flavor mixing matrix will be of the TM1 form.
(2) In the case of $A^{}_1 = B^{}_2 = B^{}_3 =0$ (or equivalently $ A^{}_2 = A^{}_3 = B^{}_1 =0$), one has
\begin{eqnarray}
M^{}_{\rm D}= \pmatrix{
\vspace{0.1cm}
\displaystyle \frac{a^{}_2 -a^{}_3}{2} & 2 b^{}_3 &  \cr
a^{}_2 & - b^{}_3 \cr
a^{}_3 & b^{}_3 \cr
} \; ,
\label{eq:6.2.8}
\end{eqnarray}
and then obtains
\begin{eqnarray}
M^\prime_\nu = - \pmatrix{
\displaystyle \frac{6 b^2_3}{M^{}_2} & 0 & 0 \cr
\vspace{0.1cm}
0 & \displaystyle \frac{3(a^{}_2 -a^{}_3)^2}{4 M^{}_1} &  \displaystyle \frac{3(a^{2}_2-a^{2}_3)}{2\sqrt{6} M^{}_1} \cr
0 & \displaystyle \frac{3(a^{2}_2-a^{2}_3)}{2\sqrt{6} M^{}_1} &  \displaystyle \frac{(a^{}_2+a^{}_3)^2}{2 M^{}_1} \cr
} \; .
\label{eq:6.2.9}
\end{eqnarray}
So we arrive at the TM1 mixing pattern with $m^{}_3 = 0$ instead of $m^{}_1 = 0$.
(3) In the case of $A^{}_1 = A^{}_3 = B^{}_2 =0$ (or equivalently $A^{}_2 = B^{}_1 = B^{}_3 =0$),
we are left with
\begin{eqnarray}
M^{}_{\rm D}= \pmatrix{
- a^{}_3 & b^{}_3 -b^{}_2 \cr
- a^{}_3 & b^{}_2  \cr
a^{}_3 & b^{}_3 \cr
} \; .
\label{eq:6.2.10}
\end{eqnarray}
As a result,
\begin{eqnarray}
M^\prime_\nu = - \pmatrix{
\vspace{0.1cm}
\displaystyle \frac{3 (b^{}_2 - b^{}_3)^2}{2 M^{}_2} & 0 & \displaystyle \frac{\sqrt{3} (b^2_3 - b^2_2)}{2 M^{}_2}  \cr
\vspace{0.1cm}
0 & \displaystyle \frac{3 a^2_3}{M^{}_1} & 0  \cr
\displaystyle \frac{\sqrt{3} (b^2_3 - b^2_2)}{2 M^{}_2} & 0 &  \displaystyle  \frac{ (b^{}_2 + b^{}_3)^2}{2 M^{}_2} \cr
} \; ,
\label{eq:6.2.11}
\end{eqnarray}
which can be diagonalized by a (1,3) rotation matrix, rendering the lepton flavor mixing matrix to be of the TM2 form. In this case either $m^{}_1 =0$ or $m^{}_3 =0$ is allowed.
We can therefore draw the conclusion that, in order to get the TM1 (or TM2) mixing pattern, one needs to either take both columns of $M^{}_{\rm D}$ to be orthogonal to $(2, -1, 1)^{T}$ (or $(1, 1, -1)^{T}$) or take one column to be orthogonal to $(2, -1, 1)^{T}$ (or $(1, 1, -1)^{T}$) and the other column to be proportional to it. As mentioned in section~\ref{section 3.4}, this result can also be understood with the help of the triangular parametrization of $M^{}_{\rm D}$.
\begin{figure*}[h]
\centering
\includegraphics[width=6in]{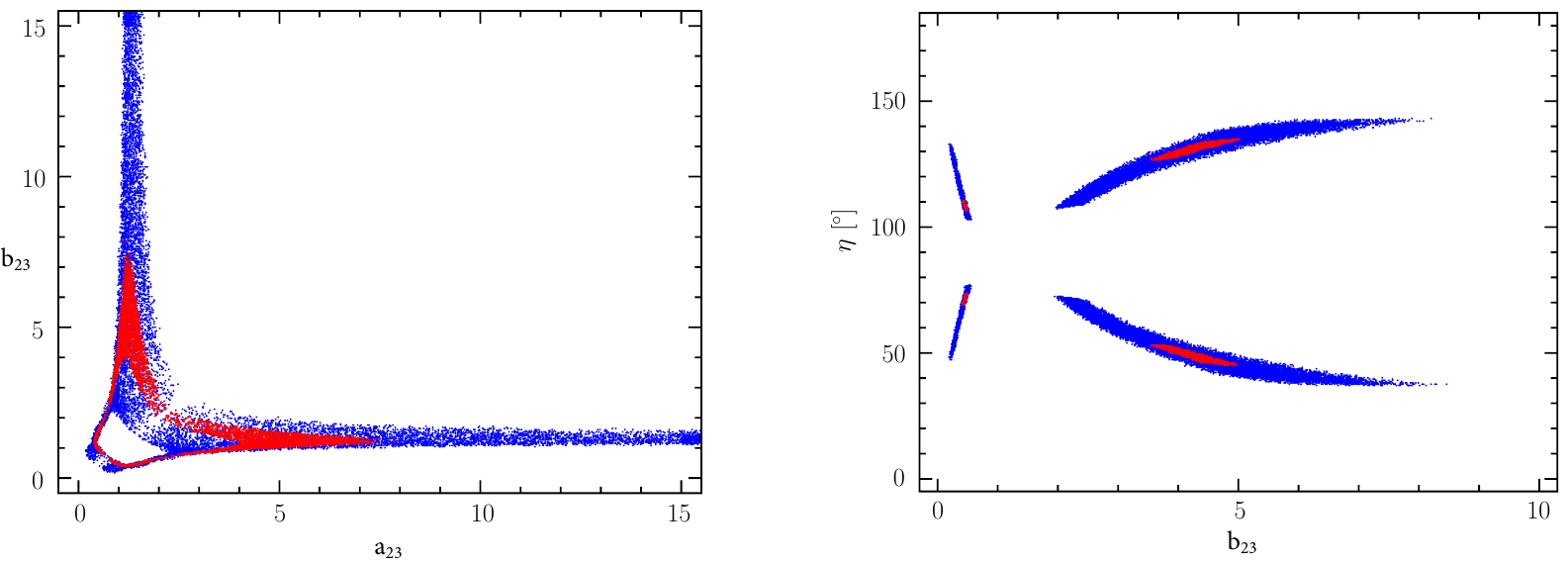}
\caption{In the TM1 case with $m^{}_1=0$, possible values of $(a^{}_{23}, b^{}_{23})$ (left) and $(b^{}_{23}, \eta)$ for $a^{}_{23}=1$ (right) that can be consistent with current experimental results at the $3 \sigma$ (blue) or $1\sigma$ (red) level. }
\label{Fig:6-3}
\end{figure*}

Now we confront the textures of $M^{}_{\rm D}$ in Eqs.~(\ref{eq:6.2.6}), (\ref{eq:6.2.8}) and (\ref{eq:6.2.10}) with current experimental data. Let us first consider $M^{}_{\rm D}$ in Eq.~(\ref{eq:6.2.6}) which can be used to realize the TM1 mixing pattern in the $m^{}_1 =0$ case. Given that only the phase differences ${\rm arg}(b^{}_i) - {\rm arg}(a^{}_i)$ have the physical meaning, without loss of generality we take $a^{}_{2, 3}$ and $b^{}_{2, 3}$ to be real and complex, respectively. Furthermore, only a simplified case in which the phase difference between $b^{}_2$ and $b^{}_3$ is trivially 0 or $\pi$ will be considered. Then we are only left with one phase (i.e., the relative phase between two columns of $M^{}_{\rm D}$), denoted as $\eta$, which is responsible for both the CP-violating effects at low energies and leptogenesis at a superhigh energy scale. Note that $\eta$ works with a period of $\pi$ in determining the low-energy flavor parameters. Now we explore possible values of $a^{}_{23} \equiv a^{}_2/a^{}_3$ and $b^{}_{23} \equiv b^{}_2/b^{}_3$ for the texture of $M^{}_{\rm D}$ in Eq.~(\ref{eq:6.2.6}) to be phenomenologically viable.
The results are shown in the left panel of Fig.~\ref{Fig:6-3}. Apparently, there is a symmetry with respect to the interchange $a^{}_{23} \leftrightarrow b^{}_{23}$. And $a^{}_{23}$ and $b^{}_{23}$ can only take positive values. Note that the equality between $a^{}_{23}$ and $b^{}_{23}$ is never allowed, because otherwise two columns of $M^{}_{\rm D}$ would be proportional to each other in which case only one light neutrino can acquire a nonzero mass.
Furthermore, the predictions for neutrino masses and lepton flavor mixing angles keep invariant under the transformation $\eta \to \pi-\eta$, and those for the CP phases undergo a sign reversal (see Table~\ref{Table:7}). This observation can be easily understood from the fact that $M^\prime_\nu$ in Eq.~(\ref{eq:6.2.7}) becomes $M^{\prime*}_\nu$ under such a transformation.

The model-building exercises in the literature have repeatedly shown that a texture of $M^{}_{\rm D}$ with zero entries or simple entry ratios (e.g., some linear equalities) can relatively easily find a justification from some discrete Abelian or non-Abelian flavor symmetries \cite{Altarelli:2010gt, King:2013eh}. Hence we pay particular attention to such possibilities for which the texture of $M^{}_{\rm D}$ in Eq.~(\ref{eq:6.2.6}) can be further simplified. We first note that $a^{}_{23}$ and $b^{}_{23}$ have no chance to reach either 0 or $\infty$, implying that $M^{}_{\rm D}$ cannot take a column pattern like $(-1, 0, 2)^{T}$ or $(1, 2, 0)^{T}$. In comparison, $a^{}_{23} =1$ (or $b^{}_{23} =1$) is possible, corresponding to a simple but interesting column pattern $(0, 1, 1)^{T}$. In Table~\ref{Table:7} we list some particular combinations of $(a^{}_{23}, b^{}_{23})$ that can be consistent with current experimental results to a good degree of accuracy (measured by the corresponding $\chi^2_{\rm min}$ values), together with their predictions for the low-energy flavor parameters (corresponding to minimalizations of the $\chi^2$ function defined in Eq.~(\ref{eq:5.1.15})) \cite{Chen:2019oey}. Taking $a^{}_{23} = 1/3$, $1/2$, $2$, $3$ or $5$ (or the same value for $b^{}_{23}$) corresponds to a column pattern like $(-1, 1, 3)^{T}$, $(-1, 2, 4)^{T}$,  $(1, 4, 2)^{T}$, $(1, 3, 1)^{T}$ or $(2, 5, 1)^{T}$, respectively. In obtaining the numerical results in Table~\ref{Table:7}, we have taken $\eta$ to be a free parameter and determined its value by fitting current experimental data. It is interesting to notice that for some particular combinations of $(a^{}_{23}, b^{}_{23})$ the best-fit values of $\eta$ are also close to certain special values (see also the right panel of Fig.~\ref{Fig:6-3}).
This tempts us to consider the possibility of $\eta$ taking a special value, which may relatively easily get a symmetry justification in the model-building exercises. The possible cases of this kind, together with their predictions for the low-energy flavor parameters, are listed in Table~\ref{Table:8}.
One can see that the compatibilities of these cases with current experimental data
are only slightly worsened as compared with the corresponding cases in Table~\ref{Table:7}.
\begin{table}[t]
\caption{ The predictions of some particular combinations of $(a^{}_{23}, b^{}_{23})$ for the low-energy flavor parameters, where the units of $\Delta m^2_{21}$ and $\Delta m^2_{31}$ are $10^{-5}$ eV$^2$ and $10^{-3}$ eV$^2$, respectively.}
\label{Table:7}
\vspace{0.1cm}
\centering
\begin{tabular}{ccccccccccc} \br
$a^{}_{23}$ & $b^{}_{23}$ & $\eta/\pi$ & $\chi^2_{\rm min}$ & $\Delta m^2_{21}$ & $\Delta m^2_{31}$ & $s^2_{12}$ & $s^2_{13}$ & $s^2_{23}$ & $\delta/\pi$ & $\sigma/\pi$  \\
\mr
\vspace{0.1cm}
1 & 1/3 & $\pm 0.345$ & 23.7 & 7.66 & 2.50 & 0.319 & 0.0212 & 0.509 & $\pm 0.488$ & $\pm 0.218$  \\
\vspace{0.1cm}
1 & 1/2 & $\pm 0.414$ & 3.44 & 7.28 & 2.53 & 0.318 & 0.0229 & 0.605 & $\pm 0.347$ & $\pm 0.271$  \\
\vspace{0.1cm}
1 & 3 & $\pm 0.325$ & 12.2 & 7.17 & 2.54 & 0.318 & 0.0231 & 0.528 & $\mp 0.462$ & $\pm 0.189$ \\
\vspace{0.1cm}
1 & 5 & $\pm 0.248$ & 2.37 & 7.47 & 2.53 & 0.318 & 0.0220 & 0.601 & $\mp 0.351$ & $\pm 0.157$  \\
\ 2 & 3 & $\pm 0.467$ & 1.80 & 7.38 & 2.51 & 0.318 & 0.0223 & 0.600 & $\mp 0.354$ & $\pm 0.365$  \\
\br
\end{tabular}
\end{table}
\begin{table}[t]
\caption{ The predictions of some particular combinations of $(a^{}_{23}, b^{}_{23}, \eta)$ for the low-energy flavor parameters, where the units of $\Delta m^2_{21}$ and $\Delta m^2_{31}$ are $10^{-5}$ eV$^2$ and $10^{-3}$ eV$^2$, respectively.}
\label{Table:8}
\vspace{0.1cm}
\centering
\begin{tabular}{ccccccccccc} \br
$a^{}_{23}$ & $b^{}_{23}$ & $\eta$ & $\chi^2_{\rm min}$ & $\Delta m^2_{21}$ & $\Delta m^2_{31}$ & $s^2_{12}$ & $s^2_{13}$ & $s^2_{23}$ & $\delta/\pi$ & $\sigma/\pi$  \\
\mr
\vspace{0.1cm}
1& 1/3 & $\pm \pi/3$  & 29.5 & 7.36 & 2.53 & 0.318 & 0.0221 & 0.489 & $\pm 0.516$ & $\pm 0.202$ \\
\vspace{0.1cm}
1 & 1/2 & $\pm 2\pi/5$ & 13.5 & 6.88 & 2.56 & 0.317 & 0.0240 & 0.575 & $\pm 0.398$ & $\pm 0.238$ \\
\vspace{0.1cm}
1 & 3 & $\pm \pi/3$ & 16.0 & 7.46 & 2.52 & 0.318 & 0.0225 & 0.513 & $\mp 0.482$ & $\pm 0.200$ \\
\ 1 & 5 & $\pm \pi/4$ & 2.37 & 7.52 & 2.52 & 0.318 & 0.0220 & 0.599 & $\mp 0.354$ & $\pm 0.158$ \\
\br
\end{tabular}
\end{table}

\begin{figure*}[h]
\centering
\includegraphics[width=6in]{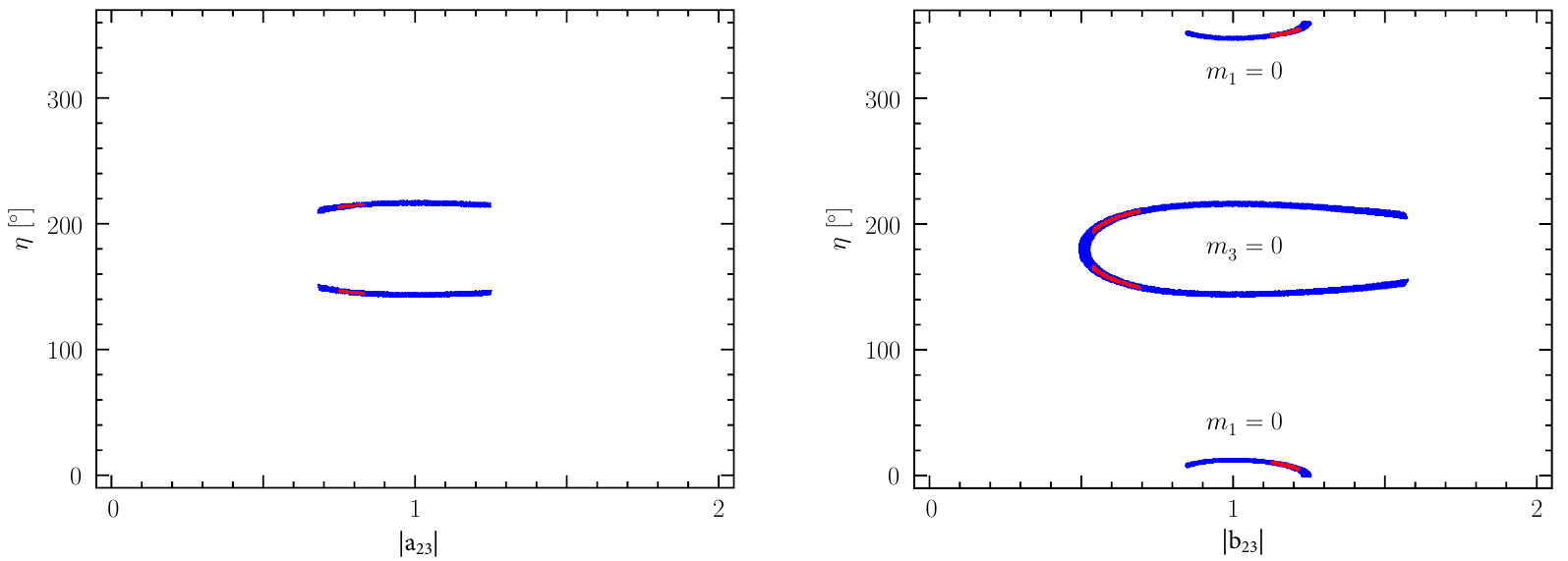}
\caption{Left panel: In the TM1 case with $m^{}_3=0$, possible values of $(|a^{}_{23}|, \eta)$ that can be consistent with current experimental data at the $3\sigma$ (blue) and $1\sigma$ (red) levels.
Right panel: In the TM2 case with $m^{}_1=0$ or $m^{}_3=0$, possible values of ($|b^{}_{23}|, \eta$) that can be consistent with current experimental data at the $3\sigma$ (blue) and $1\sigma$ (red) levels. }
\label{Fig:6-4}
\end{figure*}

As for the texture of $M^{}_{\rm D}$ in Eq.~(\ref{eq:6.2.8}) which can be used to realize the TM1 mixing pattern in the $m^{}_3 =0$ case, we take $a^{}_{2, 3}$ and $b^{}_3$ to be complex and real parameters, respectively. In this case the resulting lepton flavor mixing is determined by $a^{}_{23} \equiv a^{}_2/a^{}_3$. It is found that $a^{}_{23}$ should be a complex parameter $|a^{}_{23}| e^{{\rm i} \eta}$ in order for the texture of $M^{}_{\rm D}$ under consideration to be phenomenologically viable. The allowed values of $(|a^{}_{23}|, \eta)$ are shown in the left panel of Fig.~\ref{Fig:6-4}. One can see that $|a^{}_{23}|$ and $\eta$ are around 1 and $ \pm 0.8 \pi$, respectively. For the texture of $M^{}_{\rm D}$ in Eq.~(\ref{eq:6.2.10}) which can be used to realize the TM2 mixing pattern in either the $m^{}_1 =0$ case or the $m^{}_3 =0$ case, we take $a^{}_3$ and $b^{}_{2, 3}$ to be real and complex parameters, respectively. In this case the resulting lepton flavor mixing is determined by $b^{}_{23} \equiv b^{}_2/b^{}_3$, which also needs to be a complex parameter $|b^{}_{23}| e^{{\rm i}\eta}$ in order for the texture of $M^{}_{\rm D}$ under discussion to be phenomenologically viable. The allowed values of $(|b^{}_{23}|, \eta)$ are shown in the right panel of Fig.~\ref{Fig:6-4}. It turns out that $\eta$ is around 0 (or $\pi$) in the $m^{}_1 =0$ (or $m^{}_3 =0$) case and $|b^{}_{23}|$ is around 1.

Given the above textures of  $M^{}_{\rm D}$, the calculations of leptogenesis are straightforward \cite{Shimizu:2017vwi}. Here we shall not go into the details of this issue but just make some immediate comments. For the structure of $M^{}_{\rm D}$ shown in Eq.~(\ref{eq:6.2.6}), under the assumption that there is only one physical phase $\eta$ (i.e., the relative phase between two columns of $M^{}_{\rm D}$), a correlation between signs of the baryon number asymmetry and the low-energy CP-violating effects can be established. In the particular case of $a^{}_{23}= 1$ and $b^{}_{23}= 3$, for example, a numerical calculation yields \cite{Bjorkeroth:2014vha}
\begin{eqnarray}
Y^{}_{\rm B} \simeq 2.5 \times 10^{-11} \sin 2\eta \left( \frac{M^{}_1}{10^{10} \ \rm GeV} \right) \; .
\label{eq:6.2.12}
\end{eqnarray}
If $\eta$ is taken to be $\pi/3$, for which the observed value of $Y^{}_{\rm B}$ can be reproduced with $M^{}_1 \simeq 4 \times 10^{10}$ GeV, then one will arrive at $\delta \simeq - \pi/2$. As for the textures of $M^{}_{\rm D}$ shown in Eqs.~(\ref{eq:6.2.8}) and (\ref{eq:6.2.10}), the CP-violating asymmetry for leptogenensis vanishes due to the orthogonality of two columns of $M^{}_{\rm D}$. This is a generic  consequence of the seesaw models in which the columns of $M^{}_{\rm D}$ are simply proportional to those of $U$ (i.e., the form dominance scenario \cite{King:2003jb} mentioned in section~\ref{section 3.4}) \cite{Antusch:2006cw, King:2006hn, Jenkins:2008rb, Bertuzzo:2009im, AristizabalSierra:2009ex, Choubey:2010vs}.

Let us take the so-called {\it littlest seesaw} model as an example to illustrate how to realize a particular texture of $M^{}_{\rm D}$ like those listed in Eqs.~(\ref{eq:6.2.6}), (\ref{eq:6.2.8}) and (\ref{eq:6.2.10}) with the help of a kind of flavor symmetry. The littlest seesaw model refers to the texture of $M^{}_{\rm D}$ in Eq.~(\ref{eq:6.2.6}) with $a^{}_{23} =1$. A more specific littlest seesaw model with $b^{}_{23} =1$ and $\eta =\pi/3$ (or $b^{}_{23} =1/3$ and $\eta =-\pi/3$) deserves particular attention due to its highly-constrained form and interesting phenomenological consequences (see Table~\ref{Table:8}) \cite{King:2005bj, Antusch:2011ic, King:2013iva, King:2013xba, King:2013hoa, Bjorkeroth:2014vha, King:2015dvf, King:2016yvg, Ballett:2016yod, King:2018fqh, Geib:2017bsw, King:2016yef}.
Since the TM1 mixing pattern is a simple variant of the TBM mixing pattern, it is expected that a flavor symmetry capable of realizing the latter can also be used to realize the former. Hence our strategy is to first explore a possible flavor symmetry that can naturally accommodate the TBM mixing pattern. In principle, a flavor symmetry $\rm G^{}_{\rm F}$ in the lepton sector should be initially applied to the charged-lepton and neutrino sectors on an equal footing (because the left-handed charged-lepton and neutrino fields jointly constitute the $\rm SU(2)^{}_{\rm L}$ doublets) and then spontaneously broken down to different residual symmetries ${\rm G}^{}_l$ and $\rm G^{}_\nu$ corresponding to the two sectors in order to produce a nontrivial lepton flavor mixing pattern. Conversely, one may get a hold of $\rm G^{}_{\rm F}$ by studying ${\rm G}^{}_l$ and $\rm G^{}_\nu$.
In this connection it is useful to notice that the light Majorana neutrino mass matrix $M^{}_\nu = U^{}_{\rm TBM} D^{}_\nu U^{T}_{\rm TBM}$ reconstructed from $U^{}_{\rm TBM}$ is invariant under the following order-two (i.e., $S^2 = G^2 = I$) transformations \cite{Lam:2006wm,Lam:2007qc,Lam:2008rs}:
\begin{eqnarray}
S = \frac{1}{3} \pmatrix{
-1 & 2 & 2 \cr 2 & -1 & 2 \cr 2 & 2 & -1
} \; , \hspace{1cm} G = \pmatrix{
1 & 0 & 0 \cr 0 & 0 & 1 \cr 0 & 1 & 0
} \; .
\label{eq:6.2.13}
\end{eqnarray}
As for the charged-lepton sector, the diagonal Hermitian matrix $M^{}_l M^\dagger_l$ possesses the symmetry $T^\dagger M^{}_l M^\dagger_l T = M^{}_l M^\dagger_l$ with $T = {\rm Diag}\{e^{{\rm i}\phi^{}_1}, e^{{\rm i}\phi^{}_2}, e^{-{\rm i}(\phi^{}_1 + \phi^{}_2)} \}$.
If the above $S$, $G$ and $T$ symmetries of $M^{}_\nu$ and $M^{}_l M^\dagger_l$ are identified as the residual ones (i.e., $\rm G^{}_\nu$ and ${\rm G}^{}_l$) of $\rm G^{}_{\rm F}$, then the latter can be generated from their exhaustive multiplications. In order for the resulting $\rm G^{}_{\rm F}$ to be finite, there must exist a positive integer $n$ for $T^n = I$ to hold. Note that $n$ cannot be smaller than 3 so that $T$ is capable of distinguishing the three charged-lepton fields. A detailed analysis shows that $T = {\rm Diag} \{ 1, \omega^2, \omega \}$ (for $\omega \equiv e^{{\rm i}2\pi/3}$, corresponding to $n=3$) together with the above forms of $S$ and $G$ constitutes the generators of the $\rm S^{}_4$ group:
\begin{eqnarray}
S^2 = G^2 = T^3 = (SG)^2 = (ST)^3= (TG)^2 = (STG)^4 = I \;.
\label{eq:6.2.14}
\end{eqnarray}
The $\rm S^{}_4$ group has five irreducible representations: ${\bf 1}$, ${\bf 1^\prime}$, ${\bf 2}$, ${\bf 3}$ and ${\bf 3^\prime}$. The above explicit forms of $T$, $S$ and $G$ give their representation matrices in the ${\bf 3^\prime}$ representation, while those in the ${\bf 3}$ representation just differ by a sign for $G$. The Kronecker products of two representations relevant for our study are given by
\begin{eqnarray}
\bf 1^\prime \times 1^\prime = 1  \; , \hspace{0.5cm}  1^\prime \times 3 = 3^\prime  \; , \hspace{0.5cm}  1^\prime \times 3^\prime = 3  \; , \nonumber \\
\bf 3^{(\prime)} \times 3^{(\prime)} = 1 + 2 + 3 + 3^\prime \; , \hspace{0.5cm} 3 \times 3^\prime = 1^\prime + 2 + 3 + 3^\prime \; .
\label{eq:6.2.15}
\end{eqnarray}
More mathematical details about the $\rm S^{}_4$ group can be found in Ref.~\cite{Ishimori:2010au}.
As the unique group for naturally realizing the TBM mixing pattern \cite{Lam:2008sh}, the $\rm S^{}_4$ group will be employed here to realize the littlest seesaw model. As a matter of fact, the light Majorana neutrino mass matrix $M^{}_\nu = U^{}_{\rm TM1} D^{}_\nu U^{T}_{\rm TM1}$ reconstructed in terms of $U^{}_{\rm TM1}$ retains the $SG$ symmetry of $\rm S^{}_4$ \cite{King:2016yvg}.

Once $\rm G^{}_{\rm F}$ is specified, an immediate question will be how to break this symmetry while preserving the desired residual symmetries. To answer this question, one needs to introduce the so-called flavon fields, which do not carry the SM quantum numbers but may constitute nontrivial representations of $\rm G^{}_{\rm F}$. They break $\rm G^{}_{\rm F}$ to the desired pattern by acquiring proper vacuum expectation value (VEV) alignments. As discussed in Ref.~\cite{King:2009ap}, the flavor-symmetry breaking can proceed in two distinct approaches, based on how the symmetries for $M^{}_l M^\dagger_l$ and $M^{}_\nu$ come about. In the {\it direct} approach, the residual symmetries ${\rm G}^{}_l$ and $\rm G^{}_\nu$ are preserved by the relevant flavon VEVs and subsequently by $M^{}_l M^\dagger_l$ and $M^{}_\nu$. In the {\it indirect} approach, the flavon VEVs do not necessarily preserve any flavor symmetry, but their particular alignments accidentally give rise to the desired forms of $M^{}_l M^\dagger_l$ and $M^{}_\nu$. Here we present a concrete model to illustrate the indirect approach. Some discussions about the direct approach will be given in section~\ref{section 6.3}.
\begin{table}
\caption{ The transformation properties of the lepton, Higgs and flavon superfields
under the $\rm S^{}_4 \times Z^{}_2 \times Z^{}_3$ symmetries and their $\rm U(1)^{}_{\rm R}$ charges \cite{King:2015dvf}. }
\label{Table:9}
\vspace{-0.28cm}
$$
\begin{array}{||c||ccccccc||c||cccccc||}
\hline \hline
 & e^c & \mu^c & \tau^c
& \phi^{}_e & \phi^{}_\mu & \phi^{}_\tau
& H_{d} & l & H^{}_{u} & \phi^{}_{1} & \phi^{}_{2} &N^c_{1} & N^c_{2}
& \xi  \\  \hline
\hline
\rm  S^{}_4 &  {\bf 1} & {\bf 1} & {\bf 1} & {\bf 3} & {\bf 3} & {\bf 3}  & {\bf 1}
& {\bf 3} & {\bf 1} & {\bf 3'} & {\bf 3} & {\bf 1'} & {\bf 1} & {\bf 1}
\\ \hline  \hline
\rm  Z^{}_2 & 1 & 1 & 1 & 1 & 1  & 1 & 1 & 1 & 1 & -1 & -1 & -1
& -1 & 1  \\\hline
\rm Z^{}_3 & \omega  & \omega^2 & 1 & \omega^2 & \omega & 1 & 1
& 1 & 1 & \omega & 1 & \omega^2 & 1  & \omega^2 \\ \hline
\rm U(1)^{}_{\rm R} &  1 & 1 & 1 & 0 & 0 & 0 & 0 & 1 & 0 & 0 & 0 & 1 & 1 & 0  \\
\hline
\end{array}
$$
\end{table}

Now we present an indirect model that can reproduce the specific littlest seesaw model with $b^{}_{23} =1$ and $\eta = \pi/3$ \cite{King:2015dvf}.
The model employs the $\rm S^{}_4 \times Z^{}_2 \times Z^{}_3$ symmetries, under which the transformation properties of the related fields are listed in Table~\ref{Table:9}. To facilitate the following realization of the desired flavon VEV alignments through the F-term alignment mechanism \cite{Altarelli:2005yx}, which takes advantage of the $\rm U(1)^{}_{\rm R}$ symmetry (under which the superpotential terms should carry a total charge of 2) of the supersymmetric theories, this model is embedded in the supersymmetry framework.
Given that the flavor symmetry is broken in different manners in the charged-lepton and neutrino sectors, the $\rm Z^{}_2$ symmetry is introduced to distinguish the flavon fields for these two sectors. And the
$\rm Z^{}_3$ symmetry is used to further distinguish the flavon fields for different flavors. Furthermore, this $\rm Z^{}_3$ symmetry can also help us realize $\eta = \pi/3$, as will be seen below. Under our setup, the superpotential $W$ invariant under the SM gauge symmetry and flavor symmetries appears as \cite{King:2015dvf}
\begin{eqnarray}
W & = & \frac{y^{}_1}{\Lambda} H^{}_d \left({l}.\phi^{}_1\right) N^c_{1}
+ \frac{y^{}_2}{\Lambda} H^{}_d \left({l}.\phi^{}_2\right) N^c_{2}
+ \xi N^c_{1} N^c_{1}
+ M^{}_2 N^c_{2} N^c_{2} \nonumber \\
&& +\frac{y^{}_e}{\Lambda}  H^{}_u \left( {l}.\phi^{}_e\right) e^c
+ \frac{y^{}_\mu}{\Lambda}  H^{}_u \left({l}.\phi^{}_{\mu}\right) \mu^c
+\frac{y^{}_\tau}{\Lambda} H^{}_u \left({l}.\phi^{}_{\tau}\right) \tau^c \; ,
\label{eq:6.2.16}
\end{eqnarray}
where $(\alpha.\beta) = \alpha^{}_1 \beta^{}_1 + \alpha^{}_2 \beta^{}_2 + \alpha^{}_3 \beta^{}_3 $ denotes the contraction of two triplet representations of $\rm S^{}_4$ into a singlet representation in the basis of Ref.~\cite{Ishimori:2010au}, $\Lambda$ is the cutoff scale of the flavor-symmetry physics, and the ratios of the flavon VEVs to $\Lambda$ are typically assumed to be small so that the contributions of higher-dimension terms are naturally suppressed. When the flavon fields acquire the VEV alignments
\begin{eqnarray}
\langle \phi^{}_1 \rangle  = v^{}_1 \pmatrix{ 0 \cr 1 \cr 1 } \; , \hspace{0.5cm}
\langle \phi^{}_2 \rangle = v^{}_2 \pmatrix{ 1 \cr 3 \cr 1 } \; ,  \hspace{0.5cm}
\langle \xi \rangle = M^{}_1 \; , \nonumber \\
\langle \phi^{}_e \rangle = v^{}_e \pmatrix{ 1 \cr 0 \cr 0 } \; , \hspace{0.5cm}
\langle \phi^{}_\mu \rangle = v^{}_\mu \pmatrix{ 0 \cr 1 \cr 0 } \; , \hspace{0.5cm}
\langle \phi^{}_\tau \rangle = v^{}_\tau \pmatrix{ 0 \cr 0 \cr 1 } \; ,
\label{eq:6.2.17}
\end{eqnarray}
the specific littlest seesaw model with $b^{}_{23} =1$ will be successfully reproduced. It is also possible to explain the mass hierarchies of three charged leptons by simply including an additional Froggatt-Nielsen symmetry \cite{Froggatt1979Hierarchy}.

Finally, we give a brief account of how to achieve the desired flavon VEV alignments via the F-term alignment mechanism \cite{Altarelli:2005yx}. For this purpose, we introduce some driving fields $\psi$ which carry a $\rm U(1)^{}_{\rm R}$ charge of 2 and thus can linearly couple with the flavon fields to form certain superpotential terms. Then the minimization requirement of the potential energy $V(\phi) = \sum |\partial W/\partial \psi|^2$ brings about the constraint $\partial W/\partial \psi =0$ for the flavon VEVs.
For example, the superpotential terms
\begin{eqnarray}
\psi \left(g \phi^{}_2 \phi^{}_2 + g^\prime \xi^\prime \phi^{}_2 + g^{}_\mu \xi^{}_\mu \phi^{}_\mu  \right) \; ,
\label{eq:6.2.18}
\end{eqnarray}
where $\psi$, $\xi^\prime$ and $\xi^{}_\mu$ have the transformation properties $({\bf 3}, 1, 1)$, $({\bf 1}, -1, 1)$ and $({\bf 1}, 1, \omega^2)$ under $\rm S^{}_4 \times Z^{}_2 \times Z^{}_3$,
will lead to the following constraints on the VEV alignment of $\phi^{}_2$:
\begin{eqnarray}
2 g \pmatrix{
\langle \phi^{}_{2} \rangle^{}_2  \langle \phi^{}_{2} \rangle^{}_3 \cr
\langle \phi^{}_{2} \rangle^{}_3  \langle \phi^{}_{2} \rangle^{}_1 \cr
\langle \phi^{}_{2} \rangle^{}_1  \langle \phi^{}_{2} \rangle^{}_2 }
+ g^\prime \langle \xi^\prime  \rangle
\pmatrix{
\langle \phi^{}_{2} \rangle^{}_1 \cr
\langle \phi^{}_{2} \rangle^{}_2 \cr
\langle \phi^{}_{2} \rangle^{}_3 }
+ g^{}_{\mu} \langle \xi^{}_{\mu}  \rangle
\pmatrix{
\langle \phi^{}_{\mu} \rangle^{}_1 \cr
\langle \phi^{}_{\mu} \rangle^{}_2 \cr
\langle \phi^{}_{\mu} \rangle^{}_3 }
= \pmatrix{ 0 \cr 0 \cr 0 } \; .
\label{eq:6.2.19}
\end{eqnarray}
Taking account of the VEV alignment of $\phi^{}_\mu$ in Eq.~(\ref{eq:6.2.17}), we arrive at $\langle \phi^{}_{2} \rangle \propto (1, n, 1)^T$ with $n$ unspecified. Provided there is also a superpotential term $\psi^\prime \phi^{}_2 \phi^{}_3$ with $\psi^\prime$ being a singlet representation of $\rm S^{}_4$ and $\phi^{}_3$ being a triplet representation of $\rm S^{}_4$, the constraint $\langle \phi^{}_2 \rangle^{}_1 \langle \phi^{}_3 \rangle^{}_1 + \langle \phi^{}_2 \rangle^{}_2 \langle \phi^{}_3 \rangle^{}_2 + \langle \phi^{}_2 \rangle^{}_3 \langle \phi^{}_3 \rangle^{}_3 =0$ together with $\langle \phi^{}_{3} \rangle \propto (2, -1, 1)^T$ will result in $\langle \phi^{}_{2} \rangle \propto (1, n, n-2)^T$. The combination of $\langle \phi^{}_{2} \rangle \propto (1, n, n-2)^T$ and $\langle \phi^{}_{2} \rangle \propto (1, n, 1)^T$ then yields the desired $\langle \phi^{}_{2} \rangle \propto (1, 3, 1)^T$. The VEV alignments of other flavon fields can be achieved in a similar way \cite{King:2015dvf}.

To understand the origin of $\eta = \pi/3$, one needs to impose the CP symmetry and then break it in a particular way \cite{King:2013iva,King:2013xba,Antusch:2013wn,Antusch:2011sx}.
In the present model the $\rm Z^{}_3$ symmetry can help us fulfill this role: the superpotential term $\psi^{\prime \prime} ( \xi^3/\Lambda - M^2)$ (with $\psi^{\prime \prime}$ being a singlet under the flavor symmetries and $M$ being real due to the CP symmetry) leads to the constraint $\langle \xi \rangle^3/\Lambda - M^2 =0$, which can give $M^{}_1 = \langle \xi \rangle = e^{{\rm i} 2\pi/3} M$. Such an $M^{}_1$ is equivalent to $M^{}_1$ being real but $\eta = \pi/3$. A straightforward generalization of the above tactics allows us to achieve a phase of $2\pi/n$ with the help of a ${\rm Z}^{}_n$ symmetry.

\subsection{The tri-direct CP approach}
\label{section 6.3}

Now let us give an introduction of the so-called tri-direct CP approach proposed in Refs.~\cite{Ding:2018fyz,Ding:2018tuj}, which is dedicated to the minimal seesaw mechanism, to illustrate the direct approach for flavor symmetry breaking. In the literature a popular and successful way of implementing some flavor symmetry $\rm G^{}_{\rm F}$ is to impose the CP symmetry $\rm H^{}_{\rm CP}$ simultaneously so that both the lepton flavor mixing angles and CP-violating phases can be predicted. However, the canonical CP transformation may not be consistent with $\rm G^{}_{\rm F}$. In order for $\rm H^{}_{\rm CP}$ to be compatible with $\rm G^{}_{\rm F}$, the following consistency condition must be satisfied \cite{Feruglio:2012cw,Holthausen:2012dk,Chen:2014tpa}:
\begin{eqnarray}
X \rho^*(g) X^{-1} = \rho(g^\prime)  \; , \hspace{1cm} g , g^\prime \in {\rm G^{}_{F}} \; ,
\label{eq:6.3.1}
\end{eqnarray}
where $\rho(g)$ and $\rho(g^\prime)$ are the representation matrices of $g$ and $g^\prime$, and $X$ is the generalized CP transformation matrix of $\rm H^{}_{CP}$. In general, $g$ and $g^\prime$ are different from each other, in which case the full symmetry is a semi-direct product of $\rm G^{}_F$ and $\rm H^{}_{CP}$: $\rm G^{}_F \rtimes H^{}_{CP}$. Of course, the semi-direct product will be reduced to a direct product if $g = g^\prime$ holds.
In the generic direct approach for flavor symmetry breaking, $\rm G^{}_F \rtimes H^{}_{CP}$ is typically assumed to be spontaneously broken down to ${\rm G}^{}_l \rtimes {\rm H}^l_{\rm CP}$ and $\rm G^{}_\nu \rtimes H^{\nu}_{CP}$ in the charged-lepton and neutrino sectors, respectively. In the specific tri-direct CP approach discussed here, it is further assumed that the two right-handed neutrino fields possess different residual symmetries: $\rm G^{}_1 \rtimes H^{1}_{CP}$ for $N^{}_1$ and $\rm G^{}_2 \rtimes H^{2}_{CP}$ for $N^{}_2$.
Since ${\rm G}^{}_i$ (for $i=1$ and 2) only have an order of 2, ${\rm G}^{}_i \rtimes {\rm H}^{i}_{\rm CP}$ actually reduce to ${\rm G}^{}_i \times {\rm H}^{i}_{\rm CP}$. In comparison with the indirect approach, here the flavon VEV alignments are required to preserve the corresponding residual symmetries. As one will see, the combination of these residual symmetries can severely constrain the model parameters and lead to some testable predictions for the low-energy flavor parameters.

We first formulate such a tri-direct approach \cite{Ding:2018fyz,Ding:2018tuj}. In the mass basis of two right-handed neutrinos, the Lagrangian relevant for the charged-lepton and neutrino masses in Eq.~(\ref{eq:1.11}) can be rewritten as
\begin{eqnarray}
- {\cal L}^{}_{\rm mass} & = & \frac{Y^{}_l}{\Lambda} \overline{\ell^{}_{\rm L}}
H \phi^{}_l E^{}_{\rm R} + \frac{y^{}_i}{\Lambda} \overline{\ell^{}_{\rm L}}
\widetilde{H} \phi^{}_i N^{}_{i \rm R} + \frac{1}{2} M^{}_i \overline{N^{c}_{i\rm R}} N^{}_{i \rm R} + {\rm h.c.} \; ,
\label{eq:6.3.2}
\end{eqnarray}
where $\phi^{}_l$ and $\phi^{}_{i}$ are the flavon fields. After spontaneous symmetry breaking, invariance of the charged-lepton mass matrix $M^{}_l$ under the residual flavor symmetry ${\rm G}^{}_l$ reads
\begin{eqnarray}
g^\dagger_l M^{}_l M^\dagger_l g^{}_l = M^{}_l M^\dagger_l  \; , \hspace{1cm} g^{}_l \in {\rm G}^{}_{l} \; .
\label{eq:6.3.3}
\end{eqnarray}
Substituting the reconstruction relation $M^{}_l M^\dagger_l = U^{}_l D^{2}_l U^\dagger_l$ with $D^{2}_l = {\rm Diag} \{ m^2_e, m^2_\mu, m^2_\tau \}$ into this equation, we find that $g^{}_l$ can also be diagonalized by $U^{}_l$:
\begin{eqnarray}
U^\dagger_l g^{}_l U^{}_l = {\rm Diag} \{ e^{{\rm i} \varphi^{}_e}, e^{{\rm i} \varphi^{}_\mu} ,
e^{{\rm i} \varphi^{}_\tau} \} \; ,
\label{eq:6.3.4}
\end{eqnarray}
where $\varphi^{}_\alpha$ are some roots of unity because of the finite order of $g^{}_l$. This means that the unitary matrix $U^{}_l$ for diagonalizing $M^{}_l M^\dagger_l$ can be directly calculated from ${\rm G}^{}_l$ itself with no need of the concrete form of $M^{}_l M^\dagger_l$. As for the neutrino sector, for any given ${\rm G}^{}_i$, the corresponding ${\rm H}^{i}_{\rm CP}$ can be derived from the consistency conditions
\begin{eqnarray}
X^{}_i \rho^*(g^{}_i) X^{-1}_i = \rho(g^{}_i)  \; , \hspace{1cm} g^{}_i  \in {\rm G}^{}_{i} \; .
\label{eq:6.3.5}
\end{eqnarray}
Once ${\rm G}^{}_i \times {\rm H}^{i}_{\rm CP}$ are specified, the VEV alignments of $\phi^{}_i$ will be fixed by the requirement that they preserve the corresponding residual symmetries
\begin{eqnarray}
g^{}_i \langle \phi^{}_i \rangle = \langle \phi^{}_i \rangle \; , \hspace{1cm} g^{}_i \in {\rm G}^{}_{i} \; ; \nonumber \\
X^{}_i \langle \phi^{}_i \rangle^* = \langle \phi^{}_i \rangle \; , \hspace{1cm} X^{}_i \in {\rm H}^{i}_{\rm CP} \; .
\label{eq:6.3.6}
\end{eqnarray}
Subsequently, the texture of the Dirac neutrino mass matrix $M^{}_{\rm D}$ can be directly read out by exploiting the invariance of $\overline{\ell^{}_{\rm L}}
\widetilde{H} \phi^{}_i N^{}_{i \rm R}$ under $\rm G^{}_F$. The lepton flavor mixing matrix is then given by $U = U^\dagger_l U^{}_\nu$ with $U^{}_\nu$ being the unitary matrix for diagonalizing the effective mass matrix of three light Majorana neutrinos obtained via the seesaw formula.

We now show that the above tri-direct approach can help us reproduce the littlest seesaw model discussed in section~\ref{section 6.2}. Just as before, $\rm G^{}_F$ is chosen to be $\rm S^{}_4$, under which the transformation properties of the related fields are the same as in Table~\ref{Table:9}.
We recall that the order of ${\rm G}^{}_l$ cannot be smaller than 3 in order to be able to distinguish the three charged leptons, and that of ${\rm G}^{}_i$ is 2.
This means that only the group elements having an order $\geq 3$ and $2$ can be identified as the generators of ${\rm G}^{}_l$ and ${\rm G}^{}_i$, respectively.
For our purpose, ${\rm G}^{}_l$ is taken to be the $\rm Z^{}_3$ symmetry generated by $T$ (denoted as ${\rm Z}^{T}_3$) for which $M^{}_l M^\dagger_l$ is diagonal. On the other hand, the choice of ${\rm G^{}_1} = {\rm Z}^{G}_2$ and ${\rm G^{}_2} = {\rm Z}^{SG}_2$ with $X^{}_1 = X^{}_2 = I$ will fix the VEV alignments of $\phi^{}_i$ to the forms
\begin{eqnarray}
\langle \phi^{}_1 \rangle  = v^{}_1 \pmatrix{ 0 , 1 , 1 }^T \; , \hspace{1cm}
\langle \phi^{}_2 \rangle = v^{}_2 \pmatrix{ 1 , n , n-2 }^T \; ,
\label{eq:6.3.7}
\end{eqnarray}
where $v^{}_i$ and $n$ are real due to the CP symmetry. Then the texture of $M^{}_{\rm D}$ can be directly read out from Eq.~(\ref{eq:6.3.2}) by taking into account such VEV alignments of $\phi^{}_i$. It is straightforward to check that the littlest seesaw model can really be reproduced. Note that there is only one physical phase, $\arg(y^{}_2) - \arg(y^{}_1)$, as a direct consequence of the imposed CP symmetry.

Next, we consider alternative possibilities of ${\rm G}^{}_l$ and ${\rm G}^{}_i$. As a particular example, ${\rm G}^{}_l$, $\rm G^{}_1$ and $\rm G^{}_2$ are chosen to be ${\rm Z}^T_3$, ${\rm Z}^{TST^2}_2$ and ${\rm Z}^{G}_2$, respectively. So ${\rm G}^{}_l = {\rm Z}^T_3$ implies a diagonal $M^{}_l M^\dagger_l$, just like before. As for the neutrino sector, one needs to first figure out the residual CP symmetries compatible with ${\rm Z}^{TST^2}_2$ and ${\rm Z}^{G}_2$ by using Eq.~(\ref{eq:6.3.5}). A detailed analysis shows that $X^{}_1$ can be one of the following eight elements:
\begin{eqnarray}
SG, \hspace{0.2cm} T^2, \hspace{0.2cm} ST^2S, \hspace{0.2cm} T^2STG, \hspace{0.2cm} G, \hspace{0.2cm} ST^2, \hspace{0.2cm} T^2S, \hspace{0.2cm} TST^2G \; .
\label{eq:6.3.8}
\end{eqnarray}
When $X^{}_1$ is one of the first four elements in Eq.~(\ref{eq:6.3.8}), the VEV alignment of $\phi^{}_1$ will be fixed to the form $\langle \phi^{}_1 \rangle = v^{}_1 (1, \omega^2, \omega)^T$. The result of $\langle \phi^{}_1 \rangle$ for $X^{}_1$ being one of the last four elements in Eq.~(\ref{eq:6.3.8}) just differs by an overall factor $\rm i$, which can be absorbed by a redefinition of $y^{}_1$.
On the other hand, $X^{}_2$ can be one of the four elements $I, G, S$ and $SG$. When $X^{}_2$ is one of the first (or last) two of these elements, the VEV alignment of $\phi^{}_2$ will be fixed to the form $\langle \phi^{}_2 \rangle = v^{}_2 (1, x, x)^T$ or to the form $\langle \phi^{}_2 \rangle = v^{}_2 (1 + 2{\rm i}x, 1-{\rm i}x, 1-{\rm i}x)^T$ with $x$ being real due to the CP symmetry. A numerical calculation finds that $\langle \phi^{}_2 \rangle = v^{}_2 (1 + 2{\rm i}x, 1-{\rm i}x, 1-{\rm i}x)^T$ is unable to result in the phenomenologically viable consequences, so we are left only with $\langle \phi^{}_2 \rangle = v^{}_2 (1, x, x)^T$. Given the contraction rule $(\alpha.\beta) = \alpha^{}_1 \beta^{}_1 + \alpha^{}_2 \beta^{}_3 + \alpha^{}_3 \beta^{}_2 $ of two triplet representations of $\rm S^{}_4$ into a singlet representation in the basis of Ref.~\cite{Ding:2013hpa}, $M^{}_{\rm D}$ is obtained as
\begin{eqnarray}
M^{}_{\rm D}= \frac{v}{\Lambda} \pmatrix{
y^{}_1 v^{}_1 & y^{}_2 v^{}_2  \cr
\omega y^{}_1 v^{}_1 & x y^{}_2 v^{}_2  \cr
\omega^2 y^{}_1 v^{}_1 & x y^{}_2 v^{}_2 \cr
} \; .
\label{eq:6.3.9}
\end{eqnarray}
Again, only the phase $\arg(y^{}_2) - \arg(y^{}_1)$ is relevant in physics. Such a highly-constrained form of $M^{}_{\rm D}$ will lead to some testable predictions for the low-energy flavor parameters, as shown in Ref.~\cite{Ding:2018fyz}.
A similar and exhaustive analysis of all the possible patterns of $\rm S^{}_4 \rtimes H^{}_{CP}$ symmetry breaking has been done in Ref.~\cite{Ding:2018tuj}.

\setcounter{equation}{0}
\setcounter{figure}{0}
\section{Some other aspects of the minimal seesaw model}
\label{section 7}

\subsection{Lepton-number-violating processes}
\label{section 7.1}

\begin{figure*}[t]
\centering
\includegraphics[width=5in]{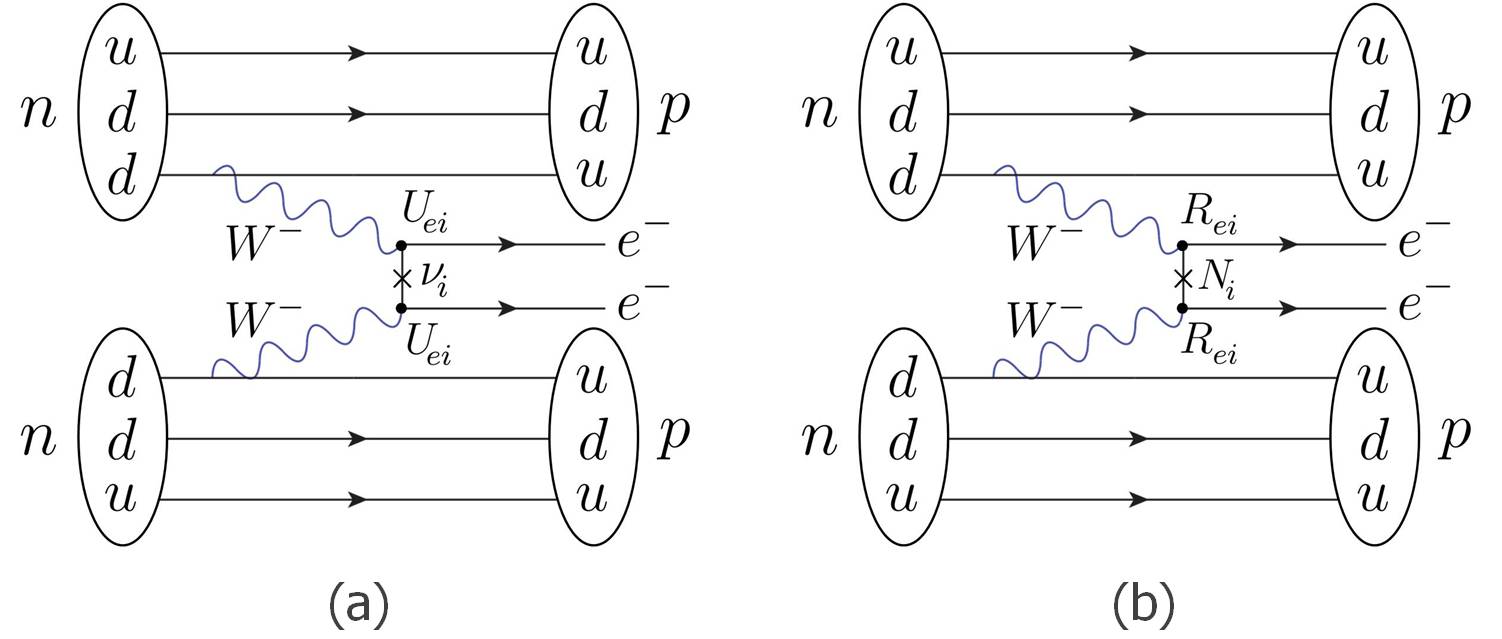}
\caption{The Feynman diagrams for a benchmark $0\nu 2\beta$ decay mediated by (a) three light Majorana neutrinos $\nu^{}_i$ (for $i=1,2,3$) and (b) two heavy Majorana neutrinos $N^{}_i$ (for $i=1,2$) in the minimal seesaw mechanism.}
\label{Fig:7-1}
\end{figure*}

In the minimal seesaw mechanism both three light neutrinos $\nu^{}_i$ (for $i=1,2,3$)
and two heavy neutrinos $N^{}_i$ (for $i=1,2$) are of the Majorana nature. Hence both
of them can mediate the lepton-number-violating $0\nu 2\beta$ decays of some nuclei,
$(A, Z) \to (A, Z+2) + 2 e^-$, where the atomic mass number $A$ and the atomic number $Z$ are both even \cite{Furry:1939qr}. The Feynman diagrams for a benchmark $0\nu 2\beta$ process mediated by
$\nu^{}_i$ and $N^{}_i$ are illustrated in Fig.~\ref{Fig:7-1}(a) and Fig.~\ref{Fig:7-1}(b), respectively. The corresponding amplitudes of these two diagrams are expected to be proportional to $m^{}_i$ (for $q^2 \gg m^2_i$) and $-1/M^{}_i$ (for $q^2 \ll M^2_i$), respectively, where $q \sim 0.1 ~{\rm GeV}$ measures the energy scale or momentum transfer of such a $0\nu 2\beta$ transition \cite{Rodejohann:2011mu}. Therefore, the overall width of a $0\nu 2\beta$ decay in the minimal seesaw scenario can be approximately expressed as \cite{Xing:2009ce}
\begin{eqnarray}
\Gamma^{}_{0\nu 2\beta} & \propto & \left|\sum^3_{i=1} m^{}_i U^2_{e i}
- M^2_A\sum^2_{i=1} \frac{R^2_{e i}}{M^{}_i} {\cal F}(A, M^{}_i)\right|^2
\nonumber \\
& = & \left|\sum^2_{i=1} M^{}_i R^2_{e i} \left[1 + \frac{M^2_A}{M^{2}_i}
{\cal F}(A, M^{}_i)\right] \right|^2 \; ,
\label{eq:7.1.1}
\end{eqnarray}
where $U^{}_{e i}$ and $R^{}_{e i}$ are the corresponding elements of $U$ and $R$ which show up in the weak charged-current interactions of $\nu^{}_i$ and $N^{}_i$ as described by Eq.~(\ref{eq:3.1.9}), $A$ stands for the atomic number of the isotope, ${\cal F}(A, M^{}_i) \simeq 0.1$ is a dimensionless factor depending mildly on the decaying nucleus, and $M^{}_A \sim q \sim 0.1$ GeV \cite{Haxton:1985am,Blennow:2010th}. In obtaining the second equality of
Eq.~(\ref{eq:7.1.1}) we have used the exact seesaw relation $(U D^{}_\nu U^T)^{}_{ee} =
-(R D^{}_N R^T)^{}_{ee}$ shown in Eq.~(\ref{eq:3.1.13}). Since a seesaw model is in general expected to naturally work only when its mass scale is far above the electroweak scale, $M^{}_i \gg M^{}_A$ holds and leads us to the following excellent approximation \cite{Xing:2009ce,Rodejohann:2009ve}:
\begin{eqnarray}
\Gamma^{}_{0\nu 2\beta} \propto \left|\sum^2_{i=1} M^{}_i R^2_{e i}\right| = \left|\sum^3_{i=1} m^{}_i U^2_{e i}\right|^2 \; .
\label{eq:7.1.2}
\end{eqnarray}
In other words, the contribution of $N^{}_i$ to a $0\nu 2\beta$ decay mode must be
negligible in all the reasonable parameter space, unless the contribution of
$\nu^{}_i$ is vanishing or vanishingly small as a result of significant cancellations
among the three different $m^{}_i U^2_{e i}$ components in Eq.~(\ref{eq:7.1.1}) \cite{Xing:2003jf}. This interesting observation implies that the $0\nu 2\beta$ decays
are not {\it directly} sensitive to the heavy degrees of freedom in the seesaw mechanism.
In view of the fact that the PMNS matrix $U$ is not exactly unitary in the presence of
slight flavor mixing between the light and heavy Majorana neutrinos (i.e.,
$UU^\dagger = I - RR^\dagger$), however, we stress that a small impact of $N^{}_i$ on
the $0\nu 2\beta$ transitions is {\it indirectly} reflected by the non-unitarity of $U$
in Eq.~(\ref{eq:7.1.2}).

In the Euler-like parametrization of $U$ and $R$ described in section~\ref{section 3.1}, one can see that possible deviations of $U$ from its exactly unitary limit $U^{}_0$ are measured by $\sin^2\theta^{}_{i4}$ and $\sin^2\theta^{}_{i5}$ (for $i=1,2,3$) which are expected to be at most of ${\cal O}(10^{-2})$ \cite{Antusch:2006vwa,Antusch:2014woa,Blennow:2016jkn}. It is therefore reasonable to assume $U$ to be exactly unitary for the time being. That is to say, the lepton-number-violating $0\nu 2\beta$ decays are not expected to serve as a sensitive playground to test the minimal seesaw mechanism (or the type-I seesaw mechanism in general) in the foreseeable future.

\begin{figure*}[t]
\centering
\includegraphics[width=5.5in]{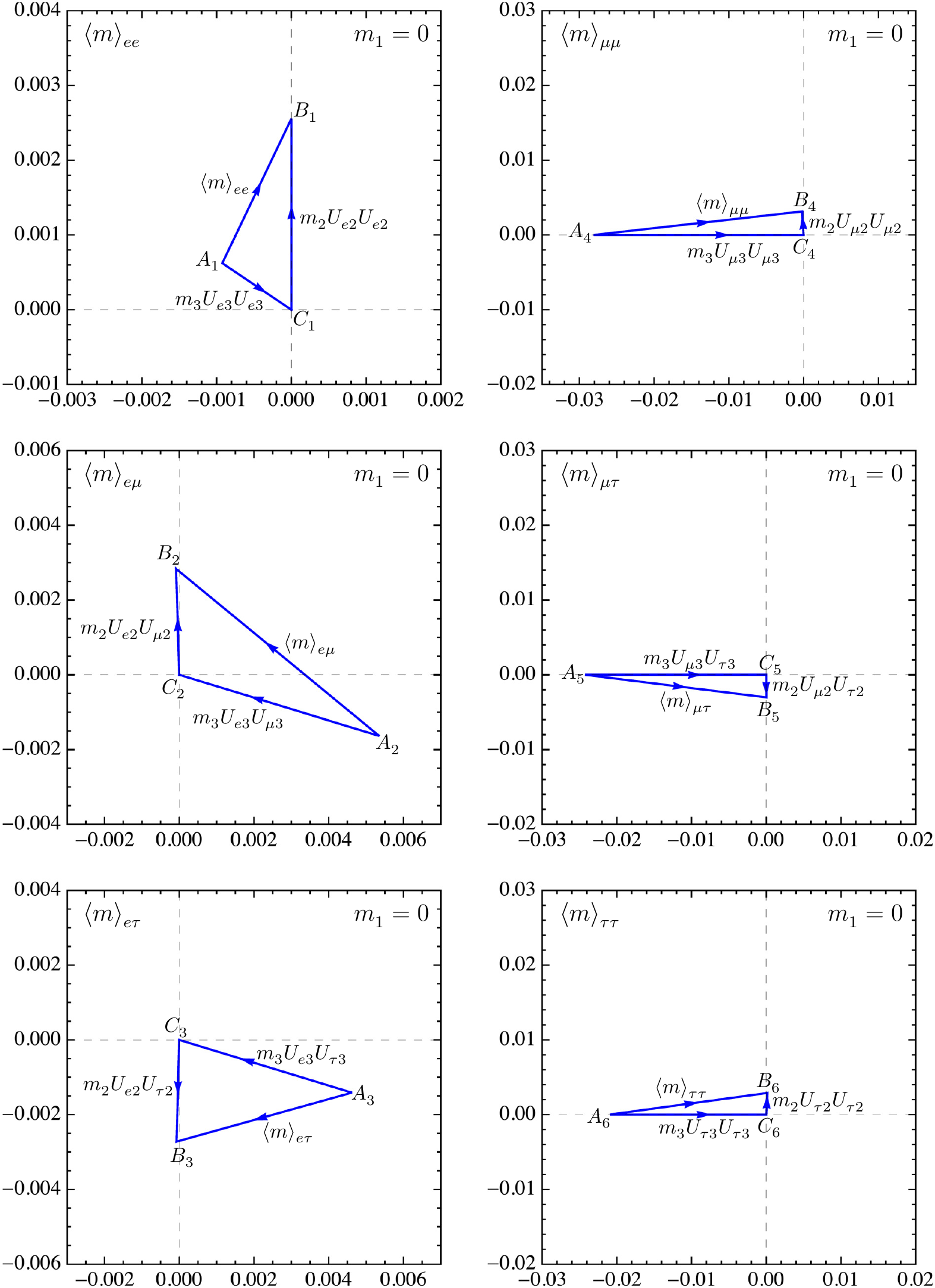}
\caption{Six effective mass triangles $\triangle A^{}_i B^{}_i C^{}_i$ (for $i=1,2, \cdots, 6$) of three light Majorana neutrinos with $m_1 =0$ in the complex plane, plotted by assuming the Majorana CP phase $\sigma = \pi/4$ and inputting the best-fit values of $\Delta m^2_{21}$, $\Delta m^2_{31}$, $\theta^{}_{12}$, $\theta^{}_{13}$, $\theta^{}_{23}$ and $\delta$ in the normal neutrino mass ordering case.}
\label{Fig:7-2}
\end{figure*}

\begin{figure*}[t]
\centering
\includegraphics[width=5.5in]{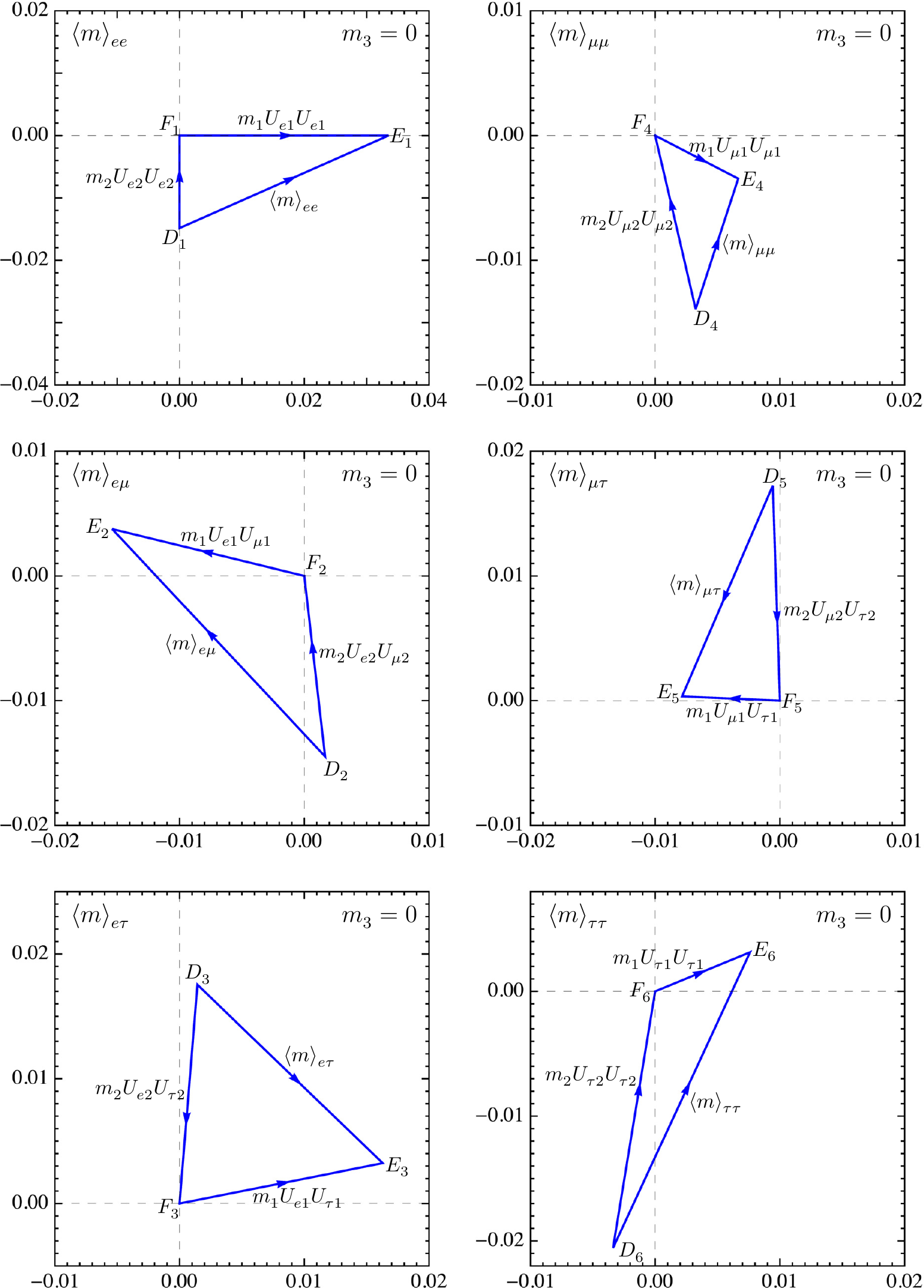}
\caption{Six effective mass triangles $\triangle D^{}_i E^{}_i F^{}_i$ (for $i=1,2, \cdots, 6$) of three light Majorana neutrinos with $m_3 =0$ in the complex plane, plotted by assuming the Majorana CP phase $\sigma = \pi/4$ and inputting the best-fit values of $\Delta m^2_{21}$, $\Delta m^2_{31}$, $\theta^{}_{12}$, $\theta^{}_{13}$, $\theta^{}_{23}$ and $\delta$ in the inverted neutrino mass ordering case.}
\label{Fig:7-3}
\end{figure*}

As a straightforward consequence of the minimal seesaw mechanism, the effective Majorana neutrino mass matrix elements $\langle m\rangle^{}_{\alpha\beta}$ defined in Eq.~(\ref{eq:2.4}) can be simplified to six effective mass triangles in the complex plane \cite{Xing:2015uqa,Xing:2015wzz}:
\begin{eqnarray}
m^{}_1 = 0: \hspace{0.5cm} \langle m\rangle^{}_{\alpha\beta} = \sqrt{\Delta m^2_{21}} \hspace{0.08cm} U^{}_{\alpha 2} U^{}_{\beta 2} + \sqrt{\Delta m^2_{31}} \hspace{0.08cm} U^{}_{\alpha 3} U^{}_{\beta 3} \;;
\nonumber \\
m^{}_3 = 0: \hspace{0.5cm} \langle m\rangle^{}_{\alpha\beta} = \sqrt{\left|\Delta m^2_{31}\right|} \hspace{0.08cm} U^{}_{\alpha 1} U^{}_{\beta 1} + \sqrt{\left|\Delta m^2_{32}\right|} \hspace{0.08cm} U^{}_{\alpha 2} U^{}_{\beta 2} \;,
\label{eq:7.1.3}
\end{eqnarray}
where $\alpha$ and $\beta$ run over $e$, $\mu$ and $\tau$. The size and shape of each triangle are illustrated in Fig.~\ref{Fig:7-2} with $m^{}_1 =0$ or Fig.~\ref{Fig:7-3} with $m^{}_3 =0$, where the best-fit values of two neutrino mass-squared differences, three neutrino mixing angles and the Dirac CP phase have been taken as the typical inputs \cite{Esteban:2020cvm}. If such effective mass triangles are finally established from the measurements of some lepton-number-violating processes, it will be possible to determine or constrain the Majorana CP phase $\sigma$ or its combination with the Dirac CP phase $\delta$.

Even if the $0\nu 2\beta$ decays are experimentally observed in the future, one can only get some information on the effective Majorana neutrino mass term $|\langle m\rangle^{}_{ee}|$. To determine or constrain the other five effective mass terms (i.e., $|\langle m\rangle^{}_{e\mu}|$, $|\langle m\rangle^{}_{e\tau}|$, $|\langle m\rangle^{}_{\mu\mu}|$, $|\langle m\rangle^{}_{\mu\tau}|$ and $|\langle m\rangle^{}_{\tau\tau}|$), whose small magnitudes have been shown by Fig.~\ref{Fig:5-1} in the minimal seesaw framework, we have to explore other relevant lepton-number-violating processes which are certainly much more challenging.
\begin{figure*}[t]
\centering
\includegraphics[width=3.6in]{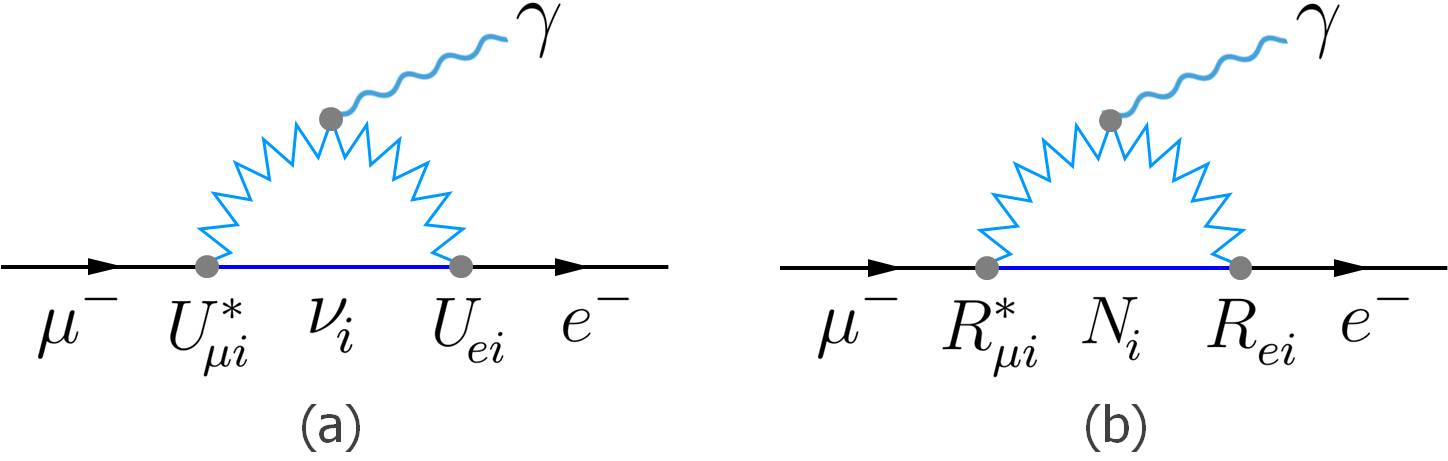}
\caption{The one-loop Feynman diagrams for the radiative $\mu^- \to e^- + \gamma$ decay mediated by (a) three light Majorana neutrinos $\nu^{}_i$ (for $i=1,2,3$) and (b) two heavy Majorana neutrinos $N^{}_i$ (for $i=1,2$) in the minimal seesaw mechanism.}
\label{Fig:7-4}
\end{figure*}

\subsection{Lepton-flavor-violating processes}
\label{section 7.2}

Needless to say, the lepton-flavor-violating decay modes of charged leptons can also be mediated by both the light Majorana neutrinos $\nu^{}_i$ and the heavy Majorana neutrinos $N^{}_i$ in the minimal seesaw mechanism. Fig.~\ref{Fig:7-4} illustrates the one-loop Feynman diagrams for the radiative $\mu^- \to e^- + \gamma$ transition of this kind, where the elements of $U$ and $R$ matrices determine the strengths of weak charged-current interactions of light and heavy Majorana neutrinos with charged leptons, respectively. The rates of radiative $\alpha^- \to \beta^- + \gamma$ decays (for $\alpha, \beta = e, \mu, \tau$ and $m^{}_\alpha > m^{}_\beta$) against those of the SM-allowed $\alpha^- \to \beta^- + \nu^{}_\alpha + \overline{\nu}^{}_\beta$ decays can be approximately expressed as
\begin{eqnarray}
\frac{\Gamma (\alpha^- \to \beta^- + \gamma)}{\Gamma (\alpha^- \to \beta^- + \nu^{}_\alpha + \overline{\nu}^{}_\beta)} & \simeq &
\frac{3 \alpha^{}_{\rm em}}{2 \pi} \left|\sum^3_{i=1} U^*_{\alpha i}
U^{}_{\beta i} G^{}_\gamma(x^{}_i) + \sum^2_{i=1} R^*_{\alpha i} R^{}_{\beta i}
G^{}_\gamma(x^{\prime}_i)\right|^2 \; ,
\nonumber \\
& \simeq & \frac{3 \alpha^{}_{\rm em}} {32 \pi} \left|\sum^3_{i=1} U^*_{\alpha i} U^{}_{\beta i}
\frac{m^2_i}{M^2_W} + 2\sum^2_{i=1} R^*_{\alpha i} R^{}_{\beta i} \right|^2 \; ,
\label{eq:7.2.1}
\end{eqnarray}
where $\alpha^{}_{\rm em} \simeq 1/137$ denotes the fine-structure constant of quantum electrodynamics, $x^{}_i \equiv m^2_i/M^2_W \ll 1$ and $x^\prime_i \equiv M^2_i/M^2_W
\gg 1$ (for $i = 1,2,3$) have been taken into account for a {\it natural} version of the minimal seesaw mechanism, and $G^{}_\gamma (x)$ is the loop function which approaches $x/4$ for $x\ll 1$ or $1/2$ for $x \gg 1$ \cite{Ilakovac:1994kj,Alonso:2012ji,Lindner:2016bgg}. Given the exact correlation
$RR^\dagger = I - UU^\dagger$ between $U$ and $R$ below Eq.~(\ref{eq:3.1.9}), one may simplify Eq.~(\ref{eq:7.2.1}) to the following expression:
\begin{eqnarray}
\frac{\Gamma (\alpha^- \to \beta^- + \gamma)}{\Gamma (\alpha^- \to \beta^- + \nu^{}_\alpha + \overline{\nu}^{}_\beta)} & \simeq & \frac{3 \alpha^{}_{\rm em}} {32 \pi} \left|\sum^3_{i=1} U^*_{\alpha i} U^{}_{\beta i} \left(\frac{m^2_i}{M^2_W} - 2\right) \right|^2
\nonumber \\
& \simeq & \frac{3 \alpha^{}_{\rm em}} {8 \pi} \left|\sum^3_{i=1} U^*_{\alpha i} U^{}_{\beta i} \right|^2 = \frac{3 \alpha^{}_{\rm em}} {8 \pi} \left|\sum^2_{i=1} R^*_{\alpha i} R^{}_{\beta i} \right|^2 \;
\label{eq:7.2.2}
\end{eqnarray}
with the terms proportional to $m^2_i/M^2_W$ being safely neglected in the second row of Eq.~(\ref{eq:7.2.2}). Note that in the absence of heavy degrees of freedom the neglected $m^2_i/M^2_W$ terms are actually the standard contributions of three light Majorana neutrinos to radiative $\alpha^- \to \beta^- + \gamma$ decays \cite{Petcov:1976ff,Bilenky:1977du,Cheng:1976uq,Marciano:1977wx,Lee:1977qz,Lee:1977tib,Bernstein:2013hba,
Albrecht:2013wet}, but they are so suppressed by $\Delta m^2_{21}/M^2_W \sim {\cal O}(10^{-26})$ and $\Delta m^2_{31}/M^2_W \sim {\cal O}(10^{-25})$ that it is hopeless to measure such rare lepton-flavor-violating processes in any realistic experiments (e.g., the branching ratio of $\mu^- \to e^- + \gamma$ is expected to be of ${\cal O}(10^{-54})$ or smaller as constrained by current neutrino oscillation data \cite{Xing:2019vks}).

If the contributions of two heavy Majorana neutrinos to radiative $\alpha^- \to \beta^- + \gamma$ decays
are dominant, Eq.~(\ref{eq:7.2.2}) tells us that their magnitudes are measured by
\begin{eqnarray}
\left|R^*_{\alpha 1} R^{}_{\beta 1} + R^*_{\alpha 2} R^{}_{\beta 2}\right| \simeq
\left|\hat{s}_{i4} \hat{s}^*_{j4} + \hat{s}^{}_{i5} \hat{s}^*_{j5}\right| \lesssim {\cal O}(10^{-2}) \; ,
\label{eq:7.2.3}
\end{eqnarray}
where the Euler-like parametrization of $R$ in Eq.~(\ref{eq:3.1.10}) has been used (for $i, j = 1, 2, 3$ and $i \neq j$). As a result, the ratio of $\Gamma (\alpha^- \to \beta^- + \gamma)$ to $\Gamma (\alpha^- \to \beta^- + \nu^{}_\alpha + \overline{\nu}^{}_\beta)$ is naively expected to be of ${\cal O}(10^{-8})$ or smaller. In practice, one usually follows the opposite way to model-independently constrain
the unknown active-sterile neutrino mixing angles $\theta^{}_{i4}$ and $\theta^{}_{i5}$ or their combinations (for $i=1,2,3$) from current experimental data on the upper bounds of $\alpha^- \to \beta^- + \gamma$ decay modes \cite{Antusch:2006vwa}. Given \cite{Tanabashi:2018oca}
\begin{eqnarray}
\frac{\Gamma (\mu^- \to e^- + \gamma)}{\Gamma (\mu^- \to e^- + \nu^{}_\mu + \overline{\nu}^{}_e)} < 4.2 \times 10^{-13} \; ,
\nonumber \\
\frac{\Gamma (\tau^- \to e^- + \gamma)}{\Gamma (\tau^- \to e^- + \nu^{}_\tau + \overline{\nu}^{}_e)} < 1.9 \times 10^{-7} \; ,
\nonumber \\
\frac{\Gamma (\tau^- \to \mu^- + \gamma)}{\Gamma (\tau^- \to \mu^- + \nu^{}_\tau + \overline{\nu}^{}_\mu)} < 2.5 \times 10^{-7} \; ,
\label{eq:7.2.4}
\end{eqnarray}
at the $90\%$ confidence level, one may make use of Eqs.~(\ref{eq:7.2.2}) and (\ref{eq:7.2.3}) to get the following preliminary constraints:
\begin{eqnarray}
\left|R^*_{\mu 1} R^{}_{e 1} + R^*_{\mu 2} R^{}_{e 2}\right| \simeq
\left|\hat{s}_{24} \hat{s}^*_{14} + \hat{s}^{}_{25} \hat{s}^*_{15}\right|
< 2.2 \times 10^{-5} \; ,
\nonumber \\
\left|R^*_{\tau 1} R^{}_{e 1} + R^*_{\tau 2} R^{}_{e 2}\right| \simeq
\left|\hat{s}_{34} \hat{s}^*_{14} + \hat{s}^{}_{35} \hat{s}^*_{15}\right|
< 1.5 \times 10^{-2} \; ,
\nonumber \\
\left|R^*_{\tau 1} R^{}_{\mu 1} + R^*_{\tau 2} R^{}_{\mu 2}\right| \simeq
\left|\hat{s}_{34} \hat{s}^*_{24} + \hat{s}^{}_{35} \hat{s}^*_{25}\right|
< 1.7 \times 10^{-2} \; .
\label{eq:7.2.5}
\end{eqnarray}
Because of the phase parameters are entangled with the active-sterile neutrino mixing angles in the above combinations, it is still impossible to constrain any of the individual parameters even in the minimal seesaw scheme.

To enhance the rates of rare $\alpha^- \to \beta^- + \gamma$ decays to a level close to the
present experimental sensitivity, one may consider to give up the assumption $M^2_{1,2} \gg M^2_W$ by lowering the conventional minimal seesaw scale down to
the TeV regime or even lower such that the active-sterile neutrino mixing angles
$\theta^{}_{ij}$ (for $i = 1,2,3$ and $j = 4,5$) of $R$ can be as large
as possible \cite{Antusch:2006vwa,Ilakovac:1994kj,
Alonso:2012ji,Abada:2007ux,Deppisch:2010fr,Ibarra:2011xn}. If one prefers to go beyond the SM
framework by incorporating the minimal seesaw scenarios with some supersymmetric
models, for instance, it will certainly be possible to achieve much richer
phenomenology of lepton flavor violation in the charged-lepton sector
\cite{Raidal:2002xf,Ibarra:2003up,Dutta:2003ps,Cao:2003zv}. Since this aspect has been well  reviewed in Refs.~\cite{Lindner:2016bgg,Guo:2006qa}, here we shall not go into details.

\subsection{Low-scale seesaw models}
\label{section 7.3}

In this subsection we give some brief discussions about the low-scale seesaw models. To explain the motivation for considering this kind of models, let us recall the reasoning for considering the conventional seesaw models at a superhigh energy scale: if the Yukawa couplings $(Y^{}_{\nu})^{}_{\alpha i}$ between the left- and right-handed neutrinos take seemingly natural ${\cal{O}}(1)$ values, then the sub-eV light neutrino masses will be achieved in correspondence to the ${\cal{O}}(10^{14})$ GeV right-handed neutrino masses via the seesaw formula. But this might just be a prejudice: even within the SM, the Yukawa couplings of different fermions span many orders of magnitude, from ${\cal{O}}(10^{-6})$ (the electron) to $\simeq 1$ (the top quark). If the Yukawa couplings of the neutrinos are somewhat comparable with that of the electron, then the seesaw mechanism will allow right-handed neutrinos with just TeV-scale masses to generate the sub-eV light neutrino masses. In this sense a TeV-scale (e.g., the left-right-symmetric models \cite{Pati:1974yy, Mohapatra:1974hk, Senjanovic:1975rk, Wyler:1982dd}, see also \cite{Davidson:1987mh}) or GeV-scale seesaw model is absolutely acceptable, and even the eV-scale seesaw models are not impossible \cite{Branco:2019avf}.
Here let us focus on the minimal version of such models, which contains only two right-handed neutrinos. If one further requires that the model be also viable for leptogenesis, then the two right-handed neutrinos must be nearly degenerate in their masses so that a resonant amplification can be achieved. It is worth noting that such a mass degeneracy will not be necessary any more if more than two right-handed neutrinos are responsible for leptogenesis \cite{Drewes:2012ma}. All in all, here we shall pay our attention to the low-scale seesaw models with two nearly degenerate right-handed neutrinos.

From the experimental point of view, the most attractive feature of those low-scale seesaw models is that their new degrees of freedom are likely to be probed in the laboratory \cite{Shrock:1980ct, Shrock:1981wq, Langacker:1988ur}. Once kinematically allowed, the right-handed neutrinos participate in any processes as the left-handed neutrinos do but their amplitudes are suppressed by the mixing factors $R^{}_{\alpha i} = (Y^{}_{\nu})^{}_{\alpha i} v/M^{}_i$, where $R^{}_{\alpha i}$ are the corresponding elements of $R$ showing up in the weak charged-current interactions of $N^{}_i$ as described by Eq.~(\ref{eq:3.1.9}). If the small mass splitting $\Delta M \equiv M^{}_2 - M^{}_1$ is beyond the experimental resolution capabilities, a case in most of the leptogenesis parameter space, then the flavor-dependent (or flavor-independent) processes are only sensitive to $R^{2}_\alpha = \sum^{}_i |R^{}_{\alpha i}|^2$ (or $R^{2} = \sum^{}_\alpha R^{2}_{\alpha }$) instead of individual $|R^{}_{\alpha i}|^2$ (or $R^{2}_i = \sum^{}_\alpha |R^{}_{\alpha i}|^2$).
If there are no strong cancellations between the contributions of two right-handed neutrinos to $M^{}_\nu$, a naive seesaw expectation gives $R^2 \sim  \sqrt{|\Delta m^2_{31}|} /M^{}_0$ with $M^{}_0 \equiv (M^{}_1 + M^{}_2)/2$, suggesting that $R^2$ increase with the lowering of $M^{}_0$.
Unfortunately, even for the GeV-scale values of $M^{}_0$, $R^2$ remains too small to be accessible in most of the realistic future experiments. Much larger values of $R^2$, which are experimentally accessible but still consistent with small light neutrino masses, can only be achieved in the presence of strong cancellations that keep the total contribution of two right-handed neutrinos to $M^{}_\nu$ small enough in spite of large individual Yukawa couplings. Such strong cancellations will be a natural consequence if an approximate lepton number conservation is invoked, such as in the model illustrated by Eq.~(\ref{eq:4.4.3}): in the limit of lepton number conservation, the two right-handed neutrinos are exactly degenerate in their masses and the light neutrinos remain massless; when the lepton number is approximately conserved, the non-zero but tiny $\Delta M$ and light neutrino masses are to be protected \cite{Branco:1988ex, Shaposhnikov:2006nn, Kersten:2007vk}.
Two popular classes of specific models that implement such a scenario are the inverse seesaw models \cite{GonzalezGarcia:1988rw, Mohapatra:1986aw, Mohapatra:1986bd, Das:2012ze} and the linear seesaw models \cite{Akhmedov:1995vm, Akhmedov:1995ip, Barr:2003nn, Malinsky:2005bi}. Phenomenologically, the most appealing low-scale seesaw model seems to be the so-called  neutrino minimal Standard Model ($\nu$MSM) \cite{Asaka:2005pn, Asaka:2005an}, in which two GeV-scale nearly-degenerate right-handed neutrinos are responsible for both the generation of light neutrino masses and the baryon asymmetry of the Universe (effectively a minimal seesaw model), and one keV-scale right-handed neutrino serves as the candidate for warm dark matter \cite{Adhikari:2016bei, Boyarsky:2018tvu}.

\begin{figure*}[t]
\centering
\includegraphics[width=6in]{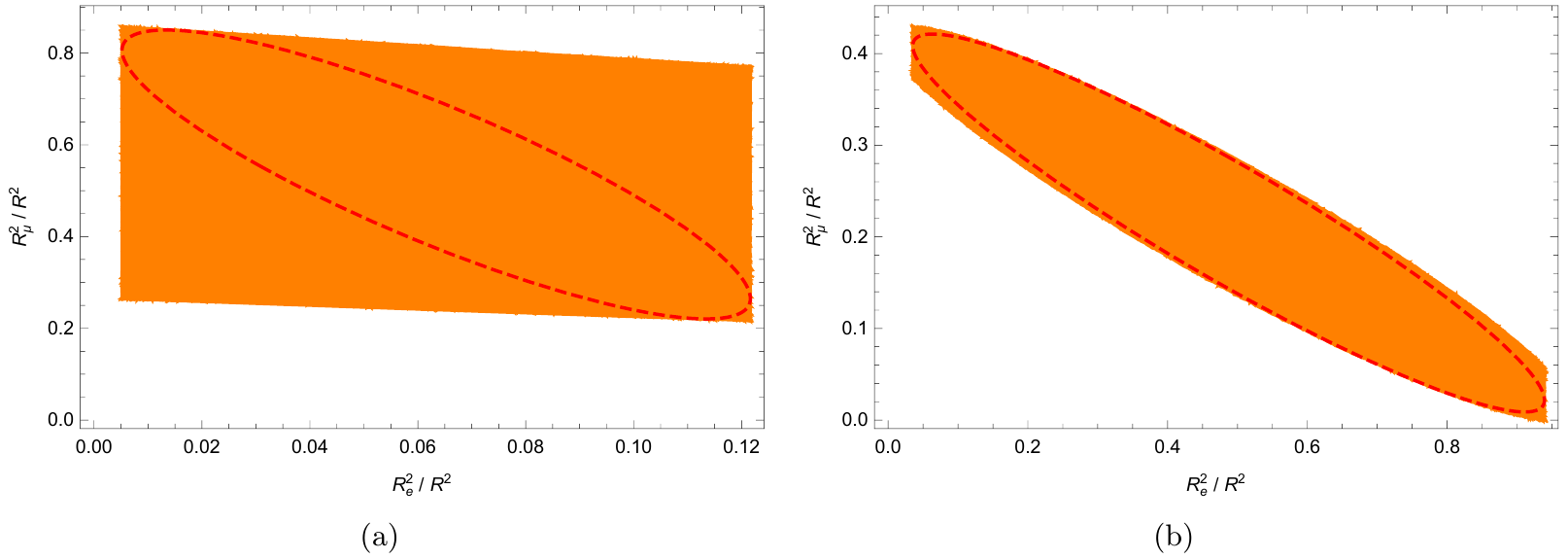}
\caption{The allowed values of $R^2_\mu/R^2$ versus $R^2_e/R^2$ in the $m^{}_1 =0$ (a) and $m^{}_3 =0$ (b) cases. In obtaining these results, we have taken the best-fit values for the neutrino mass-squared differences and the neutrino mixing angles, and allow $\sigma$ to vary from $0$ to $2\pi$. The shaded regions and dashed lines correspond to the $3\sigma$ range and best-fit value of $\delta$, respectively. }
\label{Fig:7-5}
\end{figure*}

Without going into detail about a concrete model, one can generically parameterize the Yukawa couplings in the Casas-Ibarra form which automatically guarantees a viable reproduction of the observed values of the low-energy neutrino parameters. In such a parametrization, the level of cancellations is measured by the imaginary part of $z$. The case of strong cancellations (great enhancements of the Yukawa couplings) corresponds to ${\rm Im}(z) \gg 1$. In this case, one has
\begin{eqnarray}
m^{}_1 = 0: \hspace{1cm} O \simeq \frac{1}{2}  e^{{\rm Im}(z)} e^{-{\rm i Re}(z) }  \pmatrix{ 0 & 0 \cr 1 & -  {\rm i} \cr {\rm i} &  1 } \;;  \nonumber \\
m^{}_3 = 0: \hspace{1cm} O \simeq \frac{1}{2}  e^{{\rm Im}(z)} e^{-{\rm i Re}(z) } \pmatrix{ 1 & -  {\rm i} \cr {\rm i} &  1 \cr 0 & 0 } \;,
\label{eq:7.3.1}
\end{eqnarray}
which leads respectively to
\begin{eqnarray}
R^2_{\alpha} = \frac{1}{2 M^{}_0} e^{2{\rm Im}(z)} | \sqrt{m^{}_i} U^{}_{\alpha i} + {\rm i} \sqrt{m^{}_j} U^{}_{\alpha j} |^2 \; , \nonumber \\
R^2 = \frac{1}{2 M^{}_0} e^{2{\rm Im}(z)} (m^{}_i + m^{}_j) \; ,
\label{eq:7.3.2}
\end{eqnarray}
with $i =2$ and $j=3$ (or $i =1$ and $j=2$) for the $m^{}_1 =0$ (or $m^{}_3 =0$) case.
One can see that the measurements of $R^2_{\alpha}$ will allow us to fix $M^{}_0$ (which can also be determined kinematically), ${\rm Im}(z)$ and the Majorana CP phase $\sigma$.
And the relative sizes of $R^2_{\alpha}$ among three flavors are determined by the low-energy neutrino observables alone \cite{Shaposhnikov:2008pf, Asaka:2011pb, Ruchayskiy:2011aa, Hernandez:2016kel}. In Fig.~\ref{Fig:7-5} we plot the allowed values of $R^2_\mu/R^2$ versus $R^2_e/R^2$ in the $m^{}_1 =0$ and $m^{}_3 =0$ cases, while the value of $R^2_\tau/R^2$ is fixed by the definition $R^2 = \sum^{}_\alpha R^2_{\alpha}$.
The main features of these results can be understood from the following analytical approximations.
For the $m^{}_1 =0$ case, with the neglect of $m^{}_2$ (which is actually a bad approximation), $R^2_{\alpha}$ approximate to
\begin{eqnarray}
&& R^2_{e} \simeq \frac{\sqrt{|\Delta m^2_{31}|} }{2 M^{}_0} e^{2{\rm Im}(z)} s^2_{13} \; , \nonumber \\
&& R^2_{\mu} \simeq  \frac{\sqrt{|\Delta m^2_{31}|} }{2 M^{}_0} e^{2{\rm Im}(z)} c^2_{13} s^2_{23} \;, \hspace{1cm} R^2_{\tau} \simeq  \frac{\sqrt{|\Delta m^2_{31}|} }{2 M^{}_0} e^{2{\rm Im}(z)} c^2_{13} c^2_{23} \;.
\label{eq:7.3.3}
\end{eqnarray}
It is obvious that $R^2_e$ is suppressed with respect to $R^2_\mu$ and $R^2_\tau$ due to the smallness of $\theta^{}_{13}$, and $R^2_\mu \sim R^2_\tau$ due to the closeness of $\theta^{}_{23}$ to $\pi/4$.
For the $m^{}_3 =0$ case, with the neglect of $\theta^{}_{13}$ and the deviation of $\theta^{}_{23}$ from $\pi/4$, $R^2_{\alpha}$ are approximately given by
\begin{eqnarray}
&& R^2_{e} \simeq \frac{\sqrt{|\Delta m^2_{31}|} }{2 M^{}_0} e^{2{\rm Im}(z)} (1 - \sin 2 \theta^{}_{12} \sin \sigma) \; , \nonumber \\
&& R^2_{\mu} \simeq R^2_{\tau} \simeq \frac{\sqrt{|\Delta m^2_{31}|} }{4 M^{}_0} e^{2{\rm Im}(z)} (1 + \sin 2 \theta^{}_{12} \sin \sigma) \;.
\label{eq:7.3.4}
\end{eqnarray}
In this case $R^2_e$ has a chance (for $\sigma \simeq -\pi/2$) to be much larger than $R^2_\mu$ and $R^2_\tau$.

The properties of the right-handed neutrinos are subject to constraints from many aspects, including direct searches \cite{Atre:2009rg, Deppisch:2015qwa, Cai:2017mow, Das:2017nvm}, indirect searches \cite{Gorbunov:2014ypa, Antusch:2014woa, Lindner:2016bgg, Fernandez-Martinez:2016lgt, Chrzaszcz:2019inj} and cosmological considerations \cite{Boyarsky:2009ix, Hernandez:2014fha} (for a summary, see Refs.~\cite{Canetti:2012kh, Drewes:2013gca, Drewes:2015iva, Drewes:2016jae}). Two distinct strategies can be employed for direct searches of the right-handed neutrinos. The first one is related to their production: for the mass ranges below the $K$-meson masses, between the $K$- and $D$-meson masses and between the $D$- and $B$-meson masses, they can be searched for in the $K$, $D$ and $B$ decays, respectively; for the mass range above the $B$-meson masses, they can only be produced in high energy colliders \cite{Gorbunov:2007ak, Bondarenko:2018ptm, Das:2015toa}. The second strategy is to
search for their decays inside a detector. Finally, these two strategies can be combined if their production and decays occur inside the same detector.
The negative results obtained in the past direct searches in the fixed-target experiments and colliders place some upper bounds on $|R^{2}_{\alpha i}|$ as functions of $M^{}_i$. It is noteworthy that these negative results have ruled out the mass range below about 100 MeV, when combined with the requirement that the right-handed neutrinos should have a lifetime shorter than about 0.1 second in order not to spoil the success of the Big Bang nucleosynthesis \cite{Gorbunov:2007ak, Hernandez:2014fha}.

The properties of the right-handed neutrinos can also be constrained by indirect searches, where they (as virtual particles) affect some observables or the rates of some processes such as the lepton-number-violating and lepton-flavor-violating ones discussed in last two subsections.
In particular, the rate of the $0\nu 2\beta$ decay is modified to \cite{Drewes:2016lqo, Asaka:2016zib}
\begin{eqnarray}
\Gamma^{}_{0\nu 2\beta} & \propto & \left| \sum^3_{i=1} m^{}_i U^2_{e i}
+ \sum^2_{i=1} M^{}_i R^2_{e i} f^{}_A(M^{}_i)\right|^2
\nonumber \\
& = & \left| \left[ 1 - f^{}_A (M^{}_0 ) \right] \sum^3_{i=1} m^{}_i U^2_{e i}
+  \sum^2_{i=1} M^{}_i R^2_{e i} \left[ f^{}_A(M^{}_i) -  f^{}_A (M^{}_0 ) \right] \right|^2 \nonumber \\
& = & \left| \left[ 1 - f^{}_A (M^{}_0 ) \right] \sum^3_{i=1} m^{}_i U^2_{e i}
+ f^{2}_A (M^{}_0 ) \frac{M^2_0 }{\Lambda^2} \Delta M  \left( R^2_{e 1} -  R^2_{e 2} \right) \right|^2 \; ,
\label{eq:7.3.5}
\end{eqnarray}
where the approximation
\begin{eqnarray}
f^{}_A (M ) \simeq  \frac{\Lambda^2 }{\Lambda^2 + M^2} \;
\label{eq:7.3.6}
\end{eqnarray}
with $\Lambda^2 = (159 \ {\rm MeV})^2$ in the Argonne model \cite{Faessler:2014kka}
and the exact seesaw relation $(U D^{}_\nu U^T)^{}_{ee} =
-(R D^{}_N R^T)^{}_{ee}$ shown in Eq.~(\ref{eq:3.1.13}) have been used. By using the Casas-Ibarra parametrization for ${\rm Im}(z) \gg 1$, $R^2_{e 1} -  R^2_{e 2}$ in Eq.~(\ref{eq:7.3.5}) can be recast as
\begin{eqnarray}
\hspace{-1.3cm} m^{}_1 = 0: \hspace{1cm} R^2_{e 1} -  R^2_{e 2} \simeq  \frac{\sqrt{|\Delta m^2_{31}|}}{2 M^{}_0}  s^2_{13} e^{2{\rm Im}(z)} e^{-2{\rm i} [{\rm Re}(z) + \delta] } \;;  \nonumber \\
\hspace{-1.3cm} m^{}_3 = 0: \hspace{1cm} R^2_{e 1} -  R^2_{e 2} \simeq - \frac{\sqrt{|\Delta m^2_{31}|}}{2 M^{}_0}  c^2_{13} e^{2{\rm Im}(z)} e^{-2{\rm i} {\rm Re}(z) } \left( c^{}_{12} + {\rm i} s^{}_{12} e^{ {\rm i} \sigma } \right)^2 \;.
\label{eq:7.3.7}
\end{eqnarray}
For comparably small $M^{}_0$ and comparably large $\Delta M$, the contributions from the right-handed neutrinos may become significant. For example, it is found that $M^{}_0 \sim 1$ GeV and $\Delta M > 10^{-4}$ GeV can enhance the rate of the $0 \nu 2 \beta$ decay while allowing for a viable leptogenesis \cite{Drewes:2016lqo, Asaka:2016zib, Hernandez:2016kel}. This allows us
to constrain $M^{}_0$, $\Delta M$, ${\rm Re}(z)$, ${\rm Im}(z)$ and $\sigma$.

The constraints from cosmological considerations include the aforementioned requirements for a viable leptogenesis and avoidance of spoiling the success of the Big Bang nucleosynthesis.
It should be noted that when the right-handed neutrinos have some masses below $\simeq 130$ GeV where the sphaleron processes become decoupled \cite{DOnofrio:2014rug}, the baryon asymmetry of the Universe is not generated in their decays but via the CP-violating oscillations during their production \cite{Akhmedov:1998qx, Asaka:2005an}: coherent pairs of right-handed neutrinos are constantly created from the thermal bath of the early Universe and oscillate in a CP-violating manner. As a result, both the left-handed and right-handed sectors develop some lepton asymmetries, which are of nearly equal amounts but with opposite signs, keeping the total lepton asymmetry vanishingly small. The latter fact is due to that the right-handed neutrino masses (which act as the source of lepton number violation) are small with respect to the temperature at that time. Since the sphaleron processes only act on the left-handed sector, a sufficient amount of baryon asymmetry can be generated in spite of the vanishingly small total lepton asymmetry. A detailed analysis of this mechanism has been performed in Refs.~\cite{Canetti:2012kh, Canetti:2014dka}, where it is found that leptogenesis can only be viable when the right-handed neutrino masses are degenerate at a level of $\Delta M/M^{}_0 \lesssim 10^{-3}$, and the corresponding upper and lower bounds on $R^2_i$ are identified. Here the upper bounds on $R^2_i$ (which set a benchmark goal for the sensitivities of direct-search experiments) arise because if they are too large, the right-handed neutrinos will enter in thermal equilibrium above the electroweak scale, suppressing the production of a baryon asymmetry.

With the help of the above constraints on the properties of the right-handed neutrinos, if any neutral lepton of this kind and this weight is discovered in future experiments, then one may evaluate whether it can indeed be responsible for the generations of light neutrino masses and the baryon asymmetry of the Universe \cite{Agrawal:2021dbo}.

\section{Concluding remarks}
\label{section 8}

Although the SM is extraordinarily successful in describing all the known fundamental particles and their interactions, it does not shed any light on the origin of neutrino masses even in a qualitative way, nor it provides any quantitative information on flavor mixing and CP violation. Therefore, we are well motivated to go beyond the SM by exploring all the possible ways of neutrino mass generation and lepton flavor structures. Among various ideas that have so far been proposed, the canonical seesaw mechanism remains most popular for its three salient features: (1) it allows massive neutrinos to have the Majorana nature which may make a great impact on many lepton-number-violating processes in nuclear physics, particle physics and cosmology; (2) it offers a very natural explanation of why the three active neutrinos are so light as compared with their charged counterparts; and (3) it can naturally account for the observed baryon number  asymmetry of the Universe via the thermal leptogenesis mechanism. But all the merits of the seesaw and leptogenesis mechanisms are qualitative, and hence they fail in making any quantitative predictions that can be experimentally tested.

That is why we have recurred to the use of Occam's razor to cut one species of the heavy Majorana neutrinos in the conventional seesaw picture and arrive at its simplified version --- the minimal seesaw scenario. The smoking gun of such a simplified seesaw mechanism is its two striking predictions: (a) the smallest neutrino mass $m^{}_1$ (or $m^{}_3$) is vanishing at the tree and one-loop levels, and it is vanishingly small and thus completely negligible even after quantum corrections are taken into account at the two-loop level; (b) one of the two Majorana CP phases accordingly loses its physical meaning, and hence it will have little impact on those lepton-number-violating processes.

Given its briefness and predictability, the minimal seesaw mechanism has been studied in depth and from many perspectives in the past twenty years. In this article we have made an up-to-date review of various phenomenological aspects of this simple but instructive seesaw picture and its associated leptogenesis mechanism in neutrino physics and cosmology. Our real interest has been in possible flavor structures of such benchmark seesaw and leptogenesis scenarios and in confronting their predictions with current neutrino oscillation data and cosmological observations. We have paid particular attention to the topics of lepton number violation, lepton
flavor violation, discrete flavor symmetries, CP violation and antimatter of the Universe.

Can one experimentally verify the (minimal) leptogenesis mechanism associated with the (minimal) seesaw mechanism under discussion? An immediate answer is negative, unfortunately. More accurate experimental data to be accumulated in the foreseeable future may only allow us to exclude some specific seesaw-plus-leptogenesis scenarios of this kind by examining their quantitative consequences at low energies, but it is almost impossible to single out a unique model due to the lack of experimentally accessible observables. Nevertheless, in this connection one should not be excessively pessimistic either. As argued by Hitoshi Murayama, one will probably believe thermal leptogenesis to be the correct solution to the puzzle of why primordial antimatter has disappeared in the Universe if the following ``archaeological" evidence can be finally collected \cite{Murayama:2002jq}: (1) the electroweak baryogenesis mechanism is definitely ruled out; (2) the Majorana nature of massive neutrinos is established via the $0\nu 2\beta$ decays and (or) other lepton-number-violating processes; and (3) leptonic CP violation is convincingly observed in the next-generation long-baseline neutrino oscillation experiments.

At low energies the experimental tests of some phenomenological consequences of the minimal seesaw picture will be available in the next twenty years. No matter whether this simplified seesaw mechanism can survive such tests or not, it will provide us with some valuable implications about how to proceed to theoretically understand the true origin of tiny neutrino masses, significant lepton flavor mixing effects and mysterious CP violation at low and high energy scales.

\section*{Acknowledgements}

We would like to express our deep gratitude to all of our collaborators who have worked with us on the minimal seesaw and leptogenesis models in the past twenty years. We are especially indebted to Marco Drewes, Yan-bin Sun, Di Zhang, Shun Zhou and Jing-yu Zhu for useful discussions and friendly helps during our writing of this review article. This work was supported in part by the National Natural Science Foundation of China under Grant No. 12075254 (ZZX), No. 11775231 (ZZX), No. 11835013 (ZZX), No. 11605081 (ZHZ) and No. 12047570 (ZHZ), the Chinese Academy of Sciences (CAS) Center for Excellence in Particle Physics (ZZX), and by the Natural Science Foundation of the Liaoning Scientific Committee under Grant NO. 2019-ZD-0473 (ZHZ).

\section*{References}

\bibliographystyle{iopart-num}
\bibliography{iopart-num}

\end{document}